\documentclass[12pt]{article}

\usepackage{verbatim,color,amssymb}
\usepackage[dvipsnames]{xcolor}
\usepackage{amsmath}
\usepackage{amsthm}
\usepackage{natbib}
\usepackage{multirow}
\usepackage{setspace}
\usepackage[mathscr]{euscript}
\usepackage{fancyhdr}
\usepackage{enumitem}
\usepackage{graphicx}
\usepackage{lineno}
\usepackage{array,booktabs}
\usepackage{geometry}
\usepackage[bookmarks=false]{hyperref}
\usepackage{soul}

\usepackage{sectsty}
\usepackage{lipsum}

\usepackage{dsfont}
\usepackage{tikz}
\usetikzlibrary{bayesnet,calc}
\tikzstyle{black} = [circle,fill=black,draw=black,inner sep=1pt,
minimum size=20pt, font=\fontsize{10}{10}\selectfont, node distance=1]
\usetikzlibrary{arrows}
\usetikzlibrary{patterns}
\newcommand{\convexpath}[2]{
[
    create hullnodes/.code={
        \global\edef\namelist{#1}
        \foreach [count=\counter] \nodename in \namelist {
            \global\edef\numberofnodes{\counter}
            \node at (\nodename) [draw=none,name=hullnode\counter] {};
        }
        \node at (hullnode\numberofnodes) [name=hullnode0,draw=none] {};
        \pgfmathtruncatemacro\lastnumber{\numberofnodes+1}
        \node at (hullnode1) [name=hullnode\lastnumber,draw=none] {};
    },
    create hullnodes
]
($(hullnode1)!#2!-90:(hullnode0)$)
\foreach
[
    evaluate=\currentnode as \previousnode using \currentnode-1,
    evaluate=\currentnode as \nextnode using \currentnode+1
]
\currentnode in {1,...,\numberofnodes} {
    -- ($(hullnode\currentnode)!#2!-90:(hullnode\previousnode)$)
      let \p1 = ($(hullnode\currentnode)!#2!-90:(hullnode\previousnode) -
      (hullnode\currentnode)$),
        \n1 = {atan2(\y1,\x1)},
        \p2 = ($(hullnode\currentnode)!#2!90:(hullnode\nextnode) -
        (hullnode\currentnode)$),
        \n2 = {atan2(\y2,\x2)},
        \n{delta} = {-Mod(\n1-\n2,360)}
      in
        {arc [start angle=\n1, delta angle=\n{delta}, radius=#2]}
}
-- cycle
}
\usepackage{algorithm, algorithmicx}
\usepackage[utf8]{inputenc}

\usepackage{multibib}
\newcites{latex}{References}

\usepackage{lscape}

\usepackage{subfig}
\usepackage{setspace}
\usetikzlibrary{arrows}
\usepackage{multirow}

\makeatletter
\renewcommand{\paragraph}{%
  \@startsection{paragraph}{4}%
  {\z@}{1ex \@plus 1ex \@minus 1ex}{-1em}%
  {\normalfont\normalsize\bfseries}%
}
\def\thm@space@setup{\thm@preskip=5pt
\thm@postskip=5pt}
\makeatother

\newtheorem*{Proof*}{Proof}

\def\C{{\cal C}}

\def\E{{\cal E}}

\def\G{{\cal G}}

\def\N{{\cal N}}
\def\calP{{\cal P}}

\def\V{{\cal V}}
\def\X{{\cal X}}

\def\Ind{\hbox{1}}
\def\wh{\widehat}
\def\wt{\widetilde}

\def\IG{\hbox{Inv-Ga}}

\def\MVN{\hbox{MVN}}

\def\Normal{\hbox{Normal}}

\def\P_25_ICML{{\it Proceedings of the 25th international conference on Machine
learning}}

\def\bse{\begin{eqnarray*}}
\def\ese{\end{eqnarray*}}
\def\be{\begin{eqnarray}}
\def\ee{\end{eqnarray}}
\def\bq{\begin{equation}}
\def\eq{\end{equation}}

\def\wh{\widehat}

\def\trans{^{\rm T}}

\def\calA{{\cal A}}
\def\th{^{th}}
\def\bone{{\mathbf 1}}

\def\b1e{{\mathbf e}}

\def\bI{{\mathbf I}}

\def\bP{{\mathbf P}}

\def\bW{{\mathbf W}}

\def\bX{{\mathbf X}}
\def\by{{\mathbf y}}

\def\bzero{{\mathbf 0}}

\newcommand{\bmu}{\mbox{\boldmath $\mu$}}

\newcommand{\bepsilon}{\mbox{\boldmath $\epsilon$}}

\newcommand{\bbeta}{\mbox{\boldmath $\beta$}}

\newcommand{\bSigma}{\mbox{\boldmath $\Sigma$}}

\newcommand{\abs}[1]{\left\vert#1\right\vert}

\renewcommand\footnoterule{\kern-3pt \hrule \textwidth 2in \kern 2.6pt}

\def\boxit#1{\vbox{\hrule\hbox{\vrule\kern6pt \vbox{\kern6pt
\textcolor{blue}{#1}\kern6pt}\kern6pt\vrule}\hrule}}

\def\authorfootnote#1{{\let\thefootnote\relax\footnotetext{#1}}}

\def\partition{\X^{\star}}
\def\partlite{x^{\star}}

\begin{document}
\thispagestyle{empty}
\baselineskip=28pt

\begin{center}
{\LARGE{\bf
Bayesian Semiparametric\\\vskip -10pt
Hidden Markov Random Partition
Fields\\\vskip -10pt

for Factor Collapse on Graphs:\\
A Study of Cortical Mapping of Fingertips

}}
\end{center}
\baselineskip=12pt

\vskip 2mm
\begin{center}
Blake Moya$^{1}$ (blakemoya@utexas.edu)\\
\vskip 2mm
Kevin Sitek$^{2}$ (kevin.sitek@northwestern.edu)\\
\vskip 2mm
Arkaprava Roy$^{3}$ (arkaprava.roy@ufl.edu)\\
\vskip 2mm
Bharath Chandrasekaran$^{2}$ (bchandra@northwestern.edu)\\
\vskip 2mm
Abhra Sarkar$^{1}$ (abhra.sarkar@utexas.edu) \\
\vskip 2mm
$^{1}$Department of Statistics and Data Sciences,
University of Texas at Austin\\
2317 Speedway D9800, Austin, TX 78712, USA\\
\vskip 2mm

$^{2}$Department of Communication Sciences and Disorders,
Northwestern University,\\
70 Arts Circle Drive, Evanston, IL 60208, USA
\vskip2mm
$^{3}$Department of Biostatistics,
University of Florida\\
2004 Mowry Road, Gainesville, FL  32611, USA
\end{center}

\begin{abstract}
\baselineskip=12pt
Understanding how the brain’s cortical regions respond to stimuli like
fingertip tapping is a significant challenge in neuroscience,
especially with high-resolution functional magnetic resonance imaging data.
To address this, we propose a new statistical method for evaluating how a
factor's influence locally varies across a complex graph,
such as the brain's cortex.
Our approach is designed to handle the complexities of large graph sizes and
computational demands.
Our method uses a novel Bayesian hidden Markov random field to partition the
factor's influence into collapsed states with similar effects on the outcome at
each node. This unique model promotes sparsity in the partitions and penalizes
large variations across adjacent nodes.
The result is a highly detailed local influence map that captures subtle
changes in the patterns and magnitudes of a factor's influence across the graph.
To ensure efficient analysis of large datasets,
we developed a specialized Markov chain Monte Carlo algorithm.
Our simulation experiments demonstrate significant improvements over existing
techniques in both accuracy and scalability. Ultimately,
our method provides a powerful new framework for exploring the intricate
relationship between stimuli and brain activity,
offering a clearer picture of the cortical mapping of fingertips.

\end{abstract}

\vskip 20pt
\baselineskip=12pt
\noindent\underline{\bf Key Words}:
Cortical mapping,
Dependent random partitions,
Factor collapse,
Local inference,
Markov chain Monte Carlo,
Markov random fields

\clearpage\pagebreak\newpage
\pagenumbering{arabic}
\newlength{\gnat}
\setlength{\gnat}{25pt}
\baselineskip=\gnat

\section{Introduction} \label{sec:intro}
\vspace*{-1ex}

\paragraph{Scientific Motivation.}

Neural responses are typically localized, enabling functional brain mapping.
However, overlapping cortical representations within these activated regions
complicate precise differentiation.
At the single-neuron level,
firing rates across distinct tasks can be indistinguishable. At a macro scale,
many regions serve multiple functions,
and inter-individual heterogeneity further causes task-associated regions to
vary across subjects.
These challenges hinder a unified model of cortical mapping -- a complex
pursuit central to modern neuroscience since seminal works like
\cite{1937Penfield} and \cite{1909Brodmann} (translated in
\citealp{2006BrodmannGary}).

Recent advances in ultra-high-resolution functional magnetic resonance imaging
(fMRI) provide the precision needed to differentiate response regions,
enabling the mapping of even fine sensorimotor representations.
Studies on fingertip use have specifically suggested
that they are not cleanly parcellated across the brain but overlap in
clinically relevant ways \citep{2015Ejaz,2014Besle,2019Huber}.
Conversely, dominance maps of fingertips across the cortical surface
\citep{2020ONeill,2019Huber} and the cerebellum \citep{2013vanderZwaag} can
still be constructed by assigning regions where the response to stimulation of
a single fingertip exceeds that of all others.
Nevertheless, achieving a finer,
participant-specific understanding of the cortical mapping of fingertips
remains a challenge.

To address this, we introduce a Bayesian semiparametric local hidden Markov
random field partition model (lhMRPF) that
can characterize cortical topography by identifying regions of both unique and
simultaneous activation, as well as areas of equivalent activation,
enabling the delineation of cortical zones where the fingertips may be
represented in the brain, distinctly or functionally undifferentiably.
{\ul{While our proposed framework broadly supports average-level parameter
estimation (whether at the population level, treating subjects as replicates,
or the individual level, using repeated runs),
this work focuses specifically on the inference of individual maps,
leaving population-level inference for future exploration.
This scope is motivated by high inter-individual variability in fingertip
topography, expanded access to ultra-high-resolution replicates,
and a growing emphasis in the scientific literature on such individual-level
``precision maps"} \citep{laumann2015functional,gordon2017precision}.
}

\paragraph{General Problem Statement.}

In broader statistical terms,
our method applies to scenarios in which different factor levels
(\emph{fingertips}) can influence the outcome differently at different
locations on a graph (\emph{cortex}).
Our formulation naturally incorporates local variable selection,
allowing the factor to be important in some regions while being entirely
irrelevant in others (\emph{a fingertip may activate specific regions and not
others}).
Moreover, the nature of the influence, when important,
can vary arbitrarily across such locations.
For instance, factor levels important at two different locations on the graph
may influence the outcome differently at those locations
(\emph{a fingertip may activate two different voxels,
including in adjacent areas, with similar or different strengths});
conversely, several factor levels may influence the same location similarly or
differently
(\emph{multiple fingertips may activate the same voxel,
with similar or different strengths}).
We overcome significant methodological and computational challenges to achieve
this in a principled, flexible and realistic manner,
promoting smoothness (\emph{often adjacent voxels show similar activation
patterns}) and sparsity (\emph{often only small areas get activated})
in ultra-high-resolution graph settings (\emph{precision fMRI}).

\paragraph{Our Proposed Approach.}

The basic working principle of our method relies on the idea of `factor
collapse'.
Cortical mapping studies often
treat stimuli as factors to construct effect contrasts.
However, not all factor levels have unique effects,
leading to over-parameterization.
In Tukey's words, ``We wish to separate the varieties into distinguishable
groups as often as we can without too frequently separating varieties which
should stay together" \citep{1949Tukey}.
Partitioning (or collapsing) the factor levels into fewer unique levels thus
allows the estimation of fewer unique effects,
each with a greater number of representative data points.

However, unlike traditional factor-collapsing methods,
the factors are expected to collapse differentially over space.
Furthermore, they tend to form local clusters as large spatial processes,
such as brain activation profiles, exhibit heterogeneities of local features,
with a small subset of significant features often controlling the overall
spatial variations.
Our aim is to address these two key challenges motivated by the problem of
mapping the brain activation patterns from a finger-tapping task-based fMRI
dataset.

Specifically, consider a factor $x$ with $p$ levels
$\X=(x_{1},\dots,x_{p})\trans$ and its associated effects $\bbeta =
(\beta_{1},\dots,\beta_{p})\trans$ on some continuous outcome $y$
which can be modeled in terms of a collapsed latent factor $\partlite$ with $q$
levels $\partition = (\partlite_{1},\dots,\partlite_{q})\trans$ and associated
unique collapsed effects $\bbeta^{\star} =
(\beta_{1}^{\star},\dots,\beta_{q}^{\star})\trans$ or ``atoms",
where $q \leq p$.
This can be done via a mapping $\X=\bP\partition$,
and correspondingly $\bbeta = \bP\bbeta^{\star}$,
where $\bP$ is a $p \times q$ partition matrix with exactly one 1 and $p - 1$
0's in each row,
inducing a partition of $\X$ and $\bbeta$ in terms of their collapsed unique
values $\partition$ and $\bbeta^{\star}$, respectively.
In particular, when $\abs{\partition} = \abs{\bbeta^{\star}} = q=1$,
all factor effects collapse together,
implying that the factor $x$ is irrelevant.

Our work is focused on the development of such a framework
for data distributed on an undirected graph $\G=(\V,\E)$,
where $v\in \V$ denotes the nodes and $e \in \E$ denotes the edges connecting
the nodes, determining a neighborhood structure.
We propose a novel lhMRPF framework,
integrating additional structural components to enable detailed in-depth
inference about a factor $x$'s varying influences on an outcome across $\G$.

We do this by (i) defining the mappings $\X_{v} \mapsto \partition_{v}$ and
$\bbeta_{v} \mapsto \bbeta_{v}^{\star}$ for all $v \in \V$, and
(ii) allowing them to vary across $v \in \V$,
favoring sparsity and smoothness across neighboring locations connected by the
edges $e\in\E$.
In doing so, we allow \emph{all} of the following to vary across locations $v$
in a systematic manner:
(1) the associated partitions $\partition_{v}$ based on the differential
influences of the factor levels on the response,
(2) the corresponding effect values $\bbeta_{v}^{\star}$, and hence also
(3) the number of relevant collapsed factor levels and effects $q_{v}$.
Importantly, as before,
when $\abs{\partition_{v}} = \abs{\bbeta_{v}^{\star}} = q_{v} = 1$,
the factor $x$ becomes unimportant at location $v$,
thereby allowing us to perform local variable selection.
To model such an organization, we use two Markov random fields in concert:
(a) a discrete hMRF on the underlying factor level partitions,
(b) and then conditional on these partitions, a continuous MRF for the effects.
Moreover, our model follows three guiding principles:
(A) ``\emph{sparsity}" of the partitions across $v$ to favor small values of
$q_{v}$ facilitating the collapse of the factor levels,
(B) ``\emph{similarity}" of the partitions $\partition_{v}$ across adjacent
$v$'s to favor the factor having similar patterns of influence on the response
at neighboring locations, and
(C) ``\emph{smoothness}" of the collapsed effects $\bbeta_{v}^{\star}$ across
adjacent $v$'s to favor similar values of these effects at contiguous locations.
These properties are well-motivated by cortical fingertip mapping applications
since, as discussed above,
typically only very small brain regions are activated by task stimuli (A),
and both their influence patterns (B) and their activation magnitudes (C)
exhibit strong spatial continuity across brain regions.

We adopt a Bayesian estimation framework using Markov chain Monte Carlo (MCMC),
which enables hierarchical model construction,
enforcement of key principles through prior regularization,
and finite-sample uncertainty quantification using posterior samples.

\paragraph{Existing Methods.}

The statistical literature on clustering methods and related random partition
models is extensive
\citep[e.g.,][etc.]{binder1978bayesian,hartigan1990partition,pitman1995exchangeable,crowley1997product,quintana2003bayesian,holmes2005bayesian};
see~\cite{wade2023bayesian,grazian2023review} for recent reviews.
Clustering methods for data distributed on structured domains have also
received attention; \cite{zhang2023review} provides a recent review.

Most existing approaches to clustering structured data, however,
focus on partitioning the outcome values or processes globally across the
entire domain,
which fails to capture the often nuanced, region-specific local effects.
Carefully designed local clustering approaches can  address these limitations
and are starting to receive more attention in the literature.

Early contributions in this area focused on functional data and relied on
various adaptations of the Dirichlet process
\citep{petrone2009hybrid,dunson2009nonparametric,nguyen2011dirichlet,nguyen2014bayesian}.
\cite{suarez2016bayesian} introduced a method to leverage locally clustered
features through wavelet basis expansions.
More recently, \cite{paulon2024bayesian,fan2024bayesian,toto2025bayesian}
proposed Markov partition models for locally clustering time-indexed functional
data exploiting B-spline basis expansions.
\cite{yao2024flexible} proposed a product of Dirichlet process priors on
wavelet-based basis coefficients to enable resolution-specific local clustering.
These local clustering approaches provide enhanced flexibility and refined
inferential capabilities by capturing the local heterogeneity in the data
across their domain.
Recent work has also advanced dependent random partition models for discrete
domains, with particular emphasis on capturing time-varying structures
\citep{page2022dependent, paganin2023informed, giampino2024local,
dahl2025dependent}.

Early work on factor collapse methods,
which cluster factor levels (as opposed to values of the outcome variable),
dates back to \cite{1949Tukey}, with further developments by \cite{1958Fisher,
1974Scott}.
More recent developments,
using regularization-based methods,
are proposed in \cite{2009Bondell, 2013Post, 2020DiStefano}.
\cite{paulon2024bayesian} introduced Markov
random partition models that enable time-varying factor collapse,
facilitating local inference in longitudinal functional data by flexibly
clustering their levels according to their influences on the outcome process.

A large body of work also exists on clustering spatially distributed data
\citep{knorr2000bayesian,lawson2002spatial,denison2001bayesian,reich2007multivariate,papageorgiou2015bayesian,page2016spatial,2021Luo, 2023Criscuolo}, including methods based on hMRFs \citep{franccois2006bayesian,robinson2010change}.
These works mainly focus on partitioning the spatial domain itself,
forming contiguous homogeneous regions that share the same atoms across all
locations within them (as opposed to constructing partitions that vary across
spatial locations).

Separately, the literature on neuroimaging data analysis methods is also
extensive.
Here, we provide a brief non-exhaustive review of functional MRI (fMRI) data
analysis techniques,
our main application focus in this paper.
For comprehensive reviews of fMRI-relevant statistical methods,
we refer readers to
\cite{lindquist2008statistical,lazar2008statistical,2019Ashby}.
fMRI studies are broadly categorized into resting-state,
where data are collected while the subject is at rest, and task-based,
where specific tasks or stimuli are used to elicit neural responses.
Statistical analysis of fMRI data aims to elucidate patterns of brain activity
across regions and their interactions under various conditions.
Mass univariate analysis (MUA) \citep{groppe2011mass} remains a popular
technique, independently modeling the relationship between brain activity and
experimental conditions at each voxel,
while spatial dependence has also been incorporated through specialized
regression models \citep{zhu2014spatially, li2015spatial, reiss2015wavelet,
morris2015functional}.
A popular class of alternatives is to use MUA estimates in a second-stage
analysis, where voxel-level $p$-values are corrected by applying a random field
accounting for smoothness \citep[e.g.,][etc.]{worsley2004unified,2018Tansey}.
Additionally, approaches relying on hMRFs to segment the brain into homogeneous
regions have also been used with success
\citep[e.g.,][]{1998Descombes_a,1998Descombes_b,2001Zhang,robinson2010change,2013Johnson,2015Shu}.

Nevertheless, existing methods are limited in their capacity to achieve our
goal of local, voxel-wise inference of factor influences.
Univariate approaches can identify active regions during specific tasks
\citep{friston1994assessing},
but are not designed to distinguish events that elicit similar voxel-level
responses.
Multivariate techniques,
such as MVPA \citep{haxby2001distributed} and RSA
\citep{kriegeskorte2008representational},
can detect distributed response patterns but do not reveal their spatial
arrangement within the brain.
There is thus a need for participant-specific approaches that capture both
patterns of activation and their detailed spatial organization,
facilitating precision mapping of functional topography within individuals.

\paragraph{Key Novelties and Inferential Advantages of Our Proposed Approach.}

In this work, we introduce an approach that allows \ul{\emph{local factor
collapse}} based on their influence on an outcome across an undirected graph.
Unlike most existing methods, our approach thus infers
\ul{\emph{dependent random partitions of the factor levels across the nodes}},
rather than partitioning the nodes themselves or clustering the factor levels
globally.
Ultimately, a partition of the nodes emerges by grouping those with similar
factor partitions.
We also allow the cluster-specific collapsed factor effects to take distinct
values at different nodes while borrowing information and ensuring smoothness
across neighboring nodes.
This again contrasts with conventional clustering approaches,
where atoms are typically shared across entire contiguous regions and are
independently distributed across such regions.

\begin{figure}[!b]
    \centering
    {\small (a) Spatial Clustering of Outcomes with a Traditional HMRF}
    \vskip 5pt
    \resizebox{0.90\textwidth}{!}{
        \begin{tikzpicture}
    \node[latent] (z1) {$z_{1}$} ;
    \node[const, below=of z1] (h1) {} ;
    \node[obs, style={draw=CadetBlue, fill=CadetBlue!25!}, below=of h1,
    xshift=0cm, yshift=-0.5cm] (y1) {$\wt{\beta}_{1}$} ;
    \node[latent, right=of z1, xshift=4cm] (z3) {$z_{3}$} ;
    \node[const, below=of z3] (h3) {} ;
    \node[obs, style={draw=CadetBlue, fill=CadetBlue!25!}, below=of h3,
    xshift=0cm, yshift=-0.5cm] (y3) {$\wt{\bbeta}_{3}$} ;
    \node[latent, right=of z3, xshift=4cm] (z5) {$z_{5}$} ;
    \node[const, below=of z5] (h5) {} ;
    \node[obs, style={draw=CadetBlue, fill=CadetBlue!25!}, below=of h5,
    xshift=0cm, yshift=-0.5cm] (y5) {$\wt{\bbeta}_{5}$} ;
    \node[latent, right=of z1, xshift=0.75cm, yshift=-0.6cm] (z2) {$z_{2}$} ;
    \node[const, below=of z2] (h2) {} ;
    \node[obs, style={draw=CadetBlue, fill=CadetBlue!25!}, below=of h2,
    xshift=0cm, yshift=-0.5cm] (y2) {$\wt{\bbeta}_{2}$} ;
    \node[latent, right=of z3, xshift=0.75cm, yshift=-0.6cm] (z4) {$z_{4}$} ;
    \node[const, below=of z4] (h4) {} ;
    \node[obs, style={draw=CadetBlue, fill=CadetBlue!25!}, below=of h4,
    xshift=0cm, yshift=-0.5cm] (y4) {$\wt{\bbeta}_{4}$} ;
    \node[latent, right=of z5, xshift=0.75cm, yshift=-0.6cm] (z6) {$z_{6}$} ;
    \node[const, below=of z6] (h6) {} ;
    \node[obs, style={draw=CadetBlue, fill=CadetBlue!25!}, below=of h6,
    xshift=0cm, yshift=-0.5cm] (y6) {$\wt{\bbeta}_{6}$} ;
    \tikzset{plate caption/.style={caption, node distance=0, inner sep=0pt,
    below left=0pt and 0pt of #1.south,text height=1.25em}}
    \plate[lightgray] {} {(y1) (z1)} {$v=1$} ;
    \plate[lightgray] {} {(y2) (z2)} {$v=2$} ;
    \plate[lightgray] {} {(y3) (z3)} {$v=3$} ;
    \plate[lightgray] {} {(y4) (z4)} {$v=4$} ;
    \plate[lightgray] {} {(y5) (z5)} {$v=5$} ;
    \plate[lightgray] {} {(y6) (z6)} {$v=6$} ;
    \edge[-] {z1} {z2}
    \edge[-] {z1} {z3}
    \edge[-] {z2} {z4}
    \edge[-] {z3} {z4}
    \edge[-] {z3} {z5}
    \edge[-] {z4} {z6}
    \edge[-] {z5} {z6}
    \edge[lightgray] {z1} {y1}
    \edge[lightgray] {z2} {y2}
    \edge[lightgray] {z3} {y3}
    \edge[lightgray] {z4} {y4}
    \edge[lightgray] {z5} {y5}
    \edge[lightgray] {z6} {y6}
    \node (g0) at ($(z3)!0.4!(z4)$) {} ;
    \node[latent, below=of g0, style={draw=RoyalBlue}] (g1) {$\bbeta^{\star}$} ;
    \path (g1) edge[->, lightgray, bend right=15]  (y1) ;
    \edge[lightgray] {g1} {y2}
    \edge[lightgray] {g1} {y3}
    \edge[lightgray] {g1} {y4}
    \edge[lightgray] {g1} {y5}
    \path (g1) edge[->, lightgray, bend left=15]  (y6) ;
    \node[const, right=of z6, xshift=4cm] (l0) {} ;
    \node[const, right=of z6] (l1) {$z_{v} \sim \hbox{MRF}(\G, \eta)$} ;
    \node[const, below=of l1] (l2) {$\textcolor{RoyalBlue}{\beta^{\star}_{z}}
    \overset{iid}{\sim} \textcolor{Maroon}{\Normal}(\mu, \sigma^{2})$} ;
    \node[const, right=of y6] (l3) {$\textcolor{CadetBlue}{\wt{\beta}_{v}}
    \overset{ind}{\sim}
    \textcolor{gray}{\Normal}(\textcolor{RoyalBlue}{\beta^{\star}_{z_{v}}},
    \sigma^{2}_{\epsilon})$} ;
\end{tikzpicture}
    }
    \vskip 5pt
    {\small (b) Local Spatial Clustering of Factor Levels and Effects with Our
    Proposed lhMRPF}
    \vskip 5pt
    \resizebox{0.90\textwidth}{!}{
        \begin{tikzpicture}
    \node[latent] (z1) {$\partition_{1}$} ;
    \node[latent, style={draw=RoyalBlue}, below=of z1,
    xshift=0cm] (beta1) {$\bbeta^{\star}_{1}$} ;
    \node[obs, style={draw=CadetBlue, fill=CadetBlue!25!}, below=of beta1,
    xshift= 0cm, yshift=-0.5cm]  (y1) {$\wt{\bbeta}_{1}$} ;
    \edge[OliveGreen] {z1} {beta1} ;
    \path (z1) edge[->, OliveGreen, bend right]  (y1) ;
    \edge[lightgray] {beta1} {y1} ;
    \node[latent, right=of z1, xshift=4cm] (z4) {$\partition_{3}$} ;
    \node[latent, style={draw=RoyalBlue}, below=of z4,
    xshift=0cm] (beta4) {$\bbeta^{\star}_{3}$} ;
    \node[obs, style={draw=CadetBlue, fill=CadetBlue!25!}, below=of beta4,
    xshift= 0cm, yshift=-0.5cm]  (y4) {$\wt{\bbeta}_{3}$} ;
    \edge[OliveGreen] {z4} {beta4} ;
    \path (z4) edge[->, OliveGreen, bend right]  (y4) ;
    \edge[lightgray] {beta4} {y4} ;
    \node[latent, right=of z4, xshift=4cm] (z7) {$\partition_{5}$} ;
    \node[latent, style={draw=RoyalBlue}, below=of z7,
    xshift=0cm] (beta7) {$\bbeta^{\star}_{5}$} ;
    \node[obs, style={draw=CadetBlue, fill=CadetBlue!25!}, below=of beta7,
    xshift= 0cm, yshift=-0.5cm]  (y7) {$\wt{\bbeta}_{5}$} ;
    \edge[OliveGreen] {z7} {beta7} ;
    \path (z7) edge[->, OliveGreen, bend right]  (y7) ;
    \edge[lightgray] {beta7} {y7} ;
    \node[latent, right=of z1, xshift=0.75cm,
    yshift=-0.6cm] (z2) {$\partition_{2}$} ;
    \node[latent, style={draw=RoyalBlue}, below=of z2,
    xshift=0cm] (beta2) {$\bbeta^{\star}_{2}$} ;
    \node[obs, style={draw=CadetBlue, fill=CadetBlue!25!}, below=of beta2,
    xshift= 0cm, yshift=-0.5cm]  (y2) {$\wt{\bbeta}_{2}$} ;
    \edge[OliveGreen] {z2} {beta2} ;
    \path (z2) edge[->, OliveGreen, bend right]  (y2) ;
    \edge[lightgray] {beta2} {y2} ;
    \node[latent, right=of z4, xshift=0.75cm,
    yshift=-0.6cm] (z5) {$\partition_{4}$} ;
    \node[latent, style={draw=RoyalBlue}, below=of z5,
    xshift=0cm] (beta5) {$\bbeta^{\star}_{4}$} ;
    \node[obs, style={draw=CadetBlue, fill=CadetBlue!25!}, below=of beta5,
    xshift= 0cm, yshift=-0.5cm]  (y5) {$\wt{\bbeta}_{4}$} ;
    \edge[OliveGreen] {z5} {beta5} ;
    \path (z5) edge[->, OliveGreen, bend right]  (y5) ;
    \edge[lightgray] {beta5} {y5} ;
    \node[latent, right=of z7, xshift=0.75cm,
    yshift=-0.6cm] (z8) {$\partition_{6}$} ;
    \node[latent, style={draw=RoyalBlue}, below=of z8,
    xshift=0cm] (beta8) {$\bbeta^{\star}_{6}$} ;
    \node[obs, style={draw=CadetBlue, fill=CadetBlue!25!}, below=of beta8,
    xshift= 0cm, yshift=-0.5cm]  (y8) {$\wt{\bbeta}_{6}$} ;
    \edge[OliveGreen] {z8} {beta8} ;
    \path (z8) edge[->, OliveGreen, bend right]  (y8) ;
    \edge[lightgray] {beta8} {y8} ;
    \tikzset{plate caption/.style={caption, node distance=0, inner sep=0pt,
    below left=0pt and 0pt of #1.south,text height=1.25em}}
    \plate[lightgray]{} {(y1) (beta1) (z1)} {$v=1$} ;
    \plate[lightgray] {} {(y2) (beta2) (z2)} {$v=2$} ;
    \plate[lightgray] {} {(y4) (beta4) (z4)} {$v=3$} ;
    \plate[lightgray] {} {(y5) (beta5) (z5)} {$v=4$} ;
    \plate[lightgray] {} {(y7) (beta7) (z7)} {$v=5$} ;
    \plate[lightgray] {} {(y8) (beta8) (z8)} {$v=6$} ;
    \tikzset{plate caption/.style={caption, node distance=0, inner sep=0pt,
    below left=0pt and 0.5em of #1.south west,text height=1.25em,
    text width=0.5em}}
    \edge[-] {z1} {z2}
    \edge[-] {z1} {z4}
    \edge[-] {z2} {z5}
    \edge[-] {z4} {z5}
    \edge[-] {z4} {z7}
    \edge[-] {z5} {z8}
    \edge[-] {z7} {z8}
    \edge[Maroon, -] {beta1} {beta2}
    \edge[Maroon, -] {beta1} {beta4}
    \edge[Maroon, -] {beta2} {beta5}
    \edge[Maroon, -] {beta4} {beta5}
    \edge[Maroon, -] {beta4} {beta7}
    \edge[Maroon, -] {beta5} {beta8}
    \edge[Maroon, -] {beta7} {beta8}
    \node (e1) at ($(z1)!0.45!(z2)$) {} ;
    \node[rectangle, fill=black, minimum size=5pt,
    inner sep=0pt] at (e1.center) {} ;
    \node (e2) at ($(z1)!0.6!(z4)$) {} ;
    \node[rectangle, fill=black, minimum size=5pt,
    inner sep=0pt] at (e2.center) {} ;
    \node (e4) at ($(z2)!0.3!(z5)$) {} ;
    \node[rectangle, fill=black, minimum size=5pt,
    inner sep=0pt] at (e4.center) {} ;
    \node (e6) at ($(z4)!0.45!(z5)$) {} ;
    \node[rectangle, fill=black, minimum size=5pt,
    inner sep=0pt] at (e6.center) {} ;
    \node (e7) at ($(z4)!0.6!(z7)$) {} ;
    \node[rectangle, fill=black, minimum size=5pt,
    inner sep=0pt] at (e7.center) {} ;
    \node (e9) at ($(z5)!0.3!(z8)$) {} ;
    \node[rectangle, fill=black, minimum size=5pt,
    inner sep=0pt] at (e9.center) {} ;
    \node (e11) at ($(z7)!0.45!(z8)$) {} ;
    \node[rectangle, fill=black, minimum size=5pt,
    inner sep=0pt] at (e11.center) {} ;
    \node (f1) at ($(beta1)!0.45!(beta2)$) {} ;
    \node[rectangle, fill=Maroon, minimum size=5pt,
    inner sep=0pt] at (f1.center) {} ;
    \node (f2) at ($(beta1)!0.6!(beta4)$) {} ;
    \node[rectangle, fill=Maroon, minimum size=5pt,
    inner sep=0pt] at (f2.center) {} ;
    \node (f4) at ($(beta2)!0.3!(beta5)$) {} ;
    \node[rectangle, fill=Maroon, minimum size=5pt,
    inner sep=0pt] at (f4.center) {} ;
    \node (f6) at ($(beta4)!0.45!(beta5)$) {} ;
    \node[rectangle, fill=Maroon, minimum size=5pt,
    inner sep=0pt] at (f6.center) {} ;
    \node (f7) at ($(beta4)!0.6!(beta7)$) {} ;
    \node[rectangle, fill=Maroon, minimum size=5pt,
    inner sep=0pt] at (f7.center) {} ;
    \node (f9) at ($(beta5)!0.3!(beta8)$) {} ;
    \node[rectangle, fill=Maroon, minimum size=5pt,
    inner sep=0pt] at (f9.center) {} ;
    \node (f11) at ($(beta7)!0.45!(beta8)$) {} ;
    \node[rectangle, fill=Maroon, minimum size=5pt,
    inner sep=0pt] at (f11.center) {} ;
    \edge[Maroon, dashed] {e1} {f1}
    \edge[Maroon, dashed] {e2} {f2}
    \edge[Maroon, dashed] {e4} {f4}
    \edge[Maroon, dashed] {e6} {f6}
    \edge[Maroon, dashed] {e7} {f7}
    \edge[Maroon, dashed] {e9} {f9}
    \edge[Maroon, dashed] {e11} {f11}
    \node[const, right=of z6, xshift=4cm] (l0) {} ;
    \node[const, right=of z6] (l1) {$\partition_{v} \sim \hbox{MRPF}(\G, \phi,
    \tau)$} ;
    \node[const, below=of l1] (l2) {$\textcolor{RoyalBlue}{\bbeta^{\star}_{v}}
    \sim \textcolor{Maroon}{\hbox{GMRF}}(\textcolor{Maroon}{\G^{\star}}, \bmu,
    \sigma^{2})$} ;
    \node[const, right=of y6] (l3) {$\textcolor{CadetBlue}{\wt{\bbeta}_{v}}
    \overset{ind}{\sim}
    \textcolor{gray}{\hbox{MVN}}(\textcolor{OliveGreen}{\mathbf{P}_{v}}
    \textcolor{RoyalBlue}{\bbeta^{\star}_{v}}, \wt\bSigma_{v})$} ;
\end{tikzpicture}
    }
    \caption{Main differences and novelties of our proposed lhMRPF compared to
    a traditional hMRF.
    (a) Top panel: The traditional hMRF model for spatial clustering of
    outcomes induced via an MRF prior on the latent cluster labels $z_{v}$
    associated with the outcome values $\wt{\beta}_{v}$ in a mixture model.
    The cluster-specific unique atoms $\beta_{z}^{\star}$ are independently and
    identically distributed according to a prior;
    and they are globally shared across all locations $v$.
    (b) Bottom panel: Our proposed lhMRPF model for local clustering of factor
    effects induced via multi-layer MRF priors.
    First, the factor levels $\X$ and associated effects $\bbeta_{v}$'s are
    clustered at each location $v$ into $\X_{v}^{\star}$ and
    $\bbeta_{v}^{\star}$, respectively,
    via a novel MRF on the latent cluster labels in a partition model.
    Second, the cluster-specific unique atoms $\bbeta_{v}^{\star}$ are
    distributed according to a GMRF whose neighborhood structure is determined
    by the partitions $\X_{v}^{\star}$ in the first layer.
    Specifically, as described in Section \ref{subsec:smoothness},
    the strength of the information shared between
    $\bbeta^{\star}_{\partlite_{v}}$ and $\bbeta^{\star}_{\partlite_{u}}$ at
    neighboring nodes $v$ and $u$ is determined by the matrix $\bP_{v}^{\trans}
    \bP_{u}$, where $\bP_{v}$ and $\bP_{u}$ are partition matrices mapping
    $\X_{v}^{\star}$ and $\X_{u}^{\star}$ in the first layer;
    this property is represented here by dashed lines.
    Finally, the emission distribution for observable values $\wt{\bbeta}_{v}$
    depends both on $\bbeta^{\star}$  \textit{and} $\X^{\star}$.
    Importantly, the cluster-specific unique effects $\bbeta_{v}^{\star}$'s are
    not globally shared but locally vary across locations $v$.
    Here,
    filled and unfilled nodes represent observed and latent variables;
    directed and undirected edges
    show directional and mutual dependencies;
    squares are used to indicate that the dependencies between the nodes in the
    top layer influence the dependencies between the nodes in the middle layer;
    and the dependencies among variables are color-coded according to the
    hierarchical model on the right.
    }
    \label{fig:hmrf-comparison}
\vspace*{-10pt}
\end{figure}
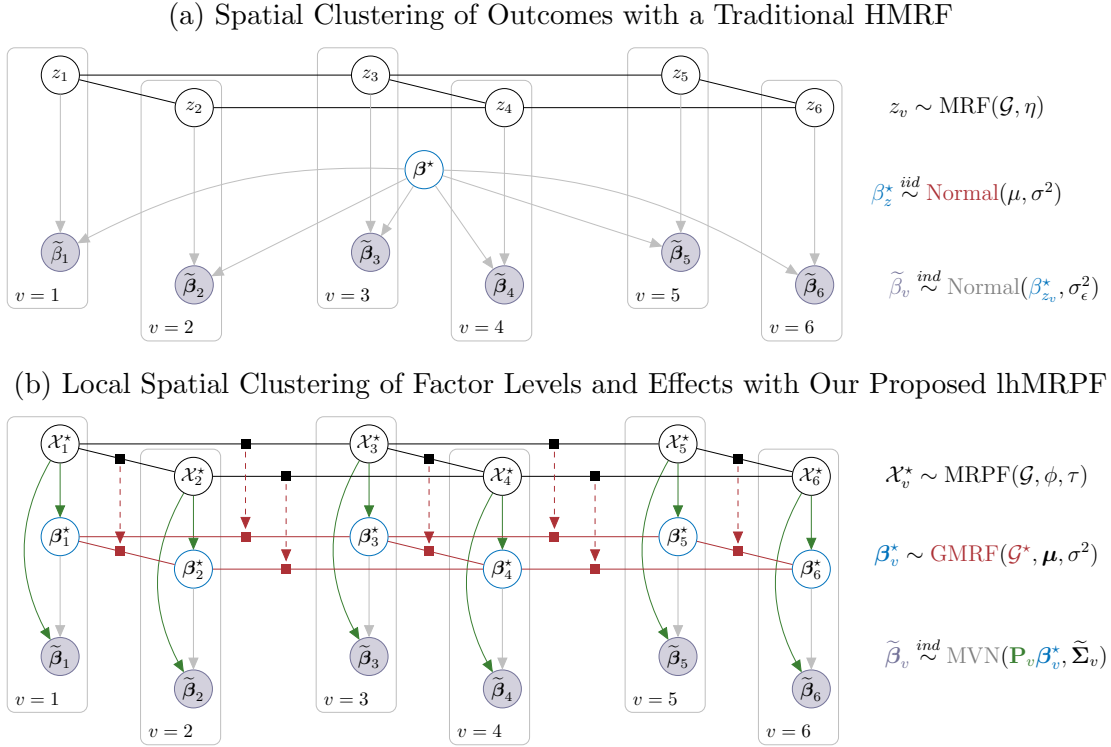

More specifically, our approach builds on the following key novelties.
{\bf \emph{First}},
local factor collapse is achieved through a novel discrete hMRPF partition
model for the collapsed factor levels $\partition_{v}$,
favoring sparsity at each location $v \in \V$ as well as similarity across
neighboring $v \in \V$.
{\bf \emph{Second}}, the collapsed factor effects $\bbeta_{v}^{\star}$,
which vary across locations $v \in \V$,
are smoothed via a novel second-layer continuous MRF,
with its neighborhood structure now informed \ul{\emph{jointly}} by the
original graph as well as the first-layer hMRF partitions.
A visual comparison of our model to conventional hMRFs is presented in Figure
\ref{fig:hmrf-comparison}.

Developing such a framework poses significant challenges on the computational
front as well -- not only due to the sheer size of the graph,
with tens of thousands of voxels,
but also because of the varying partition sizes across locations.
We address these complexities through innovative variable augmentation
strategies, combined with state-of-the-art,
locally informed MCMC methods for discrete state-space models
\citep{2017Titsias, 2019Zanella},
and parallelization techniques for hMRFs \citep{2011Gonzalez, 2020Kaplan}.

This local modeling framework supports our main scientific goal of identifying
brain regions with similar or distinct responses to fingertip tappings in an
ultra-high-resolution fMRI study.
Unlike previous studies that presuppose that finger effects are either
identical \textit{and} null or all different \textit{and} significant,
our voxel-level, graph-based approach treats fingertips as factor levels and
uncovers both \textit{distinct} and \textit{overlapping} activation patterns,
providing more granular insights into their neural representation.
Moreover, our model captures functional localization with greater nuance than
strictly volumetric approaches,
recognizing that while regions may serve multiple functions,
subsets may share identical neural pathways.

\paragraph{Article Outline.}

Section \ref{sec: background} details the fMRI dataset, preprocessing,
and scientific context.
Section \ref{sec: review} briefly reviews some relevant statistical concepts.
Section \ref{sec: method} presents the proposed lhMRPF method.
Section \ref{sec: sampling} outlines a novel MCMC-based posterior sampling
approach.
Section \ref{sec: application} reports results from our cortical mapping study.
Section \ref{sec: simulation} summarizes simulation findings.
Section \ref{sec: discussion} concludes with a discussion.
To manage space, substantive additional information -- including hyperparameter
choices, further algorithm details, additional figures,
and results -- is provided in the SM.

\vspace*{-3ex}
\section{Scientific Background} \label{sec: background}
\vspace*{-1ex}

\subsection{Cortical Digit Representation} \label{sec:digitseg}
\vspace*{-1ex}

The study of the cortical topographical representation of the digits has roots
as far back as \cite{1937Penfield}.
It has been found that distinct maps of the fingertips exist in the
sensorimotor cortex \citep{1998Gelnar}.
These representations are dynamic,
reacting differently to active and passive stimuli and spanning both
hemispheres of the sensorimotor cortex \citep{2013Diedrichsen,2019Berlot}.
Similar,
though more individually variable,
maps also exist in the cerebellum, as charted by \cite{2013vanderZwaag}.

Even with the existing techniques for standardization of neural coordinates,
high individual variability poses challenges to unified group-level study of
such regions.
That is,
even after normalization,
the regions responsible for the right index fingertip in one participant may
not overlap with the regions responsible for the same fingertip in another
participant.

Conversely,
even in the same participant,
the regions responsible for different fingertips may overlap to a high degree.
This has been investigated by \cite{2014Besle} and \cite{2019Huber} and is of
principal interest for the study of local clustering in neuroimaging.
We seek to explicitly model the overlap in cortical representation of digits by
way of a mixture model on the activation coefficients discovered by first-level
fMRI analysis.

\vspace*{-3ex}
\subsection{fMRI Data, Preprocessing, and Voxel-wise Analysis} \label{sec:VWA}
\vspace*{-1ex}

Volumes of fMRI data are recorded as participants are exposed to a stimulus of
interest over time.
Each scan $r \in \{1,\dots,n\}$ produces an array of image intensity values
$y_{t,v,r}$ for each time-point $t \in \{1,\dots,T\}$ and voxel location $v \in
\V$, capturing neural activity over time across the brain.
The onset and duration of each stimulus exposure are also recorded.
The images from each time point are then realigned to correct for the effects
of head motion during the scan and normalized to a standard template,
ensuring that corresponding anatomical regions align to the same voxels across
scans.
Preprocessing for the Flanker study and our main cortical fingertip mapping
study was done using the SPM software package \citep{SPM} and fMRIPrep
\citep{fmriprep1}.
Section S.1 in the SM presents additional details.

Let $p_{all} = (p + p_{nui})$ represent the total number of recorded
explanatory variables,
where $p$ represents the number of levels of the stimulus of interest and
$p_{nui}$ denotes the number of additional nuisance covariates.
For each voxel $v$ and each scan $r$,
a linear regression model \citep{2019Ashby} is then applied to each length-$T$
voxel-wise time series $\by_{v,r} = (y_{v,1,r},\dots,y_{v,T,r})\trans$ with a
$T \times p_{all}$ design matrix $\bX_{r}$ to estimate the contribution of each
component signal to the observed fMRI data.
An auto-regressive error structure is employed to account for temporally
dependent noise \citep{1996Bullmore,2025Goebel}.
Section S.2 in the SM presents additional details.
The regression yields a length-$p_{all}$ estimated effect
$\wt{\bbeta}_{all,v,r}$.
Using classical asymptotic theory, we then have
$\wt{\bbeta}_{all,v,r} \approx
\MVN_{p_{all}}(\bbeta_{all,v,r},\wt{\bSigma}_{all,v,r})$,
where $\bbeta_{all,v,r}$ is the underlying true unknown effect and
$\wt{\bSigma}_{all,v,r} = ( \bX_{r} \trans\wt{\bW}_{v} \bX_{r} )^{-1}$ is the
associated covariance matrix,
$\wt{\bW}_{v}$ being the error covariance estimated by the Cochrane-Orcutt
procedure \citep{1949Cochrane}.
Focusing on the effects associated with the $p$ levels of the main stimulus of
interest, denoted by $\wt{\bbeta}_{v,r}$,
and marginalizing out the nuisance variables,
we then obtain a similar result $\wt{\bbeta}_{v,r} \approx
\MVN_{p}(\bbeta_{v,r},\wt{\bSigma}_{v,r})$
which forms the basis for constructing the likelihood function used in our main
second-level analysis described in Section \ref{sec: method}.

\vspace*{-3ex}
\subsection{Our Motivating Data Set} \label{subsec:task}
\vspace*{-1ex}

The data analyzed in the present work are $n = 6$ high-resolution (2,546,073
voxels) scans of $T = 126$ time points (8 minutes, 24 seconds) from a single,
right-handed participant.
The participant underwent an auditory learning task where they used a button
pad to classify four unfamiliar monosyllabic Mandarin speech sounds
which differed in tone, the pitch trajectory of the syllable.
The participant is not a Mandarin speaker and had no prior experience
classifying the tones.
The participant pushed one of four buttons under either their right or
left-hand index or middle finger corresponding to which of the four tones they
believed they heard.
This presents an opportunity to study the representation of these fingers as an
accessory to a cognitively demanding learning task,
a task that represents an everyday use of the hands.
We use the event of button pushing as our stimulus of interest here to examine
motor processes in a real-life motor association learning scenario,
with $p = 4$ levels $\X=\{x_{1},\dots,x_{4}\} \equiv \{\text{right-index,
right-middle, left-index,
left-middle}\}$ for each of the fingers used in the task.
For additional details, see Section S.2 in the SM.

Our goal is to generate a cortical map of the fingertips using our proposed
lhMRPF method.
By adaptively partitioning fingertip effects at each voxel,
our approach provides a detailed and data-driven representation of their
cortical organization across the brain.
See Section \ref{sec: application} for further details.

\vspace*{-3ex}
\section{Review of Some Relevant Statistical Concepts} \label{sec: review}
\vspace*{-1ex}

\paragraph{Hidden Markov Random Fields.}
Let $\G =  (\V, \E)$ be an undirected graph representing random variables as
nodes and their pairwise conditional (in)dependence relationships as edges.
The absence of an edge between a pair of nodes encodes their conditional
independence given the rest.
A discrete random field $\by$ is a collection of random variables, $\{y_{v}:
v \in \V\}$ with $y_{v}$'s taking values in a finite state space.
For a given subset $\calA\subset \V$,
let $\by_{\calA}$ denote the random process restricted to $\calA$, i.e.,
$\{y_{a}: a \in \calA\}$.
Specifically, the random field $\by$ is called a Markov random field with
respect to $\G$, if it satisfies the Markov property $P(y_{v} \mid \by_{-v}) =
p(y_{v} \mid \by_{\N_{v}})$,
where $\N_{v} = \{u: (v,
u) \in \E\}$ and is the ``neighborhood" of the node $v$.

The above formulation, based on a set of conditionals,
does not necessarily guarantee the existence of a valid joint distribution.
The Hammersley-Clifford theorem states that a probability distribution having a
strictly positive density satisfies a Markov property with respect to the
undirected graph $\G$ if and only if its density can be factorized over the
cliques of the graph \citep{1971Hammersley}.
We recall that a clique $\C$ in an undirected graph $\G$ can be any single node
or a subset of nodes such that every two nodes in $\C$ are connected by an edge
from $\E$.
See also \cite[Ch. 4]{2009Koller}.

In most practical applications,
cliques of size one and two are sufficient to characterize the underlying
inter-variable dependencies flexibly,
which leads to the popular class of pairwise MRFs
\citep{2019Wainwright,2015Chen,2020Roy},
also known as `auto-models' introduced in \cite{1974Besag}.
The joint probability mass function (pmf) of $\by$ under a pairwise MRF is
\vspace{-7ex}\\
\bse
\textstyle\Pr(y_{1},\ldots,y_{v})\propto \exp\bigg\{\sum_{\{i:~ i\in \V\}}
\lambda_{1}(y_{r}) + \sum_{\{j,k:~ j,k \in \V, ~j<k\}}  \lambda_{2}(y_{j},
y_{k})\bigg\}, \label{pairMRF}
\ese
\vspace{-6ex}\\
where $\lambda_{1}(y_{r})$ is called a node potential function,
$\lambda_{2}(y_{j},
y_{\ell})$ an edge potential function and we have $\lambda_{2}(y_{j},
y_{\ell})=0$ if $(j,\ell)\notin \E$.
Thus, this distribution is pairwise Markov by construction and it satisfies the
Hammersley-Clifford theorem.

For the Potts model \citep{1952Potts,2021Izenman},
that describes a system of interacting categorical variables on a lattice,
$\lambda_{1}(y_{r})=-\sum_{k}\alpha_{k}\Ind\{y_{r}=k\}$ and
$\lambda_{2}(y_{j},y_{\ell})=-\beta\Ind\{y_{j}=y_{\ell}\}$.
For a continuous Gaussian MRF \citep{2005Rue},
they are set as $\lambda_{1}(y_{r})=-\alpha_{r} y_{r}^{2}$ and
$\lambda_{2}(y_{j},y_{\ell})=-\beta_{j,\ell}y_{j}y_{\ell}$,
where $\alpha_{j}$ and $\beta_{j,\ell}$ correspond to the diagonal and
off-diagonal elements of the precision matrix
$\bSigma^{-1}=\mathrm{cov}(\by)^{-1}$, respectively.
The intrinsic conditional autoregressive (ICAR) model is a special case that
defines each variable’s conditional distribution based on the average of its
neighbors, encouraging local smoothness.

An hMRF is obtained when an MRF is only observable through a proxy $\wt{\by}$
\citep{1995Kunsch} via a stochastic function $\wt{y}_{v} = f(y_{v})$,
called the emission function, to characterize how, at each node $v$,
the observable variable $\wt{y}_{v}$ is emitted from the hidden state $y_{v}$.

For comprehensive reviews of discrete MRFs and related graphical models,
see \cite{stoehr2017review,2021Izenman}, etc.
For detailed discussions on continuous MRFs and related graphical models,
see \cite{2005Rue}.

\begin{figure}[!ht]
    \centering
    \resizebox{0.77\textwidth}{!}{
        \input{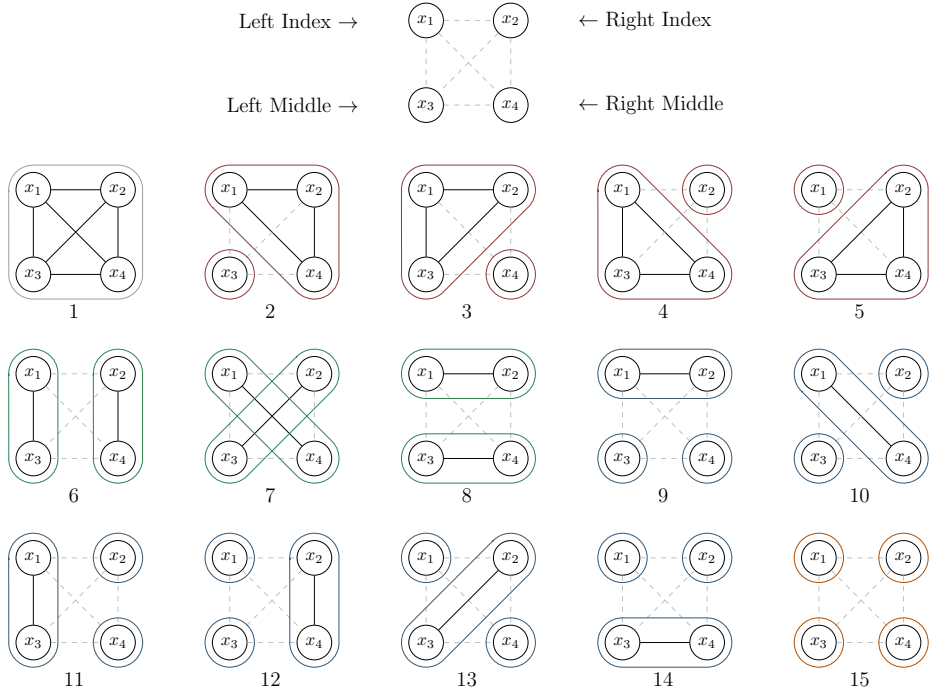}
    }
    \caption{The fifteen possible partitions of a factor with four levels $\X =
    \{x_{1},\dots,x_{4}\}$.
    For our motivating cortical fingertip mapping study,
    the levels represent the four fingers tested in the task,
    as shown at the top.
    Solid edges represent pairwise co-membership between two levels,
    and dashed edges represent pairwise distinction between two levels.
    Co-clustered levels are encircled in colored loops to clearly show the
    partition structures.}
    \label{fig:partitions-of-four}
\vspace*{-10pt}
\end{figure}

\paragraph{Partition Models.}

A partition $\partition = \{\partition_{1},\dots,\partition_{q}\}$ of a set
$\X$ is defined as a collection of nonempty disjoint subsets of $\X$ such that
$\cup_{k = 1}^{q} \partition_{k} = \X$.
We denote that two elements of $\X$ reside in the same group in $\partition$
with $x_{a} \sim_{\partition} x_{b}$.
The set of all partitions of $\X$ is denoted by
$\mathcal{P}(\X)$.
For instance, $\mathcal{P}(\{x_{1},
x_{2}\})$ comprises $\{\{x_{1}\},\{x_{2}\}\}$ and $\{x_{1}, x_{2}\}$.
The number $B_{p}$ of unique partitions of a set of $p$ elements,
referred to as the \textit{Bell number} for \cite{1934Bell},
can be computed recursively as
$B_{p} = \sum_{j = 0}^{p - 1} \binom{p - 1}{j} B_{j}$ with $B_{0} = 1$.
See also \cite{2013Mansour}.

Figure \ref{fig:partitions-of-four} shows the $B_{4} = 15$ possible partitions
of a set $\X$ with four elements $\{x_{1},\dots,x_{4}\}$.

An alternative encoding, commonly used in clustering applications,
represents a partition using allocation vectors with entries specifying the
partition sets the original elements belong to.
For example, for a set with two elements $x_{1}$ and $x_{2}$ and associated
allocation labels $\partlite_{1}$ and $\partlite_{2}$,
the set of all possible partition allocations is $\{(\partlite_{1},
\partlite_{1}), (\partlite_{1}, \partlite_{2}), (\partlite_{2}, \partlite_{1}),
(\partlite_{2}, \partlite_{2})\}$.
Under this formulation,
multiple representations can correspond to equivalent partitions.
For instance, in the above example, $(\partlite_{1},
\partlite_{1})$ and $(\partlite_{2},
\partlite_{2})$ both represent the partition $\{x_{1},x_{2}\}$;
likewise $(\partlite_{1}, \partlite_{2})$ and $(\partlite_{2},
\partlite_{1})$ both represent the partition $\{\{x_{1}\},\{x_{2}\}\}$.
While the total number of possible representations now increases from $B_{p}$
to $p^{p}$,
the latent allocation variable representation often affords computational
feasibility, especially in Bayesian settings.

A third method for encoding a set partition is via pairwise co-membership
indicators,
which has the size $2^{p(p-1)/2}$.
This encoding takes the form of a $p(p-1)/2$-length binary vector indicating
whether a pair of elements in $\X$ belong to the same set in $\partition$.
This method allows for the encoding of invalid partitions wherein it is
possible to have a representation of $\X = \{x_{1}, x_{2},
x_{3}\}$ such that $x_{1} \sim_{\partition} x_{2}$,
$x_{2} \sim_{\partition} x_{3}$, but $x_{1} \nsim_{\partition} x_{3}$.
While this can conflict with the definition of a valid partition,
this encoding can still be important for two reasons.
First, it is the only encoding that is directly available from pairwise
comparison by \textit{post hoc} testing.
Second, for valid partitions,
it is useful for constructing dissimilarity metrics \citep{1971Rand,
1973Arabie, 1985Hubert, 2015Mohlin}.

\cite{2015Mohlin} proposed a highly generalizable distance metric for
partitions.
Specifically, for partitions of discrete sets, it takes the form
\vspace{-7ex}\\
\bse
\textstyle d(\partition_{v}, \partition_{u}) = 2 \sum_{(x_{r},
x_{j}) \in \X^{2}} \Ind\{
\Ind\{x_{r} \sim_{\partition_{v}} x_{j}\}
\neq
\Ind\{x_{r} \sim_{\partition_{u}} x_{j}\}
\},
\ese
\vspace{-7ex}\\
similar to the comparison metric of \cite{1973Arabie}.
It counts the number of pairwise disagreements in the co-membership of elements
of $\X$ between the partitions $\partition_{v}$ and $\partition_{u}$ and can,
therefore, be used to penalize dissimilarities between them.

\vspace*{-3ex}
\section{Local Hidden Markov Random Partition Field} \label{sec: method}
\vspace*{-1ex}
We now propose our lhMRPF for data distributed on a graph,
using our motivating application of cortical map estimation from fMRI data as a
template for its description.

With some repetition, let $\G = (\V,
\E)$ be a graph with nodes $v \in \V$ and undirected edges $e \in \E$
determining their neighborhood structure.
We consider a factor $x$ with levels $\X=(x_{1},\dots,
x_{p})\trans$ at each node $v \in \V$ and an associated continuous outcome
$\wt\bbeta_{v}=(\wt\beta_{v,1},\dots,\wt\beta_{v,p})\trans$ with
$\wt\beta_{v,j} \equiv \wt\beta_{v,x_{v,j}}$.
Specifically, $n$ replicated measurements
$\wt\bbeta_{v,r}=(\wt\beta_{v,1,r},\dots,\wt\beta_{v,p,r})\trans, r=1,\dots,n$,
which we consider to be measurement-error-contaminated proxies of an underlying
true effect of interest $\bbeta_{v}=(\beta_{v,1},\dots,\beta_{v,p})\trans$ with
$\beta_{v,j} \equiv \beta_{v,x_{v,j}}$,
are available at each node $v \in \V$, generated according to the model
\vspace{-7ex}\\
\bse
\wt{\bbeta}_{v,r} = \bbeta_{v} + \bepsilon_{v,r},
~~~~~\bepsilon_{v,r} \overset{ind}{\sim} \MVN_{p}(\bzero, \wt\bSigma_{v,r}),
\ese
\vspace{-7ex}\\
where $\bepsilon_{v,r}=(\epsilon_{v,1,r},\dots,\epsilon_{v,p,r})\trans$'s are
multivariate normally distributed measurement errors with known covariance
matrix $\wt\bSigma_{v,r}$ (Section \ref{sec:VWA}).
Furthermore, we assume that not all factor levels in $\X$ necessarily have
distinct effects on $\wt\bbeta$,
meaning the elements of $\bbeta_{v}$ may not be unique.
Instead, at every location $v$,
$\X$ and accordingly $\bbeta_{v}$ can be clustered into a latent collapsed
factor $\partlite_{v} \in \X_{v}^{\star} =
(\partlite_{v,1},\dots,\partlite_{v,q_{v}})\trans$ with corresponding collapsed
and unique effects $\bbeta_{v}^{\star}
=(\beta_{v,1}^{\star},\dots,\beta_{v,q_{v}}^{\star})\trans$ such that
$\beta_{v,k}^{\star} \equiv \beta_{v,x_{v,k}^{\star}}^{\star}$.
This is achieved via the mapping
\vspace{-7ex}\\
\bse
\X = \bP_{v}\X_{v}^{\star},~~~\bbeta_{v} = \bP_{v}\bbeta_{v}^{\star},
\ese
\vspace{-7ex}\\
where $\bP_{v}=((P_{v,a,b}))$'s are $p \times q_{v}$ partition matrices such
that $P_{v,a,b} = \Ind\left\{ \partlite_{v}(x_{a}) = \partlite_{b} \right\}$.
These mappings are thus such that $\beta_{v,x_{1}} =
\beta_{v,x_{2}}=\beta_{v,\partlite_{v}}^{\star} \iff \partlite_{v}(x_{1}) =
\partlite_{v}(x_{2})=\partlite_{v}$ at the location $v \in \V$ in the graph.
In terms of the factor levels,
the latent collapsed levels $\partition_{v}$ of the original observed levels
$\X$ are such that $\partlite_{v}(x)$ in $\partition_{v}$ denotes the latent
cluster label associated with $x$ in $\X$ at $v \in \V$.
In terms of the factor effects,
the latent collapsed effects $\bbeta_{v}^{\star}$ of the original expanded
effects $\bbeta_{v}$ are such that $\beta_{v,\partlite_{v}(x)}^{\star}$ in
$\bbeta_{v}^{\star}$ denotes the latent true collapsed effect associated with
$\beta_{v,x}$ in $\bbeta_{v}$ at $v \in \V$.
The mappings $\bP_{v}$'s at each location $v \in \V$ in the graph,
including the number of latent clusters $q_{v}$,
are a-priori unknown and a-posteriori inferred from the data.

Importantly, when $q_{v}=1$,
all factor levels in $\X$ and their corresponding factor effects in
$\bbeta_{v}$ collapse into a single level in $\partition_{v}$ and an associated
single effect in $\bbeta_{v}^{\star}$,
implying the factor levels have no differential effect on the outcome
$\wt\bbeta$ at the location $v$.
Conversely, when $q_{v}=p$,
all factor levels in $\X$ have distinct effects in $\bbeta_{v}$.

These key features of our proposed approach offer a highly flexible and
principled framework for local inference in continuous data $\wt{\bbeta}$
distributed on a graph $\G$ generated under the influence of a stimulus
$x\in\X$.
The main methodological innovations that endow our approach with these
important properties include
(a) a prior for the partitions $\X_{v}^{\star}$ favoring sparsity while
borrowing information and ensuring smoothness in $\X_{v}^{\star}$ across
neighboring locations in $\G$,
and
(b) a prior for the collapsed effects $\bbeta_{v}^{\star}$,
conditional on the mappings $\X_{v}^{\star}$,
borrowing information and ensuring smoothness in the elements of
$\bbeta_{v}^{\star}$ characterizing the effects of the same levels in $\X$,
as determined by the mappings $\X_{v}^{\star}$,
across neighboring locations in $\G$.
These are the problems we tackle next.

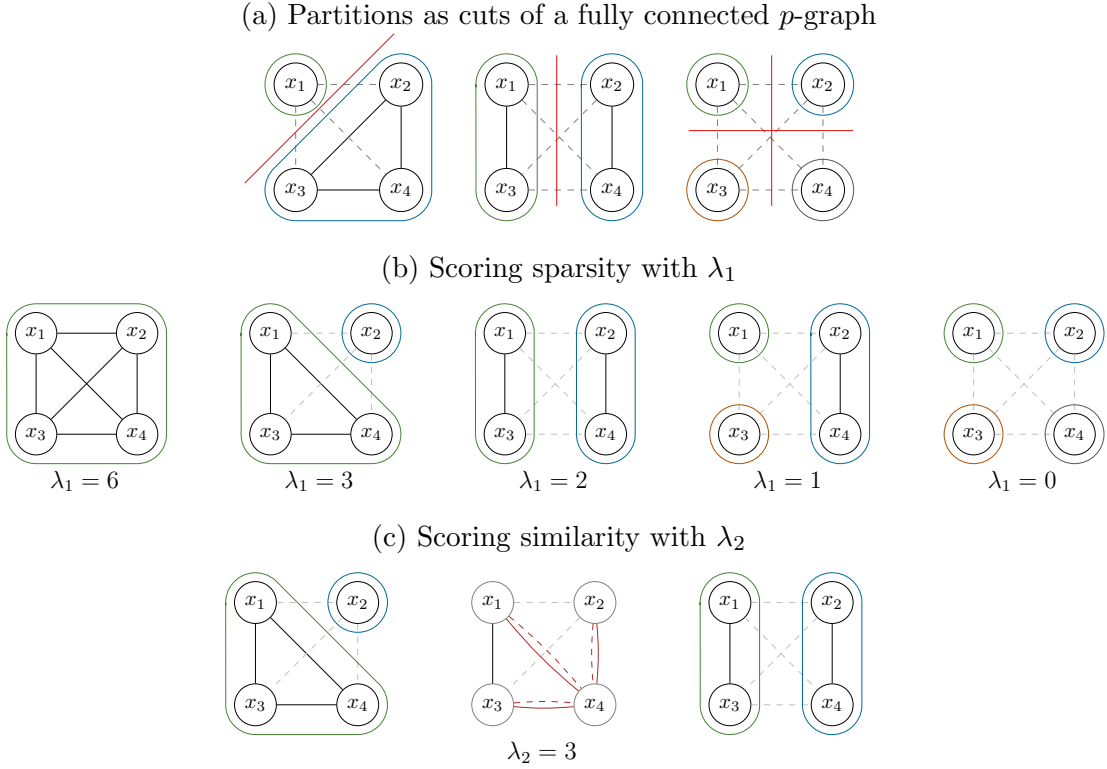
\begin{figure}[!ht]
    \centering
    {\small (a) Partitions as cuts of a fully connected $p$-graph} \\
    \vspace{0.5em}
    \resizebox{0.5\textwidth}{!}{
        \tikzset{
  cut/.style = {shorten >=-5mm, shorten <=-5mm, rounded corners=5mm,
  color={rgb:red,191;green,10;blue,10}},
}

\begin{tikzpicture}
    \node[latent] (a1) {$x_{1}$} ;
    \node[latent, right=of a1] (b1) {$x_{2}$} ;
    \node[latent, below=of a1] (c1) {$x_{3}$} ;
    \node[latent, below=of b1] (d1) {$x_{4}$} ;
    \edge[-]{b1}{c1,d1}
    \edge[-]{c1}{d1}
    \edge[-,dashed,gray]{a1}{b1,c1,d1}
    \draw[draw={rgb:red,87;green,157;blue,66}] (a1) circle (0.5cm);
    \draw[draw={rgb:red,0;green,95;blue,134}] \convexpath{c1,b1,d1}{0.5cm};
    \node[const, xshift= 1.275cm, yshift=0.5cm] (cut11) {} ;
    \node[const, yshift=-1.275cm, xshift=-0.5cm] (cut12) {} ;
    \draw[cut](cut11) -- (cut12) ;
    \node[latent, right=of b1] (a2) {$x_{1}$} ;
    \node[latent, right=of a2] (b2) {$x_{2}$} ;
    \node[latent, below=of a2] (c2) {$x_{3}$} ;
    \node[latent, below=of b2] (d2) {$x_{4}$} ;
    \edge[-]{a2}{c2}
    \edge[-]{b2}{d2}
    \edge[-,dashed,gray]{a2}{b2,d2}
    \edge[-,dashed,gray]{b2}{c2}
    \edge[-,dashed,gray]{c2}{d2}
    \draw[draw={rgb:red,87;green,157;blue,66}] \convexpath{a2,c2}{0.5cm};
    \draw[draw={rgb:red,0;green,95;blue,134}] \convexpath{b2,d2}{0.5cm};
    \node[const, xshift=4.25cm] (cut21) {} ;
    \node[const, xshift=4.25cm, yshift=-1.5cm] (cut22) {} ;
    \draw[cut](cut21) -- (cut22) ;
    \node[latent, right=of b2] (a3) {$x_{1}$} ;
    \node[latent, right=of a3] (b3) {$x_{2}$} ;
    \node[latent, below=of a3] (c3) {$x_{3}$} ;
    \node[latent, below=of b3] (d3) {$x_{4}$} ;
    \edge[-,dashed,gray]{a3}{b3,c3,d3}
    \edge[-,dashed,gray]{b3}{c3,d3}
    \edge[-,dashed,gray]{c3}{d3}
    \draw[draw={rgb:red,87;green,157;blue,66}]    (a3) circle (0.5cm);
    \draw[draw={rgb:red,0;green,95;blue,134}]     (b3) circle (0.5cm);
    \draw[draw={rgb:red,191;green,87;blue,0}]     (c3) circle (0.5cm);
    \draw[draw={rgb:red,214;green,210;blue,196}]  (d3) circle (0.5cm);
    \node[const, xshift=7.75cm] (cut31) {} ;
    \node[const, xshift=7.75cm, yshift=-1.5cm] (cut32) {} ;
    \node[const, xshift=6.875cm, yshift=-0.75cm] (cut33) {} ;
    \node[const, xshift=8.6cm, yshift=-0.75cm] (cut34) {} ;
    \draw[cut](cut31) -- (cut32) ;
    \draw[cut](cut33) -- (cut34) ;
\end{tikzpicture}
    } \\
    \vspace{0.5em}
    {\small (b) Scoring sparsity with $\lambda_{1}$} \\
    \vspace{0.5em}
    \resizebox{0.9\textwidth}{!}{
        \begin{tikzpicture}
    \node[latent]          (a1) {$x_{1}$} ;
    \node[latent, right=of a1] (b1) {$x_{2}$} ;
    \node[latent, below=of a1] (c1) {$x_{3}$} ;
    \node[latent, below=of b1] (d1) {$x_{4}$} ;
    \node[const,  right=of b1, xshift=-0.5cm] (e1) {} ;
    \node[const,  below=of c1, yshift= 0.75cm] (f1) {} ;
    \edge[-]{a1}{b1}
    \edge[-]{a1}{c1}
    \edge[-]{a1}{d1}
    \edge[-]{b1}{c1}
    \edge[-]{b1}{d1}
    \edge[-]{c1}{d1}
    \draw[draw={rgb:red,87;green,157;blue,66}] \convexpath{a1,b1,d1,c1}{0.5cm};
    \node[latent, right=of e1] (a4) {$x_{1}$} ;
    \node[latent, right=of a4] (b4) {$x_{2}$} ;
    \node[latent, below=of a4] (c4) {$x_{3}$} ;
    \node[latent, below=of b4] (d4) {$x_{4}$} ;
    \node[const,  right=of b4, xshift=-0.5cm] (e4) {} ;
    \node[const,  below=of c4, yshift= 0.75cm] (f4) {} ;
    \edge[-, dashed, lightgray]{a4}{b4}
    \edge[-]{a4}{c4}
    \edge[-]{a4}{d4}
    \edge[-, dashed, lightgray]{b4}{c4}
    \edge[-, dashed, lightgray]{b4}{d4}
    \edge[-]{c4}{d4}
    \draw[draw={rgb:red,87;green,157;blue,66}] \convexpath{a4,d4,c4}{0.5cm};
    \draw[draw={rgb:red,0;green,95;blue,134}] (b4) circle (0.5cm);
    \node[latent, right=of e4] (a6) {$x_{1}$} ;
    \node[latent, right=of a6] (b6) {$x_{2}$} ;
    \node[latent, below=of a6] (c6) {$x_{3}$} ;
    \node[latent, below=of b6] (d6) {$x_{4}$} ;
    \node[const,  right=of b6, xshift=-0.5cm] (e6) {} ;
    \node[const,  below=of c6, yshift= 0.75cm] (f6) {} ;
    \edge[-, dashed, lightgray]{a6}{b6}
    \edge[-]{a6}{c6}
    \edge[-, dashed, lightgray]{a6}{d6}
    \edge[-, dashed, lightgray]{b6}{c6}
    \edge[-]{b6}{d6}
    \edge[-, dashed, lightgray]{c6}{d6}
    \draw[draw={rgb:red,87;green,157;blue,66}] \convexpath{a6,c6}{0.5cm};
    \draw[draw={rgb:red,0;green,95;blue,134}] \convexpath{b6,d6}{0.5cm};
    \node[latent, right=of e6] (a12) {$x_{1}$} ;
    \node[latent, right=of a12] (b12) {$x_{2}$} ;
    \node[latent, below=of a12] (c12) {$x_{3}$} ;
    \node[latent, below=of b12] (d12) {$x_{4}$} ;
    \node[const,  right=of b12, xshift=-0.5cm] (e12) {} ;
    \node[const,  below=of c12, yshift= 0.75cm] (f12) {} ;
    \edge[-, dashed, lightgray]{a12}{b12}
    \edge[-, dashed, lightgray]{a12}{c12}
    \edge[-, dashed, lightgray]{a12}{d12}
    \edge[-, dashed, lightgray]{b12}{c12}
    \edge[-]{b12}{d12}
    \edge[-, dashed, lightgray]{c12}{d12}
    \draw[draw={rgb:red,87;green,157;blue,66}] (a12) circle (0.5cm);
    \draw[draw={rgb:red,0;green,95;blue,134}] \convexpath{b12,d12}{0.5cm};
    \draw[draw={rgb:red,191;green,87;blue,0}] (c12) circle (0.5cm);
    \node[latent, right=of e12] (a15) {$x_{1}$} ;
    \node[latent, right=of a15] (b15) {$x_{2}$} ;
    \node[latent, below=of a15] (c15) {$x_{3}$} ;
    \node[latent, below=of b15] (d15) {$x_{4}$} ;
    \edge[-, dashed, lightgray]{a15}{b15}
    \edge[-, dashed, lightgray]{a15}{c15}
    \edge[-, dashed, lightgray]{a15}{d15}
    \edge[-, dashed, lightgray]{b15}{c15}
    \edge[-, dashed, lightgray]{b15}{d15}
    \edge[-, dashed, lightgray]{c15}{d15}
    \draw[draw={rgb:red,87;green,157;blue,66}] (a15) circle (0.5cm);
    \draw[draw={rgb:red,0;green,95;blue,134}] (b15) circle (0.5cm);
    \draw[draw={rgb:red,191;green,87;blue,0}] (c15) circle (0.5cm);
    \draw[draw={rgb:red,214;green,210;blue,196}] (d15) circle (0.5cm);
    \node[const, below=of c1, xshift=0.825cm,
    yshift=0.75cm] (n1) {$\lambda_{1} = 6$} ;
    \node[const, below=of c4, xshift=0.825cm,
    yshift=0.75cm] (n4) {$\lambda_{1} = 3$} ;
    \node[const, below=of c6, xshift=0.825cm,
    yshift=0.75cm] (n6) {$\lambda_{1} = 2$} ;
    \node[const, below=of c12, xshift=0.825cm,
    yshift=0.75cm] (n12) {$\lambda_{1} = 1$} ;
    \node[const, below=of c15, xshift=0.825cm,
    yshift=0.75cm] (n15) {$\lambda_{1} = 0$} ;
\end{tikzpicture}
    } \\
    \vspace{0.5em}
    {\small (c) Scoring similarity with $\lambda_{2}$} \\
    \vspace{0.5em}
    \resizebox{0.55\textwidth}{!}{
        \begin{tikzpicture}
    \node[latent] (a4) {$x_{1}$} ;
    \node[latent, right=of a4] (b4) {$x_{2}$} ;
    \node[latent, below=of a4] (c4) {$x_{3}$} ;
    \node[latent, below=of b4] (d4) {$x_{4}$} ;
    \node[const,  right=of b4, xshift=-0.5cm] (e4) {} ;
    \node[const,  below=of c4, yshift= 0.75cm] (f4) {} ;
    \edge[-, dashed, lightgray]{a4}{b4}
    \edge[-]{a4}{c4}
    \edge[-]{a4}{d4}
    \edge[-, dashed, lightgray]{b4}{c4}
    \edge[-, dashed, lightgray]{b4}{d4}
    \edge[-]{c4}{d4}
    \draw[draw={rgb:red,87;green,157;blue,66}] \convexpath{a4,d4,c4}{0.5cm};
    \draw[draw={rgb:red,0;green,95;blue,134}] (b4) circle (0.5cm);
    \node[latent, draw=gray, right=of e4] (a5) {$x_{1}$} ;
    \node[latent, draw=gray, right=of a5] (b5) {$x_{2}$} ;
    \node[latent, draw=gray, below=of a5] (c5) {$x_{3}$} ;
    \node[latent, draw=gray, below=of b5] (d5) {$x_{4}$} ;
    \node[const,  right=of b5, xshift=-0.5cm] (e5) {} ;
    \node[const,  below=of c5, yshift= 0.75cm] (f5) {} ;
    \edge[-, dashed, lightgray]{a5}{b5}
    \edge[-]{a5}{c5}
    \edge[-, dashed, lightgray]{b5}{c5}
    \path (a5) edge[-, draw={rgb:red,255;green,64;blue,64},
    bend right=5]  (d5) ;
    \path (a5) edge[-, dashed, draw={rgb:red,255;green,64;blue,64},
    bend left=5]  (d5) ;
    \path (b5) edge[-, dashed, draw={rgb:red,255;green,64;blue,64},
    bend right=5]  (d5) ;
    \path (b5) edge[-, draw={rgb:red,255;green,64;blue,64}, bend left=5]  (d5) ;
    \path (c5) edge[-, draw={rgb:red,255;green,64;blue,64},
    bend right=5]  (d5) ;
    \path (c5) edge[-, dashed, draw={rgb:red,255;green,64;blue,64},
    bend left=5]  (d5) ;
    \node[latent, right=of e5] (a6) {$x_{1}$} ;
    \node[latent, right=of a6] (b6) {$x_{2}$} ;
    \node[latent, below=of a6] (c6) {$x_{3}$} ;
    \node[latent, below=of b6] (d6) {$x_{4}$} ;
    \node[const,  right=of b6, xshift=-0.5cm] (e6) {} ;
    \node[const,  below=of c6, yshift= 0.75cm] (f6) {} ;
    \edge[-, dashed, lightgray]{a6}{b6}
    \edge[-]{a6}{c6}
    \edge[-, dashed, lightgray]{a6}{d6}
    \edge[-, dashed, lightgray]{b6}{c6}
    \edge[-]{b6}{d6}
    \edge[-, dashed, lightgray]{c6}{d6}
    \draw[draw={rgb:red,87;green,157;blue,66}] \convexpath{a6,c6}{0.5cm};
    \draw[draw={rgb:red,0;green,95;blue,134}] \convexpath{b6,d6}{0.5cm};
    \node[const, below=of c5, xshift=0.825cm,
    yshift=0.75cm] (n1) {$\lambda_{2} = 3$} ;
\end{tikzpicture}
    }
    \caption{(a) Partitions as cuts of a connected graph with edges indicating
    group co-membership.
    (b) $\lambda_{1}$ counts the number of linked edges in the partition graphs
    (i.e., the number of pair co-memberships),
    shown for five different partitions of four elements.
    Our model favors larger values of $\lambda_{1}$ to promote sparsity in the
    partitions.
    (c) $\lambda_{2}$ counts the number of disagreements in linked pairs
    between two partitions,
    shown here for two partitions of four elements (left,
    right) with disagreements highlighted in red in the middle graph.
    Our model favors smaller values of $\lambda_{2}$ to promote similarity
    across adjacent partitions.
    }
    \label{fig:partition-scoring}
\vspace*{-10pt}
\end{figure}

\vspace*{-3ex}
\subsection{Sparse Factor Partitions} \label{subsec:sparsity}
\vspace*{-1ex}

We propose a prior on the partition space $\calP$
which favors parsimonious partitions across the graph $\G$.
Building on the pairwise representation of partitions (Section \ref{sec:
review}), we design a prior that penalizes the separation of element pairs,
or equivalently, rewards their co-clustering.
Specifically, defining the node potential (Section \ref{sec:
review}) to be $\lambda_{1}(\partition_{v}) = \sum_{(x_{a},
x_{b}) \in \X^{2}} \Ind\{x_{a} \sim_{\partition_{v}} x_{b}\}$,
the number of level pairs assigned to the same cluster, we let
\vspace{-7ex}\\
\bse
\Pr(\partition_{v}) \propto \exp\left\{\phi \lambda_{1}(\partition_{v})
\right\}.
\ese
\vspace{-7ex}\\
Here $\hbox{logit}^{-1}(\phi)$ is analogous to the prior probability that any
two levels of $\X$ reside in the same cluster in $\partition_{v}$.
Positive values of $\phi$ therefore encourage pairwise co-memberships,
favoring collapsed partitions of $\X$;
conversely, negative values favor expanded partitions of $\X$;
and the boundary $\phi = 0$ assigns uniform probability over $\calP$.

\vspace*{-3ex}

\subsection{Similar Factor Partitions} \label{subsec:similarity}
\vspace*{-1ex}

We next extend the prior on the partitions $\partition_{v}$'s
to favor similar configurations across neighboring locations $v\in\V$.
Specifically, we construct an MRF which penalizes differences in neighboring
partitions using the distance metric $\lambda_{2}(\partition_{v},
\partition_{u}) = d(\partition_{v}, \partition_{u})/2 = \sum_{(x_{r},
x_{j}) \in \X^{2}} \Ind[\Ind\{x_{r} \sim_{\partition_{v}} x_{j}\} \neq
\Ind\{x_{r} \sim_{\partition_{u}} x_{j}\}]$ (Section \ref{sec: review}).
When augmented with the sparsity penalty introduced above,
this now generates an MRPF prior as
\vspace{-7ex}\\
\be
\textstyle\Pr(\partition_{v} \mid \partition_{-v}) \propto \exp\left\{
\phi \lambda_{1}(\partition_{v}) - \tau \sum_{u \in \N_{v}}
\lambda_{2}(\partition_{v}, \partition_{u})
\right\},
\ee
\vspace{-7ex}\\
where the inverse-temperature parameter $\tau ~ (> 0)$ ensures that a partition
$\partition_{v}$ at node $v$ is likely to share pairwise co-membership
relations with partitions at other nodes in its neighborhood $\N_{v}$ (i.e.,
if $x_{a} \sim_{\partition_{v}} x_{b}$,
then it is likely that $x_{a} \sim_{\partition_{u}} x_{b}$ for $u \in \N_{v}$).
The prior is thus pairwise Markov and satisfies the Hammersley-Clifford
theorem, providing a valid probability distribution on the set of all
partitions $\calP(\X)$ over the graph $\G$ (Section \ref{sec: review}).

\begin{figure}[!ht]
    \centering
    \resizebox{\textwidth}{!}{
        \begin{tikzpicture}
    \node[const] (a0) {$[\{x_{1}, x_{2}, x_{3}, x_{4}\}]$};
    \node[const, above=of a0, yshift=-0.75cm] (ax) {$\mathcal{X}^{\star}_{1}$};
    \node[const, right=of a0] (b0) {$[\{x_{1}\}, \; \{x_{2}\}, \; \{x_{3}\}, \;
    \{x_{4}\}]$};
    \node[const, above=of b0, yshift=-0.75cm] (bx) {$\mathcal{X}^{\star}_{2}$};
    \node[const, right=of b0] (c0) {$[\{x_{1}, x_{2}\}, \; \{x_{3}, x_{4}\}]$};
    \node[const, above=of c0, yshift=-0.75cm] (cx) {$\mathcal{X}^{\star}_{3}$};
    \node[const, right=of c0] (d0) {$[\{x_{1}, x_{2}, x_{3}, x_{4}\}]$};
    \node[const, above=of d0, yshift=-0.75cm] (dx) {$\mathcal{X}^{\star}_{4}$};
    \edge[-] {ax} {bx}
    \edge[-] {bx} {cx}
    \edge[-] {cx} {dx}
\end{tikzpicture}
    } \\
    \vspace{1em}
    \resizebox{0.9\textwidth}{!}{
        \begin{tikzpicture}
    \node[latent] (a1) {$\beta^{\star}_{1,1}$} ;
    \node[latent, right=of a1, xshift=3cm] (b1) {$\beta^{\star}_{2,1}$} ;
    \node[latent, below=of b1, yshift=0.75cm] (b2) {$\beta^{\star}_{2,2}$} ;
    \node[latent, below=of b2, yshift=0.75cm] (b3) {$\beta^{\star}_{2,3}$} ;
    \node[latent, below=of b3, yshift=0.75cm] (b4) {$\beta^{\star}_{2,4}$} ;
    \node[latent, right=of b1, xshift=3cm] (c1) {$\beta^{\star}_{3,1}$} ;
    \node[latent, below=of c1, yshift=0.75cm] (c2) {$\beta^{\star}_{3,2}$} ;
    \node[latent, right=of c1, xshift=3cm] (d1) {$\beta^{\star}_{4,1}$} ;
    \edge[-] {a1} {b1,b2,b3,b4}
    \edge[-] {b1,b2} {c1}
    \edge[-] {b3,b4} {c2}
    \edge[-] {c1}{d1}
    \edge[-] {c2} {d1}
    \node[const, right=of a1, xshift=1cm, yshift=0.5em] (x11) {$x_{1}$};
    \node[const, right=of a1, xshift=1cm, yshift=-1em] (x12) {$x_{2}$};
    \node[const, right=of a1, xshift=1cm, yshift=-2.25em] (x13) {$x_{3}$};
    \node[const, right=of a1, xshift=1cm, yshift=-3.5em] (x14) {$x_{4}$};
    \node[const, right=of b1, xshift=0.75cm, yshift=0.5em] (x21) {$x_{1}$};
    \node[const, right=of b2, xshift=0.75cm, yshift=1.75em] (x22) {$x_{2}$};
    \node[const, right=of b3, xshift=0.75cm, yshift=2em] (x23) {$x_{3}$};
    \node[const, right=of b4, xshift=0.75cm, yshift=3em] (x23) {$x_{4}$};
    \node[const, right=of c1, xshift=0.4cm, yshift=0.5em] (x312) {$x_{1}, \;
    x_{2}$};
    \node[const, right=of c2, xshift=0.4cm, yshift=1.8em] (x33) {$x_{3}, \;
    x_{4}$};
\end{tikzpicture}
    }
    \caption{Illustration of the construction of the graph underlying the ICAR
    model in Section \ref{subsec:smoothness} for a factor with $4$ levels
    $\{x_{1},\dots,x_{4}\}$ and a linear graph $\G$ with $4$ nodes.
    Given the partition configurations $\X_{v}^{\star}$'s on $\G$ (top),
    we connect the $\beta_{v}^{\star}$'s that are associated with common levels
    of $\X$ across the adjacent nodes in $\G$ (bottom).
    }
    \label{fig:gstar}
\vspace*{-10pt}
\end{figure}
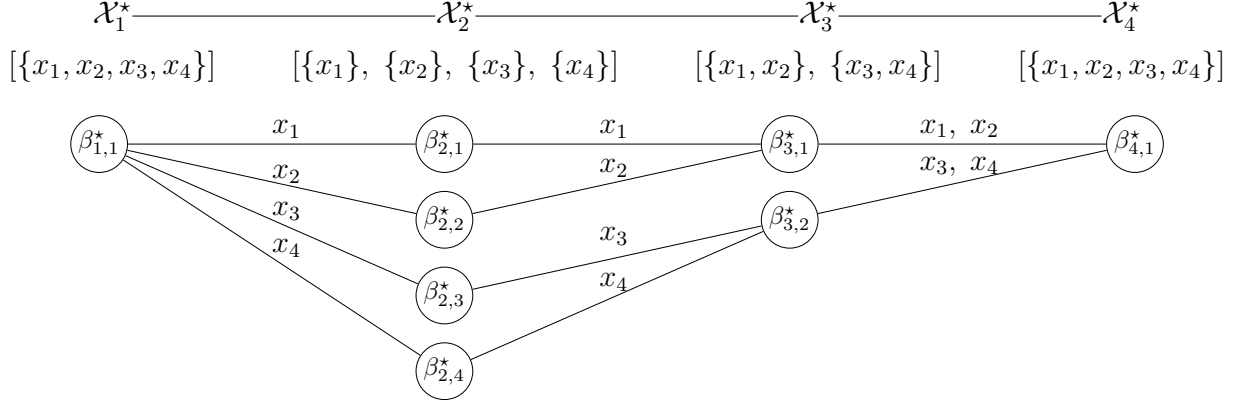

\vspace*{-3ex}

\subsection{Smooth Factor Effects}
\label{subsec:smoothness}
\vspace*{-1ex}

The final step is to design a prior for the atoms $\bbeta_{v}^{\star}$'s.
We propose a second-level continuous MRF prior that allows for sharing of
information across factor levels in $\X$ by unifying adjacency across
\textit{both} $\G$ \textit{and} $\partition$.
Specifically, for any node $v$,
we leverage $\bP_{v}$ and the $\bP_{u}$'s at neighboring nodes $u \in \N_{v}$
to connect the entries in $\bbeta_{v}^{\star}$ with those in
$\{\bbeta^{\star}_{u}: u\in\N_{v}\}$
that represent the same levels of $\X$ and are therefore considered codependent.
The strength of this co-dependence is controlled by a smoothness parameter
$\sigma^{2}_{\beta}$, which is also learned from the data.
Specifically, the prior takes the form of a novel hierarchical ICAR model
(Section \ref{sec: review}) as
\vspace{-7ex}\\
\be
\textstyle p\left( \bbeta_{v}^{\star} \mid \bP, \bbeta_{-v}^{\star},
\sigma^{2}_{\beta} \right)
= \MVN_{q_{v}}\left(\bbeta_{v}^{\star} \mid \frac{1}{|\N_{v}|} \sum_{u \in
\N_{v}}\bP_{v}^{\trans}\bP_{u}\bbeta^{\star}_{u}, \;
\frac{\sigma^{2}_{\beta}}{|\N_{v}|} \bI \right),
~~~
\sigma_{\beta}^{2} \sim \IG(a_{\sigma_{\beta}^{2}},b_{\sigma_{\beta}^{2}}),
\ee
\vspace{-7ex}\\
where the adjacency relations are based on both the original graph $\G$
\textit{and} the structure of the $q_{v} \times q_{u}$ matrices
$\bP_{v}^{\trans}\bP_{u}$ with $(a,b)\th$ element $\sum_{j = 1}^{p} \Ind\left\{
\partlite_{v}(x_{j}) = x_{a}^{\star} \right\} \Ind\left\{ \partlite_{u}(x_{j})
= x_{b}^{\star} \right\}$,
which gives the number of levels of $\X$ shared in common by the clusters
$\partlite_{a}$ in $\partition_{v}$ and $\partlite_{b}$ in $\partition_{u}$.
This creates a weighting system where a large difference between a
$\beta_{v,\partlite_{a}}^{\star}$ and a $\beta_{u,\partlite_{b}}^{\star}$ at
neighboring locations $v$ and $u$
is penalized only if they represent the collapsed effects of some common factor
levels in $\X$ at these locations,
with the penalty proportional to the number of levels they have in common.
An increased degree of shared representation between two neighboring partitions
$\partition_{v}$ and $\partition_{u}$ therefore yields additional information
about the degree to which the neighboring $\bbeta_{v}^{\star}$ and
$\bbeta_{u}^{\star}$ influence one another (Figure \ref{fig:gstar}).

\vspace*{-3ex}
\section{Posterior Inference} \label{sec: sampling}
\vspace*{-1ex}

We rely on samples drawn from the posterior using an MCMC algorithm for
finite-sample inference.
Designing such samplers for our proposed model poses significant challenges --
not only are we working with a massive graph space comprising tens of thousands
of nodes, but the model sizes $\abs{\partition_{v}} = \abs{\bbeta_{v}^{\star}}
= q_{v}$'s also vary stochastically across the nodes.
To address this, we augment each $\bbeta_v^\star$ with a vector $\bbeta_v^\circ
= (\bbeta_{v,1}^\circ, \dots, \bbeta_{v,p - q_{v}}^\circ)\trans$,
representing atoms for the $p - q_{v}$ empty clusters not linked to any factor
level at node $v$.
These are assigned independent priors $(\bbeta_v^\circ \mid \bP_v) \sim \MVN_{p
- q_{v}}(\beta_{v}^{0} \bone, \sigma_{\beta}^{2} \bI)$.
This yields a fixed-length $p$-dimensional parameter $\bbeta_{v}^\dagger =
(\bbeta_v^{\star\trans},
\bbeta_v^{\circ\trans})\trans$ with a well-defined conditional prior
$p(\bbeta_{v}^{\dagger} \mid \bP, \bbeta_{-v}^{\dagger}, \sigma_{\beta}^{2},
\beta_{v}^{0})$ for all $\bP_{v}$ configurations,
which are then easily sampled from.
For efficient exploration of the posterior,
we adapt the single-site Metropolis-Hastings (M-H) algorithm \citep{1984Geman},
combined with ``chromatic" \citep{2011Gonzalez} or ``conclique"
\citep{2020Kaplan} sampling scheme,
which allows the node-specific variables to be updated in parallel,
accelerating convergence.
Additionally, to efficiently explore the posterior over partitions,
we further incorporate the Hamming ball sampler \citep{2017Titsias} with local
likelihood-informed moves \citep{2019Zanella}.

We implemented our sampler in C++ \citep{CPP11} with an R interface
\citep{RProgramming} via Rcpp and RcppArmadillo \citep{Rcpp, RcppArmadillo},
parallelizing it using RcppThread \citep{RcppThread}. On an 8-core Apple M1,
the full run took 95 minutes (system time: 11 hours),
yielding an $85\%$ reduction and a $6.92\times$ speedup,
close to the ideal $8\times$.

\ul{Additional details about hyper-parameter selection,
posterior full conditionals, postprocessing, runtimes,
and related aspects are presented in Sections S.3 and S.4 in the SM.}

\section{Application to Cortical Mapping of Fingertips} \label{sec: application}
\vspace*{-1ex}

\begin{figure}[!ht]
    \centering
    \includegraphics[width=0.3\textwidth]{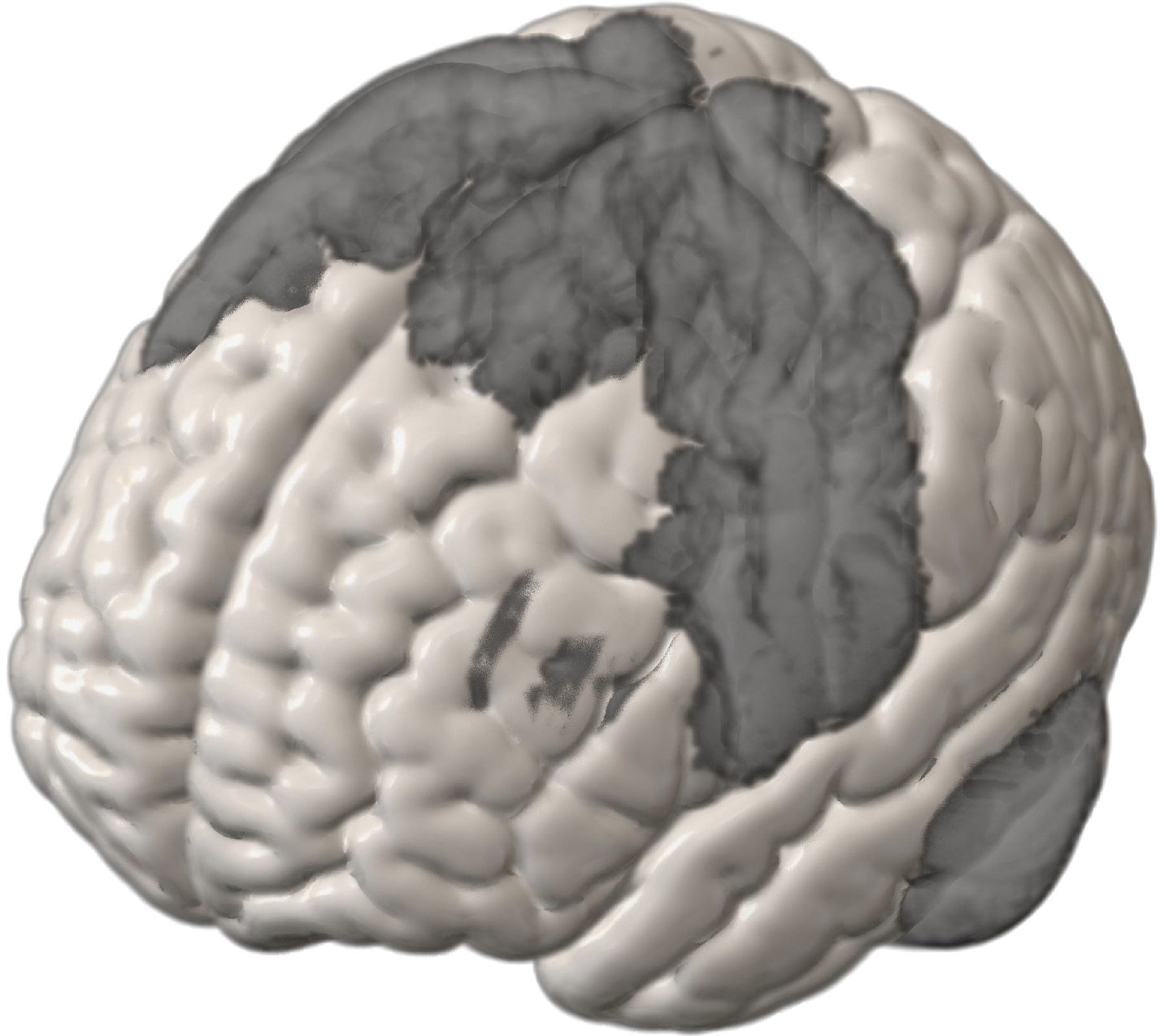}
    \hspace{25pt}
    \includegraphics[width=0.3\textwidth]{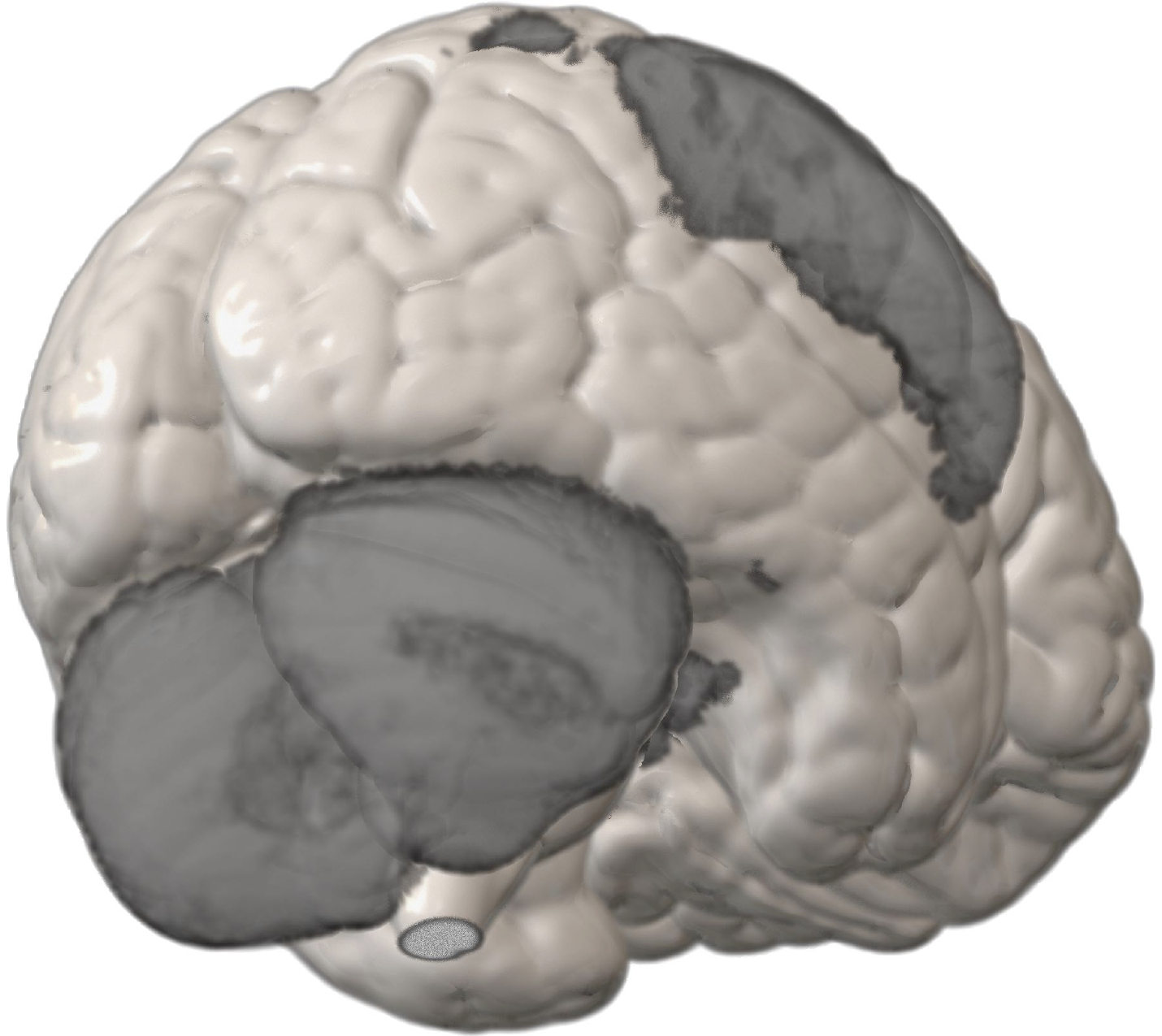}
    \caption{The inspected regions for the cortical fingertip mapping study:
    Sensorimotor areas consisting of the pre- and post-central gyri and the
    supplementary motor area, the caudate and putamen,
    and the midbrain and cerebellum.}\label{fig:mask}
\vspace*{-10pt}
\end{figure}

We apply lhMRPF to the 7T-fMRI fingertip mapping data (Section~\ref{subsec:task}).
{\ul{Given the availability of ultra-high-resolution multi-run data for each
subject, efficient individual-level `precision mapping' is feasible here,
and that is what we focus on,
leaving population-level joint modeling for future work.}}
{\ul{Results for one subject are detailed here,
with a second provided in the SM.}}
In all brain-render figures,
clusters under $64\hbox{mm}^{3}$ are hidden for clarity.

To recall, our local clustering model takes as input the results from the
first-level GLM analysis for each voxel, which arise from the distribution
$\wt\bbeta_{v,r} \sim \MVN_{p}(\bbeta_{v}, \;
\wt\bSigma_{v,r})$ (see Sections \ref{sec:VWA} and \ref{sec: method}).
We assume the underlying lattice graph $\G$ to consist of voxel nodes $v \in
\V$ and edges $e \in \E$ connecting directly adjacent voxels.
We recall that our goal is to infer the latent dependent partitions of the
effects of the $p=4$ fingertip levels $\X \equiv (\text{right index,
right middle, left index,
and left middle})\trans$ utilized in the learning task at each voxel $v\in\V$.
For computational convenience,
we focus on the most biologically relevant motor areas in the brain,
namely the cerebellum (and adjacent midbrain gray matter), caudate, putamen,
and sensorimotor cortex, as seen in Figure \ref{fig:mask}.
The corresponding graphs comprise 57,684, 6,064, and 35,893 voxels,
respectively.
Applying our lhMRPF model separately to these three regions,
which have varying neuron densities \citep{azevedo2009equal},
allows for varying smoothness in the underlying effects across them,
while voxel-specific covariance matrices $\wt\bSigma_{v}$ account for
heterogeneity within each region.

As a set of $p=4$ elements,
$\X$ has $B_{4} = 15$ possible partitions in $\calP(\X)$ at each voxel $v$
under examination (Section \ref{sec: review}),
which are shown in Figure \ref{fig:partitions-of-four}.
To simplify our analysis, we identify two classes of relevant partitions:
(A) those that exhibit digit uniqueness and (B) those that exhibit
lateralization.
Partitions exhibit digit uniqueness if any digit is partitioned into a group
with no other digits, signifying that digit (say,
$x_{1}$) to have an effect intensity ($\beta_{v,1}$) which is distinct from the
other three examined digits.
This includes any partition with a singleton cluster: partitions 2, 3, 4, 5, 9,
10, 11, 12, 13, 14, and 15.
Partitions exhibit lateralization if no group contains digits from both the
left and the right hand (which we refer to as a ``handed" partition) or if no
group contains both an index and a middle fingertip (which we call ``digited").
This includes any partition with a vertical (handed) or horizontal (digited)
cut: partitions 6, 11, 12, 15 and 8, 9, 14, 15.
Partition 15 appears in all interesting groups due to its being fully expanded,
composed of six true differences.

\begin{figure}[!ht]
    \centering
    \includegraphics[width=0.28\textwidth]{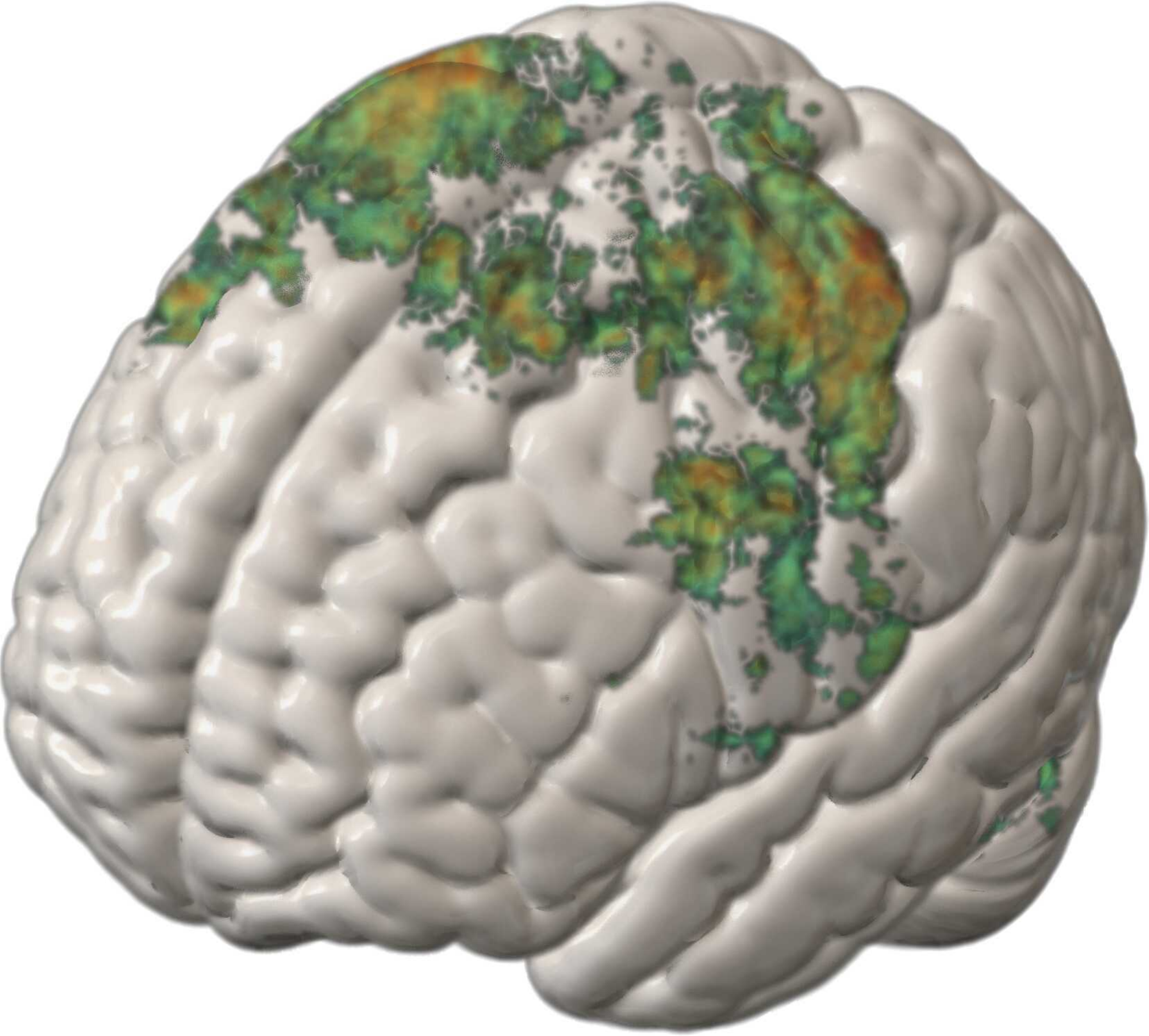}
    \hspace{15pt}
    \includegraphics[width=0.28\textwidth]{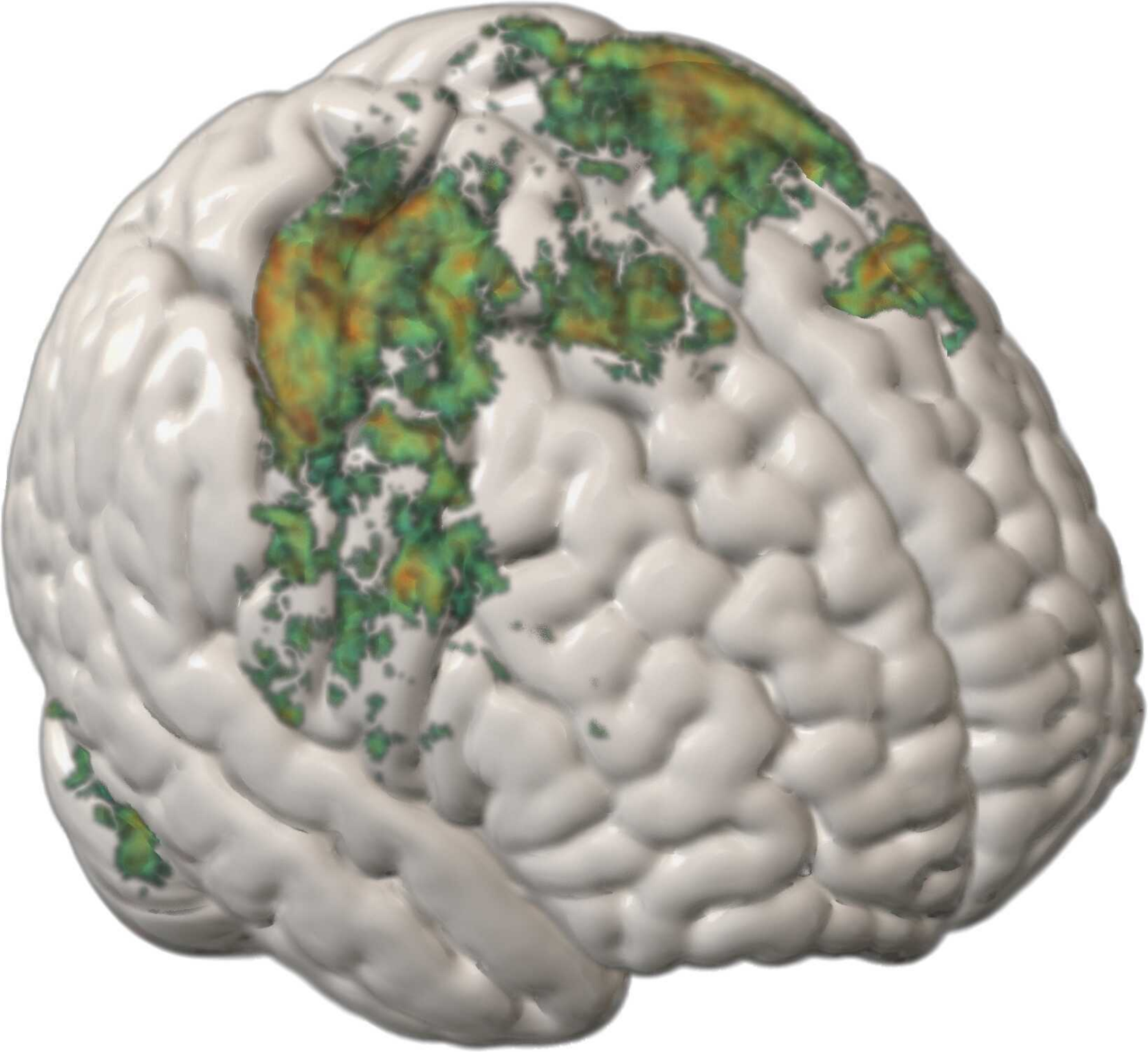}
    \hspace{15pt}
    \includegraphics[width=0.28\textwidth]{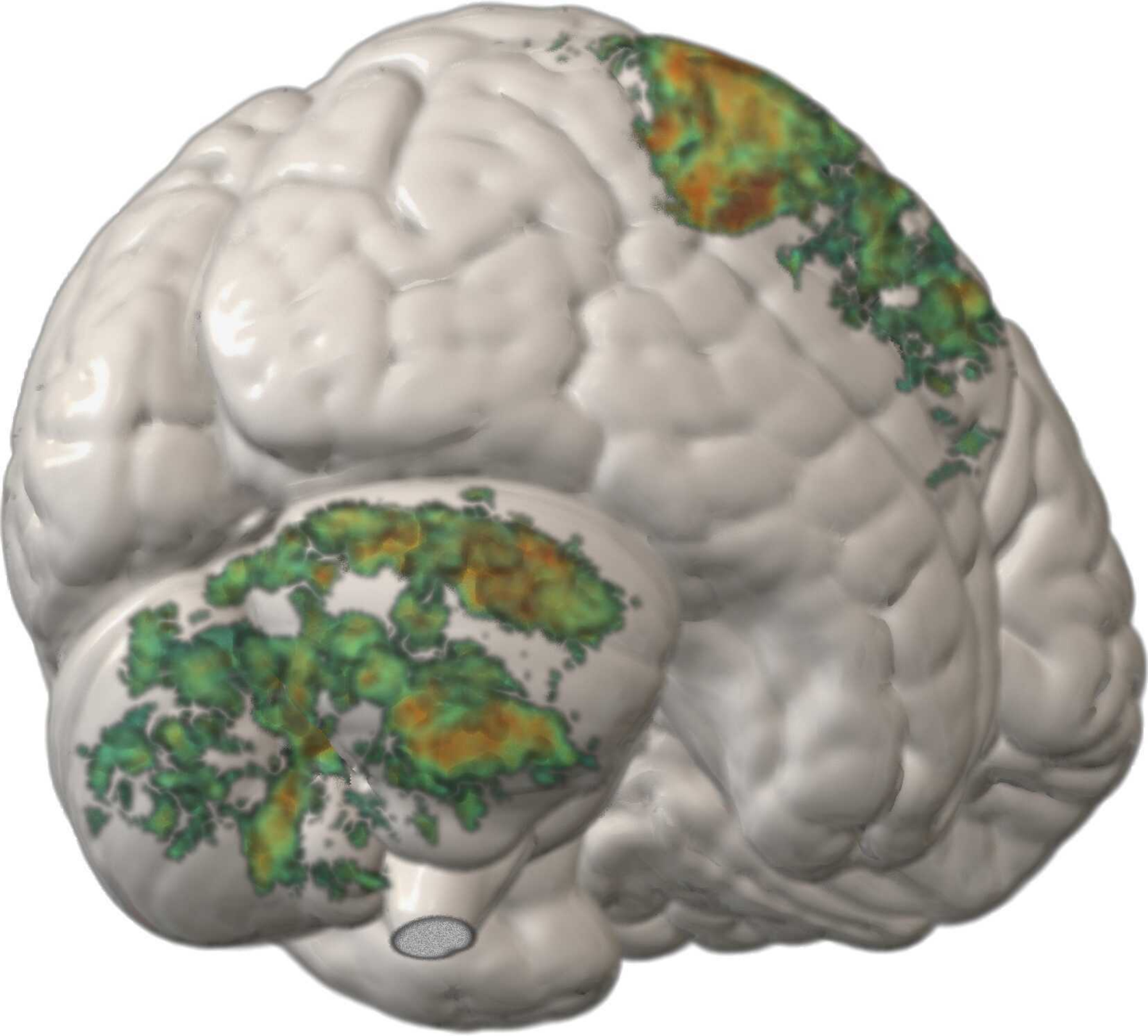}
    \newline\\
    \vspace*{-7.5pt}
    \includegraphics[width=0.75\textwidth]{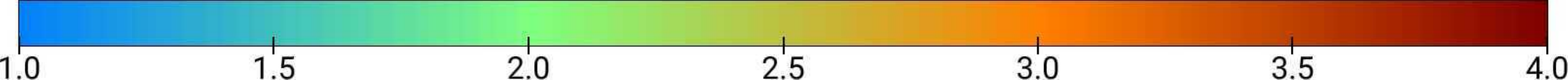}
    \vspace*{-5pt}
    \caption{
    Results for the fingertip mapping study:
    A map of the posterior expected cardinality of the partitions at each voxel.
    Only voxels with a mean estimated cardinality greater than 2 are colored.
    }\label{fig:cardinality}
\vspace*{-10pt}
\end{figure}

\paragraph{Cluster Cardinality.}
The lhMRPF method's ability to discern between the four fingers is demonstrated
in Figure \ref{fig:cardinality}.
We observe the highest degree of discernibility (high-cardinality partitions)
in the sensorimotor cortices,
aligning with established cortical mapping properties of the fingertips and
hands \citep{1998Gelnar}.
Similarly expected patterns \citep{2013vanderZwaag} are also found in the
cerebellum.

\begin{figure}[!ht]
    \centering
    {\small (a) Right-Hand}
    \vskip 10pt
    \includegraphics[width=0.25\textwidth]{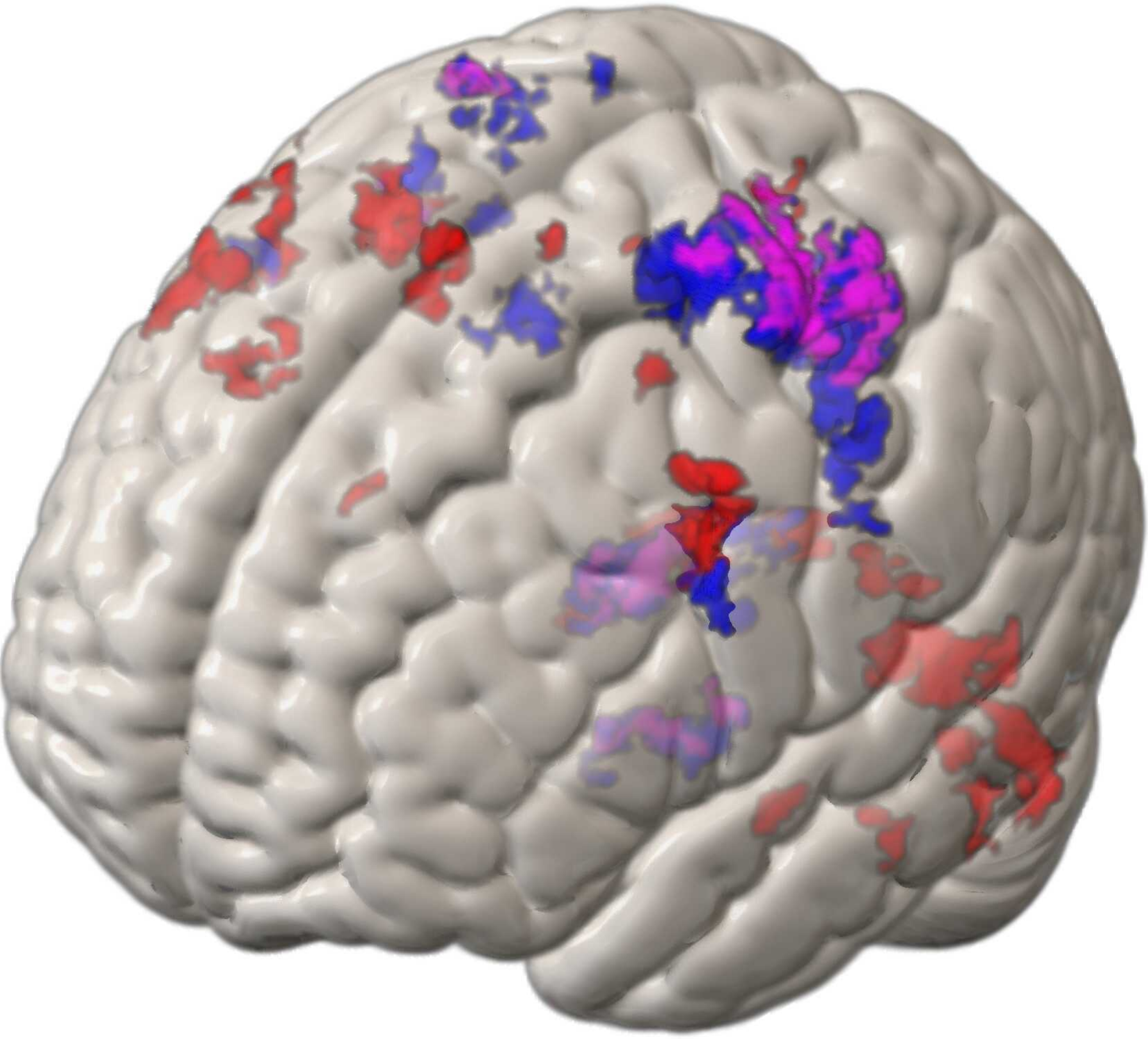}
    \hspace{15pt}
    \includegraphics[width=0.25\textwidth]{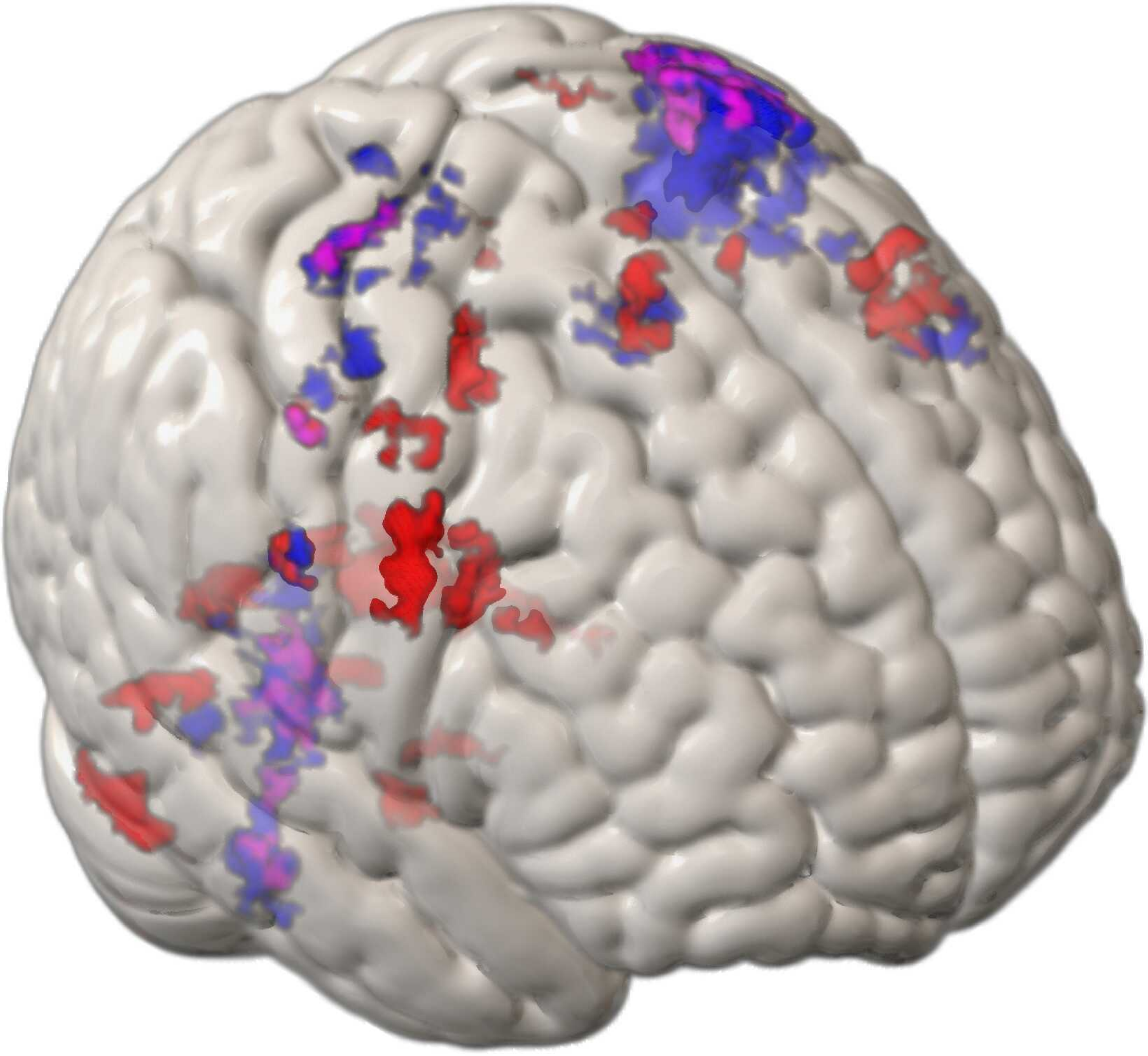}
    \hspace{15pt}
    \includegraphics[width=0.25\textwidth]{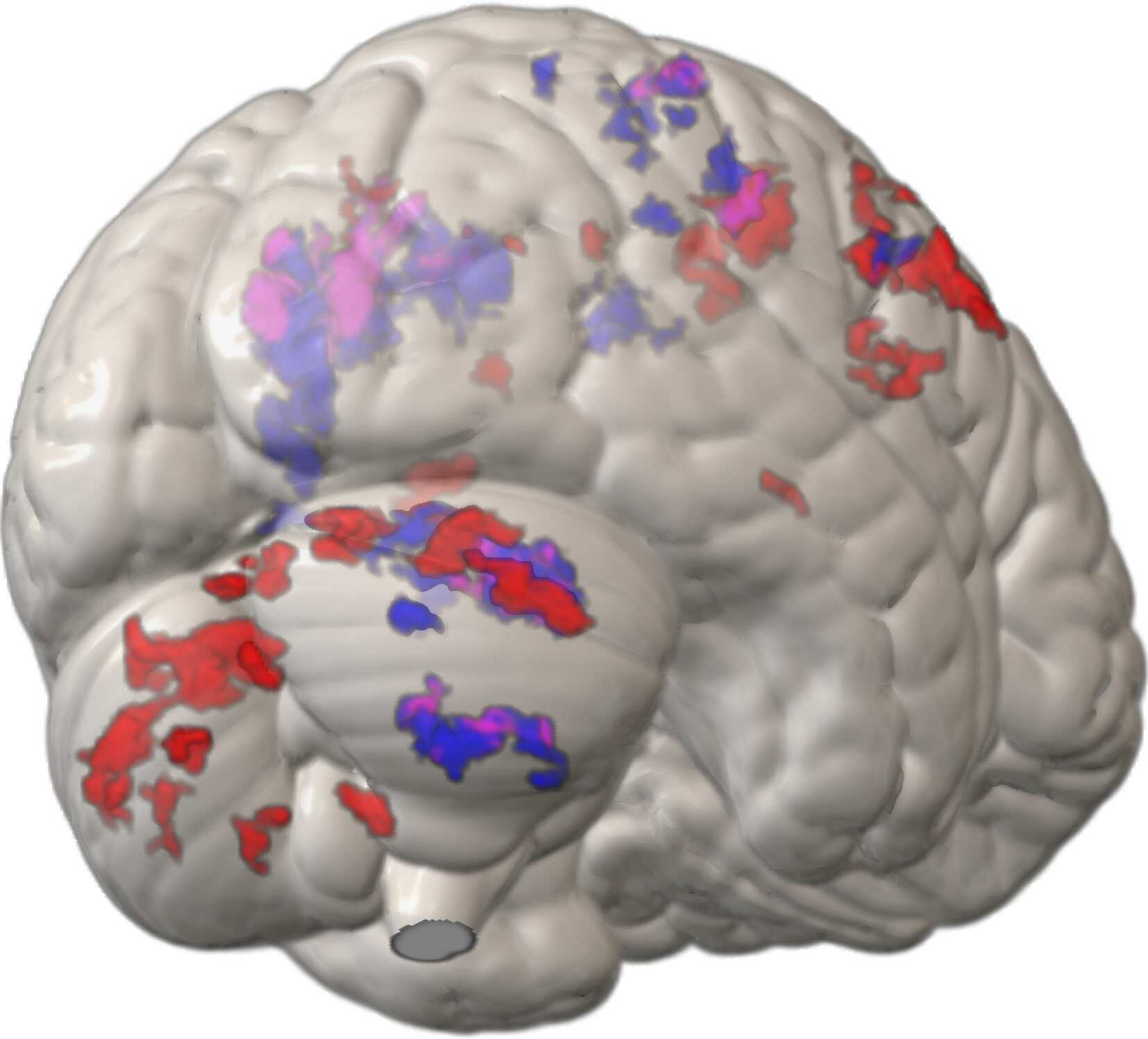}
    \vskip 10pt
    {\small (b) Left-Hand}
    \vskip 10pt
    \includegraphics[width=0.25\textwidth]{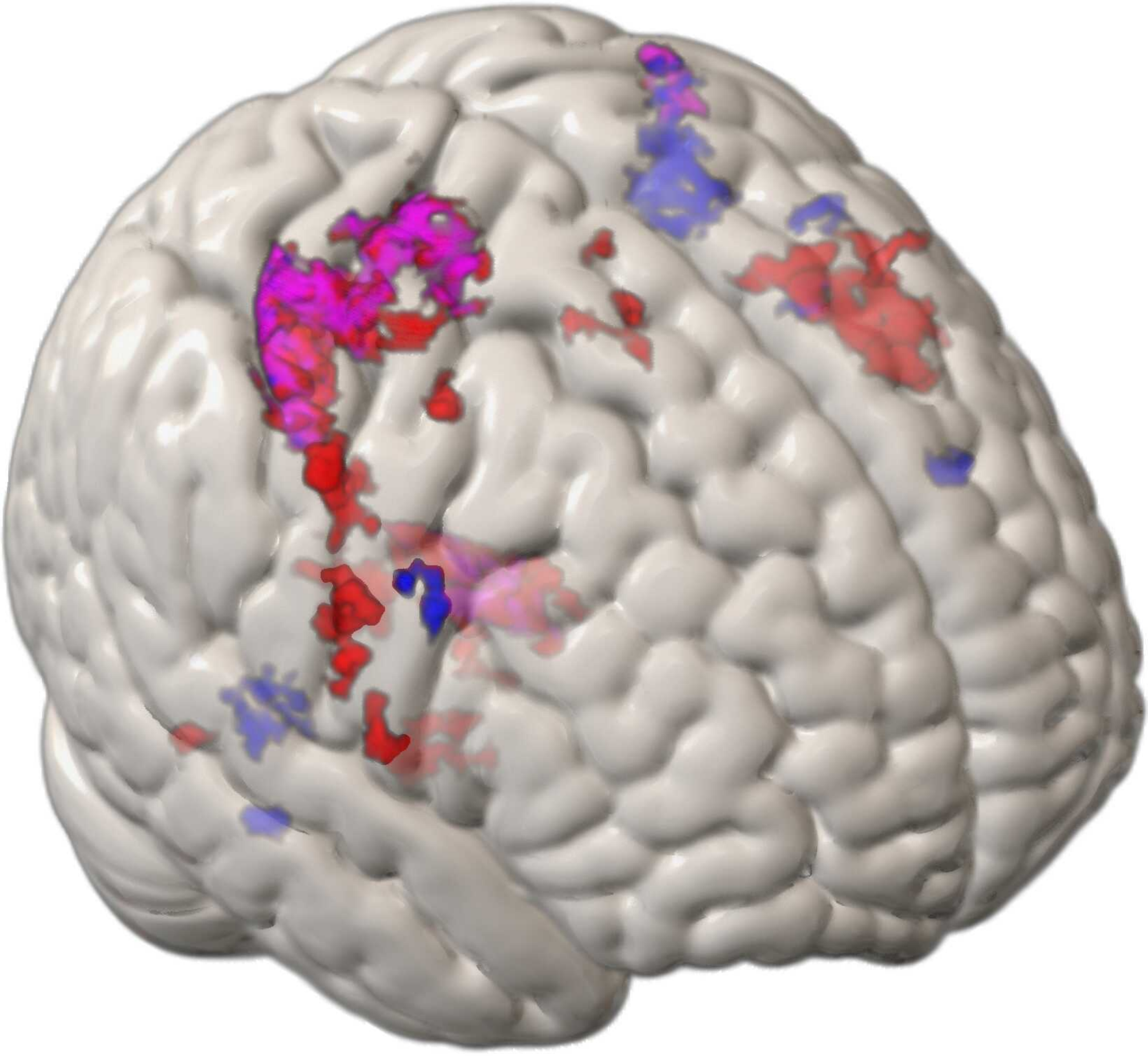}
    \hspace{15pt}
    \includegraphics[width=0.25\textwidth]{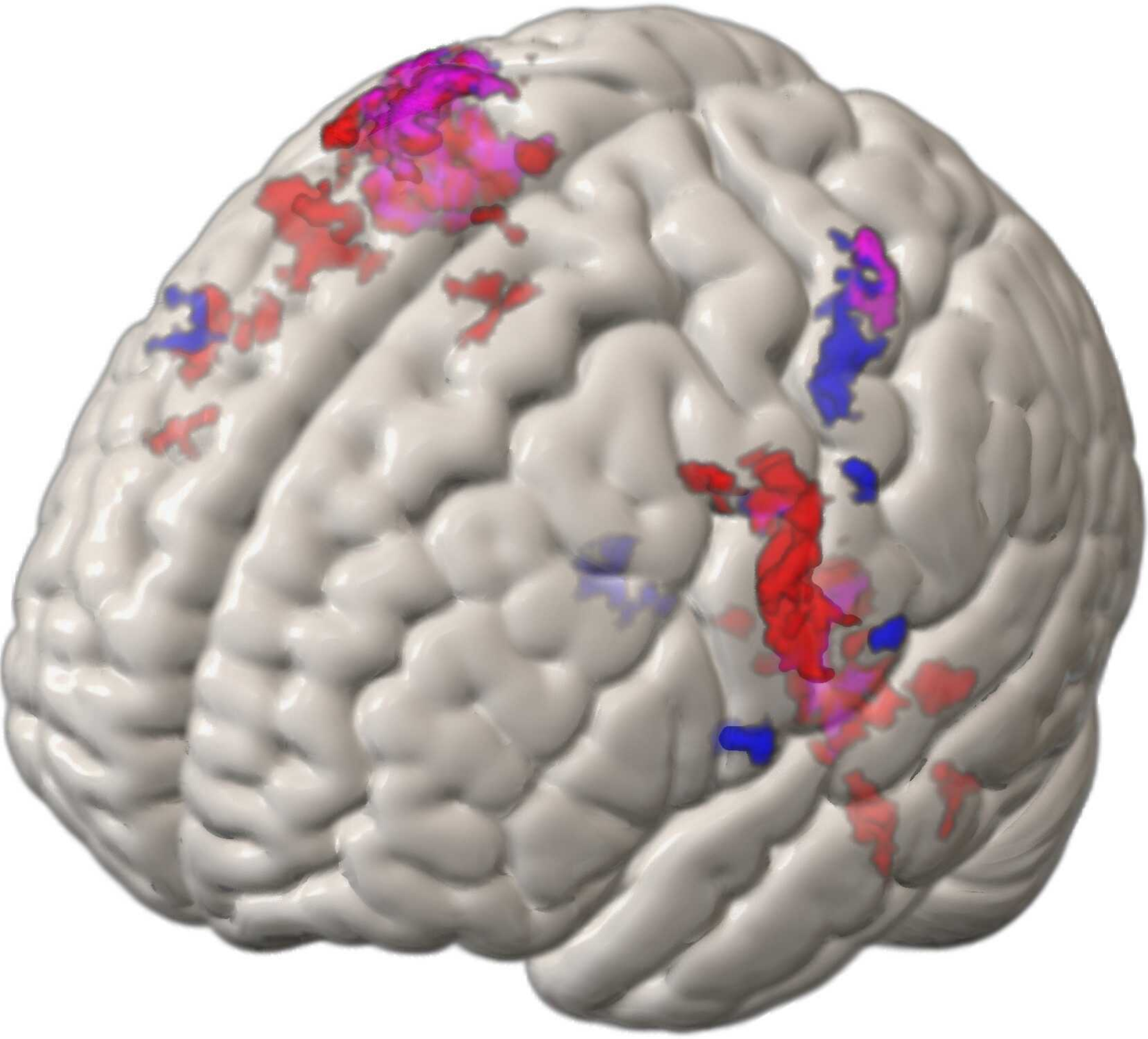}
    \hspace{15pt}
    \includegraphics[width=0.25\textwidth]{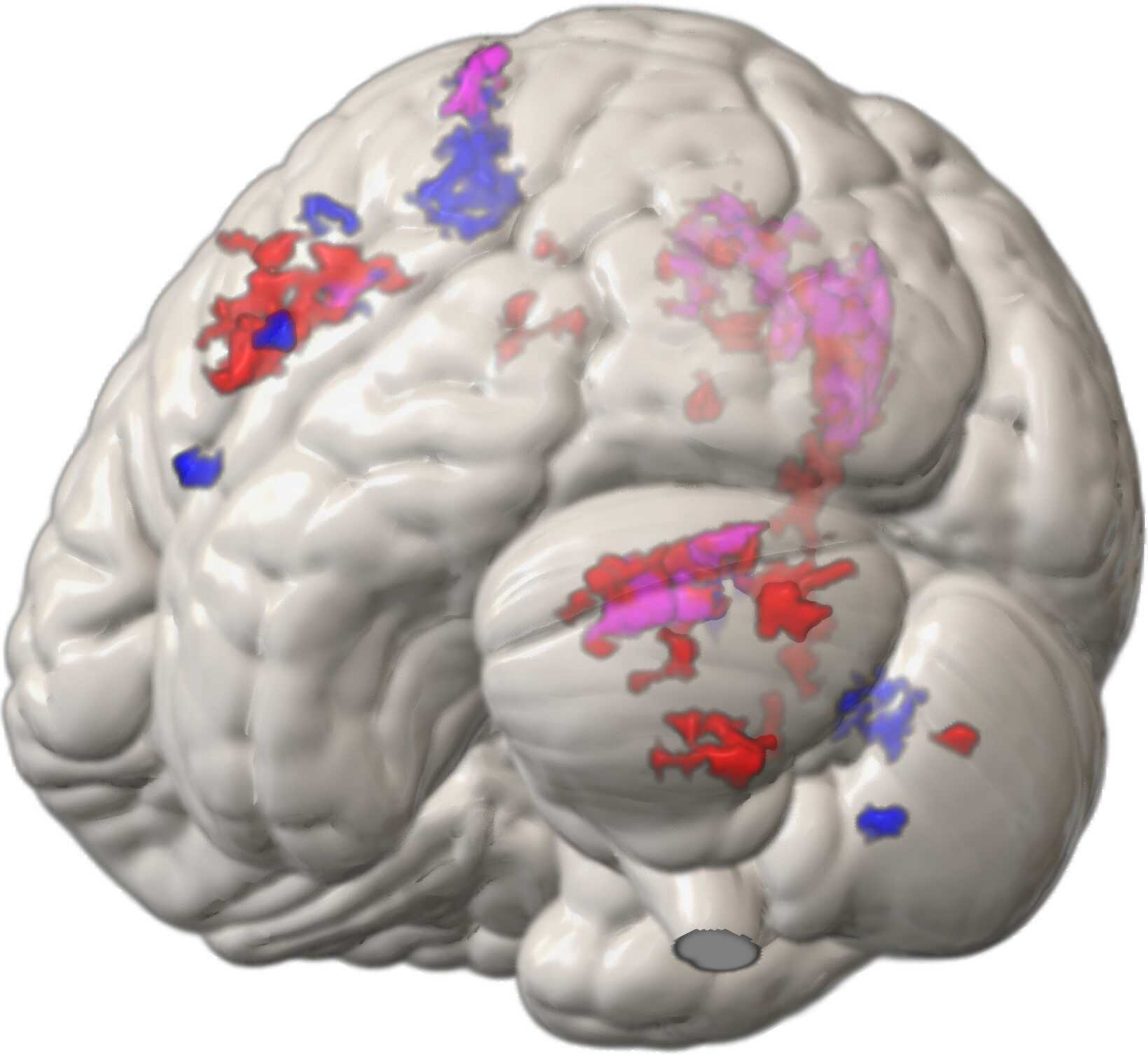}
    \caption{
    Results for the fingertip mapping study:
    Areas of unique representation of the four tested fingers for the (a) right
    and (b) left hands based on the estimated partition structures.
    Areas of unique representation for index fingers (red) and middle fingers
    (blue) distinctly overlap (purple) in the sensorimotor cortex and
    cerebellum.}\label{fig:uniqueness}
\vspace*{-10pt}
\end{figure}

\begin{figure}[!ht]
    \centering
    \includegraphics[width=0.25\textwidth]{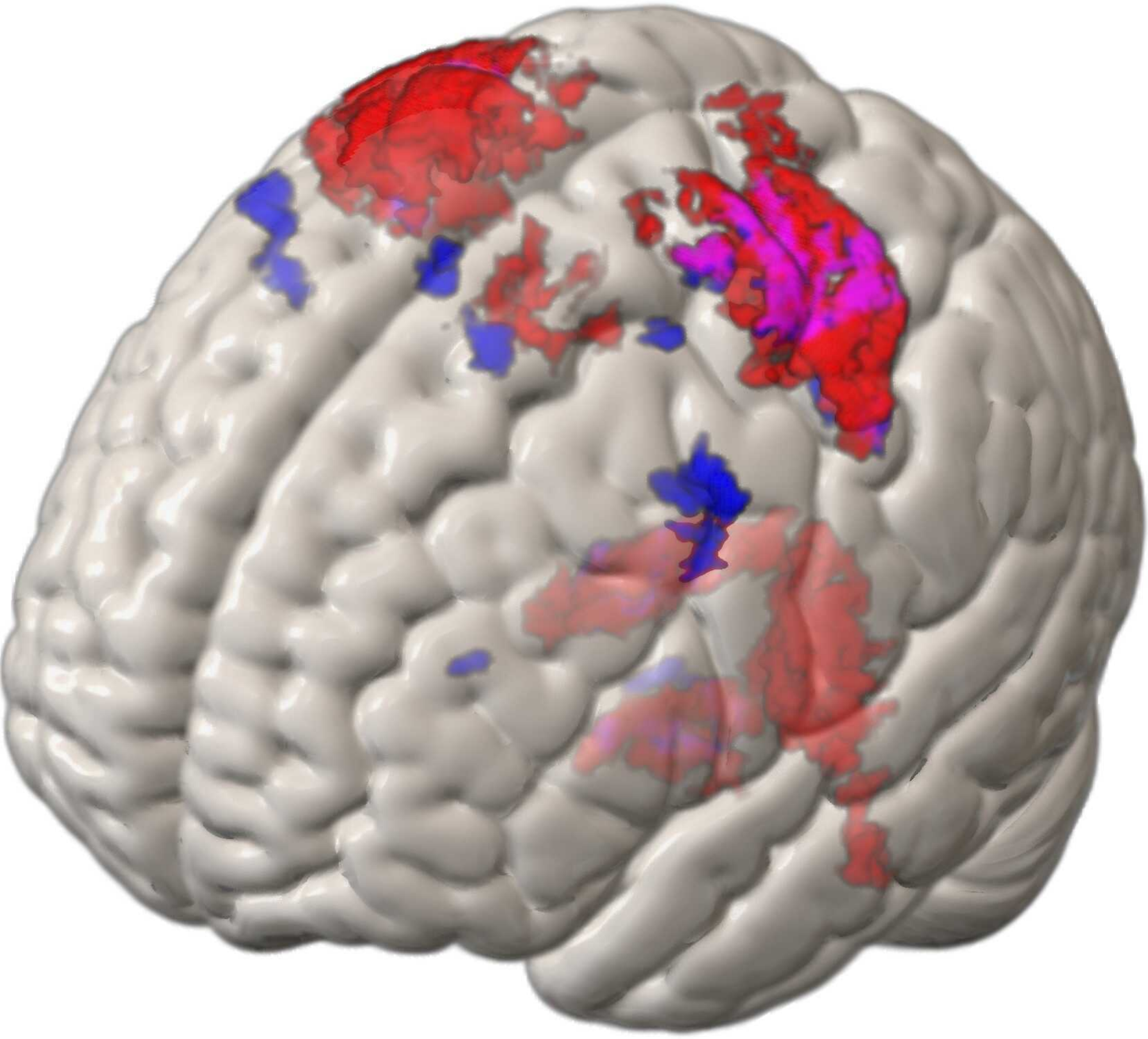}
    \hspace{15pt}
    \includegraphics[width=0.25\textwidth]{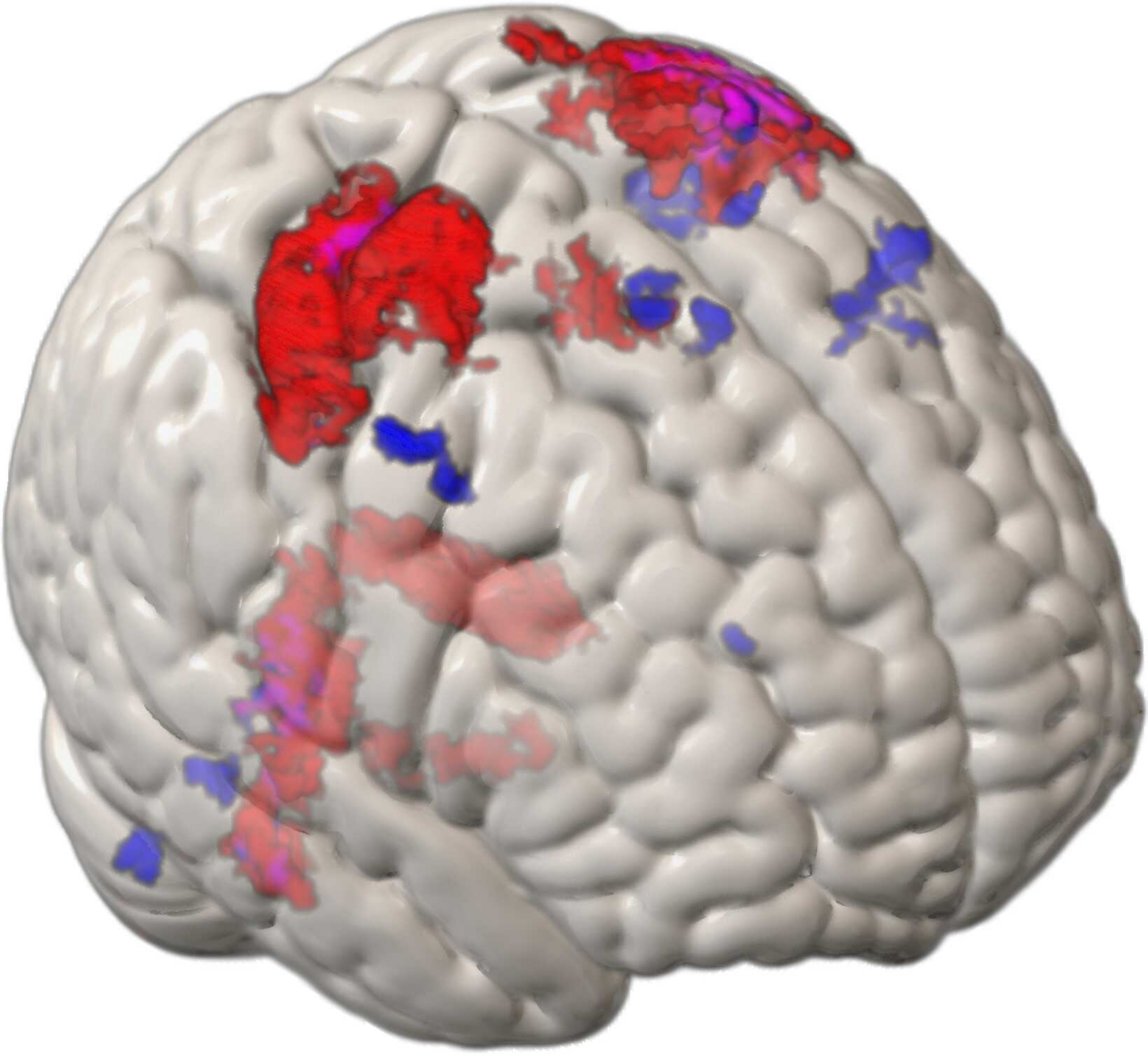}
    \hspace{15pt}
    \includegraphics[width=0.25\textwidth]{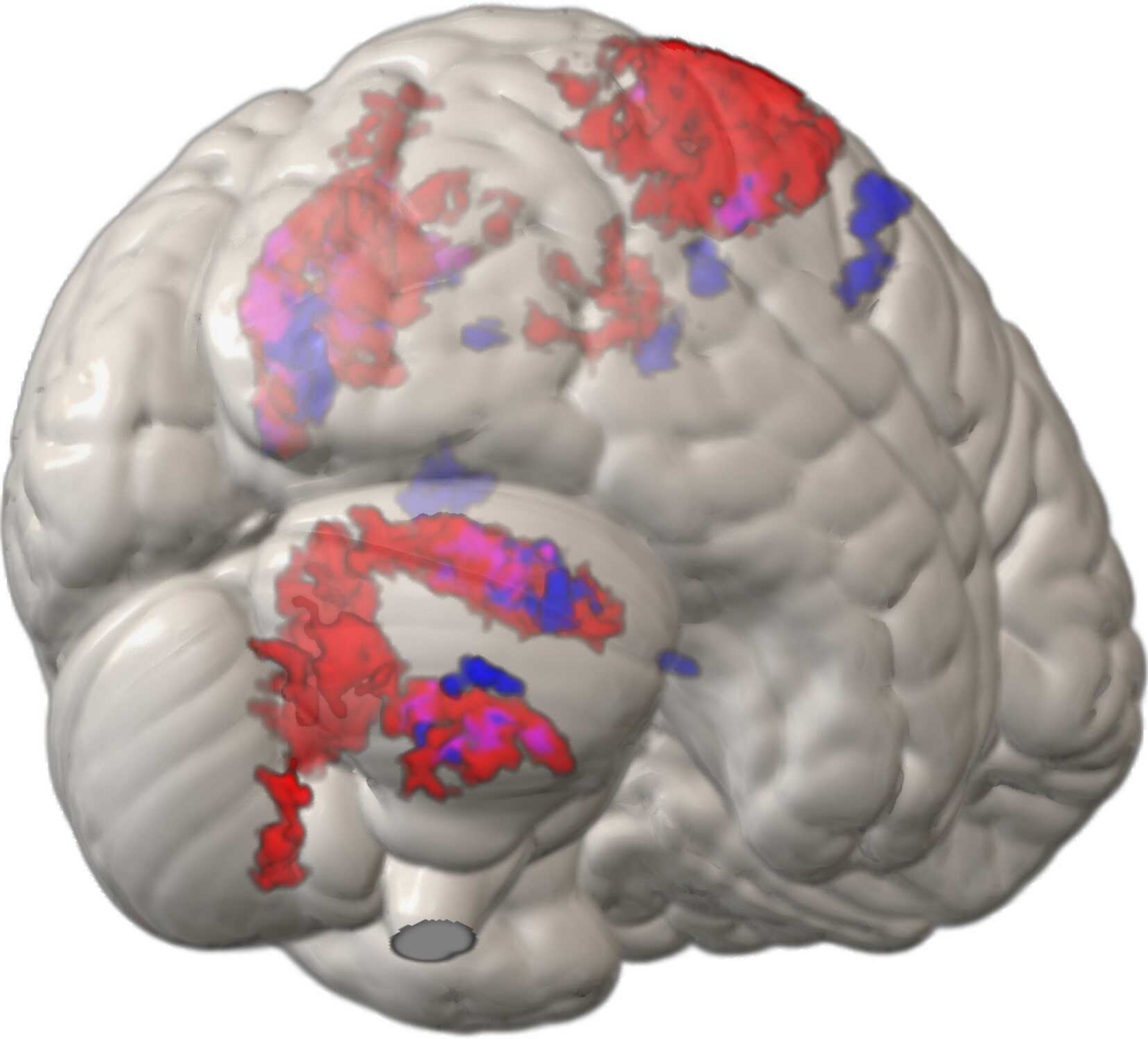}
    \caption{
    Results for the fingertip mapping study:
    Areas of split representation between the two hands (left/right,
    red) and the digits (index/middle,
    blue) based on the estimated partition structures.
    The overlap of these two partition types (purple) occurs when both the
    left/right hands and the index/middle fingers have split representations,
    i.e., when the four fingers are uniquely represented.}\label{fig:handedness}
\vspace*{-10pt}
\end{figure}

\paragraph{Cluster Organization.}
We find that the motor cortex and cerebellum appear segmented between six
overlapping regions:
four regions of unique representation for each fingertip used in the task
(Figure \ref{fig:uniqueness}),
a hand lateralization region (Figure \ref{fig:handedness}),
and a remaining volume of collapsed factor levels.
Following well-known lateralization identifying handedness regions of the
sensorimotor cortex \citep{brinkman1973cerebral,kawashima1993regional},
fingers of each hand show overlapping unique representations in these areas.
Distinct representations of the left and right-handed fingers occur in the
ipsilateral cerebellum,
with additional contralateral uniqueness shown for the right index and left
middle finger.

In addition to the primary sensorimotor areas,
the anterior ventral pre-central gyrus exhibits several compact regions of
unique representation for the index fingers,
despite the relatively low level of response intensity of this area.

\begin{figure}[!ht]
    \centering
    \includegraphics[width=0.25\textwidth]{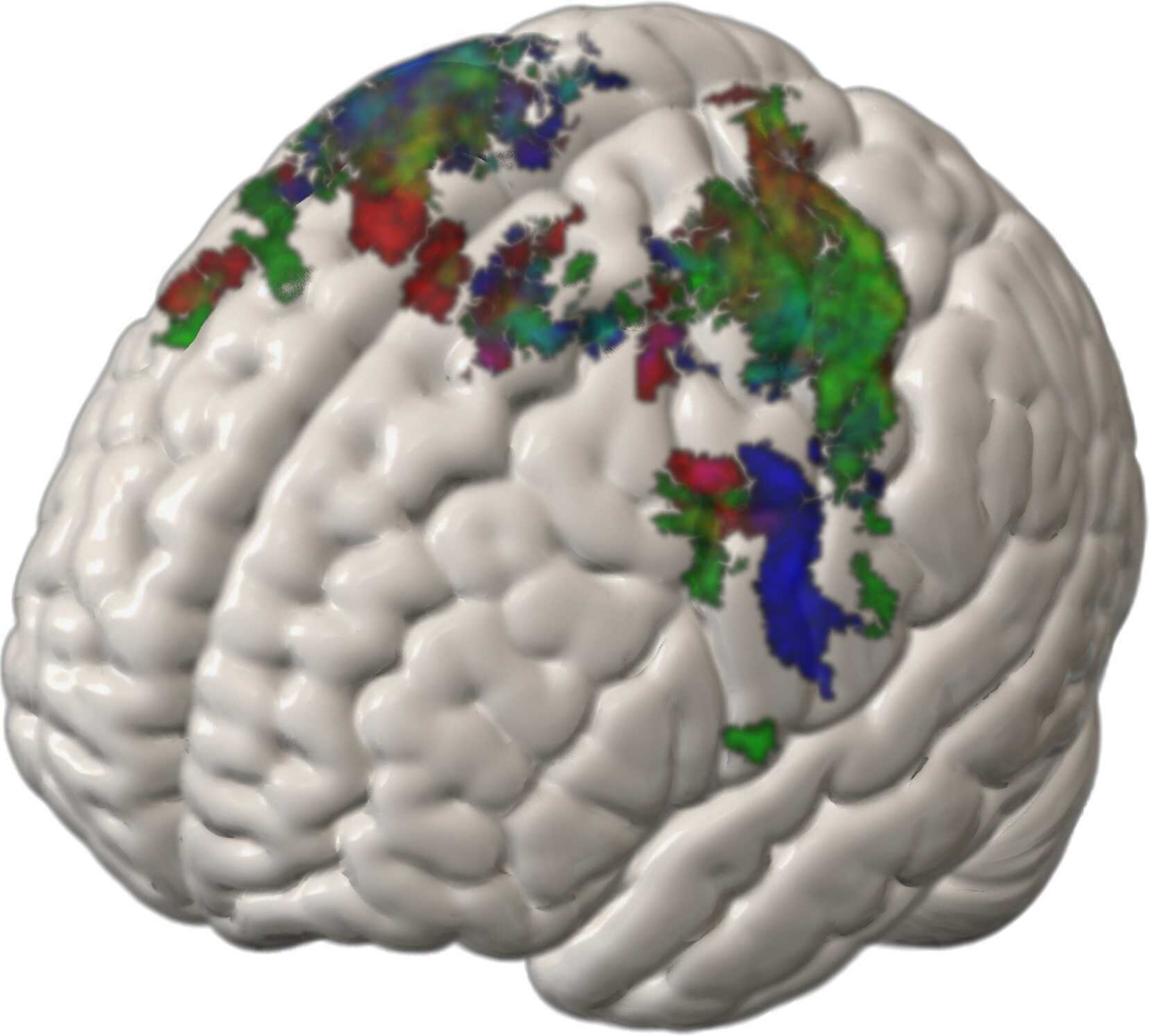}
    \hspace{15pt}
    \includegraphics[width=0.25\textwidth]{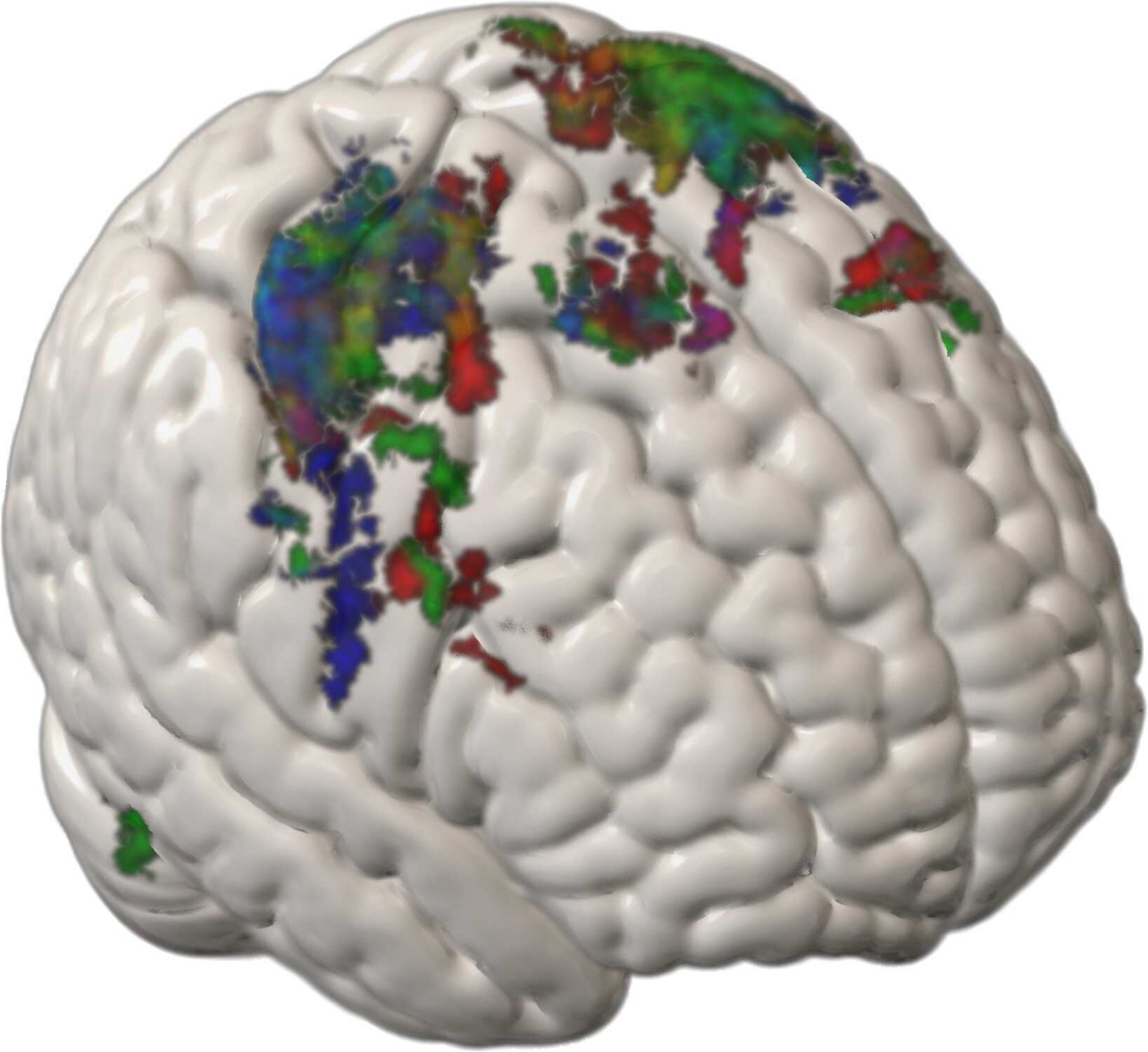}
    \hspace{15pt}
    \includegraphics[width=0.25\textwidth]{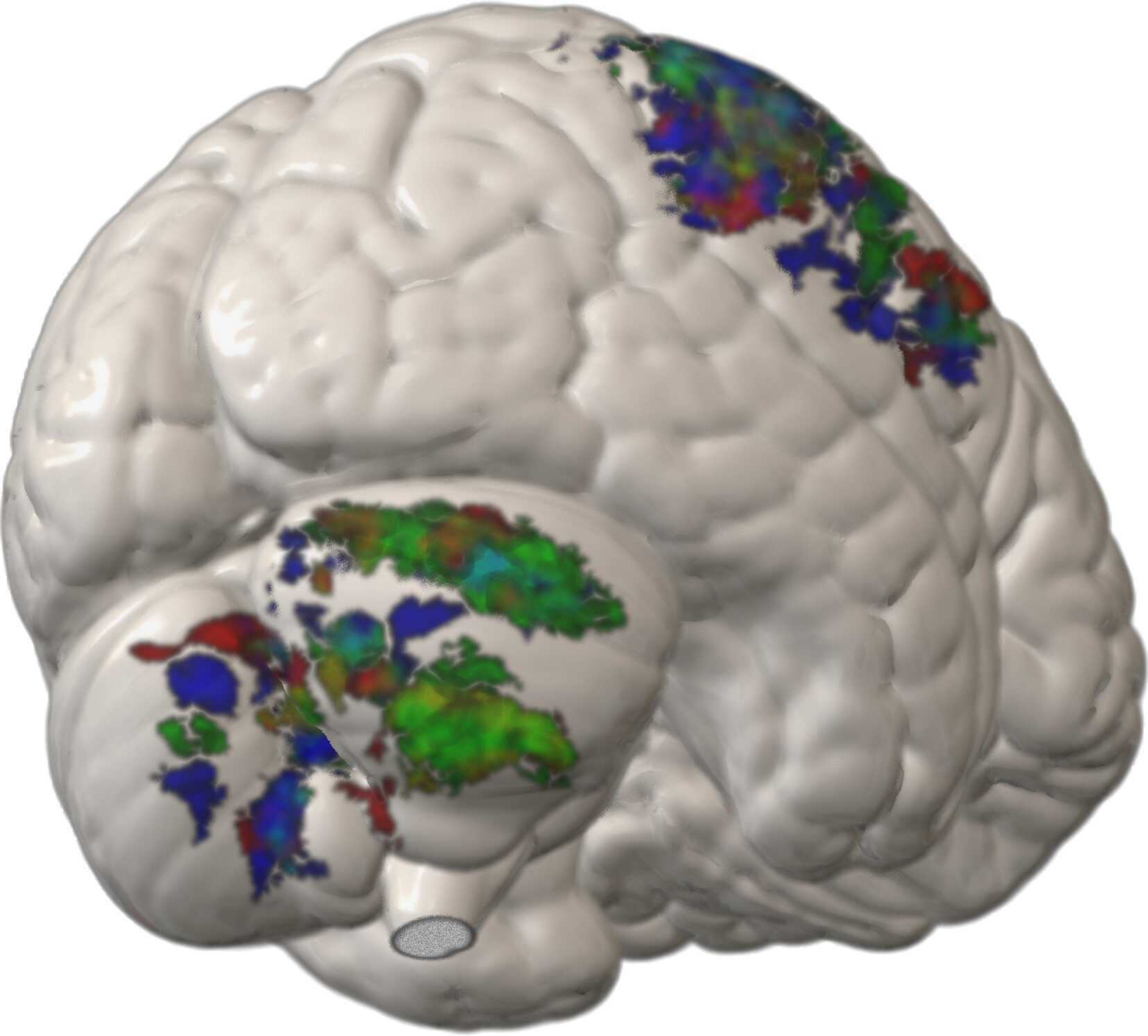}
    \caption{
    Results for the fingertip mapping study:
    Fingertip dominance across the investigated regions.
    Posterior mean dominance greater than $25\%$ is shown across voxels for the
    right index finger (red), left index (blue), and middle fingers (green)
    based on the estimated partition structures and the estimated effect
    intensities.
    Colors blend when two or more fingers exhibit greater than $25\%$ dominance
    probabilities at the same location,
    showing areas of competing dominance for the right index and either middle
    finger (yellow), left index and either middle finger (cyan),
    and the two index fingers (purple).}\label{fig:dominance}
\vspace*{-10pt}
\end{figure}

\paragraph{Dominance Maps.}
We construct fingertip dominance maps by calculating the expected posterior
probability of one finger's effect being more extreme than any of the other
tested fingers at each voxel.
That is, $\hbox{Dom}_{v}(x_{a}) = \frac{1}{M (p - 1)} \sum_{x_{b} \in \X -
x_{a}} \sum_{m = 1}^{M} \Ind\{|\beta^{\star (m)}_{v,
x_{a}}| > |\beta^{\star (m)}_{v, x_{b}}|\}$,
where $m$ is an index for the MCMC samples and $M$ is the number of
post-burn-in thinned MCMC samples used for inference.
The patterns of dominance show right-handed fingers dominating the
contralateral pre and post-central gyri and the ipsilateral cerebellum,
and vice versa for the left-handed fingers (Figure \ref{fig:dominance}).
Additionally, we found regions of dominance for the right index finger on both
the left and right pre-central gyrus (Figure \ref{fig:dominance}).

\paragraph{Unique Representation Volume as a Metric for Dexterity.}

The ranking of unique representation volume aligns well with conventional ideas
about dexterity.
The participant examined in this study is right-handed,
and the right hand exhibits a larger area of unique representation compared to
the left across all examined regions (Table \ref{tab: unique-volume1}).
Similarly, the index fingers shows a greater unique representation compared to
the middle finger (Table \ref{tab: unique-volume1}).
Furthermore, the ipsilateral cortical and contralateral cerebellar regions of
unique representation of the right index finger (Figure \ref{fig:uniqueness})
may reflect a level of focused movement beyond what is afforded to the less
dexterous fingers.
In the motor cortex,
this could reflect some of the ipsilateral circuits described in
\cite{2019Berlot}.
Although this pattern holds only approximately for individual fingers (Table
\ref{tab: unique-volume2}),
we propose that this representation could serve as a valuable complement to
dominance maps.
As \cite{2019Huber} noted,
dominance maps can be misleading in cases of representational overlap.
In contrast, the volumetric partition-based representation obtained in our
analysis offers a novel and potentially insightful perspective on functional
mapping.

\begin{table}[!ht]
\centering
\begin{tabular}{|c|cc|cc|c|}
\hline
Region & \text{Right Hand} & \text{Left Hand} & \text{Index Fingers} & \text{Middle Fingers} \\
\hline
Sensorimotor Cortex    & $\mathbf{4879}$ & $4748$ & $\mathbf{6530}$ & $5562$ \\
Cerebellum \& Midbrain & $\mathbf{2918}$ & $2109$ & $\mathbf{3612}$ & $2233$ \\
Caudate \& Putamen     & $\mathbf{247}$  & $111$  & $\mathbf{232}$  & $132$ \\
\hline
\end{tabular}
\caption{Unique volume representation for hands and digit types,
showing the number of voxels in which either of the (right,
left) handed fingers are uniquely represented or either of the (index,
middle) fingers are uniquely represented.
}\label{tab: unique-volume1}
\end{table}
\vspace*{-10pt}

\begin{table}[!ht]
\centering
\begin{tabular}{|c|c|c|c|c|c|}
\hline
Region & Right Index & Left Index & Right Middle & Left Middle  & All\\
\hline
Sensorimotor Cortex    & $2969^{3}$ & $3561^{1}$ & $3112^{2}$ & $2440^{4}$ & 1212\\
Cerebellum \& Midbrain & $2180^{1}$ & $1432^{2}$ & $1143^{3}$ & $1090^{4}$ &  405 \\
Caudate \& Putamen     & $171^{1}$  & $61^{3}$   & $81^{2}$   & $51^{4}$   &
5 \\ \hline
\end{tabular}
\caption{Unique volume representation for individual fingers,
showing the number of voxels in which each finger is uniquely represented in
each region.
The superscripts indicate their ranking.
The final column gives the count of voxels in which all four fingers have
unique representations.
}
\label{tab: unique-volume2}
\vspace*{-10pt}
\end{table}

\paragraph{Comparison with Other Works.}

We also implemented two alternative approaches,
namely spatially independent pairwise tests,
akin to standard contrasts commonly used in fMRI analysis \citep[Ind.
Tests,][]{huettel2014},
and the spatially aware probabilistic threshold-free cluster enhancement method
\citep[pTFCE,][]{2019Spisak}.

\ul{To conserve space,
here we present graphical and tabular summaries only for our proposed lhMRPF.}
\ul{Additional figures and discussions of the results obtained by the different
methods, along with results for a different subject,
are presented in Section S.5 in the SM. }

Overall,
these results suggest that the lhMRPF produces spatially compact partition
clusters directly, without the need for post-hoc conversion of pairwise results
into partition estimates, as required by Ind. Tests and pTFCE.
Additionally, lhMRPF captures partition sparsity across space more effectively,
leading to fewer small,
potentially spurious clusters of high-cardinality partitions.

Pairwise comparison approaches often yield invalid partitions (Section
\ref{sec: review}), requiring post hoc adjustments, e.g.,
splitting or merging levels, introducing sparsity but reducing sensitivity.
In contrast, our MCMC-based method directly samples valid partition
configurations, which we can summarize via voxel-wise co-membership
probabilities.
Even when using a Bayesian FDR correction \citep{2004Muller} to compare
directly with pairwise methods,
only about 5\% of our lhMRPF partitions required adjustment,
compared to 40.0\% for Ind. Tests and 64.1\% for pTFCE.
This underscores the benefits of modeling partitions directly and may explain
our method's clearer separation of left- and right-hand finger representations.

On the scientific side,
the overlap in unique representations that we detected reflects the findings of
\cite{2014Besle} and \cite{2015Ejaz},
and the general hand structure in the somatosensory cortex is supported by the
findings in \cite{2020ONeill}.
We found the cerebellum to be organized into four regions of finger
representation, which is validated by the maps charted in \cite{2024Brouwer}
and \cite{2013vanderZwaag}.

Altogether, our results provide a highly detailed, spatially smooth,
and sparse cortical map of the fingertips during an auditory-motor association
task, confirming their known sensorimotor representations,
accounting for overlap,
while also revealing novel ipsilateral finger representations in the precentral
gyrus.
Additionally, we identified a potentially useful motor dexterity metric.

Conventional voxelwise fMRI analyses primarily identify regions that respond to a task or stimulus relative to baseline. 
Our objective is different. 
Conditional on the fingertip-specific effects estimated in the first-level analysis, 
we determine which fingertips produce distinguishable responses at each voxel and how these equality relationships vary across neighboring voxels. 
A voxel may therefore respond strongly to all four fingertips while having a one-cluster partition, 
indicating a shared motor response without fingertip-specific differentiation. 
Partitions with two or more clusters identify progressively more detailed forms of local selectivity, 
including hand-specific, digit-specific, and fingertip-unique representations. 
Thus, the proposed method characterizes the organization of responses within active and potentially active regions rather than producing an overall activation map.

\vspace*{-3ex}
\section{Simulation Study} \label{sec: simulation}
\vspace*{-1ex}

Here, we briefly report the performance of our method compared to Ind.
Tests and pTFCE in simulation settings.
\ul{To save space, as with real data analysis,
the graphical summaries presented here are restricted to our proposed method.
Additional details, results,
graphical and tabular summaries are presented in Section S.6 in the SM.}

\begin{figure}[!ht]
    \centering
    {\small (a) True Effects}
    \vskip 10pt
    \includegraphics[width=0.15\textwidth]{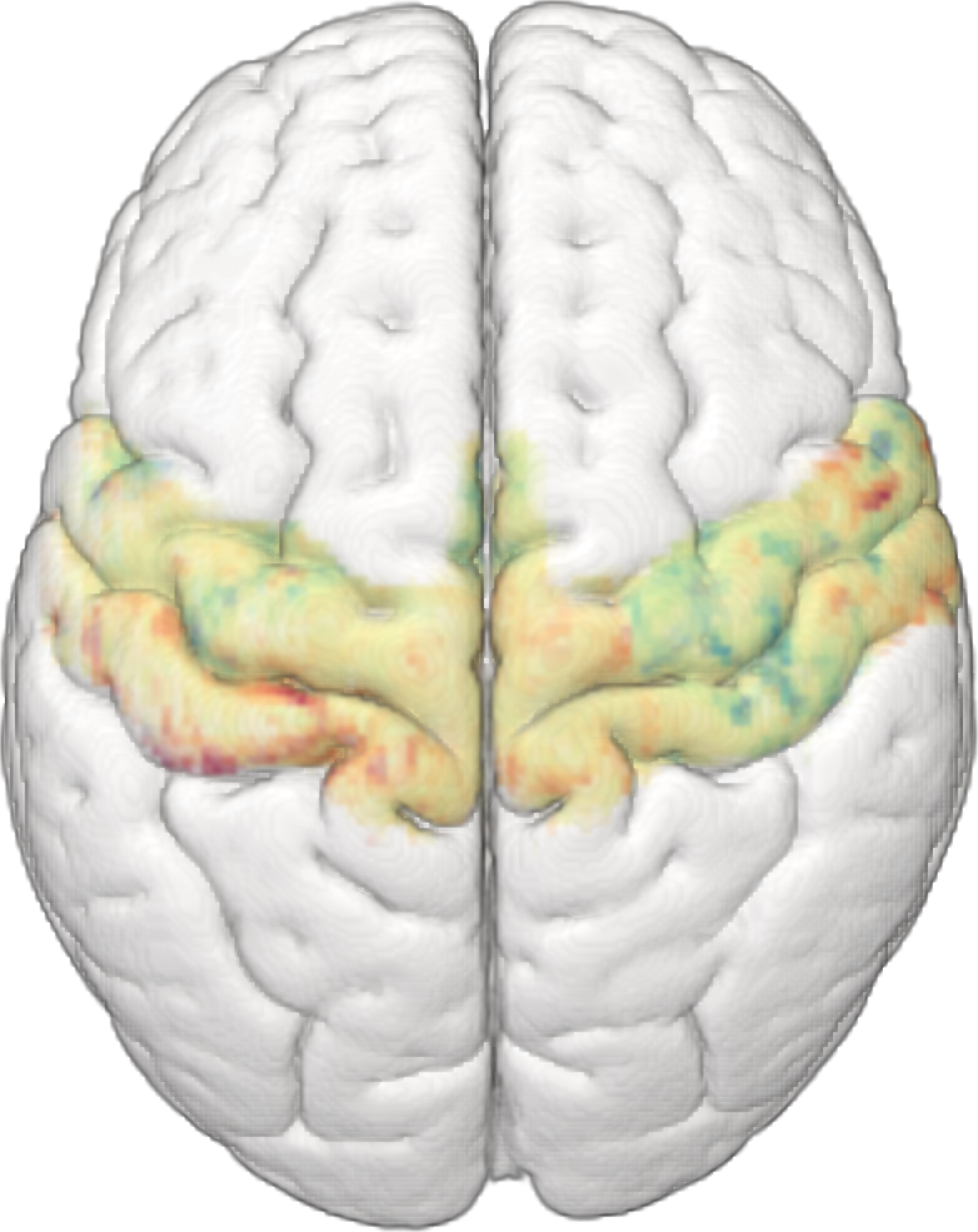}
    \hspace{15pt}
    \includegraphics[width=0.15\textwidth]{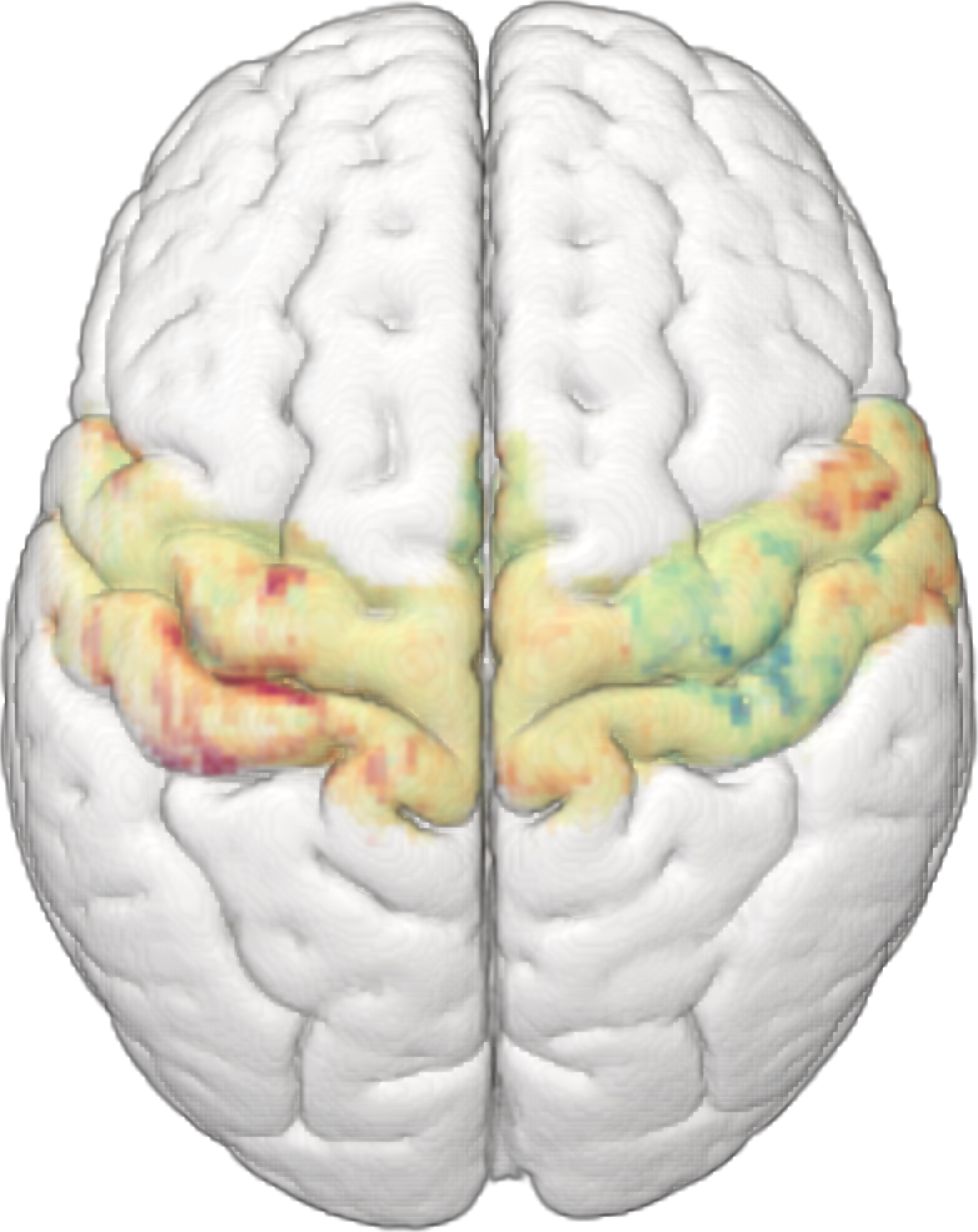}
    \hspace{15pt}
    \includegraphics[width=0.15\textwidth]{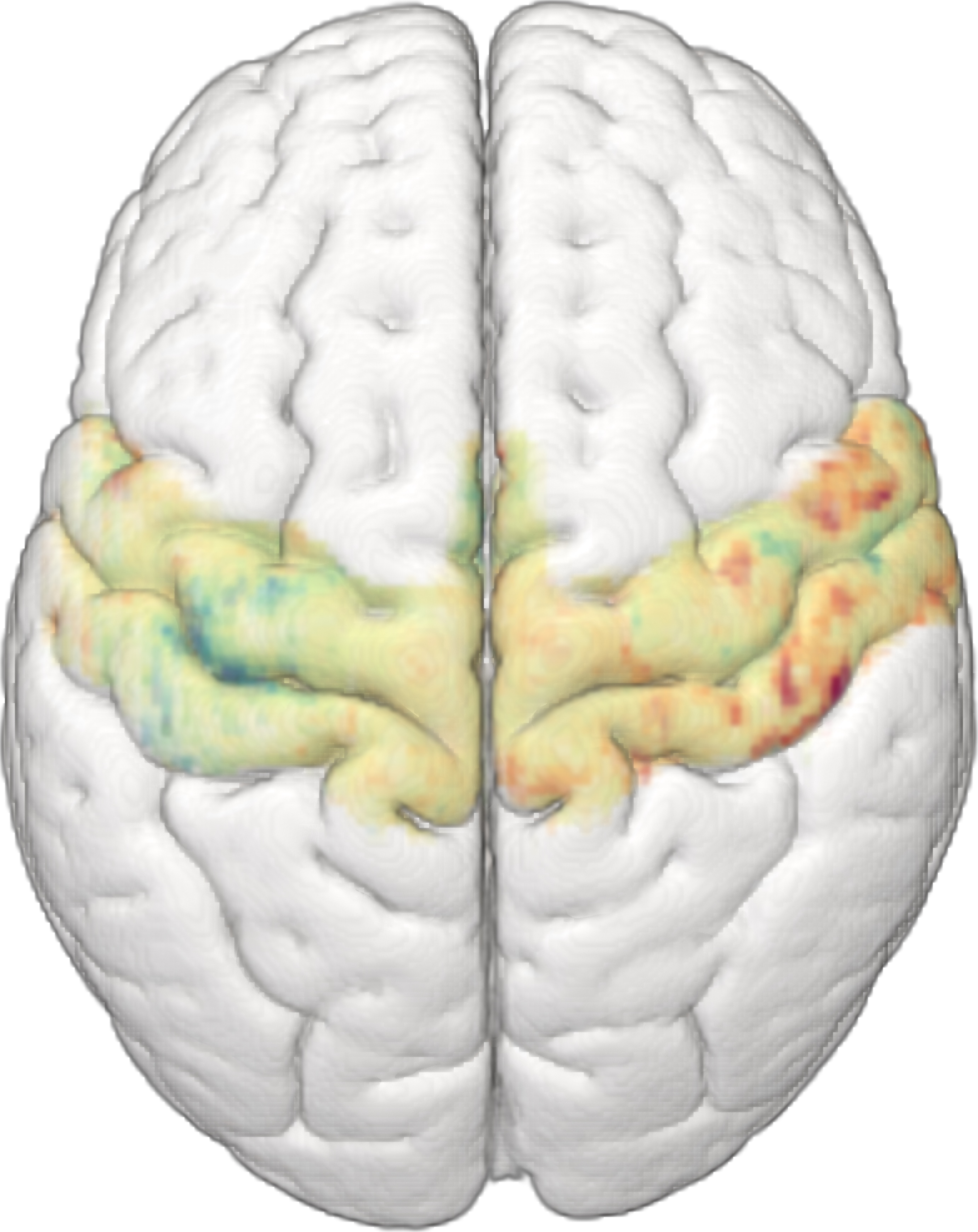}
    \hspace{15pt}
    \includegraphics[width=0.15\textwidth]{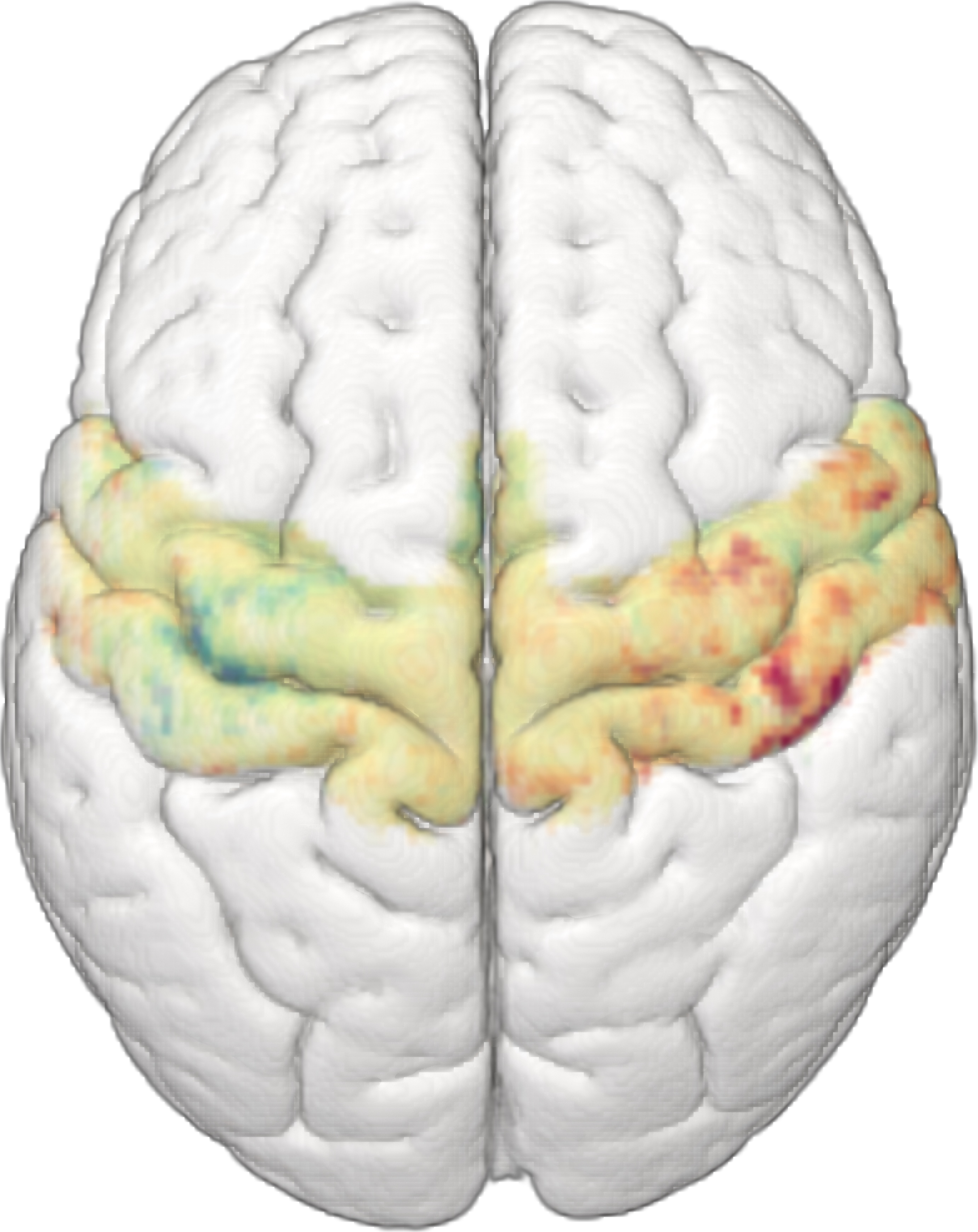}\\
    {\small (b) Recovered Effects}
    \vskip 10pt
    \includegraphics[width=0.15\textwidth]{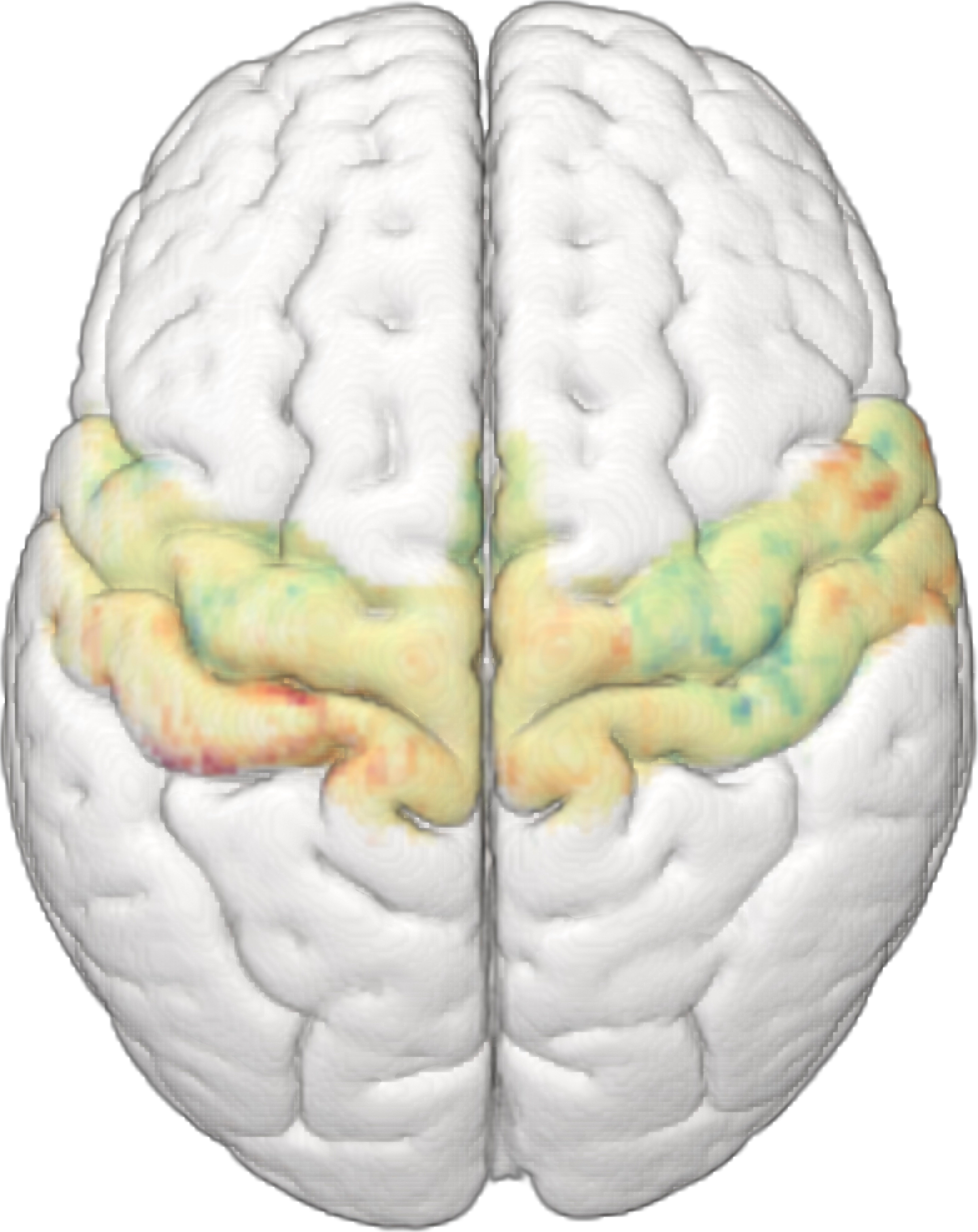}
    \hspace{15pt}
    \includegraphics[width=0.15\textwidth]{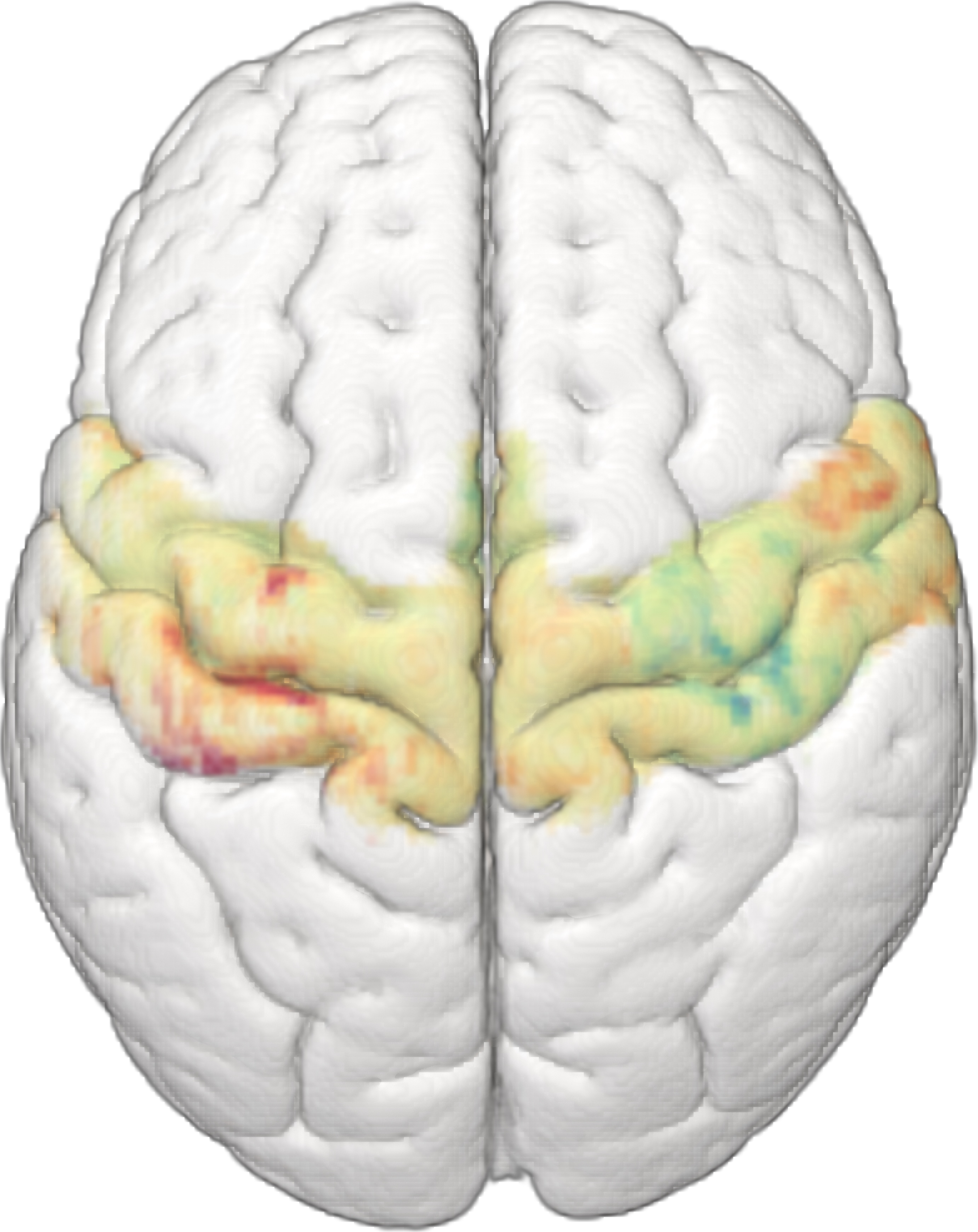}
    \hspace{15pt}
    \includegraphics[width=0.15\textwidth]{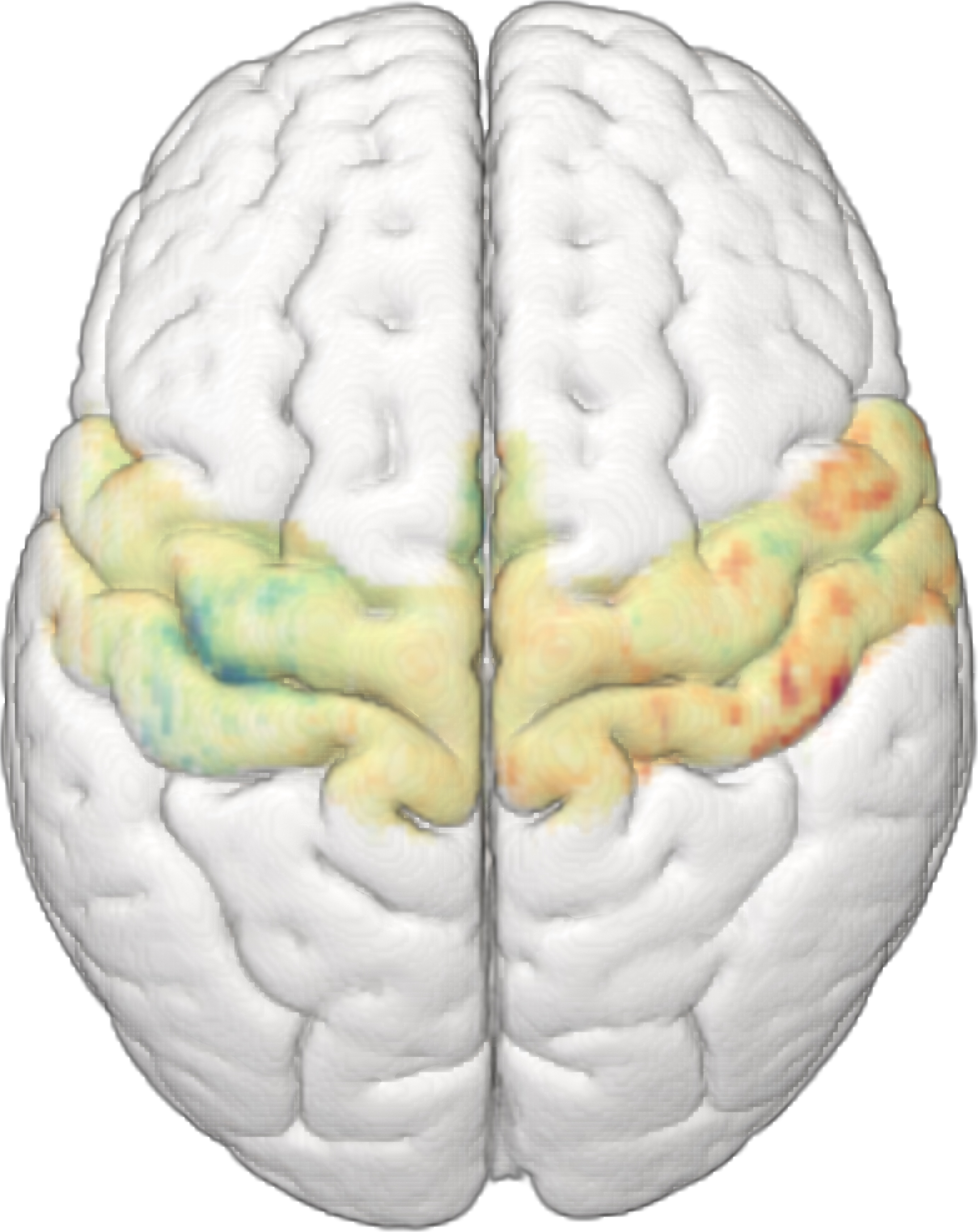}
    \hspace{15pt}
    \includegraphics[width=0.15\textwidth]{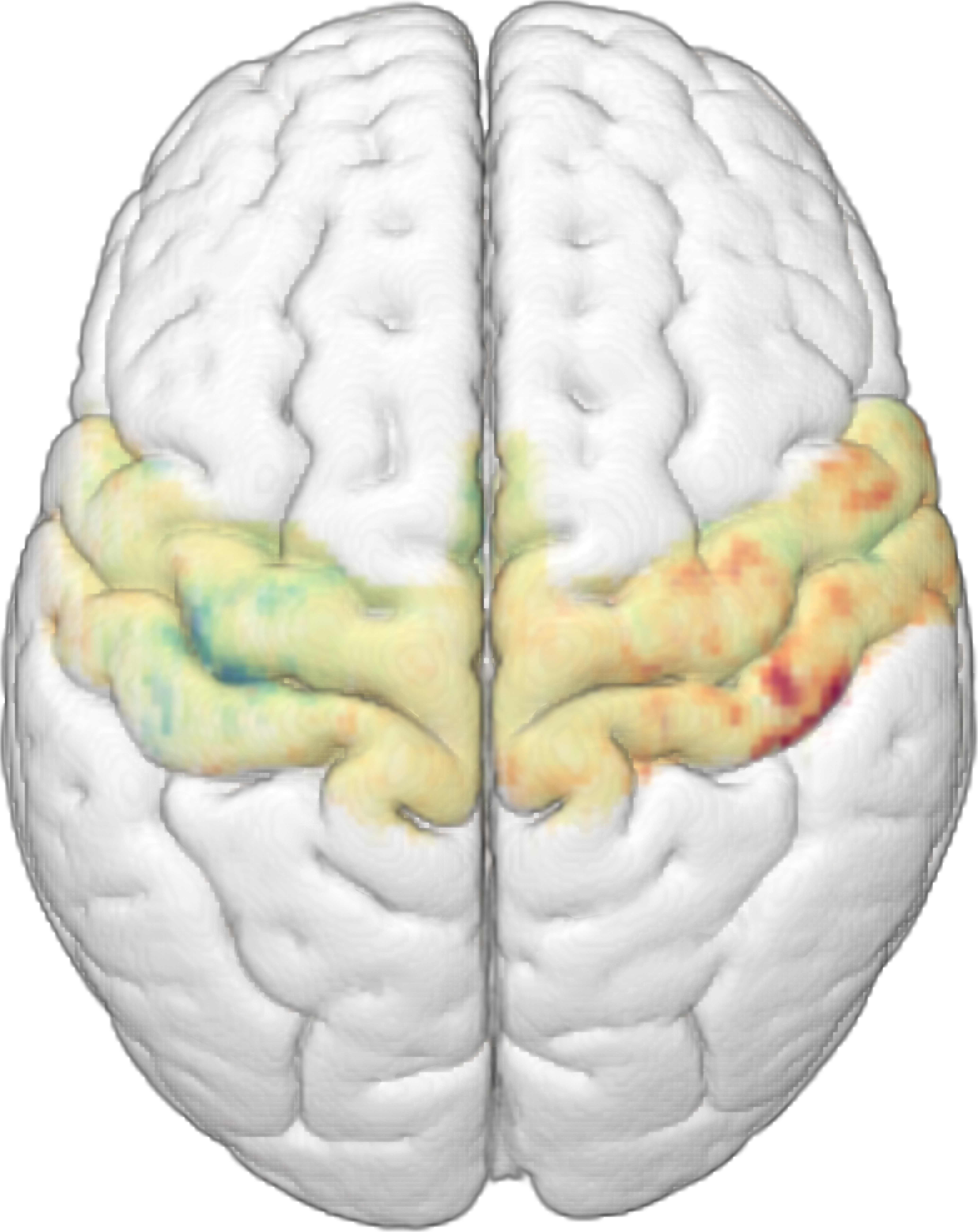}\\
    {\small (c) Recovery Errors}
    \vskip 10pt
    \hspace{20pt}
    \includegraphics[width=0.15\textwidth]{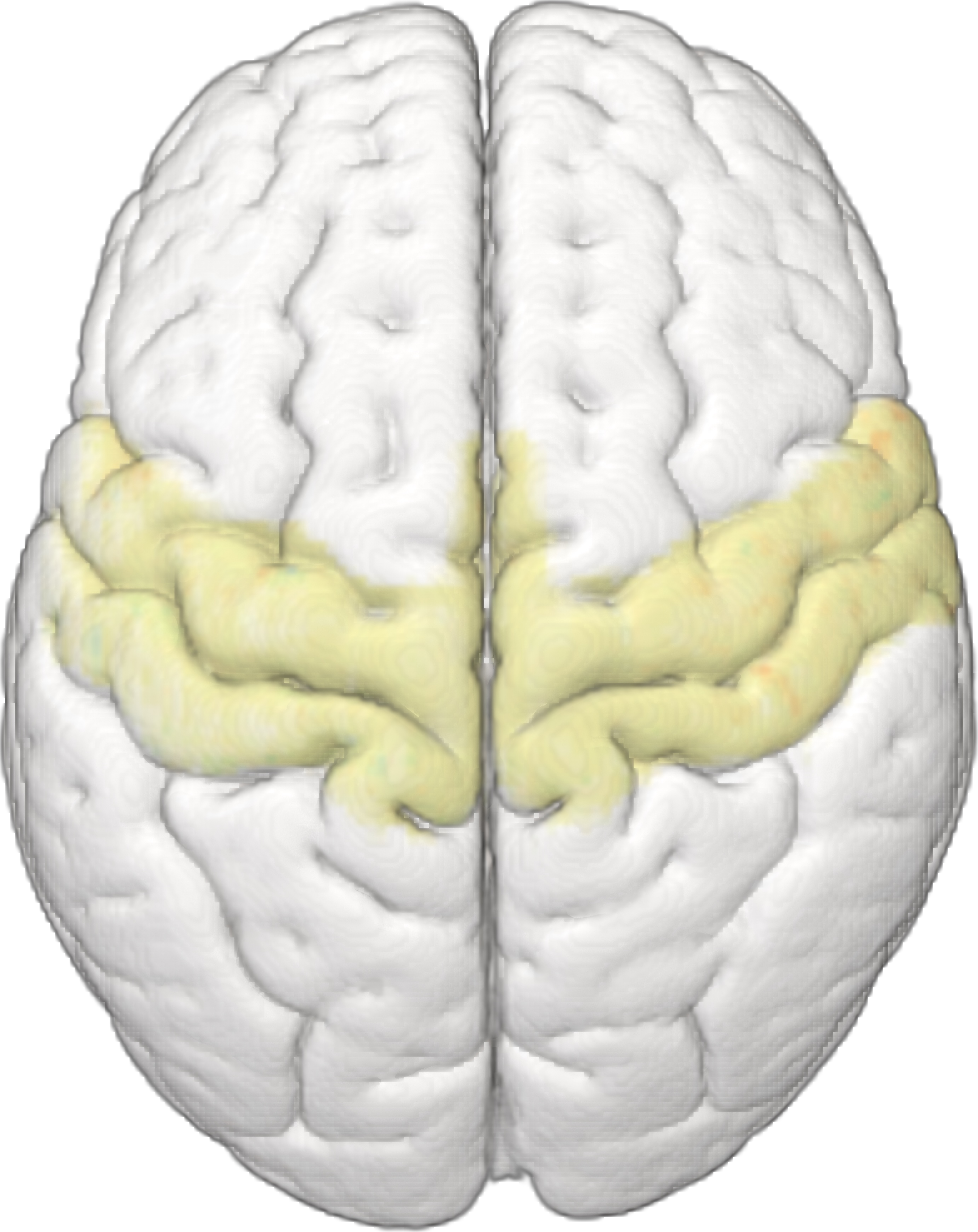}
    \hspace{15pt}
    \includegraphics[width=0.15\textwidth]{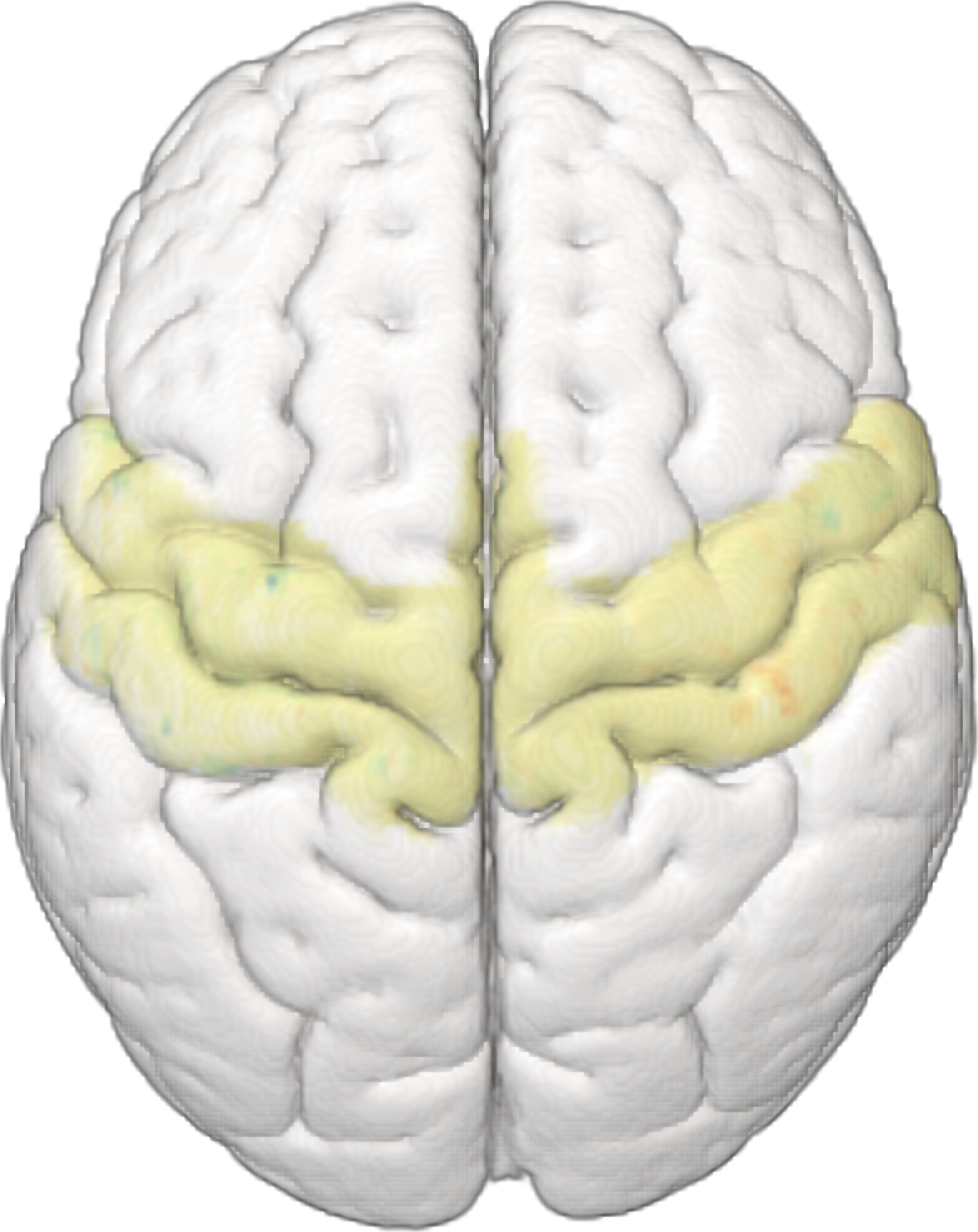}
    \hspace{15pt}
    \includegraphics[width=0.15\textwidth]{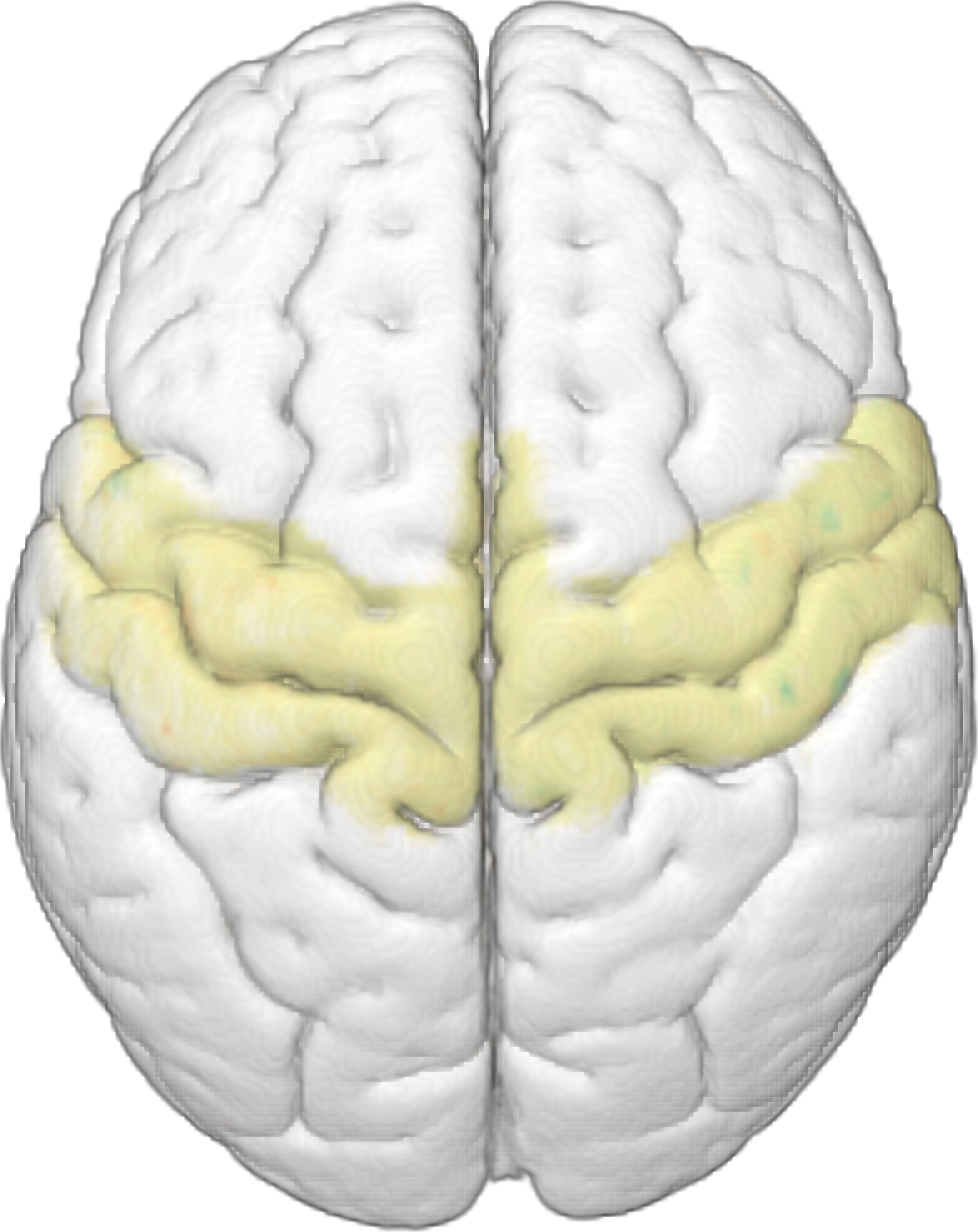}
    \hspace{15pt}
    \includegraphics[width=0.15\textwidth]{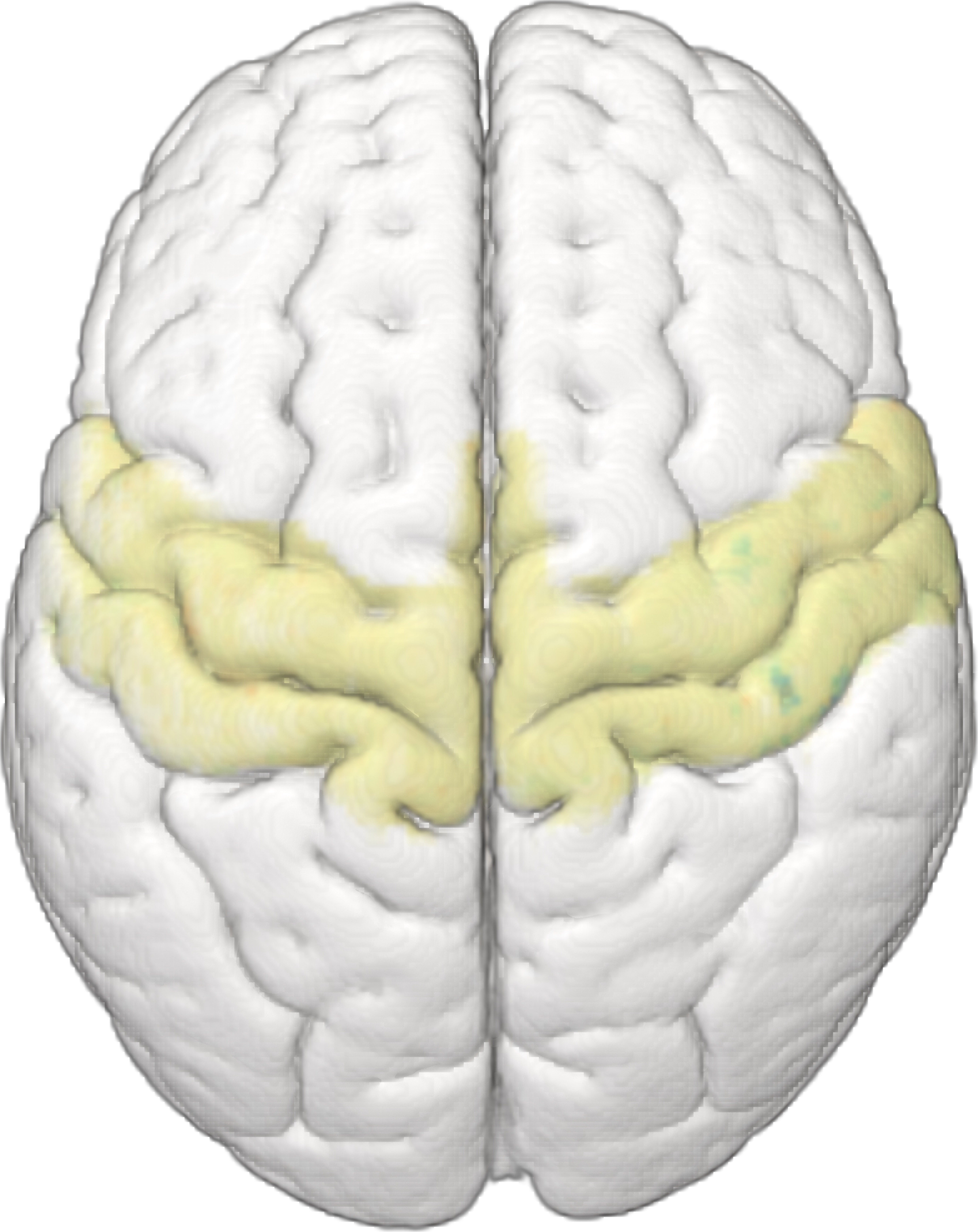}
    \newline
    \vspace{1em}
    \includegraphics[width=0.75\textwidth]{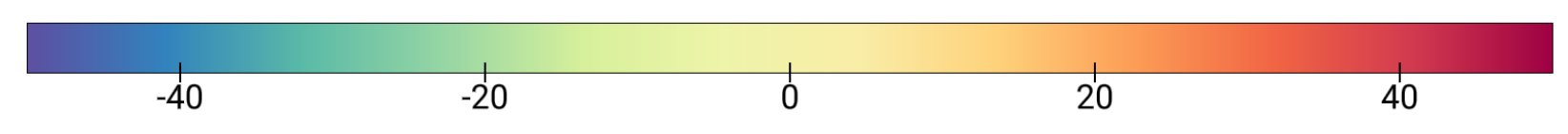}
    \vspace*{-15pt}
    \caption{
    Results for the simulation study:
    (a) True effects,
    (b) Effects recovered using our lhMRPF,
    (c) Recovery errors.
    }\label{fig:sim effects}
\vspace*{-5pt}
\end{figure}

In generating synthetic data,
we tried to closely mimic the real data application described in Section~\ref{sec: application}.
We used the final sampled MCMC state from the analysis in Section \ref{sec:
application} as the ``ground-truth" for $\partition_{v}$ and
$\bbeta^{\star}_{v}$, focusing only on the left and right motor cortices.
Then, $n=6$ synthetic images were drawn from $\mathcal{N}(\bbeta_{v}^{\star},
\widehat{\bSigma}_{v})$,
where $\widehat{\bSigma}_{v}$ were pulled from the results of the first-level
GLM analysis described in Section \ref{sec:VWA}.

The error in estimating the factor effects is measured by $\bbeta^{\star}_{v,
\text{true}} - \wh\bbeta^{\star}_{v}$
where $\wh\bbeta^{\star}_{v}$ is the posterior mean of $\bbeta^{\star}_{v}$
from our sample (Figure \ref{fig:sim effects}).
The error in recovering the partitions is calculated as the number of
incorrectly decided pairwise equalities/differences between the true and the
estimated partitions, i.e., the distance $\text{d}(\partition_{v,
\text{true}},\partition_{v})$ described in Section~\ref{subsec:similarity}
(visual summaries shown in Section S.6 in the SM).
Table \ref{tab: simulation-comparison} summarizes the partition recovery
results across the three methods.
Overall, the results show markedly improved recovery of the true partition
structure and effect values compared to alternative methods.

\begin{table}[!ht]
\centering
\begin{tabular}{|c|c|c|c|}
\hline
Method & \text{Partition Error} & \text{Partition Validity} & \text{Pairwise F1} \\
\hline
lhMRPF     & $\mathbf{1.88}$ & $\mathbf{0.75}$ & $\mathbf{0.96}$ \\
Ind. Tests & $3.14$          & $0.59$          & $0.81$ \\
pTFCE      & $2.70$          & $0.63$          & $0.43$ \\
\hline
\end{tabular}
\caption{Partition recovery metrics on simulated data for our lhMRPF method,
pTFCE, and spatially independent pairwise comparison.
The first column shows partition error as an average distance from the final
adjusted partition estimates at each voxel to the simulated ``ground-truth".
The second column shows the portion of estimated partitions that were invalid
and therefore required adjustment.
The final column shows the F1 scores for binary classification of two factor
levels as ``different" or at each voxel,
to provide a non-partition-based metric for the pairwise methods.
}\label{tab: simulation-comparison}
\end{table}
\vspace*{-10pt}

\vspace*{-1ex}
\section{Discussion} \label{sec: discussion}
\vspace*{-1ex}

\paragraph{Summary.}
In this article, we proposed a flexible hierarchical framework for evaluating
how a factor’s influence on a continuous outcome varies locally across a graph.
Our method employs a novel first-layer hidden Markov random partition field to
locally collapse factor levels into shared-effect states at each node,
promoting sparsity and discouraging abrupt spatial changes.
A second-layer novel conditional autoregressive model then induces smoothness
in the associated collapsed effects.
We induce parsimony through pairwise penalties in a graphical setting.
The resulting fine-grained influence maps can reveal spatial variation in both
the pattern and strength of factor effects.
We designed an efficient and scalable MCMC algorithm for posterior inference.
Applied to our motivating ultra-high-resolution fMRI data,
our method uncovered new insights into cortical representations of fingertips.

Current fMRI methods do not enable the joint hierarchical estimation of shared
and unique topographies of fMRI contrasts.
Our proposed method addresses this gap by providing a precision mapping of
individual functional topography.
Furthermore, while outside the scope of this article,
our framework offers potential clinical utility for identifying critical brain
areas that require preservation (e.g., during epilepsy or tumor resection).

The proposed method is not primarily motivated as an alternative procedure for selecting active fMRI voxels. 
Standard first-level analyses and spatial multiple-testing procedures are well suited to identifying responses relative to baseline. 
Our method instead addresses the subsequent question of how condition-specific effects are organized within the brain. 
In the fingertip study, 
it distinguishes voxels exhibiting a shared response to all movements from those exhibiting hand-specific, digit-specific, or fingertip-unique responses. 
These configurations cannot be fully characterized by separate activation maps or dominance maps, 
particularly when several fingertips have strong but statistically indistinguishable effects. 
Direct estimation of the local partitions also preserves the transitivity of equality relationships and provides posterior uncertainty for the inferred organization. 
The resulting maps therefore describe local functional selectivity and representational overlap, rather than activation alone.

\paragraph{Limitations and Extensions.}
Our method relies on first-level estimates;
embedding factor collapse directly into the initial additive regression could
eliminate this, though first-stage covariances offer a computationally
efficient way to capture voxel-wise heterogeneity.
Fully modeling covariances across thousands of voxels in a unified spatially
smooth framework is ongoing work.
We assumed normal errors,
with robust formulations for large deviations planned for future work.
Additionally, given multiple replicates per subject,
here we focused on efficient individual-level inference,
leaving a joint model to infer the population-level average for separate
exploration.
Finally, we also plan to extend our single-factor framework to accommodate
multiple factors.

\baselineskip=13.5pt

\vspace*{-3ex}
\section*{Supplementary Materials}
\vspace*{-1.5ex}

The Supplementary Materials provide
additional details on data preprocessing, first-level analysis,
and choice of hyperparameters;
details of the MCMC algorithm to sample from the posterior;
additional figures and results for the real data analysis;
additional details of the design of the simulation experiments and their
results;
etc.

\vspace*{-3.5ex}
\section*{Data and Code Availability Statement}
\vspace*{-1.5ex}

The raw neuroimaging data and derived data ($\wt\beta$'s and $\wt\Sigma$'s)
supporting the main paper's findings will be made publicly available at an Open
Science Framework (OSF,
https://osf.io) repository upon acceptance of this article (\ul{link excluded
now for blind review}).
Computer codes implementing our methods on the derived data are included as a
ZIP file in the SM.

\vspace*{-3.5ex}
\bibliographystyle{natbib}
\bibliography{bib}

\end{document}


\clearpage\pagebreak\newpage
\newgeometry{textheight=9in, textwidth=6.5in}
\pagestyle{fancy}
\fancyhf{}
\rhead{\bfseries\thepage}
\lhead{\bfseries Supplementary Materials}

\setcounter{equation}{0}
\setcounter{page}{1}
\setcounter{table}{1}
\setcounter{figure}{0}
\setcounter{section}{0}
\numberwithin{table}{section}
\renewcommand{\theequation}{S.\arabic{equation}}
\renewcommand{\thesubsection}{S.\arabic{section}.\arabic{subsection}}
\renewcommand{\thesection}{S.\arabic{section}}
\renewcommand{\thepage}{S.\arabic{page}}
\renewcommand{\thetable}{S.\arabic{table}}
\renewcommand{\thefigure}{S.\arabic{figure}}
\baselineskip=25pt

\begin{center}
{\LARGE{\bf
Bayesian Semiparametric\\\vskip -10pt
Hidden Markov Random Partition
Fields\\\vskip -10pt
for Factor Collapse on Graphs:\\
A Study of Cortical Mapping of Fingertips
}}
\end{center}
\vskip 20pt
\baselineskip 17pt

\vskip 10mm
{The Supplementary Materials provide additional details on data pre-processing, first-level analysis,
and choice of hyperparameters; details of the MCMC algorithm to sample from the posterior;
additional figures and results for the real data analysis; additional details of the design of the simulation experiments and their
results; etc. Computer programs implementing our methods are included in a separate ZIP file.}

\newpage

\section{Preprocessing Methods} \label{sup:preproc}

Results included in this manuscript come from preprocessing performed using
\emph{fMRIPrep} 22.1.1 (\cite{fmriprep1}; \cite{fmriprep2};
RRID:SCR\_016216), which is based on \emph{Nipype} 1.8.5 (\cite{nipype1};
RRID:SCR\_002502).

\paragraph{Preprocessing of B0 inhomogeneity mappings}
A total of 1 fieldmap was found available within the input BIDS structure for
this particular subject.
A \emph{B0}-nonuniformity map (or \emph{fieldmap}) was estimated based on two
(or more) echo-planar imaging (EPI) references with \texttt{topup}
(\cite{topup}; FSL 6.0.5.1:57b01774).
\paragraph{Anatomical data preprocessing}
A total of 1 T1-weighted (T1w) images were found within the input BIDS dataset.
The T1-weighted (T1w) image was corrected for intensity non-uniformity (INU)
with \texttt{N4BiasFieldCorrection} \citep{n4},
distributed with ANTs 2.3.3 \citep[RRID:SCR\_004757]{ants},
and used as T1w-reference throughout the workflow.
The T1w-reference was then skull-stripped with a \emph{Nipype} implementation
of the \texttt{antsBrainExtraction.sh} workflow (from ANTs),
using OASIS30ANTs as target template.
Brain tissue segmentation of cerebrospinal fluid (CSF),
white-matter (WM) and gray-matter (GM) was performed on the brain-extracted T1w
using \texttt{fast} \citep[FSL 6.0.5.1:57b01774,
RRID:SCR\_002823,]{2001Zhang}.
Brain surfaces were reconstructed using \texttt{recon-all}
\citep[FreeSurfer 7.2.0, RRID:SCR\_001847,]{fs_reconall},
and the brain mask estimated previously was refined with a custom variation of
the method to reconcile ANTs-derived and FreeSurfer-derived segmentations of
the cortical gray-matter of Mindboggle
\citep[RRID:SCR\_002438,]{mindboggle}.
Volume-based spatial normalization to one standard space (MNI152NLin2009cAsym)
was performed through nonlinear registration with \texttt{antsRegistration}
(ANTs 2.3.3), using brain-extracted versions of both T1w reference and the T1w
template.
The following template was selected for spatial normalization:
\emph{ICBM 152 Nonlinear Asymmetrical template version 2009c}
{[}\citep{mni152nlin2009casym}, RRID:SCR\_008796; TemplateFlow ID:
MNI152NLin2009cAsym{]}.
\paragraph{Functional data preprocessing}
For each of the 10 BOLD runs found per subject (across all tasks and sessions),
the following preprocessing was performed.
First, a reference volume and its skull-stripped version were generated by
aligning and averaging 1 single-band reference (SBRefs).
Head-motion parameters with respect to the BOLD reference (transformation
matrices, and six corresponding rotation and translation parameters) are
estimated before any spatiotemporal filtering using \texttt{mcflirt}
\citep[FSL 6.0.5.1:57b01774,]{mcflirt}.
BOLD runs were slice-time corrected to 0.97s (0.5 of slice acquisition range
0s-1.94s) using \texttt{3dTshift} from AFNI \citep[RRID:SCR\_005927]{afni}.
The BOLD time-series (including slice-timing correction when applied) were
resampled onto their original,
native space by applying the transforms to correct for head-motion.
These resampled BOLD time series will be referred to as \emph{preprocessed BOLD
in original space}, or just \emph{preprocessed BOLD}.
The BOLD reference was then co-registered to the T1w reference using
\texttt{bbregister} (FreeSurfer),
which implements boundary-based registration \citep{bbr}.
Co-registration was configured with six degrees of freedom.
First, a reference volume and its skull-stripped version were generated using a
custom methodology of \emph{fMRIPrep}.
Several confounding time series were calculated based on \emph{preprocessed
BOLD}: framewise displacement (FD), DVARS and three region-wise global signals.
FD was computed using two formulations following Power (absolute sum of
relative motions, \cite{power_fd_dvars}) and Jenkinson (relative root mean
square displacement between affines, \cite{mcflirt}).
FD and DVARS are calculated for each functional run,
both using their implementations in \emph{Nipype} \citep[following the
definitions by]{power_fd_dvars}.
The three global signals are extracted within the CSF, the WM,
and the whole-brain masks.
Additionally, a set of physiological regressors were extracted to allow for
component-based noise correction \citep[\emph{CompCor},]{compcor}.
Principal components are estimated after high-pass filtering the
\emph{preprocessed BOLD} time-series (using a discrete cosine filter with 128s
cut-off) for the two \emph{CompCor} variants:
temporal (tCompCor) and anatomical (aCompCor).
tCompCor components are then calculated from the top 2\% variable voxels within
the brain mask.
For aCompCor, three probabilistic masks (CSF, WM,
and combined CSF+WM) are generated in anatomical space.
The implementation differs from that of Behzadi et al.~in that instead of
eroding the masks by 2 pixels on BOLD space,
a mask of pixels that likely contain a volume fraction of GM is subtracted from
the aCompCor masks.
This mask is obtained by dilating a GM mask extracted from the FreeSurfer's
\emph{aseg} segmentation, and it ensures that components are not
extracted from voxels containing a minimal fraction of GM.
Finally, these masks are resampled into BOLD space and binarized by
thresholding at 0.99 (as in the original implementation).
Components are also calculated separately within the WM and CSF masks.
For each CompCor decomposition,
the \emph{k} components with the largest singular values are retained,
such that the retained components' time series are sufficient to explain 50
percent of variance across the nuisance mask (CSF, WM, combined, or temporal).
The remaining components are dropped from consideration.
The head-motion estimates calculated in the correction step were also placed
within the corresponding confounds file.
The confound time series derived from head motion estimates and global signals
were expanded with the inclusion of temporal derivatives and quadratic terms
for each \citep{confounds_satterthwaite_2013}.
Frames that exceeded a threshold of 0.5 mm FD or 1.5 standardized DVARS were
annotated as motion outliers.
Additional nuisance timeseries are calculated using principal components
analysis of the signal found within a thin band (\emph{crown}) of voxels around
the edge of the brain, as proposed by \citep{patriat_improved_2017}.
The BOLD time-series were resampled into standard space,
generating a \emph{preprocessed BOLD run in MNI152NLin2009cAsym space}.
First, a reference volume and its skull-stripped version were generated using a
custom methodology of \emph{fMRIPrep}.
The BOLD time-series were resampled onto the following surfaces (FreeSurfer
reconstruction nomenclature): \emph{fsnative}.
All resamplings can be performed with \emph{a single interpolation step} by
composing all the pertinent transformations (i.e.~head-motion transform
matrices, susceptibility distortion correction when available,
and co-registrations to anatomical and output spaces).
Gridded (volumetric) resamplings were performed using
\texttt{antsApplyTransforms} (ANTs),
configured with Lanczos interpolation to minimize the smoothing effects of
other kernels \citep{lanczos}.
Non-gridded (surface) resamplings were performed using \texttt{mri\_vol2surf}
(FreeSurfer).

Many internal operations of \emph{fMRIPrep} use \emph{Nilearn} 0.9.1
\citep[RRID:SCR\_001362]{nilearn},
mostly within the functional processing workflow.
For more details of the pipeline,
see \href{https://fmriprep.readthedocs.io/en/latest/workflows.html}{the section
corresponding to workflows in \emph{fMRIPrep}'s documentation}.

\hypertarget{copyright-waiver}{
\subsection{Copyright Waiver}\label{copyright-waiver}
}

The above boilerplate text was automatically generated by fMRIPrep with the
express intention that users should copy and paste this text into their
manuscripts \emph{unchanged}.
It is released under the
\href{https://creativecommons.org/publicdomain/zero/1.0/}{CC0} license.

\section{First-level Analysis} \label{sup:firstlevel}

A first stage of processing the raw BOLD time-series, each of length $t = 126$,
must be carried out to obtain the $\wt{\bbeta}_{v}$ and $\wt\bSigma_{v}$,
which will be used as the observables in our model.
We employ a generalized linear model (GLM) with an autoregressive AR(1) error
process.
The design matrix $\mathbf{X}$ was formed following standard fMRI literature as
a $t \times p_{all}$ matrix where $p_{all} = p + p_{nui}$,
a sum of the number of regressors of interest and the number of nuisance
regressors \citep{huettel2014}.
The primary regressors in this study comprised four response variables,
each corresponding to one of the four fingers participants used to indicate
their response, along with two control variables for the sound stimuli and
subsequent feedback stimuli to prevent confounding of these effects with the
participant's sound classification response.
Each regressor was encoded as a column vector reflecting the expected
hemodynamic BOLD response based on stimulus onset and duration.
Nuisance regressors included seven rigid body motion parameters (framewise
displacement, three translations, and three rotations),
seven scanner drift basis vectors, and a constant intercept.
Altogether, the design matrix consisted of 21 columns.
For further discussion on the construction of an fMRI design matrix,
see Chapters 8 and 9 of \citet{huettel2014} and the ``GLM:
First level analysis" section of the online documentation for \texttt{nilearn};
\citet{nilearn}.

The AR(1) error process was modeled by a $t \times t$ covariance matrix for
each node, namely $\bW_{v}$,
with a banded structure that summarizes the autocorrelation of successive
errors.
This let $\wt{\bbeta}_{v} = (\bX\trans \bW \bX)^{-1} \bX\trans \bW_{v} \by$.
We estimated the $\bW_{v}$'s by iterative pre-whitening of the error covariance
matrix according to the Cochrane-Orcutt procedure \citep{1949Cochrane},
for which a single iteration is considered sufficient for an AR(1) process
\citep{2001Woolrich,2023Parlak}.
These estimates are then used in the proposed lhMRPF likelihood.
Given $p$ parameters of interest,
this means we have to store $2p + p (p - 1) / 2$ values per voxel,
which number in the tens of thousands just for the regions of interest we
identify.

\section{Hyperparameter Selection: Details} \label{sec:sm hyper}

\begin{figure}
    \centering
    \includegraphics[width=0.9\textwidth,trim=0cm 3cm 0cm
    3cm,clip=true]{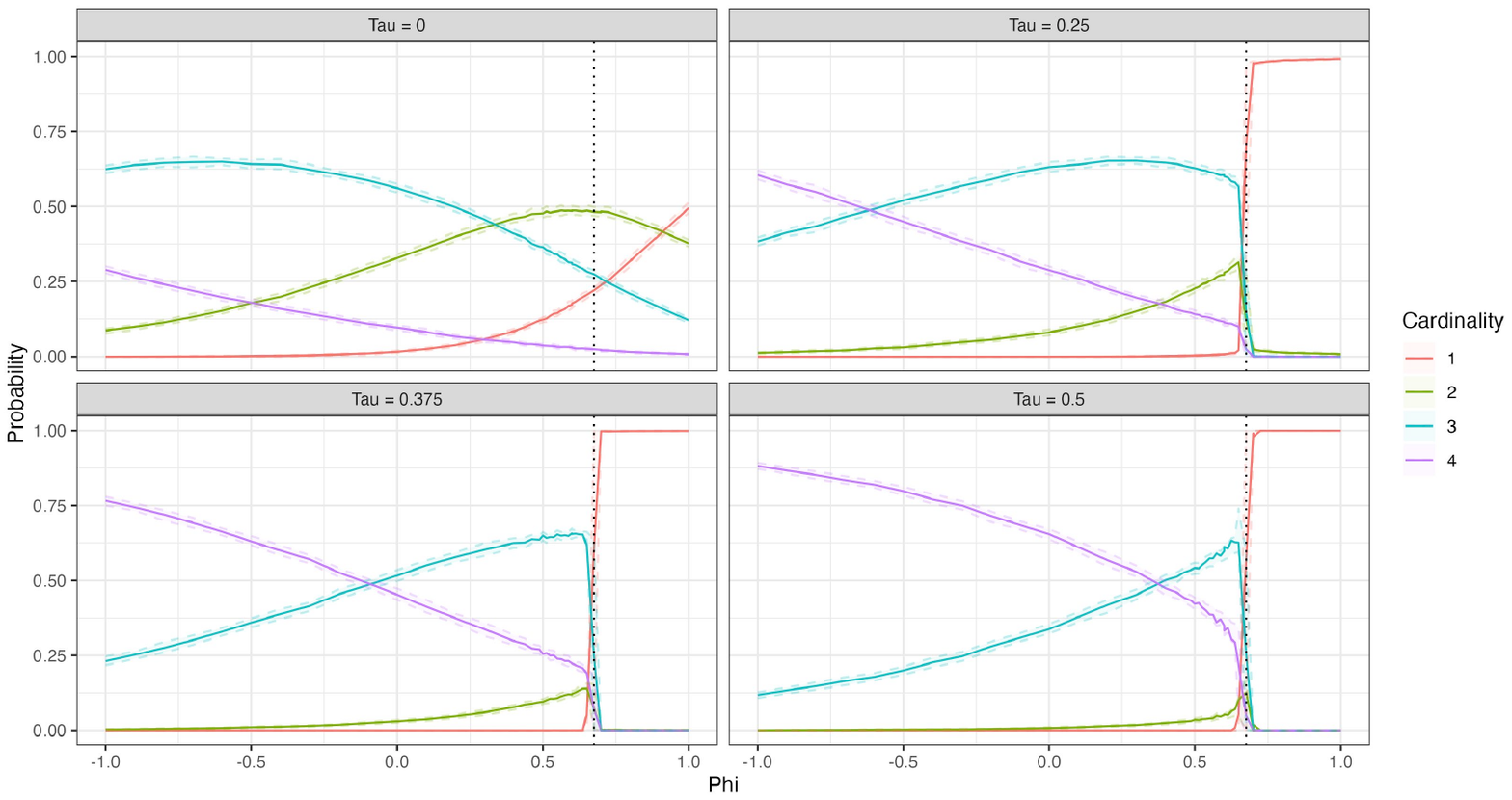}

    \caption{
    Phase transition in the expected distribution of the prior cardinality
    across configurations of $\phi$ and $\tau$.
    Each curve represents the prior probability of observing partitions of
    sizes $\abs{\X_{v}^{\star}}=\abs{\bbeta_{v}^{\star}}=q_{v}$ taking values
    $1,\dots,4$ for the given $(\phi,
    \tau)$ configuration with $\abs{\X}=p=4$ factor levels.
    The vertical dotted line shows our estimate of the critical value $\phi =
    0.675$, and the lower-left panel shows the distribution across $\phi$ at
    the second critical value $\tau = 0.375$.
    }
    \label{fig:phase-card}
\end{figure}

Inferring the hyperparameters of hMRF models, including via MCMC,
often proves prohibitively costly due to the high computational cost arising
from the normalizing constant of the MRF prior,
a product over all possible state spaces \citep{2021Izenman}.
For our model, this constant involves a sum of $\abs{\mathcal{P}(\X)}^{|\V|}$
terms, where $\V$ has tens of thousands of nodes.
We thus fixed our hyperparameters for the MRPF prior at a configuration that
induces a prior with the broadest coverage over the sample space
\citep{2015Moores}.
We used a grid search over 270 configurations for $\phi$ and $\tau$ and
simulated samples from the prior for a three-dimensional $8 \times 8 \times 8$
lattice graph with periodic boundaries where edges wrap around the graph such
that every node has six neighbors.
Such periodic boundary conditions have been traditionally widely used to
approximate MRF properties on graphs with infinite span \citep{1987Vinals}.
Here, we use it to approximate a large lattice,
such as the one that underlies our three-dimensional imaging data analyzed in
Section 6 in the main paper.
The configuration that induced the highest variance in the mean of the
$\lambda_{1}$ and $\lambda_{2}$ scores across our grid was $\phi = 0.675,
\tau = 0.375$.
The effect this configuration has on the prior mean of the $q_{v}$s across the
graph is shown in Figure \ref{fig:phase-card}.
These are then kept fixed for the rest of our analysis.
We select $a_{\sigma_{\beta}^{2}} = 0.5$ and $b_{\sigma_{\beta}^{2}} = 0.5$ to
induce a weakly informative prior on $\sigma^{2}_{\beta}$,
though the large amount of data from several thousands of nodes really drives
the posterior through the likelihood.

\newpage
\section{Posterior Inference: Details} \label{sec:sm sampling}

We rely on samples drawn from the posterior using an MCMC algorithm for
finite-sample inference.
Specifically, we adapt the single-site Metropolis-Hastings (M-H) algorithm
\citep{1984Geman},
combined with ``chromatic" \citep{2011Gonzalez} or ``conclique"
\citep{2020Kaplan} sampling scheme,
which allows the node-specific variables to be updated in parallel,
accelerating convergence.
Additionally, we adapt the Hamming ball sampler \citep{2017Titsias} with
local likelihood-informed moves \citep{2019Zanella} to efficiently explore
the posterior distribution of the partitions.

Since the cluster sizes $\abs{\partition_{v}} = \abs{\bbeta_{v}^{\star}} =
q_{v}$'s vary across $v$, it makes sampling from the posterior challenging.
To circumvent this issue,
we append each $\bbeta_{v}^{\star}$ with a $\bbeta^{\circ}_{v} =
(\bbeta^{\circ}_{v,1},\dots,\bbeta^{\circ}_{v,p-q_{v}})\trans$ which collects
the atoms for the $p - q_{v}$
empty clusters not associated with any factor level at node $v$
and is assigned a conditional prior $(\bbeta_{v}^{\circ} \mid \bP_{v}) \sim
\MVN_{p - q_{v}}(\beta^{0}_{v} \bone_{p - q_{v}},
\sigma_{\beta}^{2} \bI_{p - q_{v}})$ independent from $(\bbeta_{v}^{\star} \mid
\bP_{v})$.
This results in a convenient fixed-length-$p$ parametrization
$\bbeta^{\dagger}_{v} = (\bbeta_{v}\startrans,
\bbeta_{v}\circtrans)\trans$ with all different elements and a well-defined
multivariate normal conditional prior distribution $p(\bbeta^{\dagger}_{v} \mid
\bP, \bbeta_{-v}^{\dagger}, \sigma^{2}_{\beta}, \beta_{v}^{0})$
under any configuration of $\bP_{v}$.
Specifically,
\vspace{-4ex}\\
\bse
& \left(\bbeta^{\dagger}_{v} \mid \bP, \bbeta_{-v}^{\dagger},
\sigma^{2}_{\beta}, \beta_{v}^{0}\right)
\sim \MVN_{p}\left(
\mathbf{a}_{v,0}, \;
\bQ_{v,0}
\right),\\
& \text{where}~~\mathbf{a}_{v,0} =
\begin{pmatrix}
\abs{\N_{v}}^{-1}\sum_{u \in \N_{v}} \bP_{v}^{\trans}\bP_{u}\bbeta^{\star}_{u}
\\ \beta^{0}_{v}\mathbf{1}_{p - q}
\end{pmatrix}
~\text{and}~
\bQ_{v,0}^{-1} =
\begin{bmatrix}
 \abs{\N_{v}}\sigma_{\beta}^{-2}\bI_{q_{v}} & \mathbf{0} \\
 \mathbf{0} & \sigma^{-2}_{\beta} \bI_{p - q_{v}}
\end{bmatrix}.
\ese
\vspace{-3ex}

Next, we introduce some additional useful variables and notations to be used in
describing the sampler shortly.
We define
$\bpartlitetilde_{v}=(\partlitetilde_{v,1},\dots,\partlitetilde_{v,p})$ to be a
fixed-length-$p$ vector
whose the $j\th$ entry $\partlitetilde_{v,j} = \partlite_{v}(x_{j})$
denotes the cluster label for the $j\th$ level $x_{j}$ in $\X$.
Note that $\bpartlitetilde_{v}$ can be obtained as $\bpartlitetilde_{v} =
\bP_{v} \bpartlite_{v}$,
where we recall that $\bpartlite_{v} =
(\partlite_{v,1},\dots,\partlite_{v,q_{v}})\trans$ represents the
length-$q_{v}$ vector of unique cluster labels in $\partition_{v}$ at node $v$.
We denote the partition induced by the allocation vector $\wt\bx_{v}^{\star}$
by $\bP(\wt\bx_{v}^{\star})$.
The Hamming ball of radius $r$ about $\bpartlitetilde_{v}$,
denoted $\mathcal{H}_{r}(\bpartlitetilde_{v})$, consists of all vectors
with a Hamming distance of at most $r$ from $\bpartlitetilde_{v}$,
i.e., the set of cluster label vectors that have at most $r$ entries that are
not identical to their corresponding entries in $\bpartlitetilde_{v}$.
Finally, in what follows,
$\bzeta$ denotes a generic variable that collects the data as well as all
parameters of the model that are not explicitly mentioned.

The MCMC sampler cycles through the following steps.
We apply these steps in parallel over sets of conditionally independent nodes,
i.e., sets of nodes such that no two nodes in the same set are neighbors
\citep{2011Gonzalez,2020Kaplan},
significantly reducing the computation time.

\begin{itemize}[leftmargin=*,itemsep=0em]
\item {\bf Updating the factor effects:}
Under the specified model and prior described above,
the expanded effect parameters $\bbeta^{\dagger}_{v}$'s conveniently admit
closed-form multivariate normal posterior full conditionals and are thus
updated by straightforwardly sampling from these full conditionals as
\vspace{-6ex}\\
\bse
& \left(\bbeta^{\dagger}_{v} \mid \bzeta \right)
\sim \MVN_{p}\left[
\bOmega_{v} \left(\bQ_{v,0}^{-1} \mathbf{a}_{v,0}
+ \bQ_{v,n}^{-1} \mathbf{a}_{v,n}\right), \;
\bOmega_{v}
\right],\\
& \text{where}~~\mathbf{a}_{v,n} =
\begin{pmatrix}
(\bP_{v}^{\trans} \bP_{v})^{-1} \bP_{v}^{\trans} \left( \sum_{r=1}^{n}
\wt\bSigma_{v,r}^{-1} \right)^{-1} \left( \sum_{r=1}^{n} \wt\bSigma_{v,r}^{-1}
\wt{\bbeta}_{v,r} \right)\\
\bzero
\end{pmatrix}, \\
& \bQ_{v,n}^{-1} = \begin{bmatrix} \bP_{v}^{\trans}
\left(\sum_{r=1}^{n} \wt\bSigma_{v,r}^{-1}\right)
\bP_{v} & \bzero \\
\bzero & \bzero
\end{bmatrix},
~\text{and}~\bOmega_{v} = \left(\bQ_{v,0}^{-1} + \bQ_{v,n}^{-1} \right)^{-1}.
\ese
\vspace{-4ex}

\item {\bf Updating the partitions:}
Updating the random partitions $\partition_{v}$ across nodes $v \in \V$,
which are mapped by the cluster allocation variables $\wt\bx_{v}^{\star}$,
poses another significant challenge.
The number of possible configurations for $\wt\bx_{v}^{\star}$ at each node is
$p^{p}$ (see Section 3 in the main paper),
making a full exploration of this space computationally infeasible.
For example, our cortical fingertip mapping application has $4^4 = 256$
possible configurations per node.

We address this challenge by adapting the Hamming ball sampler
\citep{2017Titsias}, which updates a manageable,
randomly chosen subset of variables within a restricted Hamming ball radius
rather than all at once.
This improves mixing and computational efficiency,
making it well-suited for structured models with large discrete state spaces.
In our lhMRPF models,
it enhances mixing and reduces computational burden by avoiding a full scan of
all possible $\wt\bx_{v}^{\star}$ values at each iteration.

Furthermore,
we incorporate locally informed moves \citep{2019Zanella} in proposing a
new candidate partition within our Hamming ball sampler.
Instead of proposing moves that are uniformly distributed on the space of
target distribution,
locally informed moves adjust the proposal probabilities using gradient-like
information, increasing the likelihood of selecting beneficial moves while
maintaining a detailed balance, which facilitates convergence,
especially in high-dimensional discrete spaces.
In adapting this idea to our proposed lhMRPF models,
we leverage the likelihood information from the Hamming ball neighborhood
around the currently instantiated partition $\bP(\wt\bx_{v}^{\star})$.

Let $\bz_{v}$ denote an intermediate transient cluster allocation state.
Instead of taking an exploration step uniformly at random as $\bz_{v} \sim
\hbox{Unif}\{\mathcal{H}_{r}(\bpartlitetilde_{v})\}$ \citep[as originally
proposed in]{2017Titsias},
we employ a locally-balanced likelihood-informed proposal following the
approach of \cite{2019Zanella}
as
\vspace{-4ex}\\
\bse
\bz_{v} \sim \hbox{Multi}_{\bz_{v} \in \mathcal{H}_{r}(\bpartlitetilde_{v})}
\left(
g\left[
p\{\bbeta^{\dagger}_{v} \mid \bP_{v} = \bP(\bz_{v}), \bzeta\} \;
\prod_{r=1}^{n} p(\wt{\bbeta}_{v,r} \mid \bbeta_{v}, \wt\bSigma_{v,r})
\right]
\right),
\ese
\vspace{-4ex}\\
where
$g(x) = \sqrt{x}$ is the chosen balancing function.
This allows us to guide the perturbation phase of the Hamming ball sampler to
regions of high likelihood of the emission distribution.
Then,
the new allocation vector $\bpartlitetilde_{v, \text{new}}$
is sampled according to its full conditional distribution restricted to only
the vectors in $\mathcal{H}_{r}(\bz_{v})$ as
\vspace{-4ex}\\
\bse
\bpartlitetilde_{v, \text{new}}
\sim \hbox{Multi}_{
\bpartlitetilde_{v, \text{new}}
\in \mathcal{H}_{r}(\bz_{v})}
\left[
\Pr(
\bpartlitetilde_{v, \text{new}}
\mid \bpartlitetilde_{-v}, \phi, \tau)
p\{\bbeta^{\dagger}_{v} \mid \bP_{v} = \bP(
\bpartlitetilde_{v, \text{new}}
), \, \bzeta\} \;
\prod_{r=1}^{n} p(\wt{\bbeta}_{v,r} \mid \bbeta_{v}, \wt\bSigma_{v,r})
\right].
\ese
\vspace{-4ex}

\item {\bf Updating the factor smoothness parameter:}
Finally, the smoothness parameter $\sigma_{\beta}^{2}$ of the
$\beta_{v}^{\star}$'s is updated by sampling from its closed form full
conditional
\vspace{-4ex}\\
\bse
& \left(\sigma_{\beta}^{2} \mid \bzeta \right) \sim \IG\left(
a_{\sigma_{\beta}^{2}} + \frac{p|\V|-1}{2}, \,
b_{\sigma_{\beta}^{2}} + \frac{
\sum_{v \in \V} r^{2}_{v} +
\sum_{<v, u> \in \E} r^{2}_{v,u}
}{2}
\right),
\\
&\text{where}~~
r^{2}_{v} = (\bbeta_{v}^{\circ} - \beta^{0}_{v}\mathbf{1})^{\trans}
(\bbeta_{v}^{\circ} - \beta^{0}_{v}\mathbf{1}),~~\text{and}~~r^{2}_{v,u} =
(\bP_{v} \bbeta_{v}^{\star} - \bP_{u} \bbeta^{\star}_{u})^{\trans} (\bP_{v}
\bbeta_{v}^{\star} - \bP_{u} \bbeta^{\star}_{u}).
\ese
\vspace{-4ex}\\
Here, $<v,u>$ indicates the presence of an edge between the nodes $v$ and $u$.
\end{itemize}

To derive the full-conditional for $\sigma_{\beta}^{2}$,
define $n_{q} = \sum_{v\in\V} q_{v}$ and the concatenated vectors
\bse
\bB^{\star}
=
\begin{bmatrix}
\bbeta_1^{\star}\\
\vdots\\
\bbeta_{|\V|}^{\star}
\end{bmatrix},
\qquad
\bB^\circ
=
\begin{bmatrix}
\bbeta_1^\circ\\
\vdots\\
\bbeta_{|\V|}^\circ
\end{bmatrix}.
\ese
For the ICAR component,
define the block precision matrix $\bQ^{\star} = \bD-\bW$,
where the $(v,u)\th$ blocks of $\bQ^{\star}$ and $\bW$ are given respectively by
\bse
\bQ^{\star}_{v,u}
=
\begin{cases}
\bD_{v}, & v=u,\\
-\bW_{v,u}, & v\neq u,
\end{cases}
\quad\quad\quad
\text{with}
\quad\quad\quad
\bW_{v,u}
=
\begin{cases}
\bP_{v}^\top \bP_{u}, & v\sim_\G u,\\
\mathbf 0, & \text{otherwise}.
\end{cases}
\ese

Here, $\bQ^{\star}_{v,u}$ and
$\bW_{v,u}$ are both $q_{v} \times q_{u}$.
Since $\bD = \text{diag}(\bW \mathbf{1}_{n_{q}})$,
$\bQ^{\star} = \bD - \bW$ is guaranteed to satisfy the constraint $\bQ^{\star}
\mathbf{1}_{n_{q}} = \mathbf{0}_{n_{q}}$.
Also, $\bD_{v}$ is simply $\text{diag}([\bW_{v,1} , \dots ,
\bW_{v,|\V|}] \mathbf{1}_{n_{q}})$, where the block matrix $[\bW_{v,1} ,
\dots , \bW_{v,|\V|}] $ is of dimension $q_{v} \times
(q_{1}+\dots+q_{\abs{\V}}) = q_{v} \times n_{q}$,
and hence $\bD_{v}$ is $q_{v} \times q_{v}$.
Also, note that
\bse
\bW =
\begin{bmatrix}
\bW_{1,1} & \cdots & \bW_{1,\abs{\V}} \\
\vdots & \ddots & \vdots \\
\bW_{\abs{\V},1} & \cdots & \bW_{\abs{\V},\abs{\V}} \\
\end{bmatrix},
\quad
\bD =
\begin{bmatrix}
\bD_{1} & \cdots & \mathbf{0} \\
\vdots & \ddots & \vdots \\
\mathbf{0} & \cdots & \bD_{|\V|} \\
\end{bmatrix},
\ese
with the $\bW_{v,u}$'s being $\bzero$ whenever $v \not\sim_{\G} u$.

Then the intrinsic GMRF density is
\[
\pi(\bB^{\star}\mid \sigma_{\beta}^{2})
\propto
(\sigma_{\beta}^{2})^{-(n_{q}-1)/2}
\exp\left(
-\frac{1}{2\sigma_{\beta}^{2}}
\bB^{\star\top}\bQ^{\star}\bB^{\star}
\right).
\]

Now,
\bse
\bB^{\star\top}\bQ^{\star}\bB^{\star}
&=
\sum_{v\in\V}
\bbeta_{v}^{\star\top}\bD_{v}\bbeta_{v}^{\star}
-
\sum_{\substack{v,u\in\V\\v\neq u}}
\bbeta_{v}^{\star\top}\bW_{v,u}\bbeta_{u}^{\star}.
\ese

Since $\bW_{v,u}=\bP_{v}^\top\bP_{u}$ for neighboring vertices,
\begin{align*}\bB^{\star\top}\bQ^{\star}\bB^{\star}
&=
\sum_{v\sim_\G u}
\left(
\bbeta_{v}^{\star\top}\bP_{v}^\top\bP_{v}\bbeta_{v}^{\star}
+
\bbeta_{u}^{\star\top}\bP_{u}^\top\bP_{u}\bbeta_{u}^{\star}
-
2\bbeta_{v}^{\star\top}\bP_{v}^\top\bP_{u}\bbeta_{u}^{\star}
\right) \\
&=
\sum_{v\sim_\G u}
(\bP_{v}\bbeta_{v}^{\star}-\bP_{u}\bbeta_{u}^{\star})^\top
(\bP_{v}\bbeta_{v}^{\star}-\bP_{u}\bbeta_{u}^{\star})
=
\sum_{v\sim_\G u} r_{v,u}^2.\end{align*}

Therefore,
\[
\pi(\bB^{\star}\mid \sigma_{\beta}^{2})
\propto
(\sigma_{\beta}^{2})^{-(n_{q}-1)/2}
\exp\left(
-\frac{1}{2\sigma_{\beta}^{2}}
\sum_{v\sim_\G u} r_{v,u}^2
\right).
\]

For the proper Gaussian component,
$\bbeta_{v}^\circ \sim \MVN(\beta_v^0\mathbf 1,\sigma_{\beta}^{2}\bI)$, and so
\[
\pi(\bB^\circ\mid \sigma_{\beta}^{2})
\propto
(\sigma_{\beta}^{2})^{-(p|\V|-n_{q})/2}
\exp\left(
-\frac{1}{2\sigma_{\beta}^{2}}
\sum_{v\in\V} r_v^2
\right).
\]

Multiplying the two components gives
\bse
\pi(\bB^{\star},\bB^{\circ} \mid \sigma_{\beta}^{2})
&\propto
(\sigma_{\beta}^{2})^{-(p|\V|-1)/2}
\exp\left(
-\frac{1}{2\sigma_{\beta}^{2}}
\left[
\sum_{v\in\V} r_{v}^2
+
\sum_{v\sim_\G u} r_{v,u}^2
\right]
\right).
\ese

Now with an $\IG(a_\beta,b_\beta)$ prior on $\sigma_{\beta}^{2}$
with density
$\pi(\sigma_{\beta}^{2}) \propto (\sigma_{\beta}^{2})^{-(a_\beta+1)} \exp\left(
-\frac{b_\beta}{\sigma_{\beta}^{2}} \right)$,
the posterior density is
\begin{align*}\pi(\sigma_{\beta}^{2}\mid \bzeta)
&\propto
(\sigma_{\beta}^{2})^{
-\left(
a_\beta+1+\frac{p|\V|-1}{2}
\right)
}
\exp\left(
-\frac{
b_\beta
+
\frac12
\left[
\sum_{v\in\V} r_v^2
+
\sum_{v\sim_\G u} r_{v,u}^2
\right]
}{
\sigma_{\beta}^{2}
}
\right).\end{align*}

We conclude that
\bse
\sigma_{\beta}^{2}\mid \bzeta
\sim
\IG\left(
a_\beta+\frac{p|\V|-1}{2},
\,
b_\beta+
\frac12
\left[
\sum_{v\in\V} r_v^2
+
\sum_{v\sim_\G u} r_{v,u}^2
\right]
\right).
\ese

All results discussed in this article are based on the output of this
algorithm, run for 5000 iterations with the first 2000 iterations discarded as
burn-in, and the remaining samples thinned by an interval of 10.
The $\bbeta_{v}$'s were all initialized at at $\bzero$ and $\sigma^{2}_{\beta}$
at an arbitrarily high value of $1000$.
The partitions $\partition_{v}$'s were initialized by collapsing the smallest
90\% of pairwise differences across all nodes.
Partition updates were stalled until after 500 iterations to allow the chain to
settle in.
For similar reasons,
$\tau$ was gradually increased linearly from $0$ to its final value of $0.5$
over the first 750 iterations.
Finally, to obtain results comparable to standard contrast analysis,
we based our application on the posterior probability of pairwise separation,
to be comparable to the p-values returned from pairwise testing.
All separation probability values across all six pairs of levels, i.e., (1, 2),
(1, 3), (1, 4), (2, 3), (2, 4), (3, 4),
were used as the statistic of interest in the algorithm of
\cite{2004Muller},
in which a threshold is iteratively increased to achieve the desired Bayesian
FDR.

We implemented our MCMC sampler in C++ \citep{CPP11} with a front-end
interface in R \citep{RProgramming},
integrated with the help of the Rcpp \citep{Rcpp} and RcppArmadillo
\citep{RcppArmadillo} packages and parallelized with
RcppThread \citep{RcppThread}.

The total MCMC run time for the results presented in this article was about 95
minutes.
This constituted about 60 minutes for the cerebellum and midbrain,
30 minutes for the cortical motor areas,
and 5 minutes for the putamen and caudate.
The total system run time for the results,
the aggregated running time for the parallel processes, was about 11 hours,
meaning the parallel execution of the sampler reduced the real running time by
85.54 percent.
This is a 6.92 times speedup on an 8-core Apple M1 processing chip.
A perfect acceleration on 8 parallel cores would be an 8 times speedup,
showing that we achieve near-optimal performance improvements via
parallelization.

\newpage
\section{Real Data Analysis:
Additional Results} \label{sec:sm real data additional results}

This section exhibits results of the same form as in the main paper,
but for the comparison methods (spatially independent pairwise testing and
spatially aware pTFCE) on the participant analyzed in the main paper and a
weak-signal participant for which partitions were not easily segmented into
high-cardinality partitions.

\subsection{Comparison with Alternative Methods}

\begin{figure}[!ht]
    \centering

    {\small (a) lhMRPF}
    \vskip 10pt
    \includegraphics[width=0.26\textwidth]{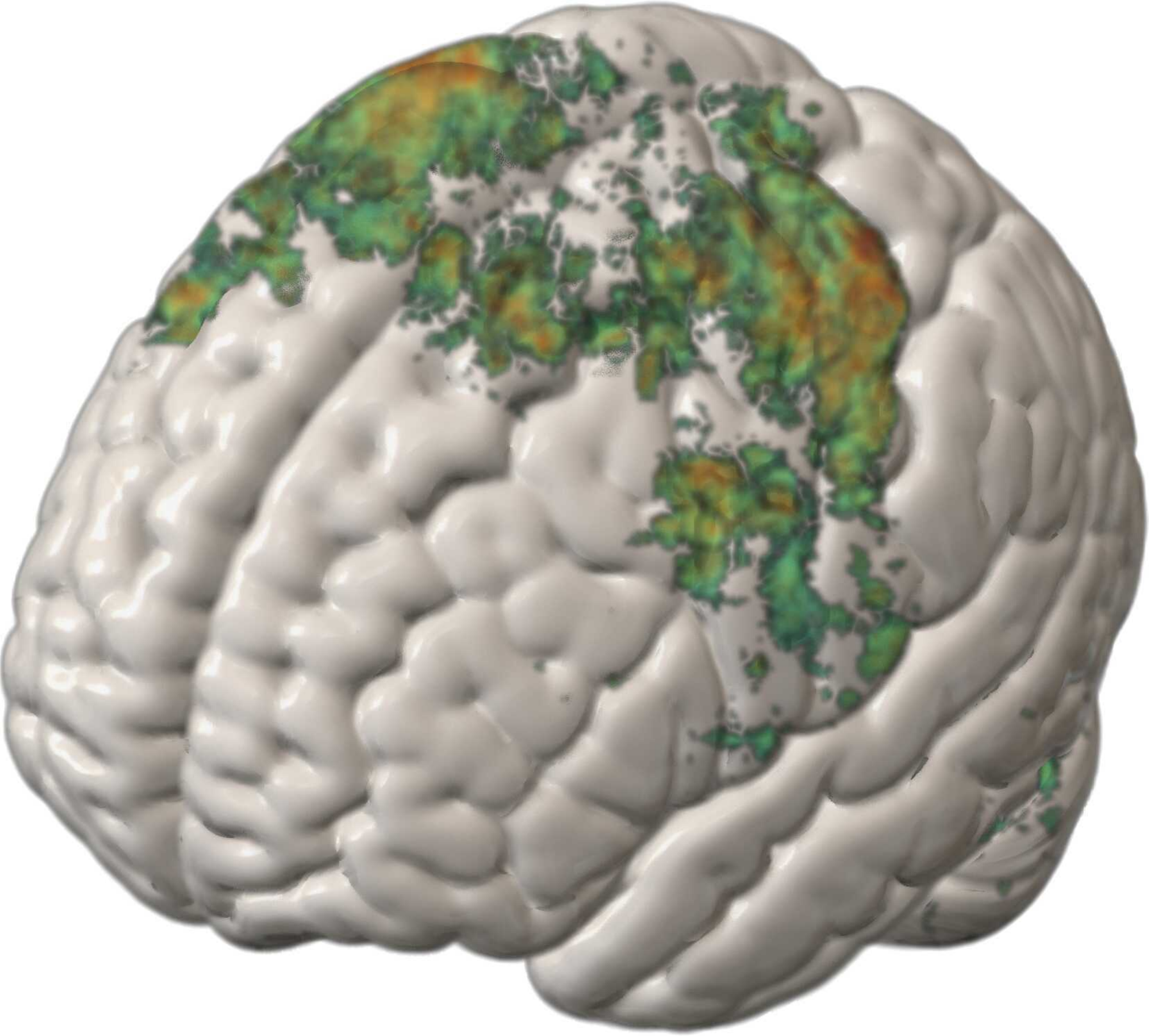}
    \hspace{15pt}
    \includegraphics[width=0.26\textwidth]{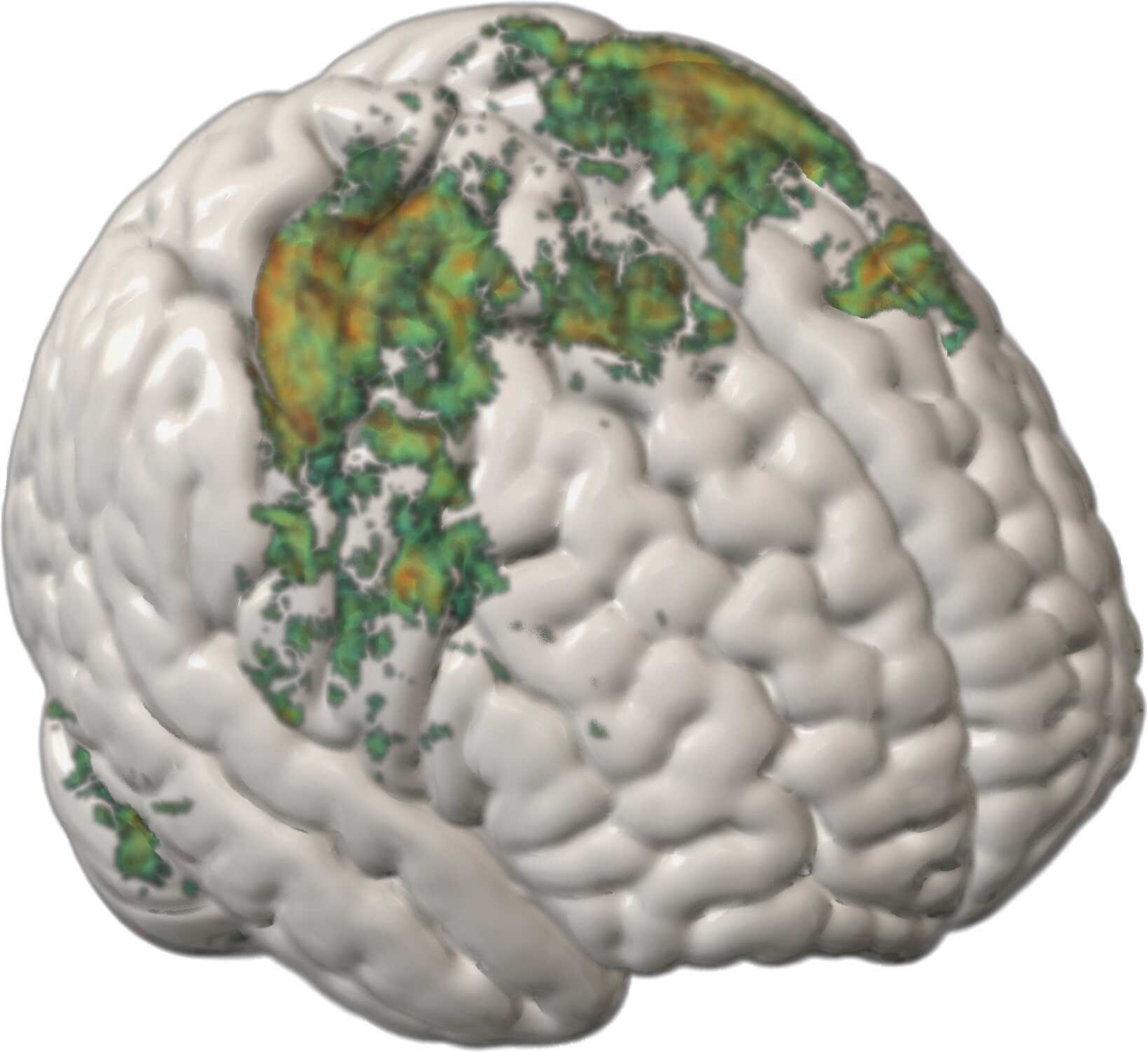}
    \hspace{15pt}
    \includegraphics[width=0.26\textwidth]{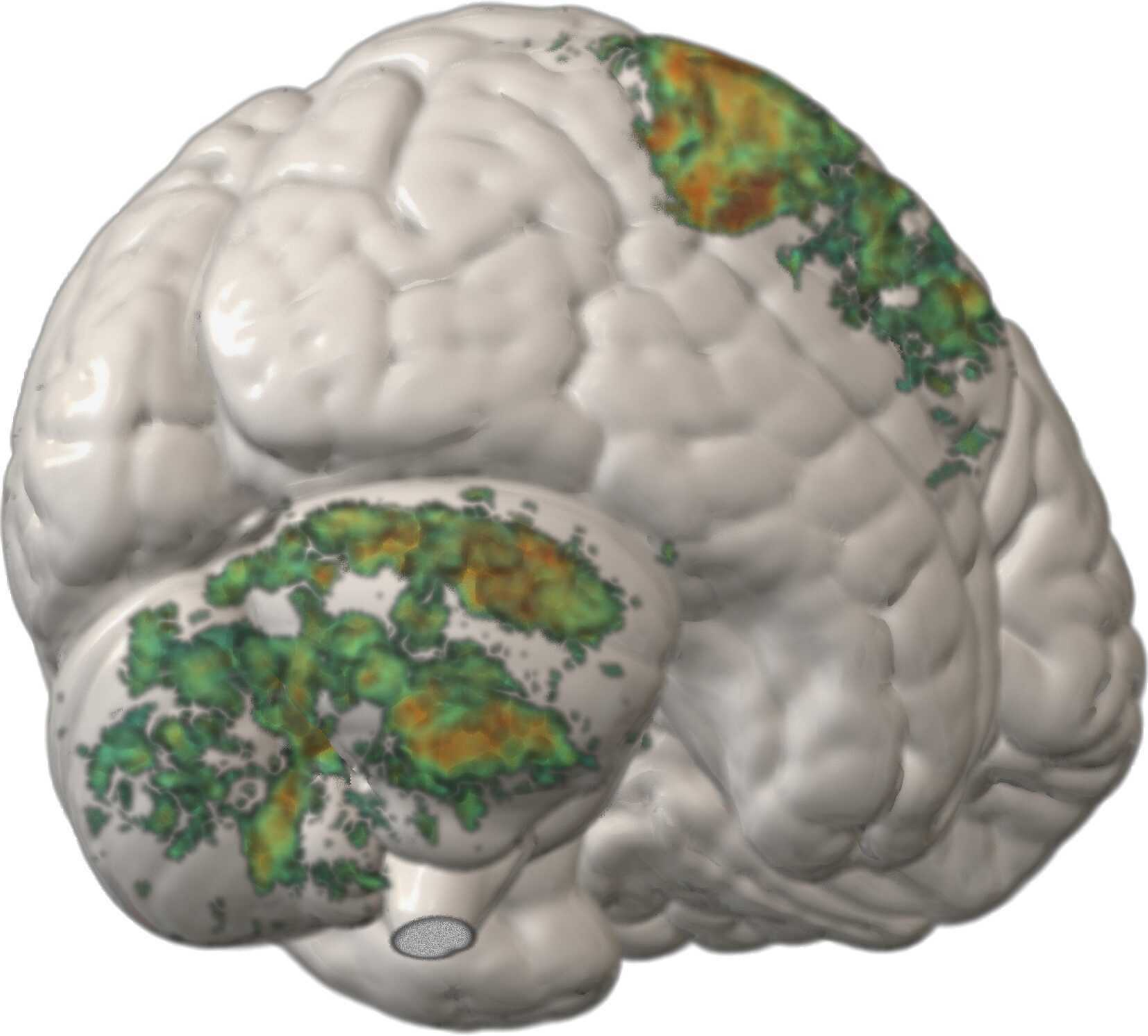}
    \vskip 10pt

    {\small (b) Ind. Tests}
    \vskip 10pt
    \includegraphics[width=0.26\textwidth]{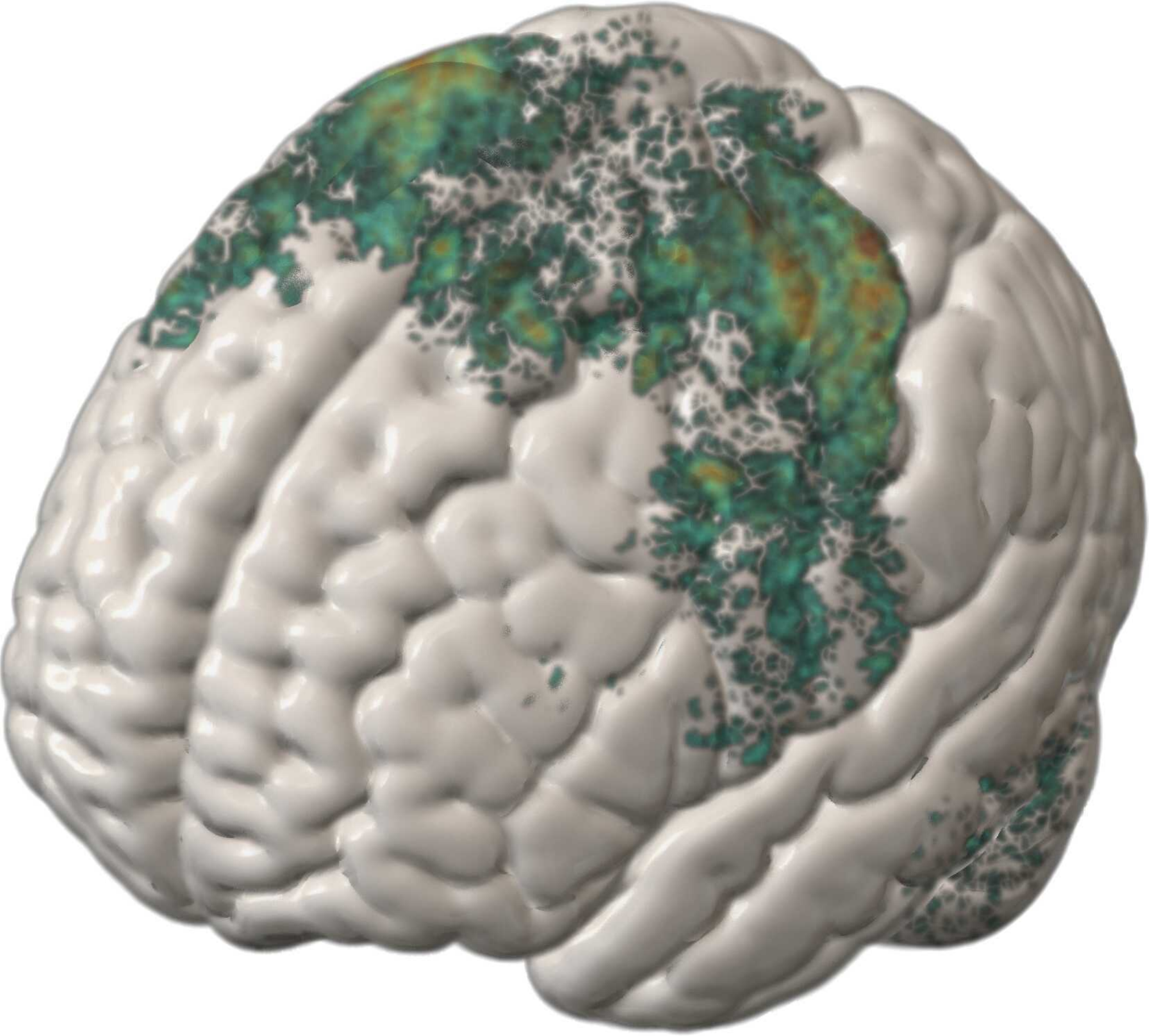}
    \hspace{15pt}
    \includegraphics[width=0.26\textwidth]{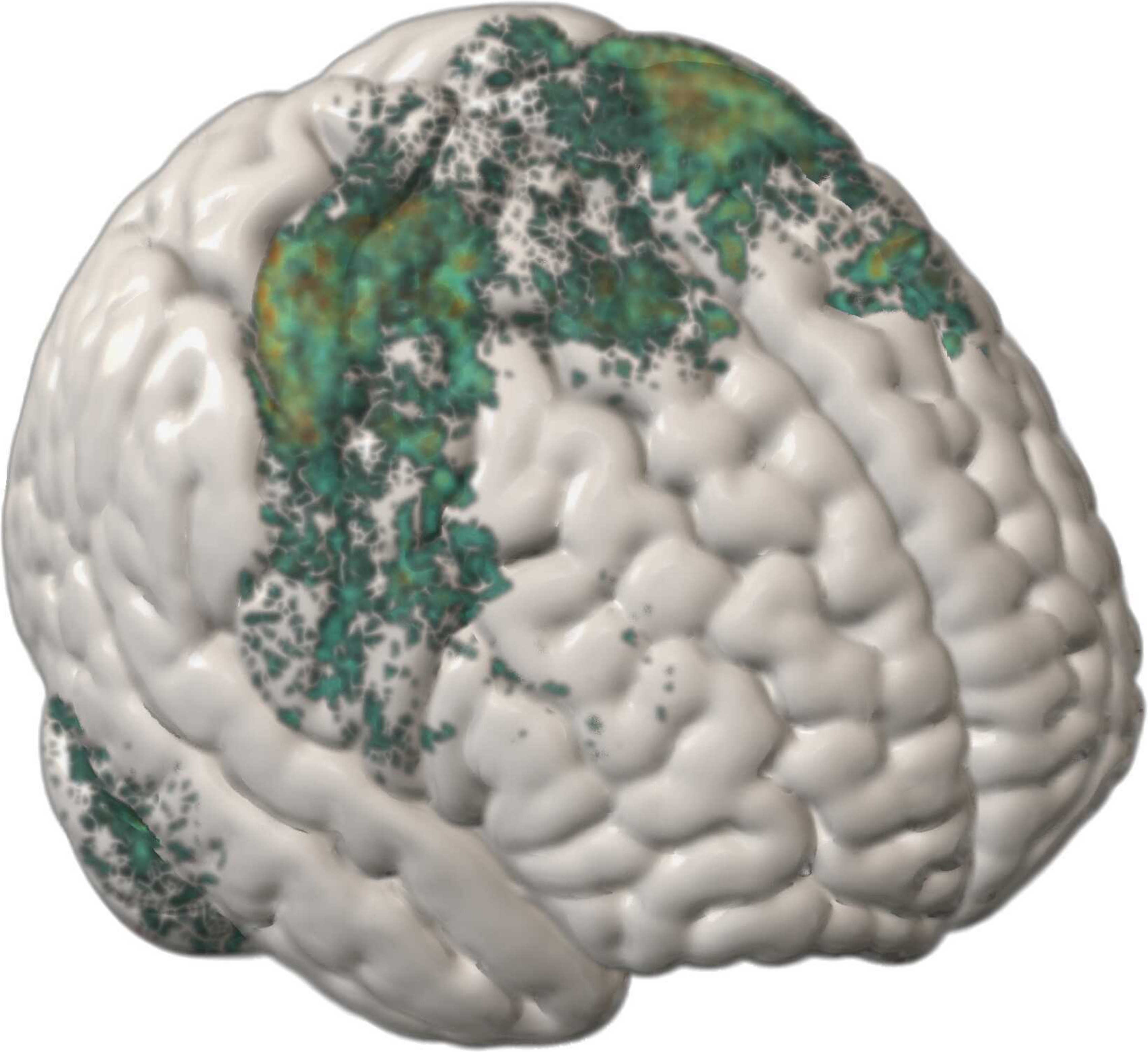}
    \hspace{15pt}
    \includegraphics[width=0.26\textwidth]{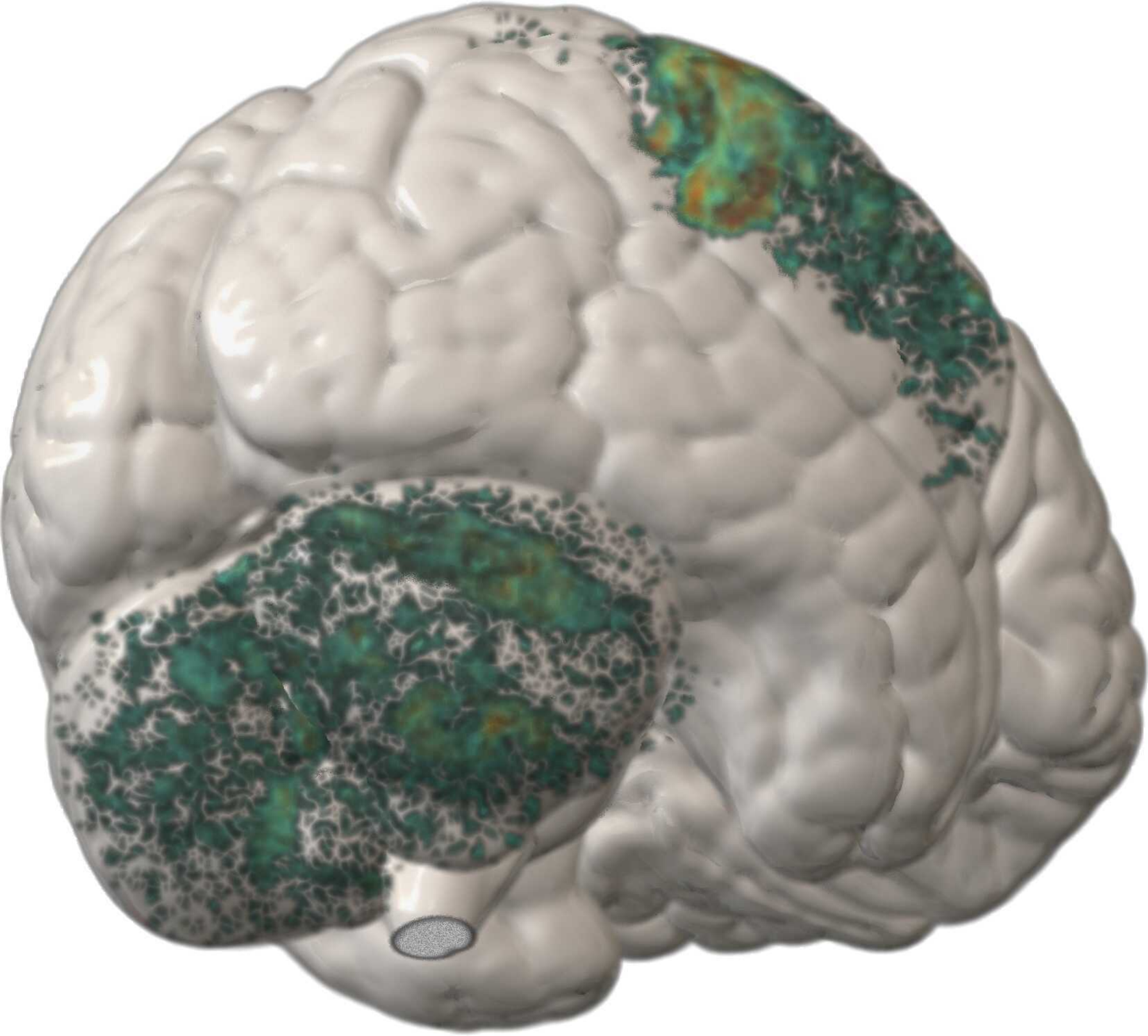}
    \vskip 10pt
    {\small (b) pTFCE}
    \vskip 10pt
    \includegraphics[width=0.26\textwidth]{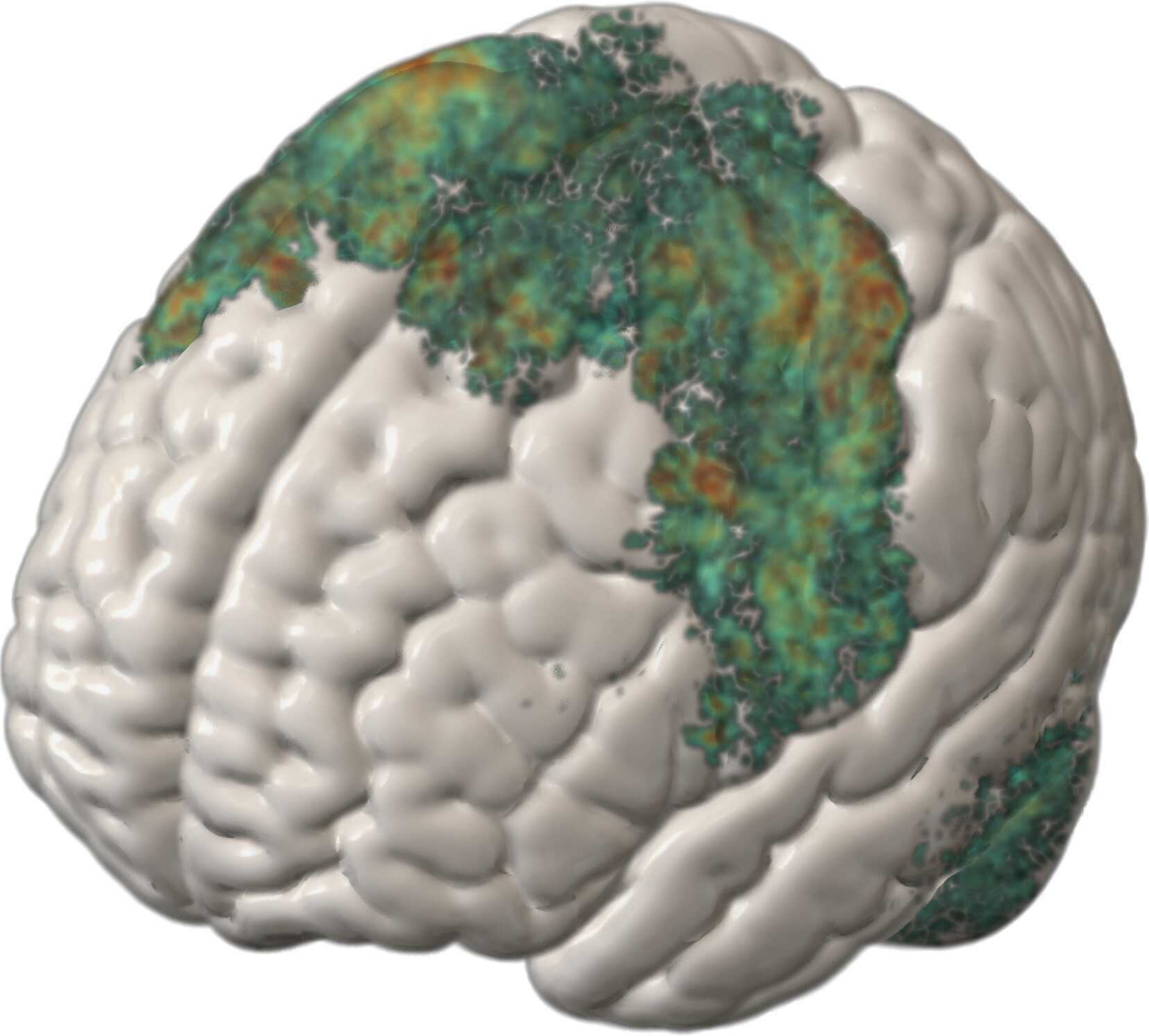}
    \hspace{15pt}
    \includegraphics[width=0.26\textwidth]{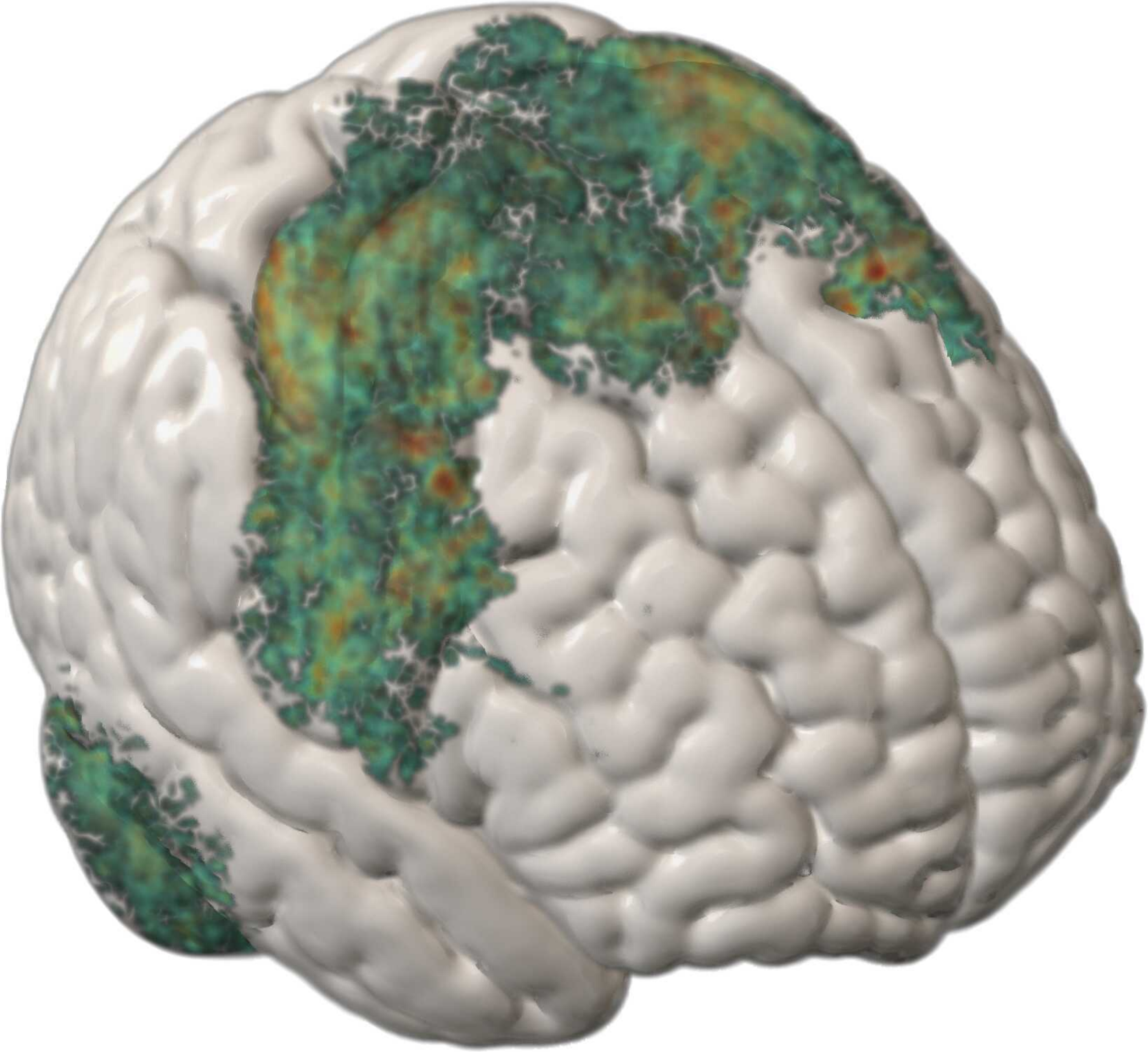}
    \hspace{15pt}
    \includegraphics[width=0.26\textwidth]{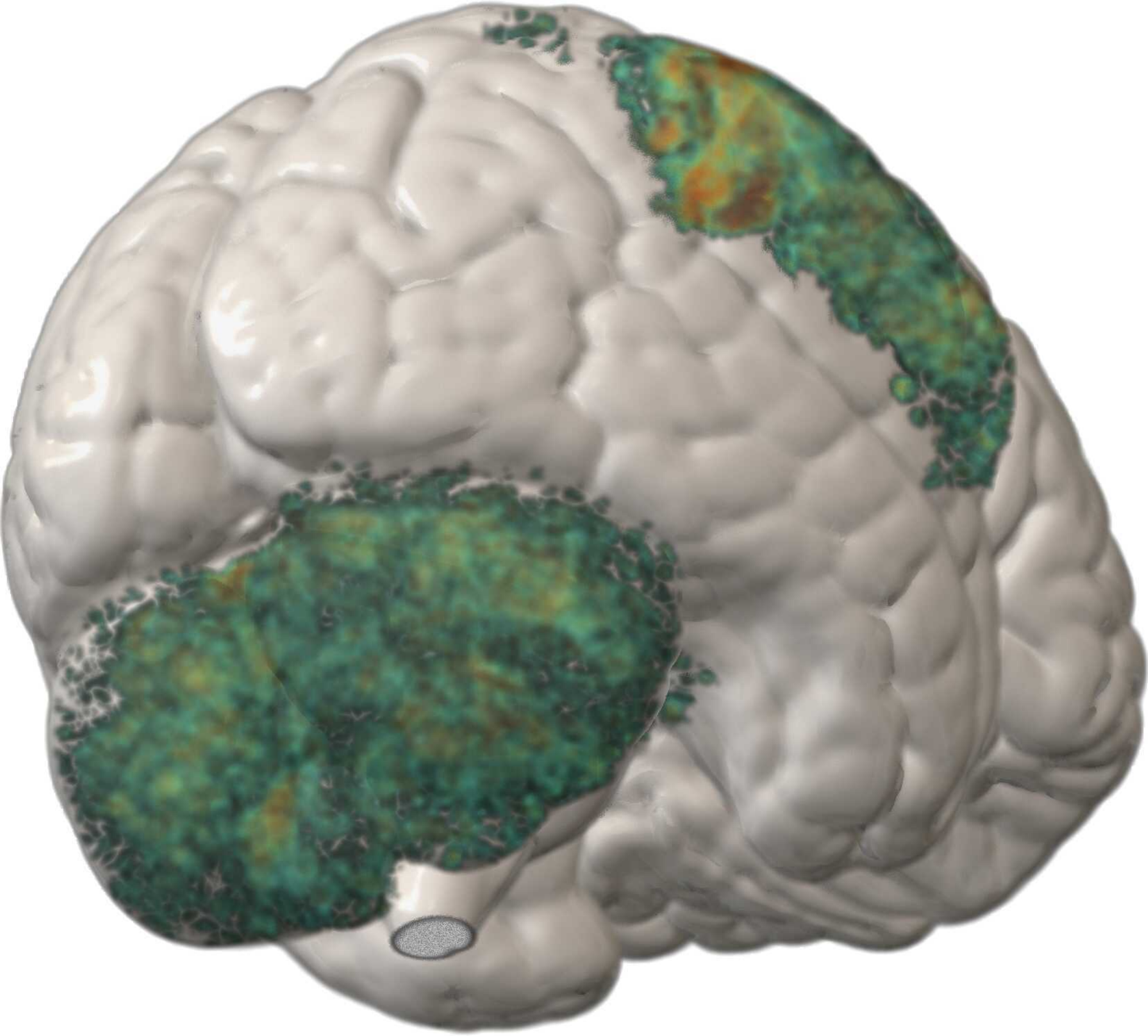}
    \newline\\
    \includegraphics[width=0.75\textwidth]{images/pdfs/colorbar-card.pdf}
    \vspace*{-0pt}
    \caption{
    Results for the fingertip mapping study:
    A map of the posterior expected cardinality of the partitions at each voxel.
    Only voxels with a mean estimated cardinality greater than 2 are colored.
    Results obtained by
    (a) our proposed lhMRPF,
    (b) spatially independent pairwise tests, and
    (c) spatially aware pairwise pTFCE.
    Unlike Figure 6 in the main paper,
    small clusters were not removed from this figure,
    highlighting differences in the methods' ability to selectively remove such
    clusters.
    See also Figure \ref{fig:sm-cardinality-sub9}.
    }
    \label{fig:sm-cardinality}
\vspace*{-10pt}
\end{figure}

\begin{figure}[!ht]
    \centering
    {\small (a) lhMRPF}
    \vskip 10pt
    \includegraphics[width=0.26\textwidth]{images/pdfs/28_halfADL.pdf}
    \hspace{15pt}
    \includegraphics[width=0.26\textwidth]{images/pdfs/28_halfADR.pdf}
    \hspace{15pt}
    \includegraphics[width=0.26\textwidth]{images/pdfs/28_halfPVR.pdf}
    \vskip 10pt
    {\small (b) Ind. Tests}
    \vskip 10pt
    \includegraphics[width=0.26\textwidth]{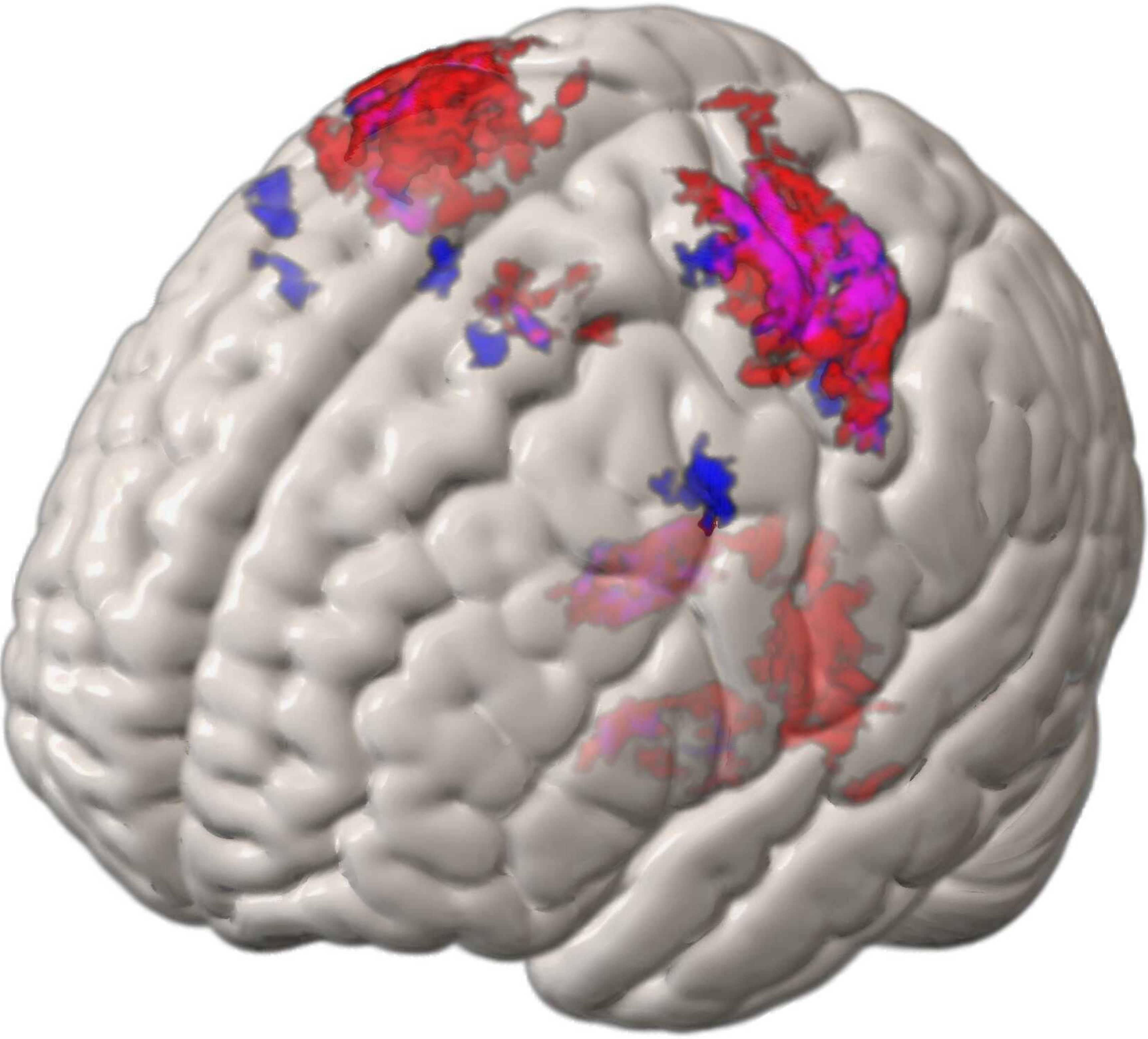}
    \hspace{15pt}
    \includegraphics[width=0.26\textwidth]{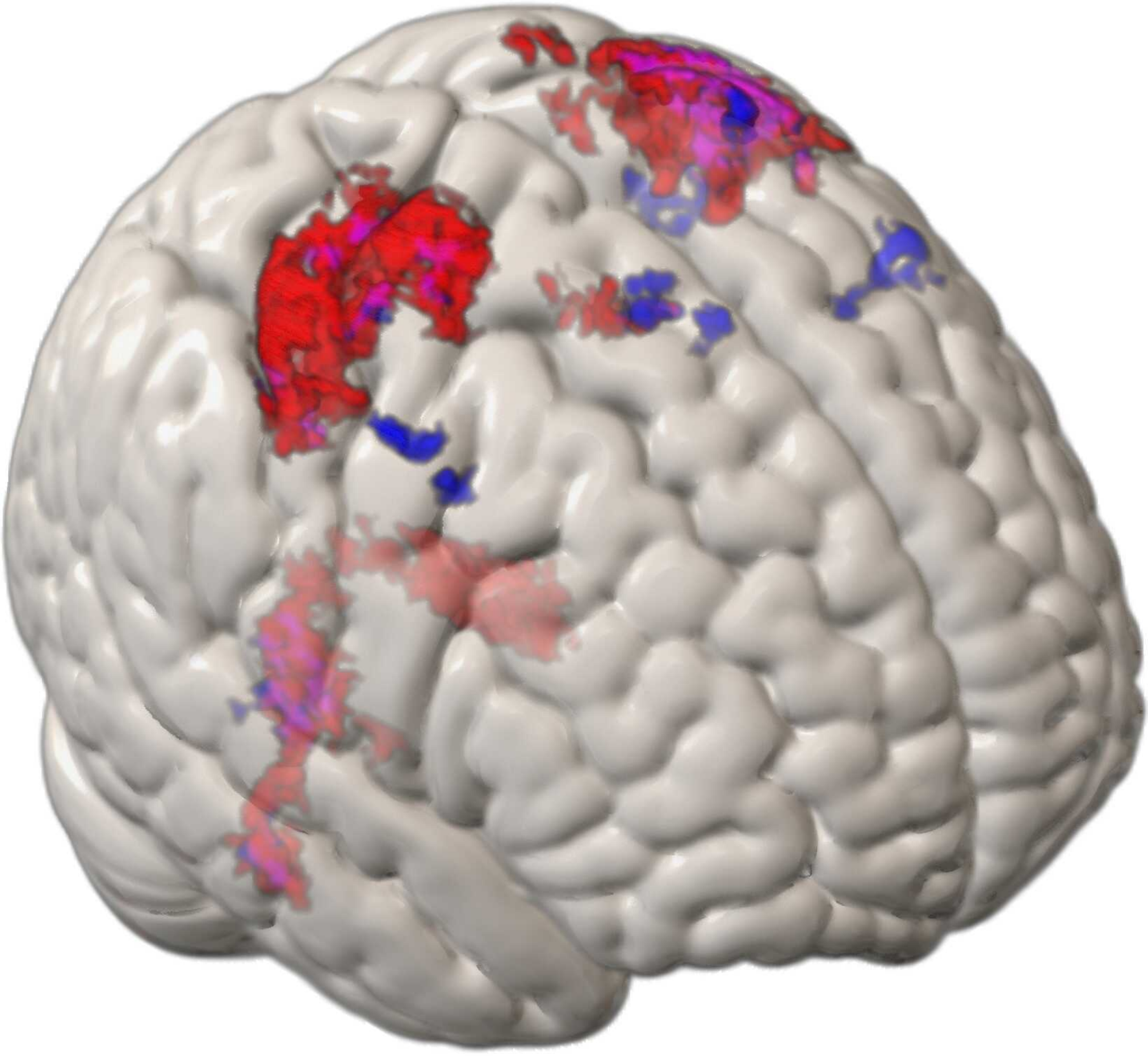}
    \hspace{15pt}
    \includegraphics[width=0.26\textwidth]{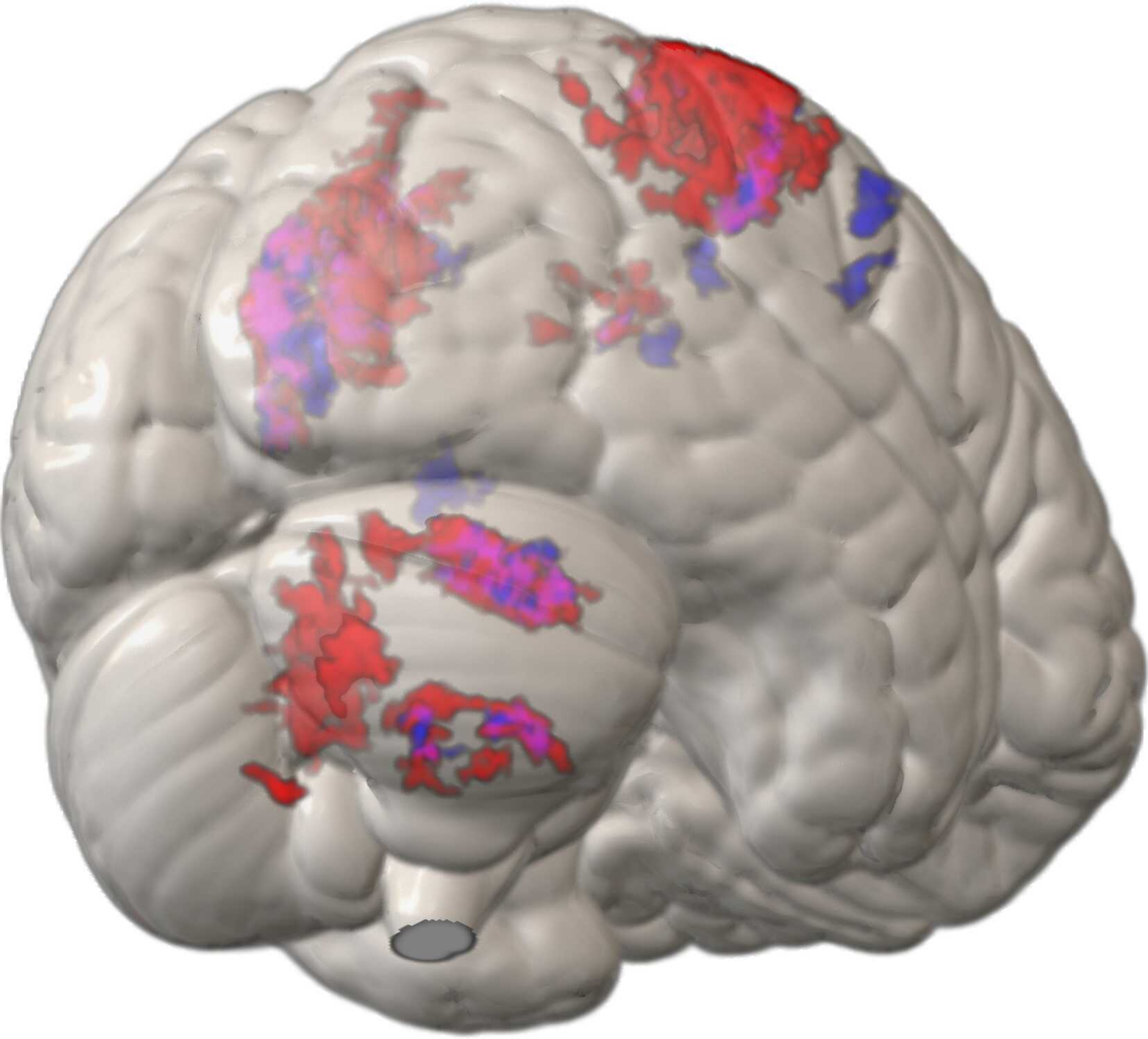}
    \vskip 10pt
    {\small (c) pTFCE}
    \vskip 10pt
    \includegraphics[width=0.26\textwidth]{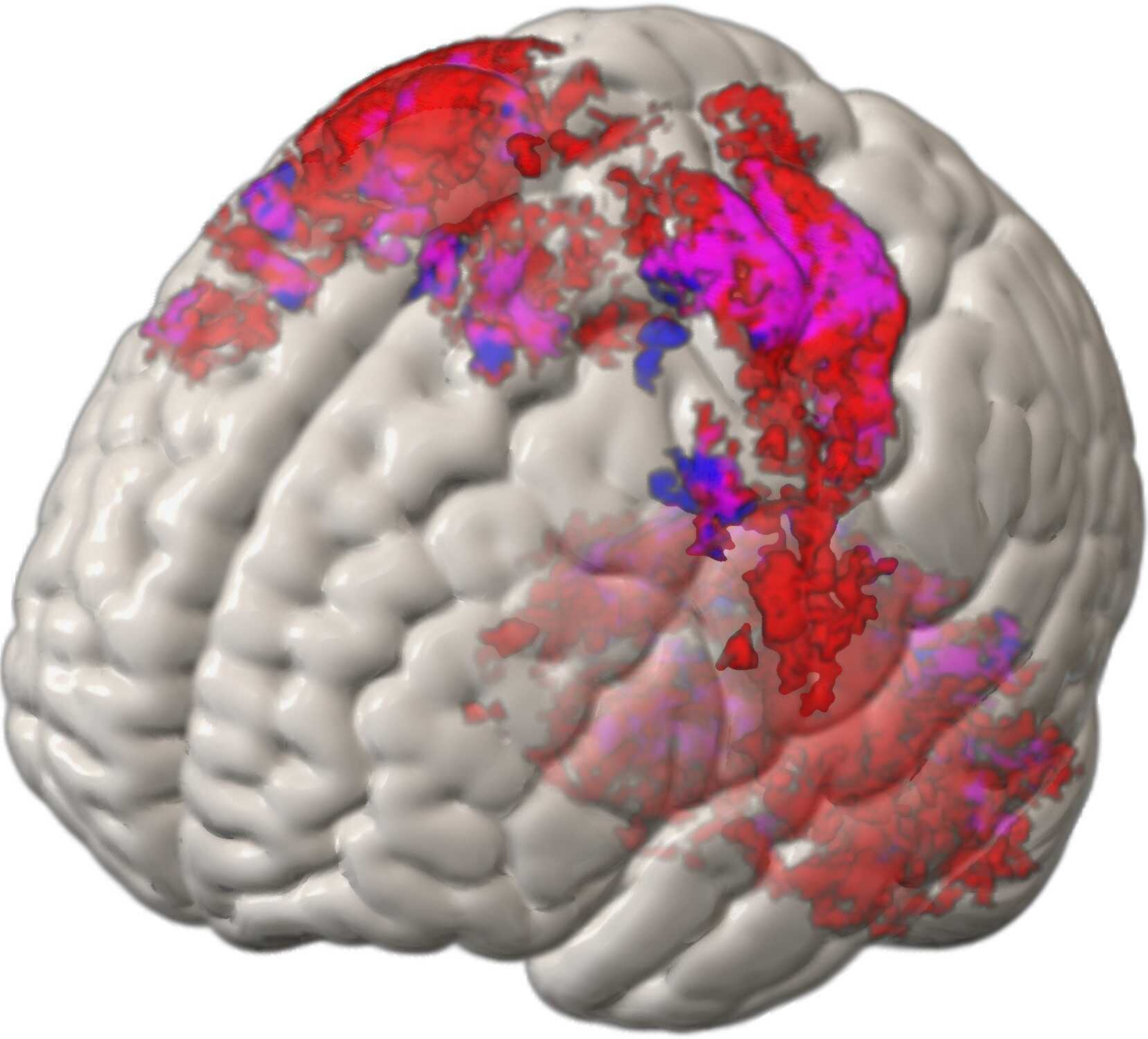}
    \hspace{15pt}
    \includegraphics[width=0.26\textwidth]{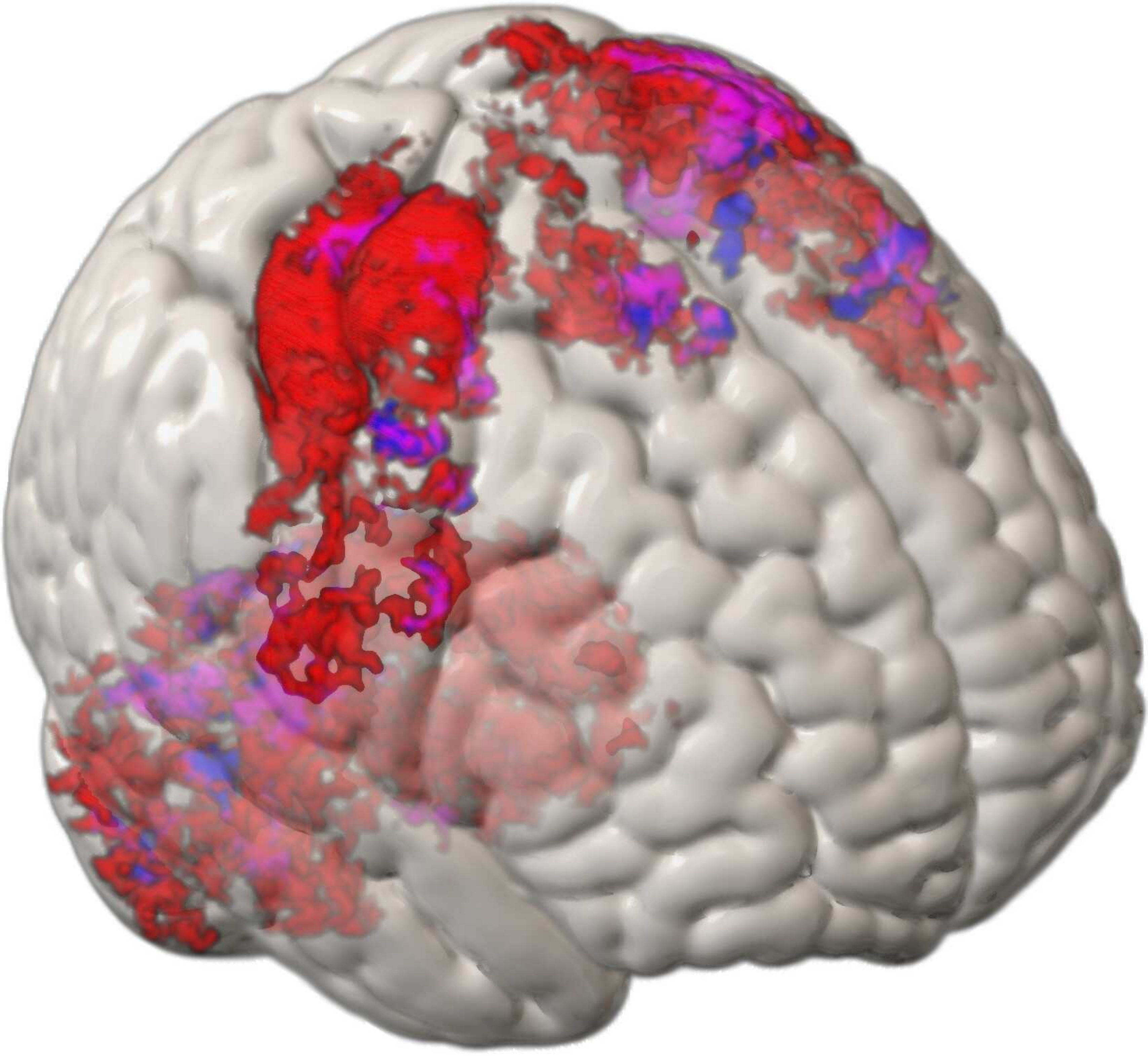}
    \hspace{15pt}
    \includegraphics[width=0.26\textwidth]{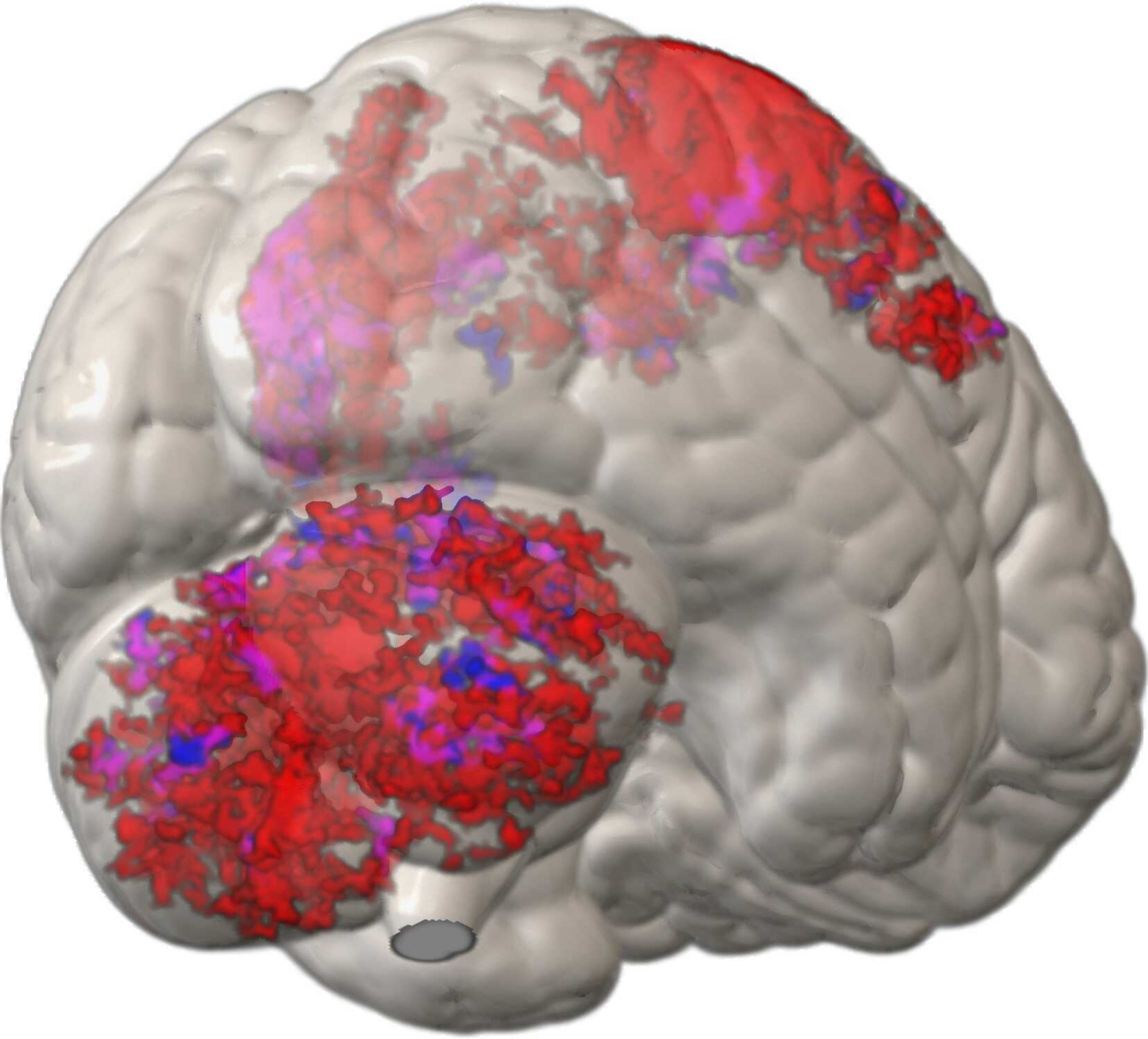}
    \caption{
    Results for the fingertip mapping study:
    Areas of split representation between the two hands (left/right,
    red) and the digits (index/middle,
    blue) based on the estimated partition structures obtained by
    (a) our proposed lhMRPF,
    (b) spatially independent pairwise tests, and
    (c) spatially aware pairwise pTFCE.
    For ease of comparison,
    lhMRPF results are repeated from Figure 8 in the main paper.
    }
    \label{fig:sm-handedness-comparison}
\end{figure}

Figure \ref{fig:sm-cardinality} displays the cardinality maps for all methods
using the main participant's data,
intentionally preserving all small clusters for maximum detail (contributing to
image "graininess").
In these sensorimotor hand areas,
lhMRPF effectively reveals the expected high-cardinality regions,
whereas the comparison methods fail distinctly: Individual Tests (Ind.
Tests) struggle to identify high-cardinality partitions due to their lack of
spatial consideration,
and pTFCE struggles to identify low-cardinality partitions,
likely caused by an overestimation of pairwise co-membership cluster size.
By explicitly representing the underlying partitions and simultaneously
considering all pairwise co-memberships,
lhMRPF achieves a superior configuration that best balances assumptions about
sparsity and voxel similarity with the empirical data.
Figure \ref{fig:sm-handedness-comparison} further demonstrates this by
highlighting differences in partition compactness and the methods' sensitivity
to the well-known lateralization of the digits.
Finally, Tables \ref{tab:unique1} and \ref{tab:unique2} provide the unique
volume representation voxel counts for all three methods,
with lhMRPF's results from the main paper repeated for ease of comparison.

\begin{table}[!ht]
\centering
\begin{tabular}{|c|c|cc|cc|}
\hline
Method & Region & \text{Right Hand} & \text{Left Hand} & \text{Index Fingers} & \text{Middle Fingers} \\
\hline
           & Sensorimotor    & $\mathbf{4879}$ & $4748$ & $\mathbf{6530}$ & $5562$ \\
lhMRPF     & Cereb. \& Midb. & $\mathbf{2918}$ & $2109$ & $\mathbf{3612}$ & $2233$ \\
           & Caud. \& Put.   &  $\mathbf{247}$ &  $111$ &  $\mathbf{232}$ &  $132$ \\
\hline
           & Sensorimotor    & $\mathbf{2026}$ & $1441$ & $\mathbf{2300}$ & $2192$ \\
Ind. Tests & Cereb. \& Midb. &  $\mathbf{543}$ &  $299$ &  $\mathbf{551}$ &  $398$ \\
           & Caud. \& Put.   &   $\mathbf{15}$ &    $5$ &   $\mathbf{19}$ &    $1$ \\
\hline
           & Sensorimotor    & $7238$ & $\mathbf{8373}$ & $\mathbf{12425}$ & $12242$ \\
pTFCE      & Cereb. \& Midb. & $\mathbf{7384}$ & $6543$ & $9854$  & $\mathbf{10532}$ \\
           & Caud. \& Put.   &  $\mathbf{278}$ &  $113$ &  $\mathbf{325}$  &    $66$ \\
\hline
\end{tabular}
\caption{Results for the fingertip mapping study:
Unique volume representation for hands and digit types of the participant
analyzed in the main paper,
showing the number of voxels in which either of the (right,
left) handed fingers is uniquely represented, or either of the (index,
middle) fingers is uniquely represented.
}
\label{tab:unique1}
\end{table}
\vspace*{-10pt}

\begin{table}[!ht]
\centering
\begin{tabular}{|c|c|cc|cc|c|}
\hline
Method & Region & \text{R. Index} & \text{L. Index} & \text{R. Middle} & \text{L. Middle} & \text{All} \\
\hline
           & Sensorimotor    & $2969^{3}$ & $3561^{1}$ & $3122^{2}$ & $2440^{4}$ & $1212$ \\
lhMRPF     & Cereb. \& Midb. & $2180^{1}$ & $1432^{2}$ & $1143^{3}$ & $1090^{4}$ & $405$ \\
           & Caud. \& Put.   &  $171^{1}$ &   $61^{3}$ &   $81^{2}$ &   $51^{4}$ &    $5$ \\
\hline
           & Sensorimotor    & $1118^{3}$ & $1182^{1}$ & $1447^{2}$ &  $745^{4}$ &  $539$ \\
Ind. Tests & Cereb. \& Midb. &  $314^{1}$ &  $237^{3}$ &  $296^{2}$ &  $102^{4}$ &   $67$ \\
           & Caud. \& Put.   &   $14^{1}$ &    $5^{2}$ &    $1^{3}$ &    $0^{4}$ &    $0$ \\
\hline
           & Sensorimotor    & $5323^{4}$ & $7102^{1}$ & $5648^{3}$ & $6594^{2}$ & $3733$ \\
pTFCE      & Cereb. \& Midb. & $4704^{4}$ & $5150^{2}$ & $5737^{1}$ & $4795^{3}$ & $3057$ \\
           & Caud. \& Put.   &  $245^{1}$ &   $80^{2}$ &   $33^{4}$ &   $33^{4}$ &    $0$ \\
\hline
\end{tabular}
\caption{Results for the fingertip mapping study:
Unique volume representation for individual fingers of the participant analyzed
in the main paper, showing the number of voxels in which each of the four
fingers is uniquely represented and where all four fingers are uniquely
represented.
}
\label{tab:unique2}
\end{table}
\vspace*{-10pt}

\clearpage\newpage
\subsection{Individual Variability}

\begin{figure}[!ht]
    \centering
    {\small (a) lhMRPF}
    \vskip 10pt
    \includegraphics[width=0.26\textwidth]{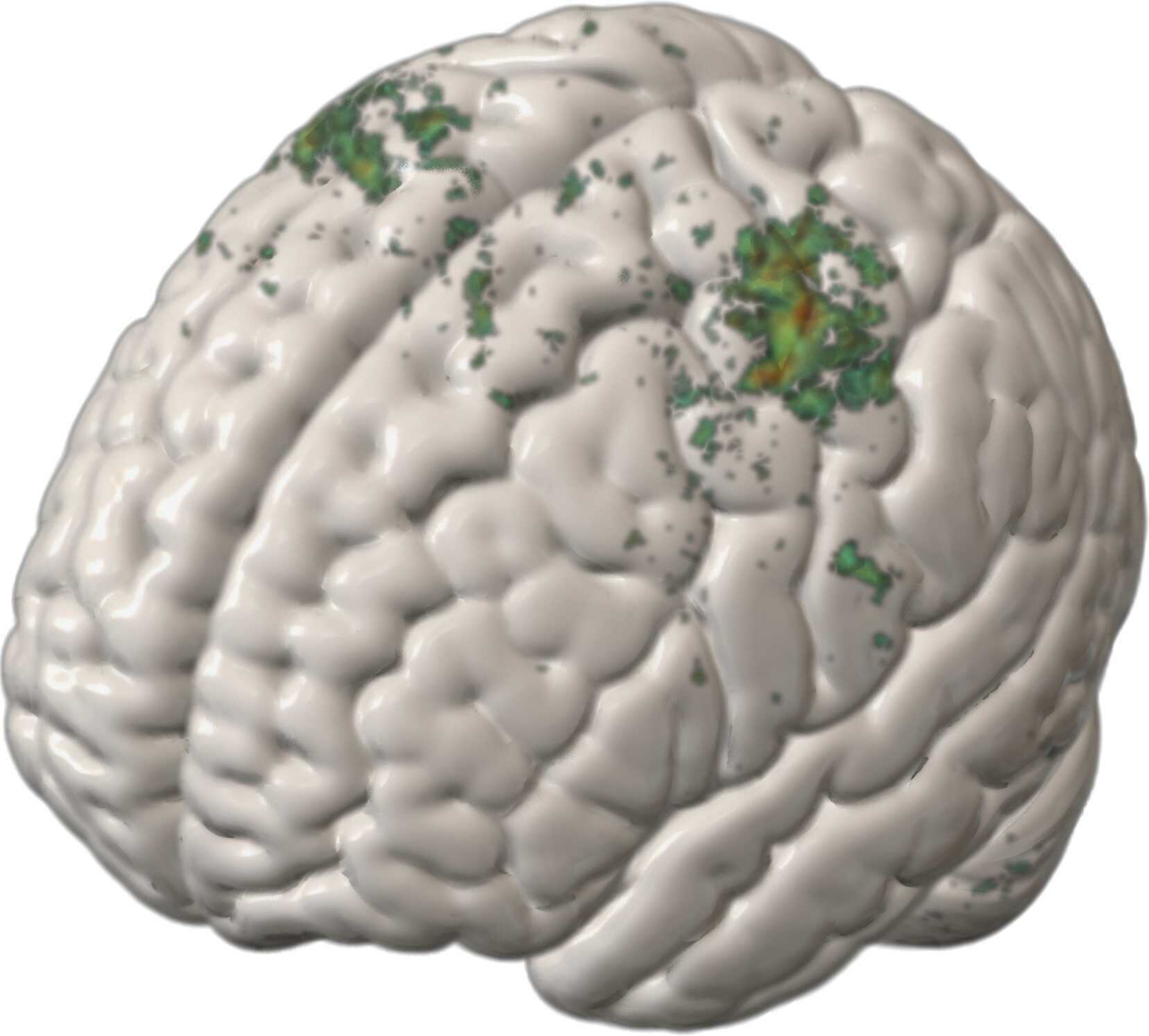}
    \hspace{15pt}
    \includegraphics[width=0.26\textwidth]{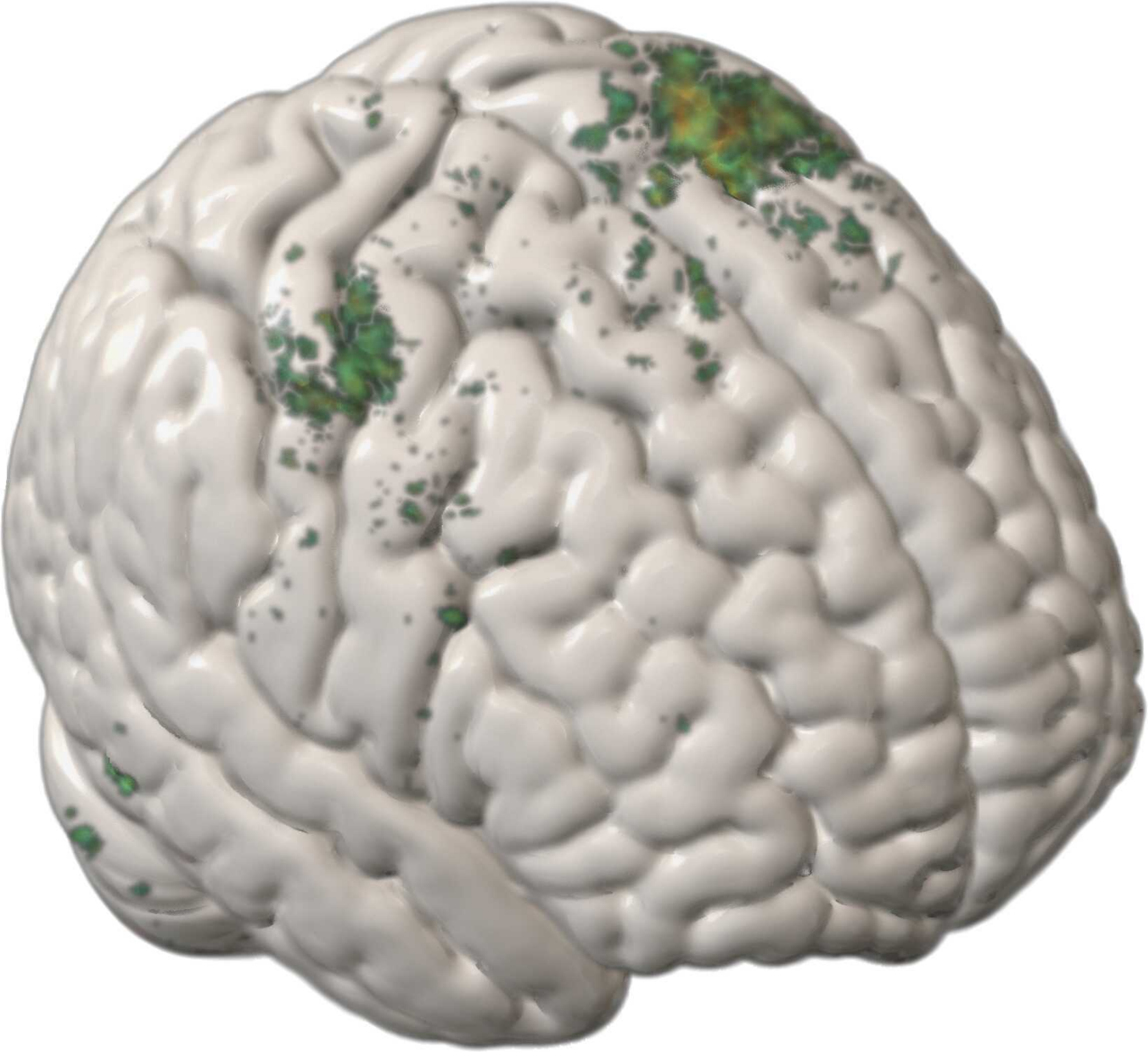}
    \hspace{15pt}
    \includegraphics[width=0.26\textwidth]{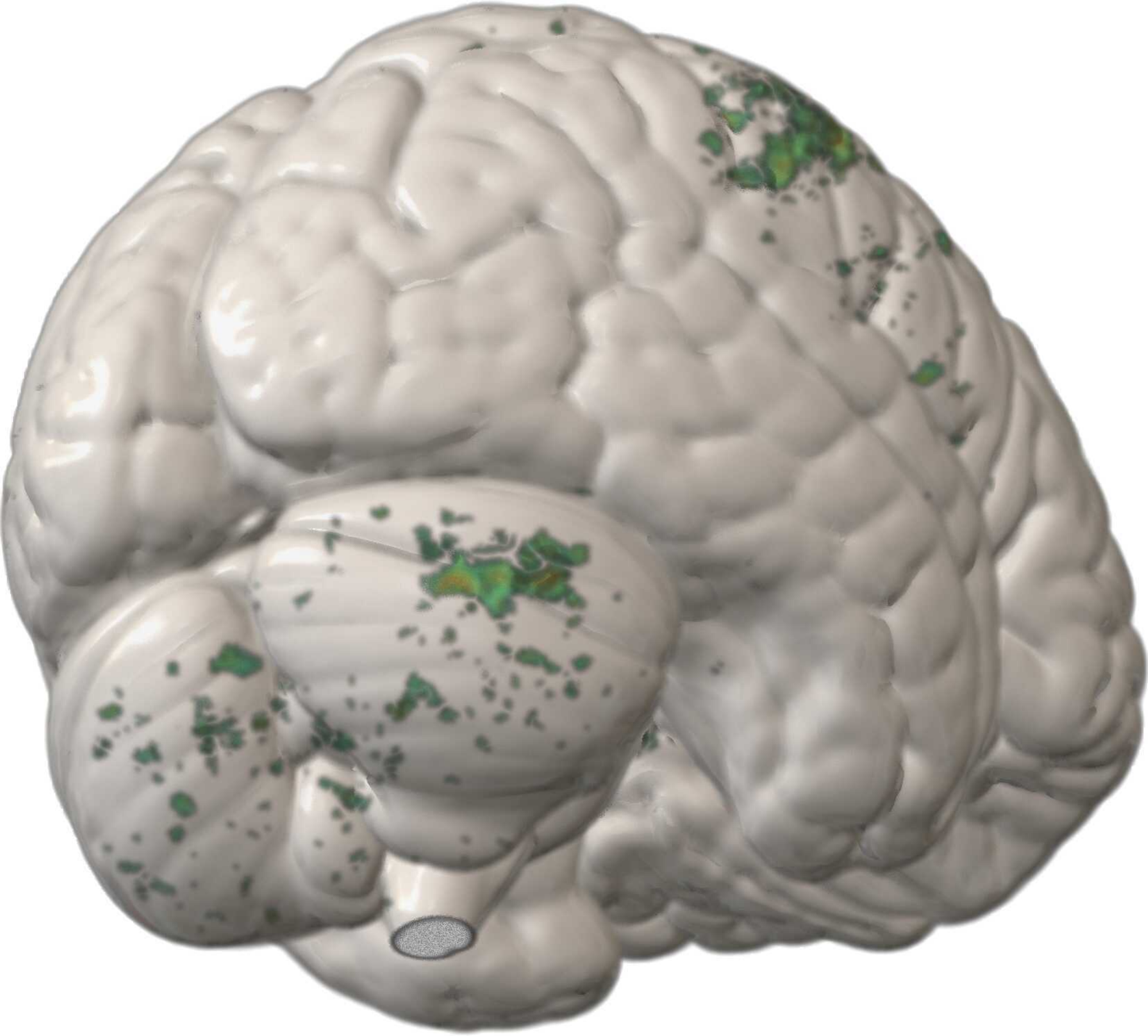}
    \vskip 10pt
    {\small (b) Ind. Tests}
    \vskip 10pt
    \includegraphics[width=0.26\textwidth]{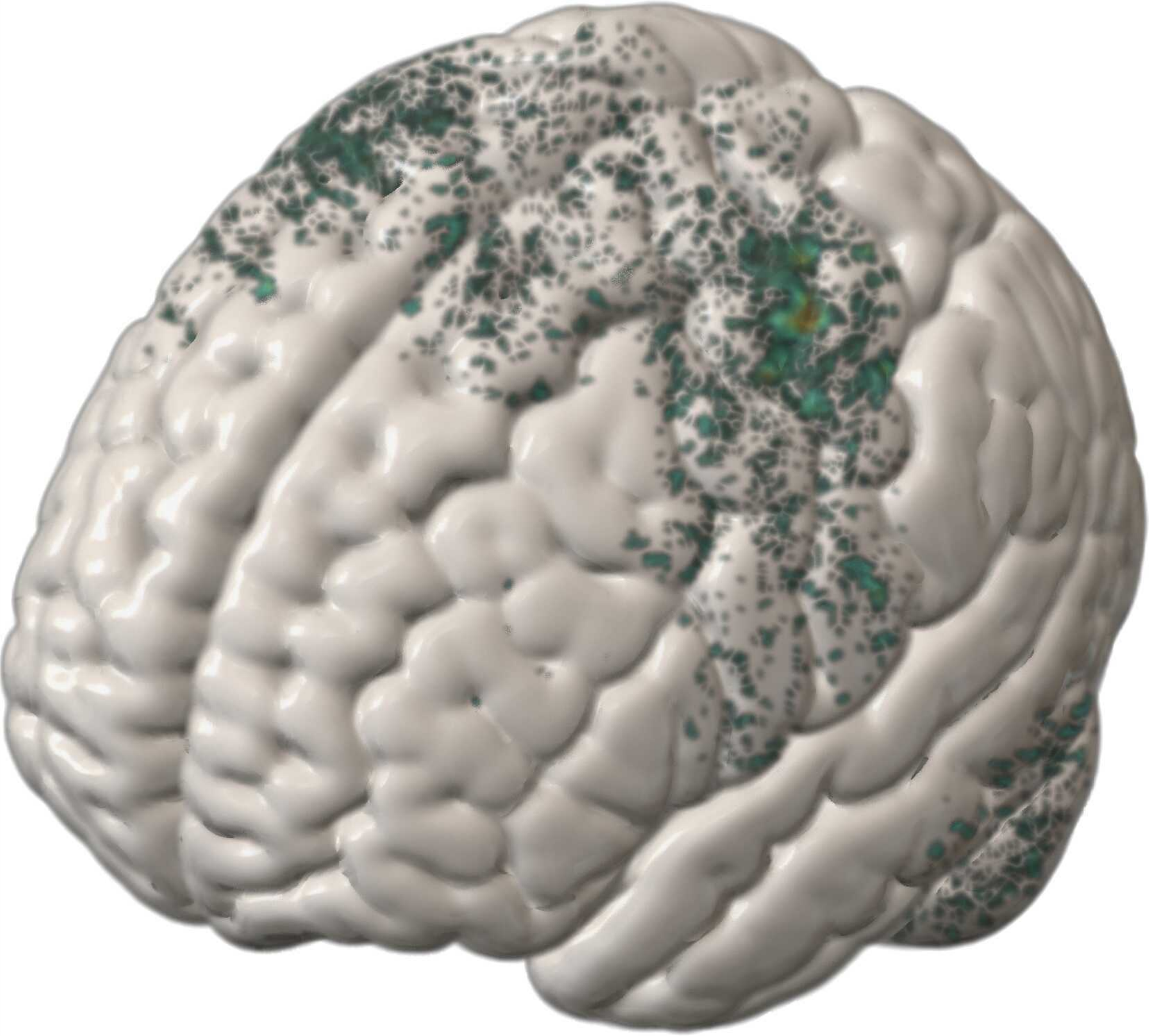}
    \hspace{15pt}
    \includegraphics[width=0.26\textwidth]{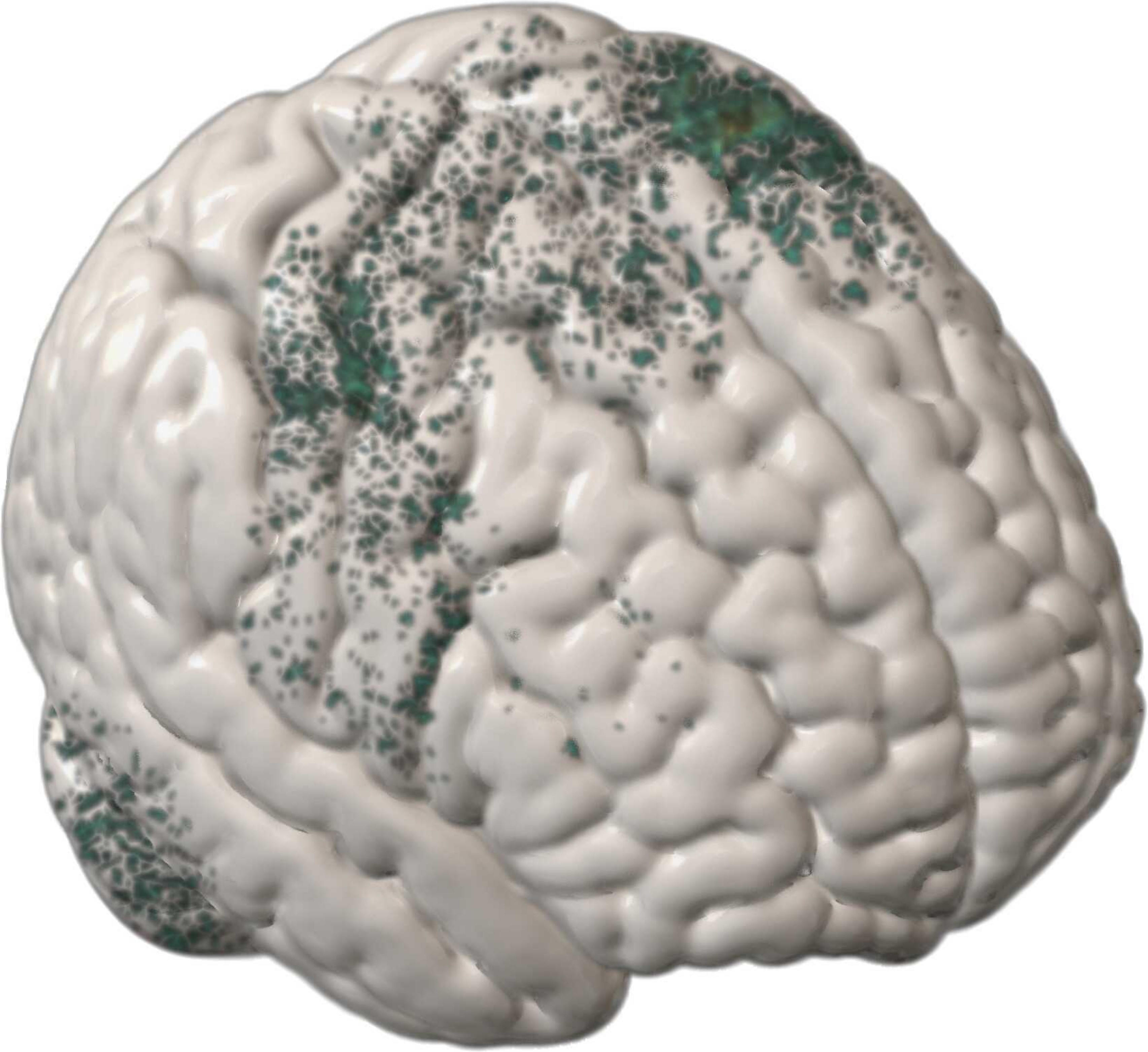}
    \hspace{15pt}
    \includegraphics[width=0.26\textwidth]{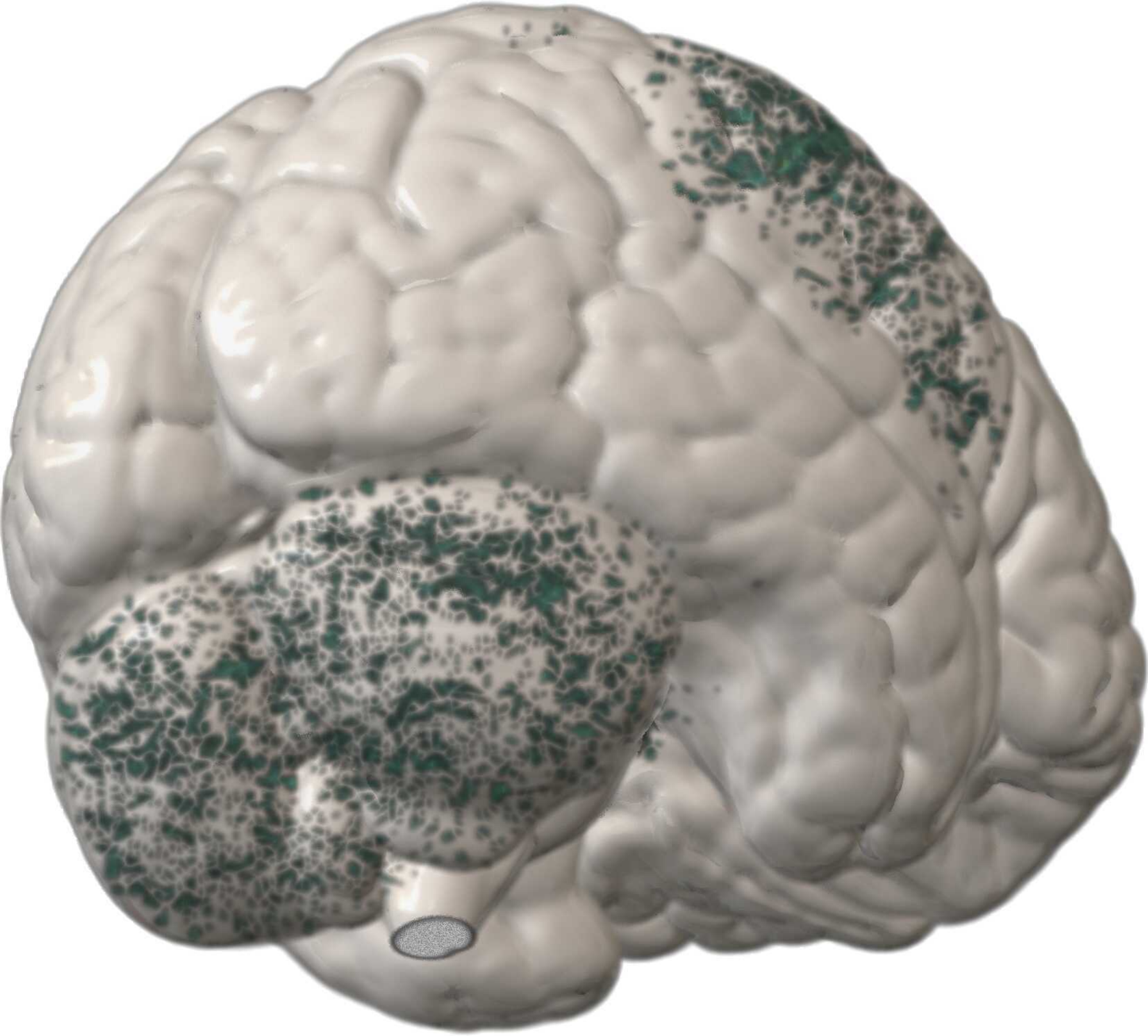}
    \vskip 10pt
    {\small (c) pTFCE}
    \vskip 10pt
    \includegraphics[width=0.26\textwidth]{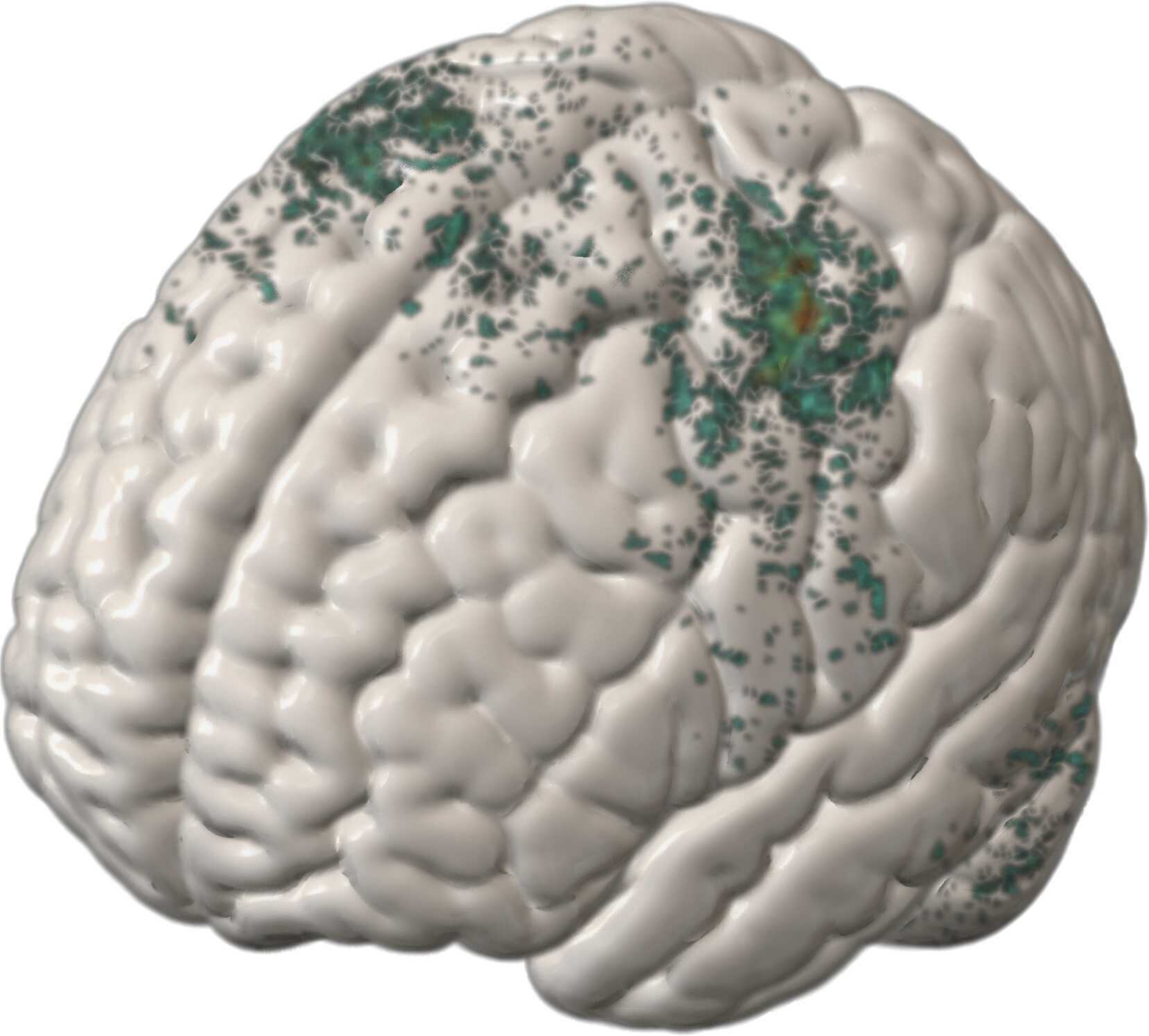}
    \hspace{15pt}
    \includegraphics[width=0.26\textwidth]{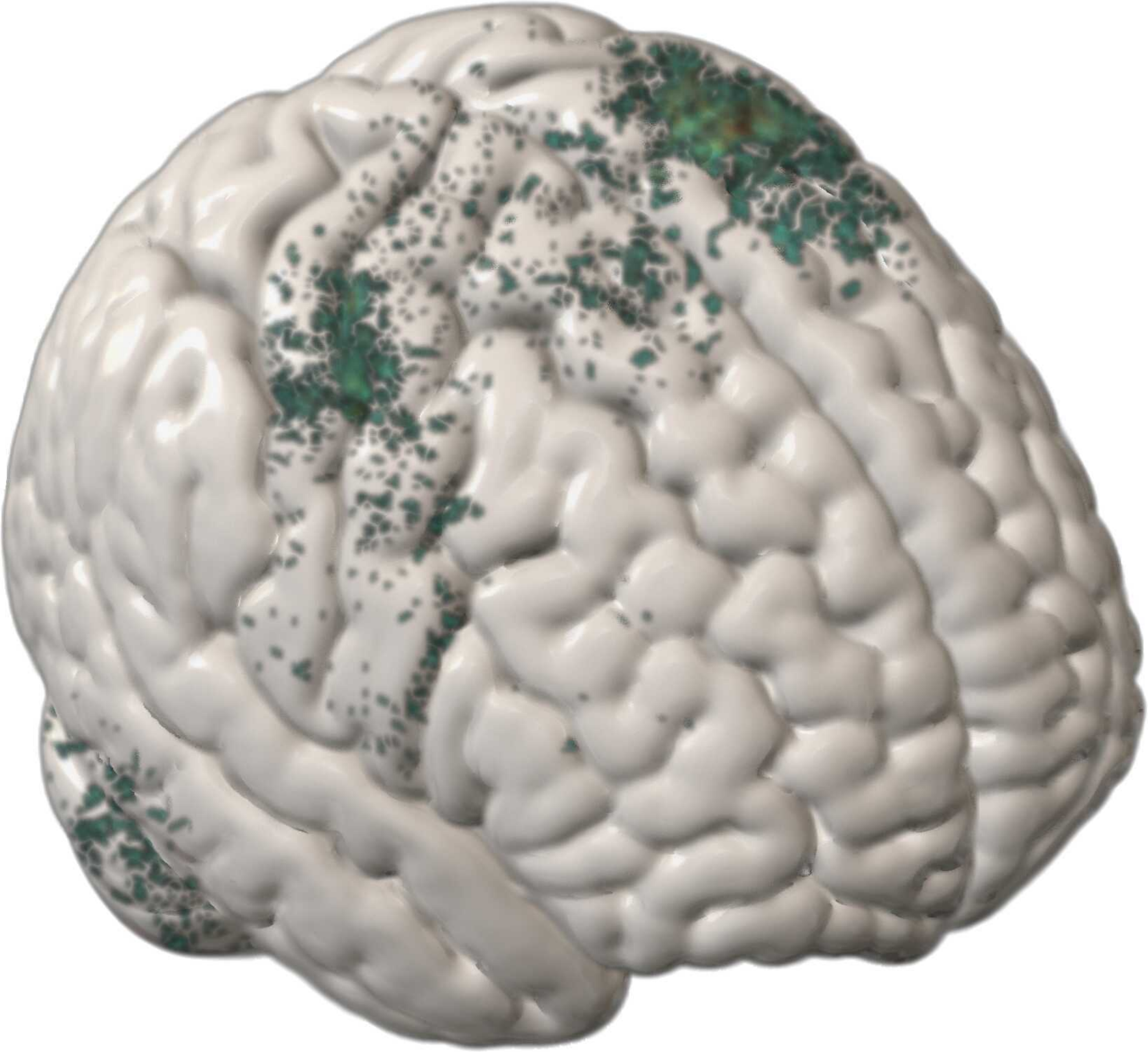}
    \hspace{15pt}
    \includegraphics[width=0.26\textwidth]{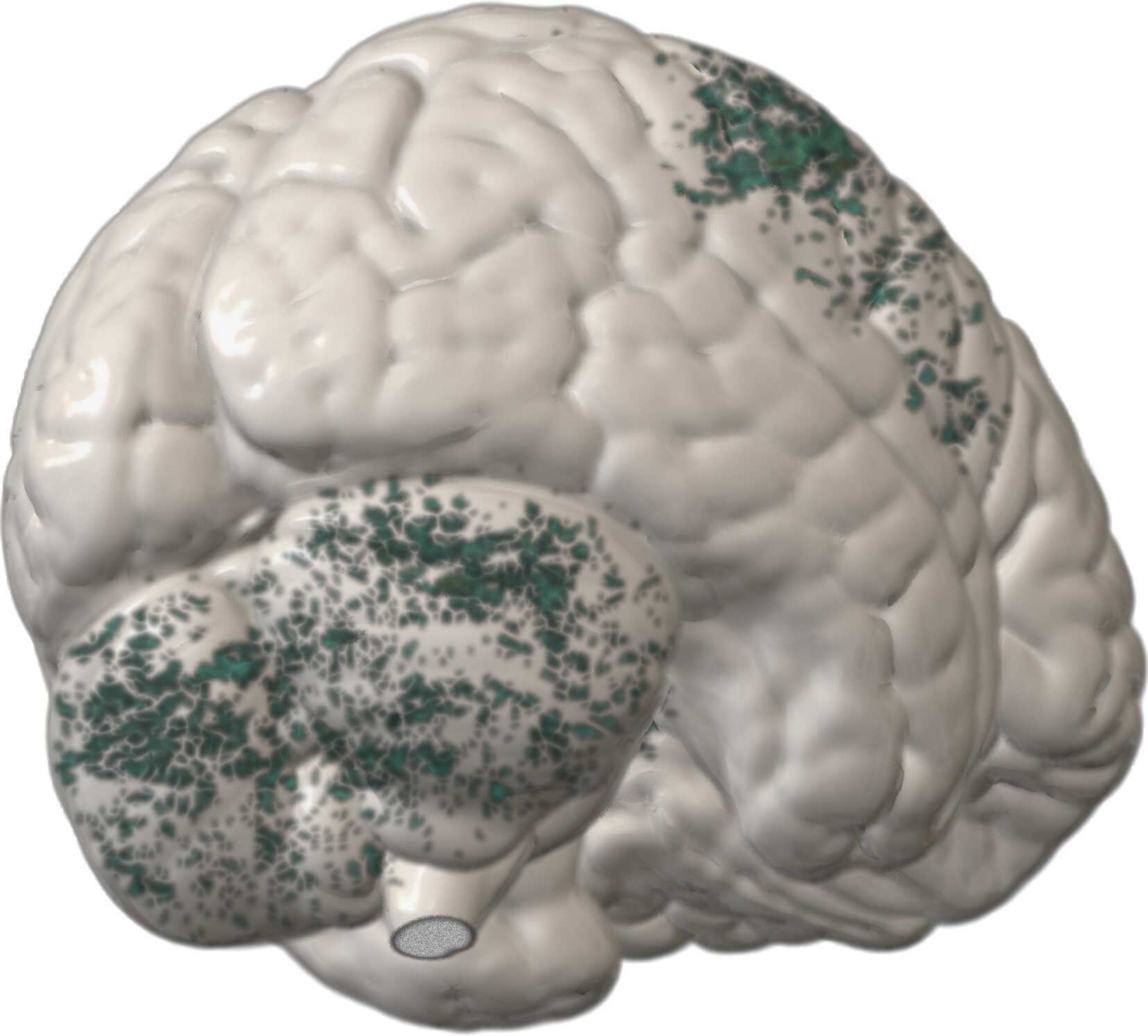}
    \newline\\
    \includegraphics[width=0.75\textwidth]{images/pdfs/colorbar-card.pdf}
    \vspace*{-0pt}
    \caption{
    Results for the fingertip mapping study for a weak-signal participant:
    A map of the posterior expected cardinality of the partitions at each voxel.
    Only voxels with a mean estimated cardinality greater than 2 are colored.
    Results obtained by
    (a) our proposed lhMRPF,
    (b) spatially independent pairwise tests, and
    (c) spatially aware pairwise pTFCE.
    Small clusters were not removed from this figure,
    highlighting differences in the methods' ability to selectively remove such
    clusters.
    See Tables \ref{tab:unique1-sub9} and \ref{tab:unique2-sub9} for more
    detailed results.}
    \label{fig:sm-cardinality-sub9}
\end{figure}

\begin{figure}[!ht]
    \centering
    {\small (a) Right-Hand}
    \vskip 10pt
    \includegraphics[width=0.25\textwidth]{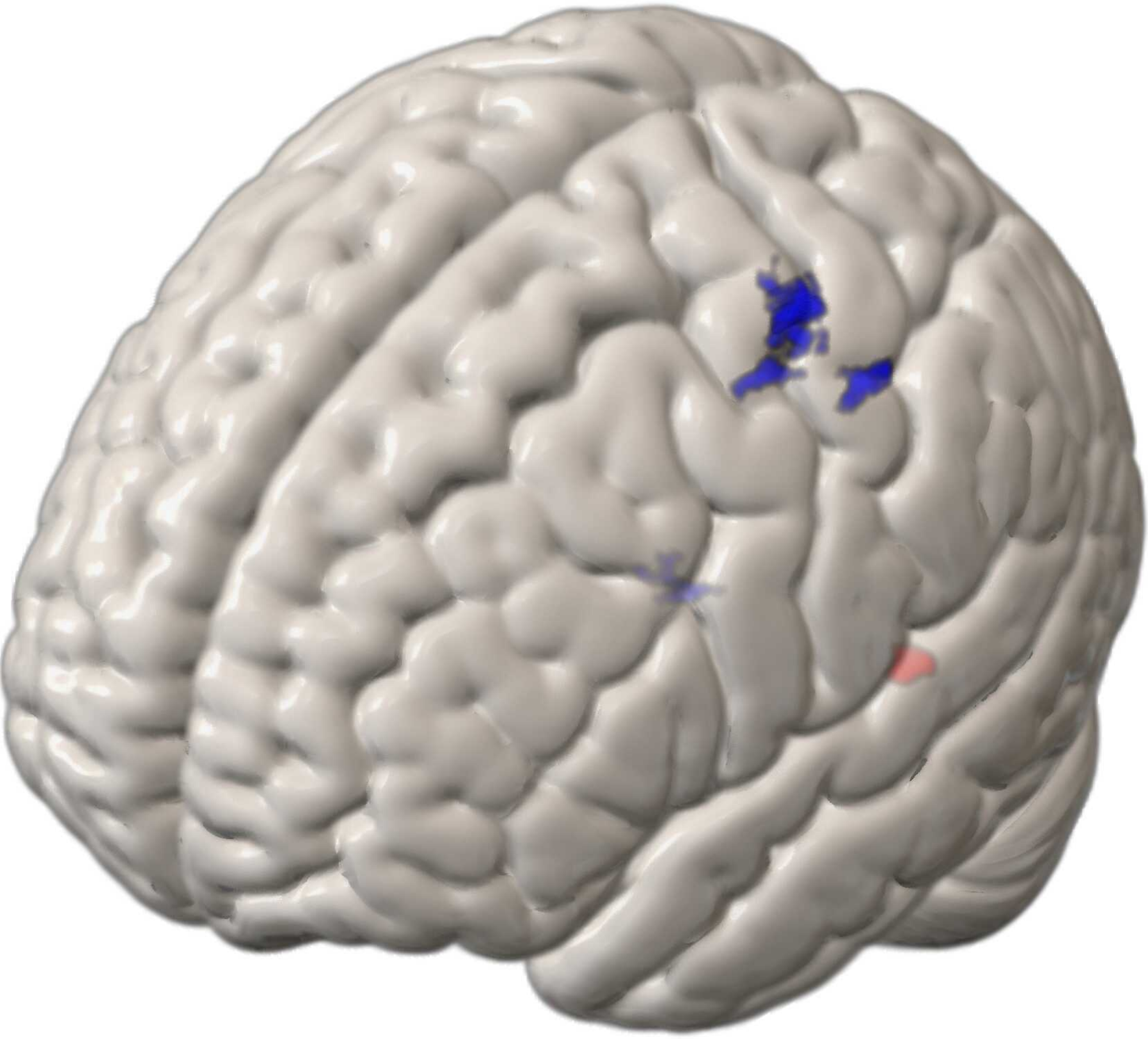}
    \hspace{15pt}
    \includegraphics[width=0.25\textwidth]{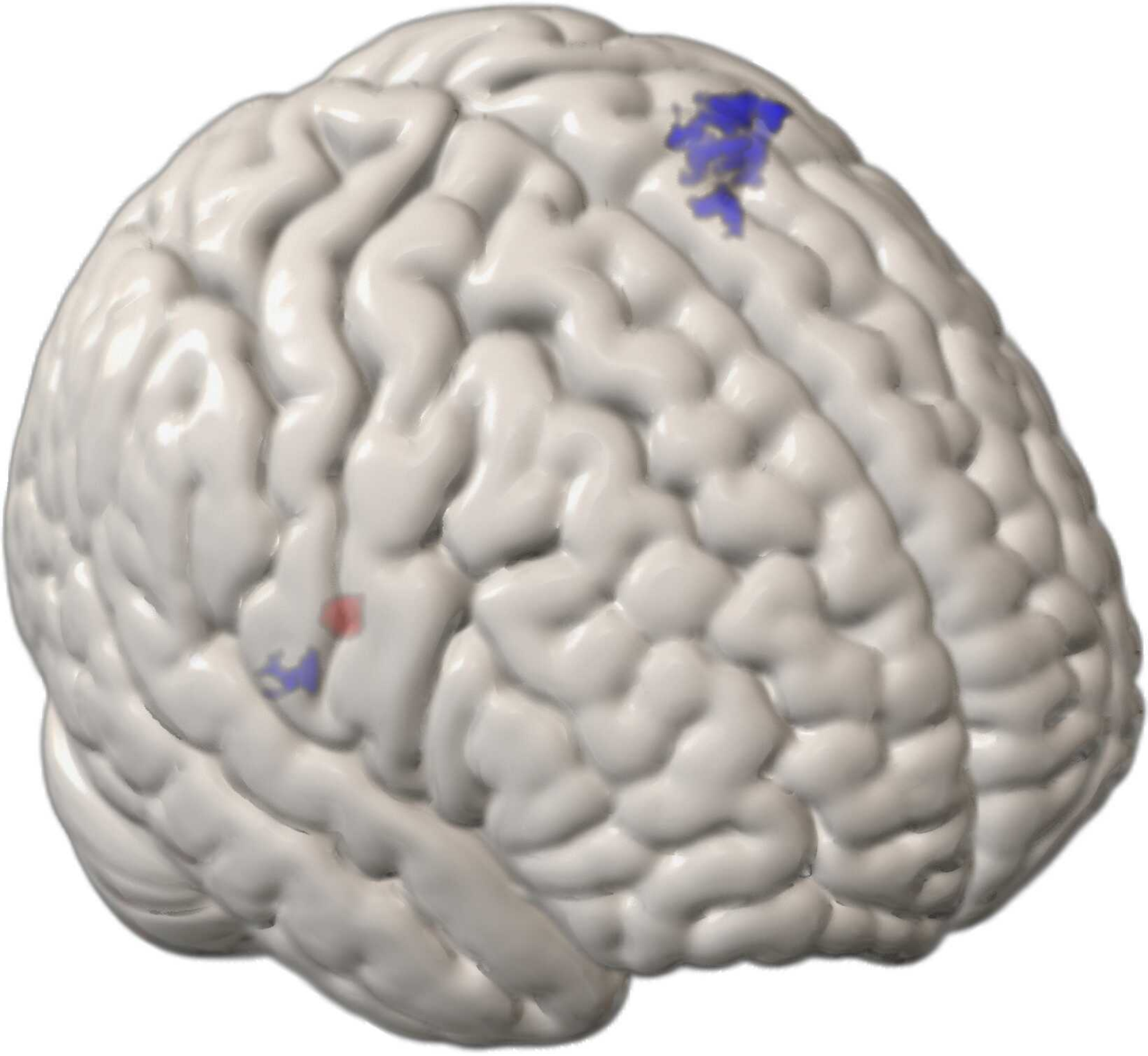}
    \hspace{15pt}
    \includegraphics[width=0.25\textwidth]{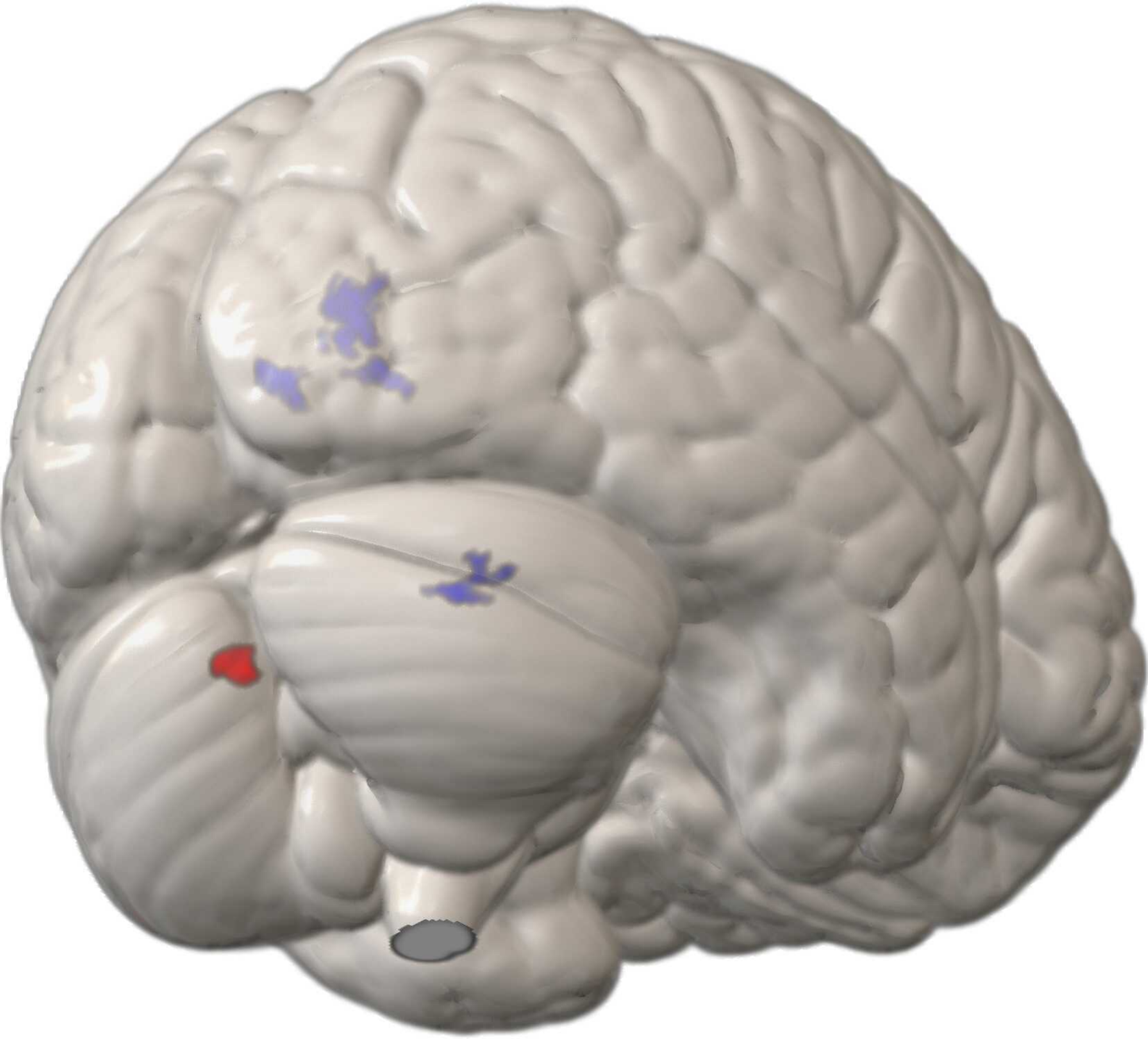}
    \vskip 10pt
    {\small (b) Left-Hand}
    \vskip 10pt
    \includegraphics[width=0.25\textwidth]{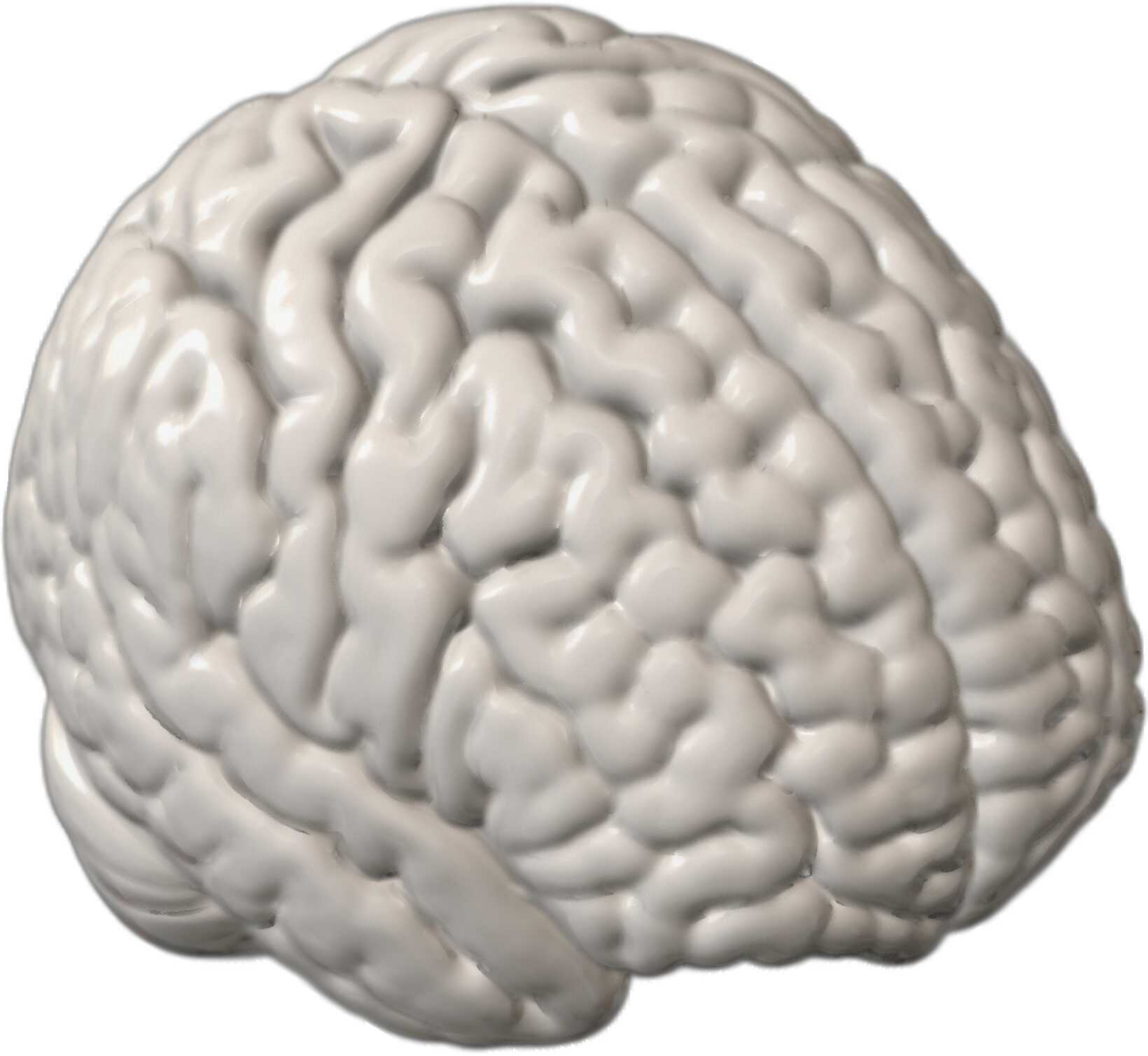}
    \hspace{15pt}
    \includegraphics[width=0.25\textwidth]{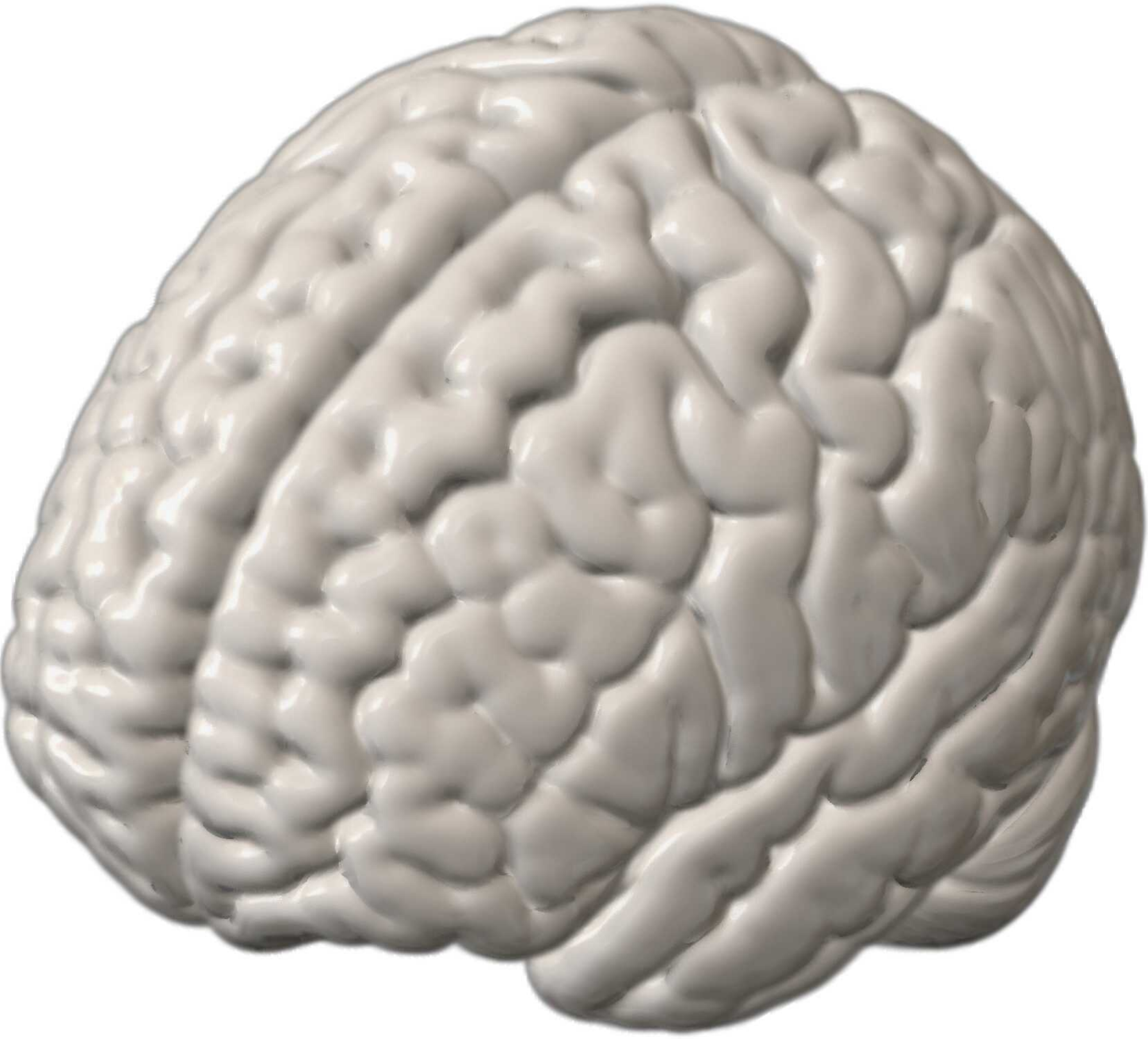}
    \hspace{15pt}
    \includegraphics[width=0.25\textwidth]{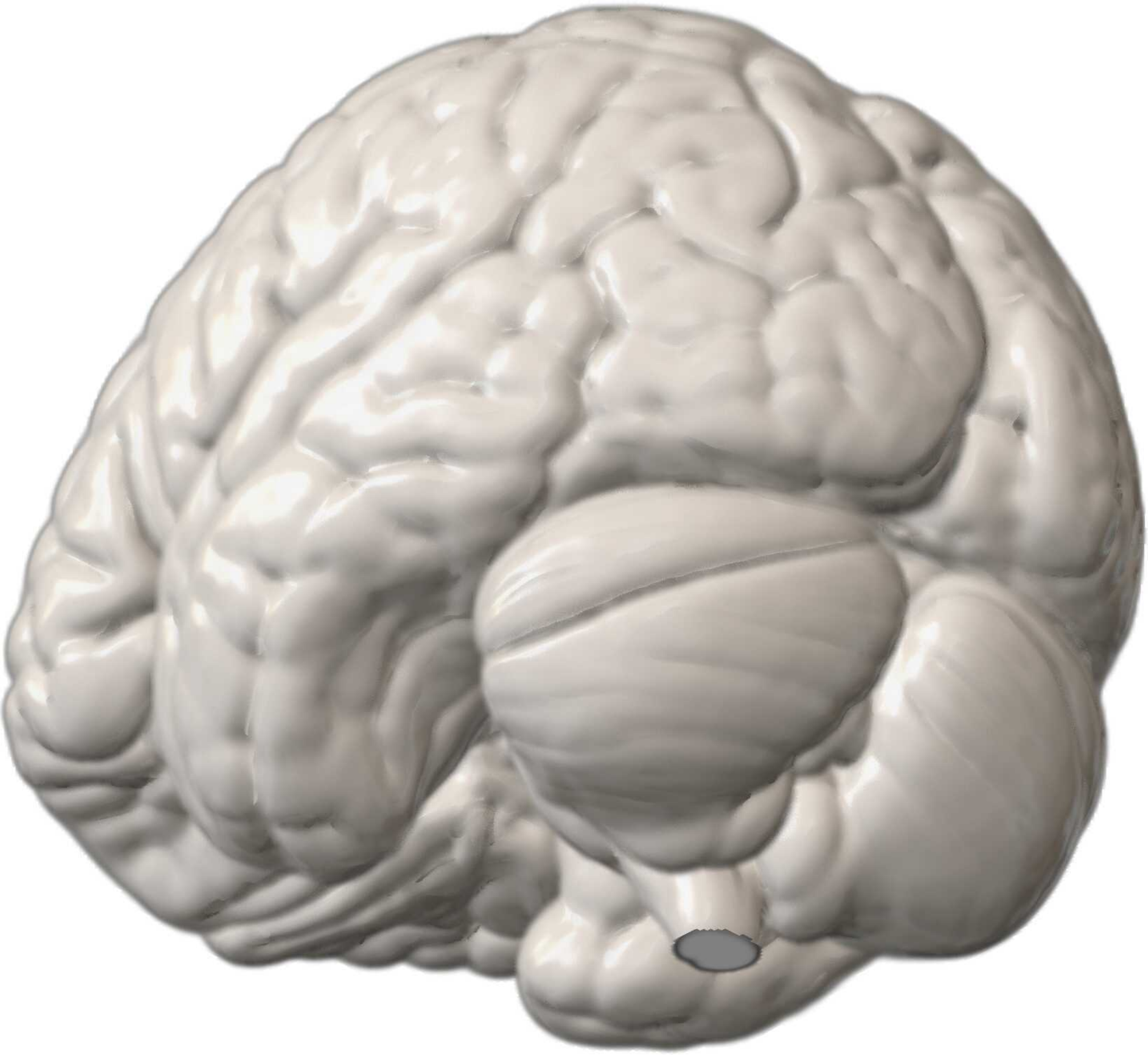}
    \caption{
    Results for the fingertip mapping study for a weak-signal participant
    obtained using lhMRPF:
    Areas of unique representation of the four tested fingers for the (a) right
    and (b) left hands based on the estimated partition structures.
    Areas of unique representation for index fingers (red) and middle fingers
    (blue) distinctly overlap (purple) in the sensorimotor cortex and
    cerebellum.}
    \label{fig:uniqueness9}
\vspace*{-10pt}
\end{figure}

\begin{figure}[!ht]
    \centering

    \includegraphics[width=0.25\textwidth]{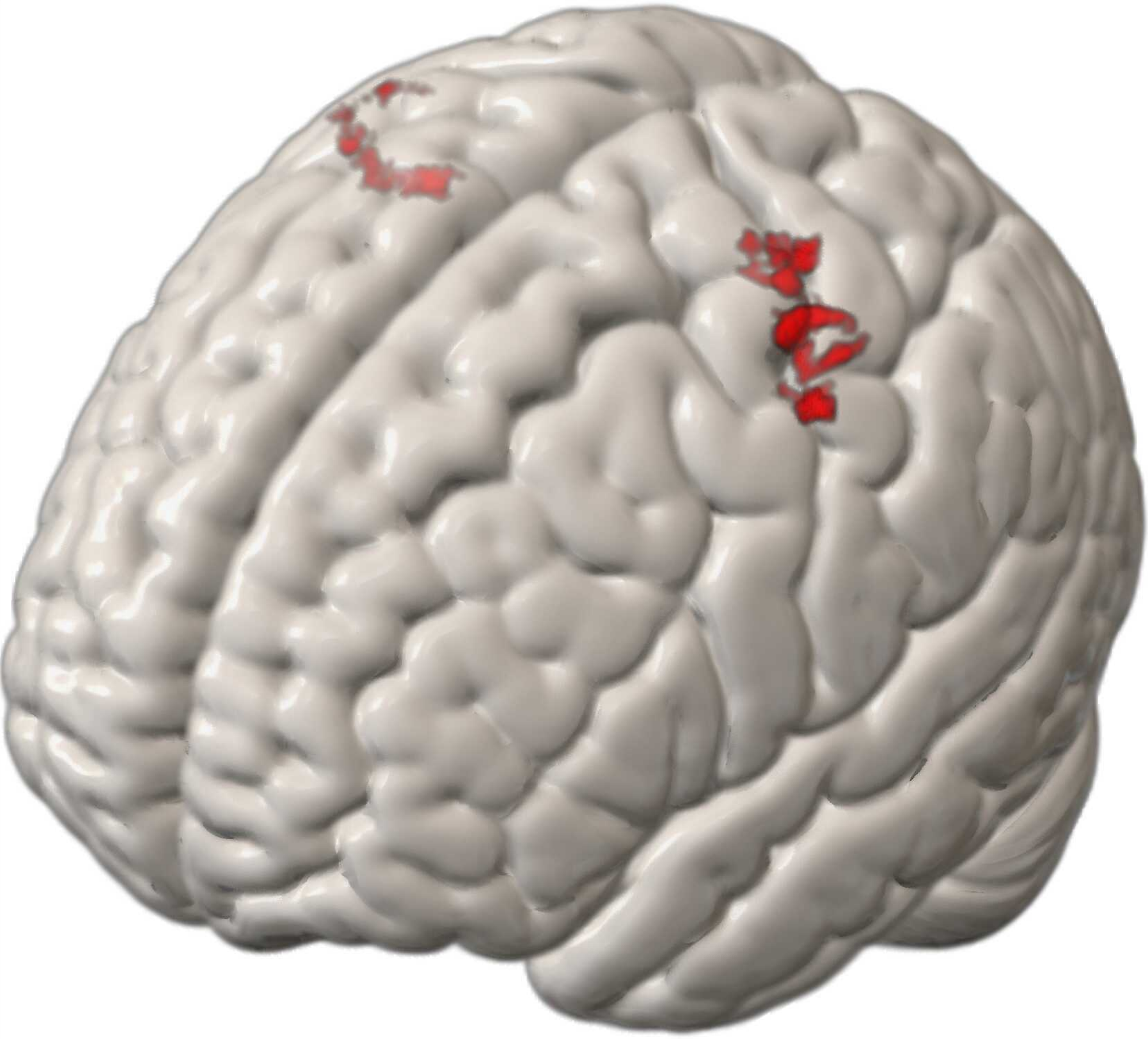}
    \hspace{15pt}
    \includegraphics[width=0.25\textwidth]{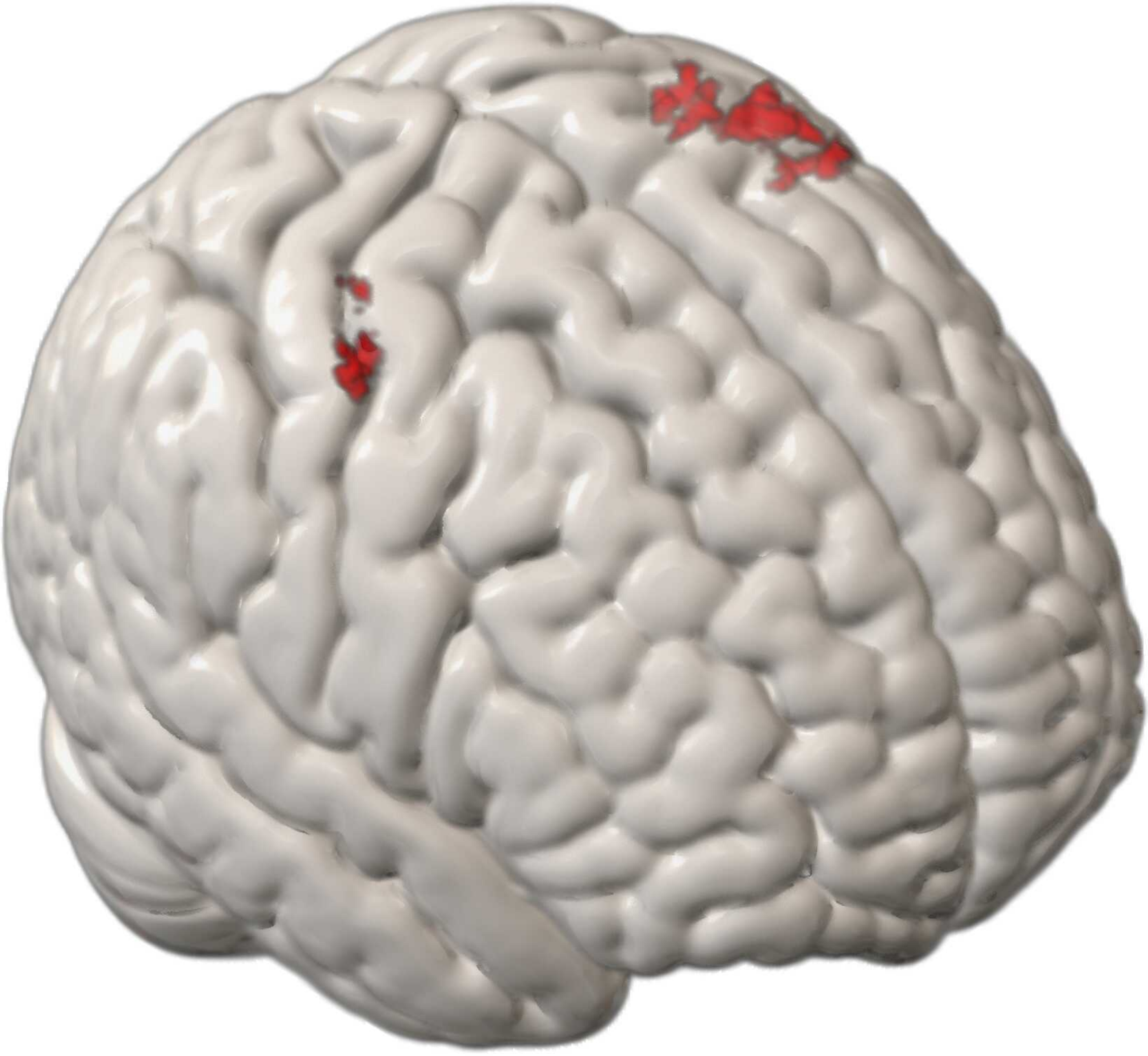}
    \hspace{15pt}
    \includegraphics[width=0.25\textwidth]{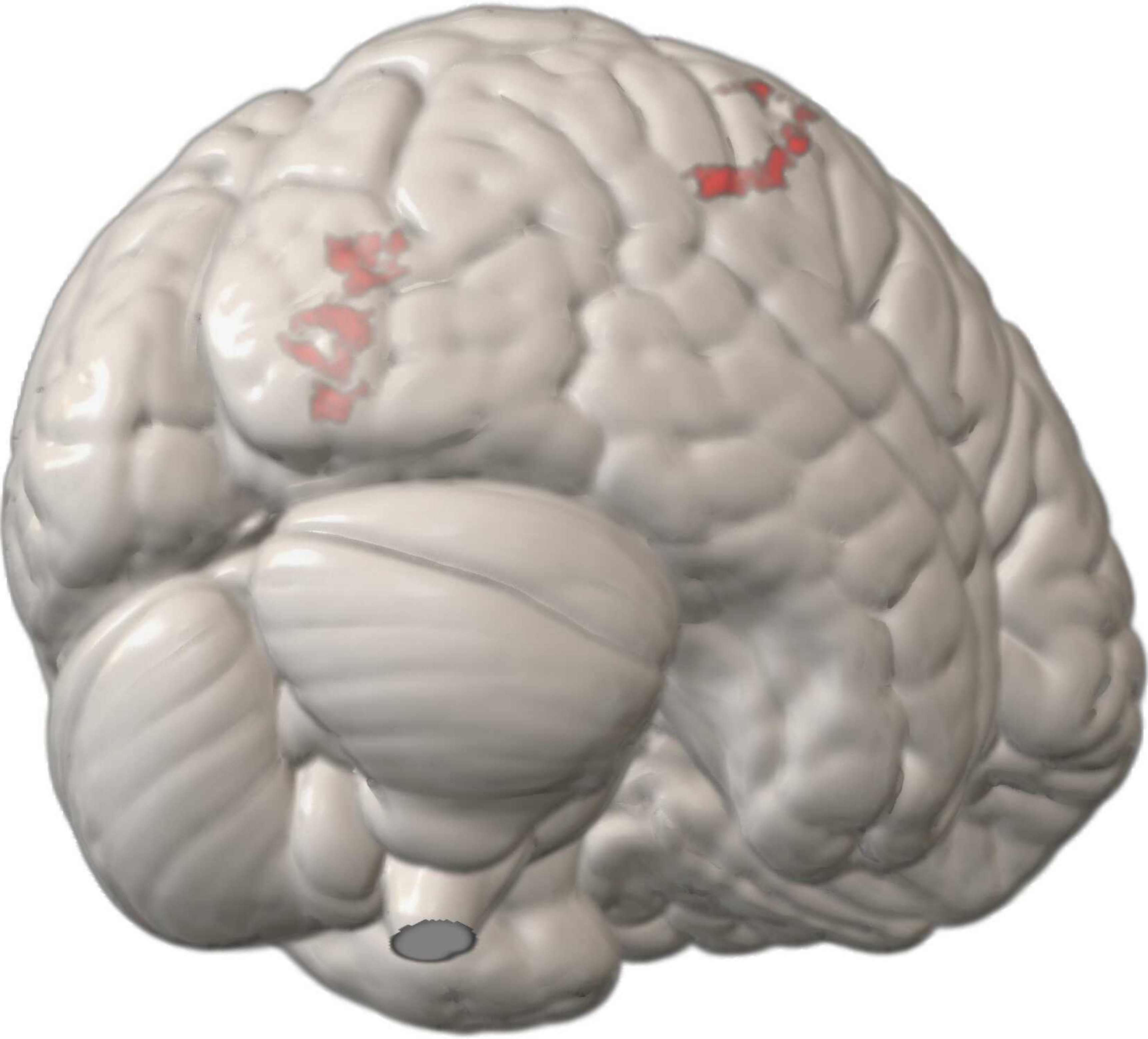}
    \caption{
    Results for the fingertip mapping study for a weak-signal participant using
    lhMRPF:
    Areas of split representation between the two hands (left/right,
    red) and the digits (index/middle,
    blue) based on the estimated partition structures.
    The overlap of these two partition types (purple) occurs when both the
    left/right hands and the index/middle fingers have split representations,
    i.e., when the four fingers are uniquely represented.}
    \label{fig:handedness9}
\vspace*{-10pt}
\end{figure}

\begin{figure}[!ht]
    \centering

    \includegraphics[width=0.25\textwidth]{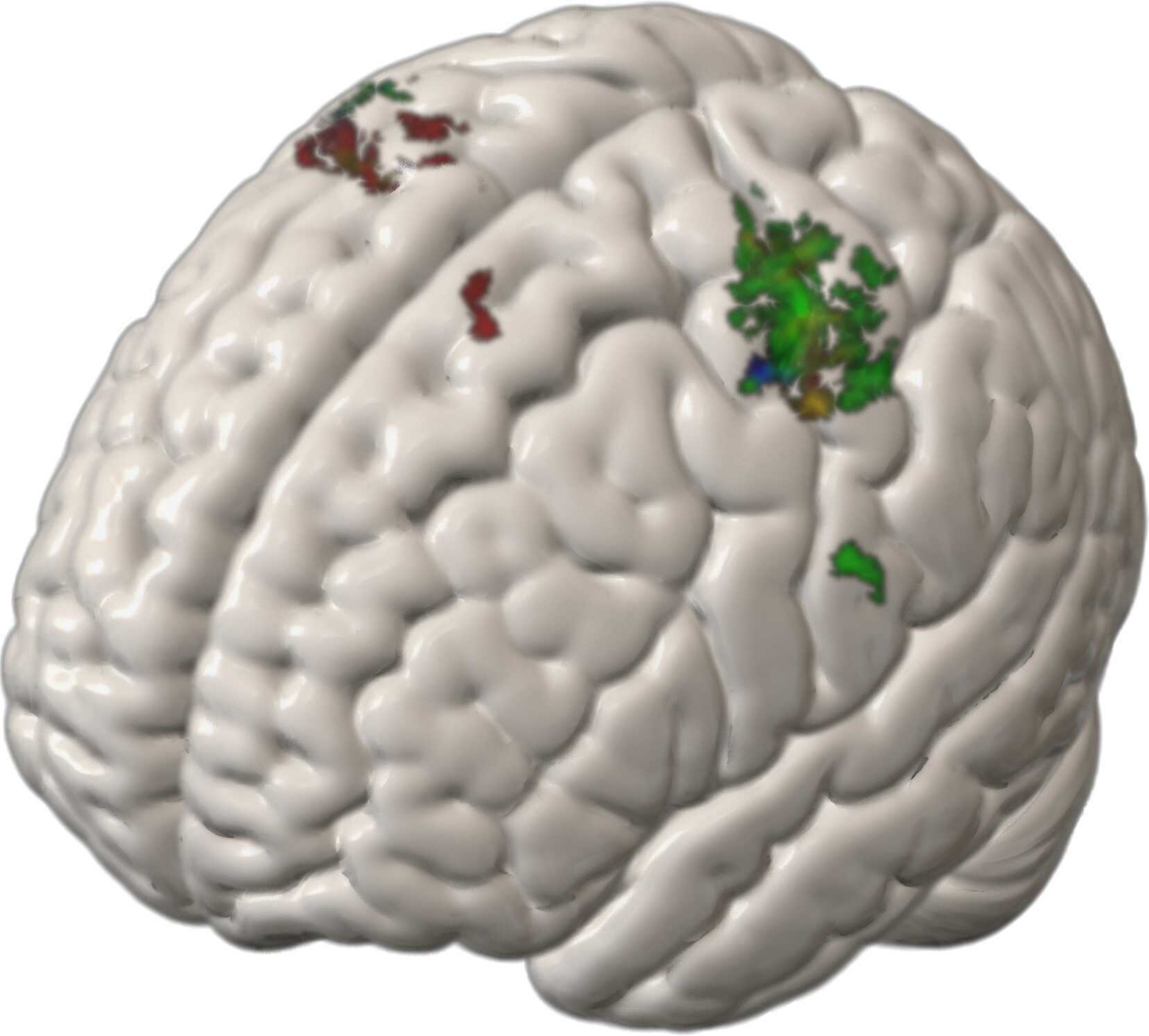}
    \hspace{15pt}
    \includegraphics[width=0.25\textwidth]{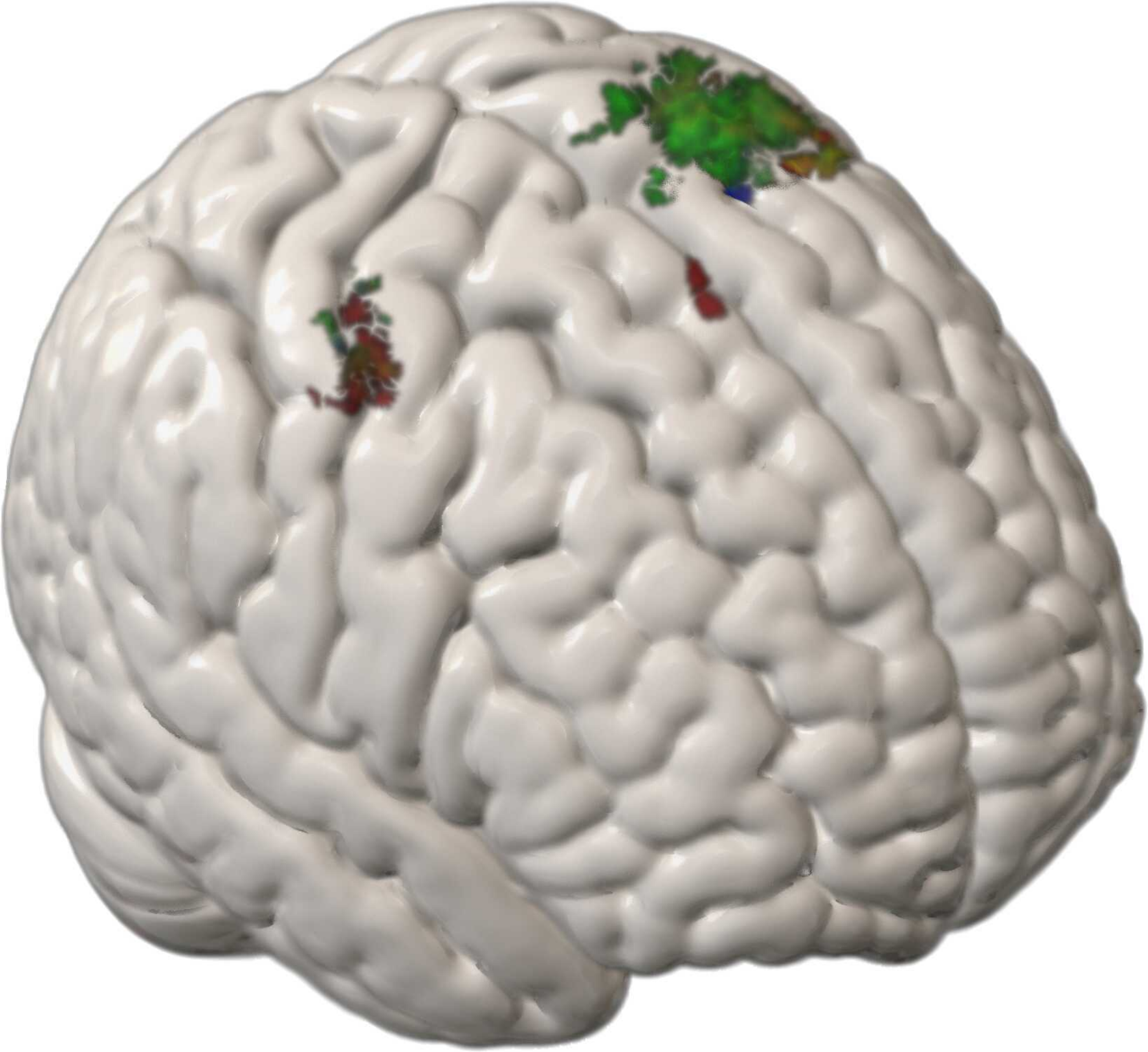}
    \hspace{15pt}
    \includegraphics[width=0.25\textwidth]{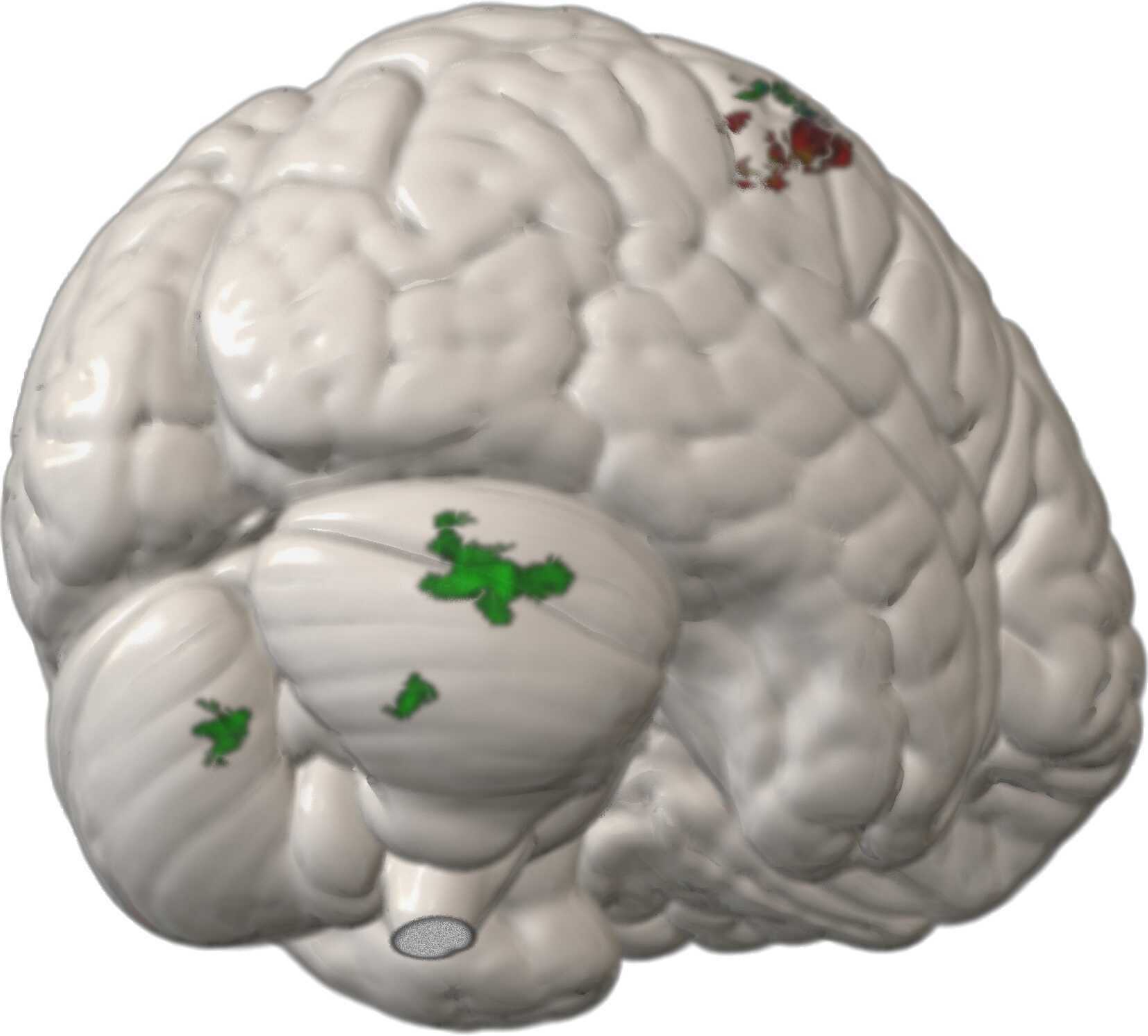}
    \caption{
    Results for the fingertip mapping study for a weak-signal participant using
    lhMRPF:
    Fingertip dominance across the investigated regions.
    Posterior mean dominance greater than $25\%$ is shown across voxels for the
    right index finger (red), left index (blue), and middle fingers (green)
    based on the estimated partition structures and the estimated effect
    intensities.
    Colors blend when two or more fingers exhibit greater than $25\%$ dominance
    probabilities at the same location,
    showing areas of competing dominance for the right index and either middle
    finger (yellow), left index and either middle finger (cyan),
    and the two index fingers (purple).}
    \label{fig:dominance9}
\vspace*{-10pt}
\end{figure}

To observe how lhMRPF performs against competing methods under a strong
sparsity assumption (where measurable differences are rare),
we analyzed a weak-signal subject.
The core benefit of lhMRPF in this regime is its ability to minimize false
positives by identifying large areas of singleton partitions where no effects
are present, while still allowing compact,
high-cardinality clusters to form in regions where some evidence exists.
In contrast, Individual Tests (Ind. Tests) struggle here,
often missing compact clusters because they identify only scattered voxels,
failing to leverage spatial information.
This advantage is evident in Figure \ref{fig:sm-cardinality-sub9},
specifically in the reduction of image `graininess' compared to other methods.
Furthermore, Tables \ref{tab:unique1-sub9} and \ref{tab:unique2-sub9}
demonstrate lhMRPF's success in capturing the expected lateralization in a
right-handed participant (left-cortical/right-cerebellar preference),
supported by its partition and dominance configurations (Figures
\ref{fig:uniqueness9}, \ref{fig:handedness9}, and \ref{fig:dominance9}).
Finally, lhMRPF surpasses the spatially-aware pTFCE method—which proved
oversensitive in high-signal data—by explicitly modeling the spatially
dependent partition field. This capability, alongside its sparsity assumptions,
prevents the spatial overflow of intense effects,
showcasing lhMRPF's advantage in weak-signal environments.

\vspace*{50pt}

\begin{table}[!ht]
\centering
\begin{tabular}{|c|c|cc|cc|}
\hline
Method & Region & \text{Right Hand} & \text{Left Hand} & \text{Index Fingers} & \text{Middle Fingers} \\
\hline
           & Sensorimotor    & $\mathbf{872}$ & $410$ & $671$ & $\mathbf{675}$ \\
lhMRPF     & Cereb. \& Midb. & $\mathbf{395}$ & $267$ & $\mathbf{363}$ & $273$ \\
           & Caud. \& Put.   & $\mathbf{202}$ &  $53$ & $\mathbf{170}$ &  $85$ \\
\hline
           & Sensorimotor    & $\mathbf{56}$ & $12$ & $\mathbf{40}$ & $29$ \\
Ind. Tests & Cereb. \& Midb. &  $\mathbf{5}$ &  $2$ &  $\mathbf{7}$ &  $0$ \\
           & Caud. \& Put.   & $\mathbf{23}$ &  $0$ & $\mathbf{23}$ &  $0$ \\
\hline
           & Sensorimotor    & $\mathbf{1486}$ &  $981$ & $\mathbf{1633}$ & $1073$ \\
pTFCE      & Cereb. \& Midb. & $\mathbf{1387}$ & $1015$ & $\mathbf{1751}$ & $660$ \\
           & Caud. \& Put.   &  $\mathbf{295}$ &   $65$ &  $\mathbf{284}$ &   $76$ \\
\hline
\end{tabular}
\caption{Results for the fingertip mapping study for a weak-signal participant:
Unique volume representation for hands and digit types,
showing the number of voxels in which either of the (right,
left) handed fingers are uniquely represented or either of the (index,
middle) fingers are uniquely represented.
}
\label{tab:unique1-sub9}
\end{table}
\vspace*{-10pt}

\begin{table}[!ht]
\centering
\begin{tabular}{|c|c|cc|cc|c|}
\hline
Method & Region & \text{R. Index} & \text{L. Index} & \text{R. Middle} & \text{L. Middle} & \text{All} \\
\hline
           & Sensorimotor    &  $416^{2}$ &  $255^{3}$ &  $499^{1}$ &  $176^{4}$ &   $43$ \\
lhMRPF     & Cereb. \& Midb. &  $184^{2}$ &  $179^{3}$ &  $213^{1}$ &   $60^{4}$ &    $2$ \\
           & Caud. \& Put.   &  $155^{1}$ &   $15^{4}$ &   $47^{2}$ &   $38^{3}$ &    $0$ \\
\hline
           & Sensorimotor    &   $33^{1}$ &    $7^{3}$ &   $24^{2}$ &    $5^{4}$ &    $1$ \\
Ind. Tests & Cereb. \& Midb. &    $5^{1}$ &    $2^{2}$ &    $0^{3}$ &    $0^{3}$ &    $0$ \\
           & Caud. \& Put.   &   $23^{1}$ &    $0^{2}$ &    $0^{2}$ &    $0^{2}$ &    $0$ \\
\hline
           & Sensorimotor    & $1045^{1}$ &  $588^{2}$ &  $569^{3}$ &  $504^{4}$ &  $128$ \\
pTFCE      & Cereb. \& Midb. &  $935^{1}$ &  $816^{2}$ &  $457^{3}$ &  $203^{4}$ &    $5$ \\
           & Caud. \& Put.   &  $268^{1}$ &   $16^{4}$ &   $27^{3}$ &   $49^{2}$ &    $0$ \\
\hline
\end{tabular}
\caption{Results for the fingertip mapping study for a weak-signal participant:
Unique volume representation for individual fingers,
showing the number of voxels in which each of the four fingers are uniquely
represented and where all four fingers are uniquely represented.
}
\label{tab:unique2-sub9}
\end{table}
\vspace*{-10pt}

\clearpage\newpage
\section{Simulation Study:
Additional Details and Results} \label{sec:sm simulation}

We conducted a simulation study on synthetic data to compare the performance of
our lhMRPF model to two existing methods:
Standard contrast analysis was conducted pairwise on factor levels and
independently across voxels \citep{huettel2014},
and probabilistic threshold-free cluster enhancement
\citep[pTFCE,]{2019Spisak} was conducted pairwise on factor levels but
accounting for spatial dependence across voxels.

We used the final MCMC sample from the real data analysis of Section 6 in the
main paper as the ``truth" for a simulated study.
The $\wh{\bbeta}_{v}$ values for six runs were generated by using the
$\bbeta^{\star}_{v}$ estimates and the $\wh{\bSigma_{v}}$ values from the
first-level analysis.
This approach allowed us to produce synthetic datasets that closely resemble
actual observations.
We focus on the motor cortex only.

\begin{figure}[!ht]
    \centering

    \vskip 10pt
    \includegraphics[width=0.2\textwidth]{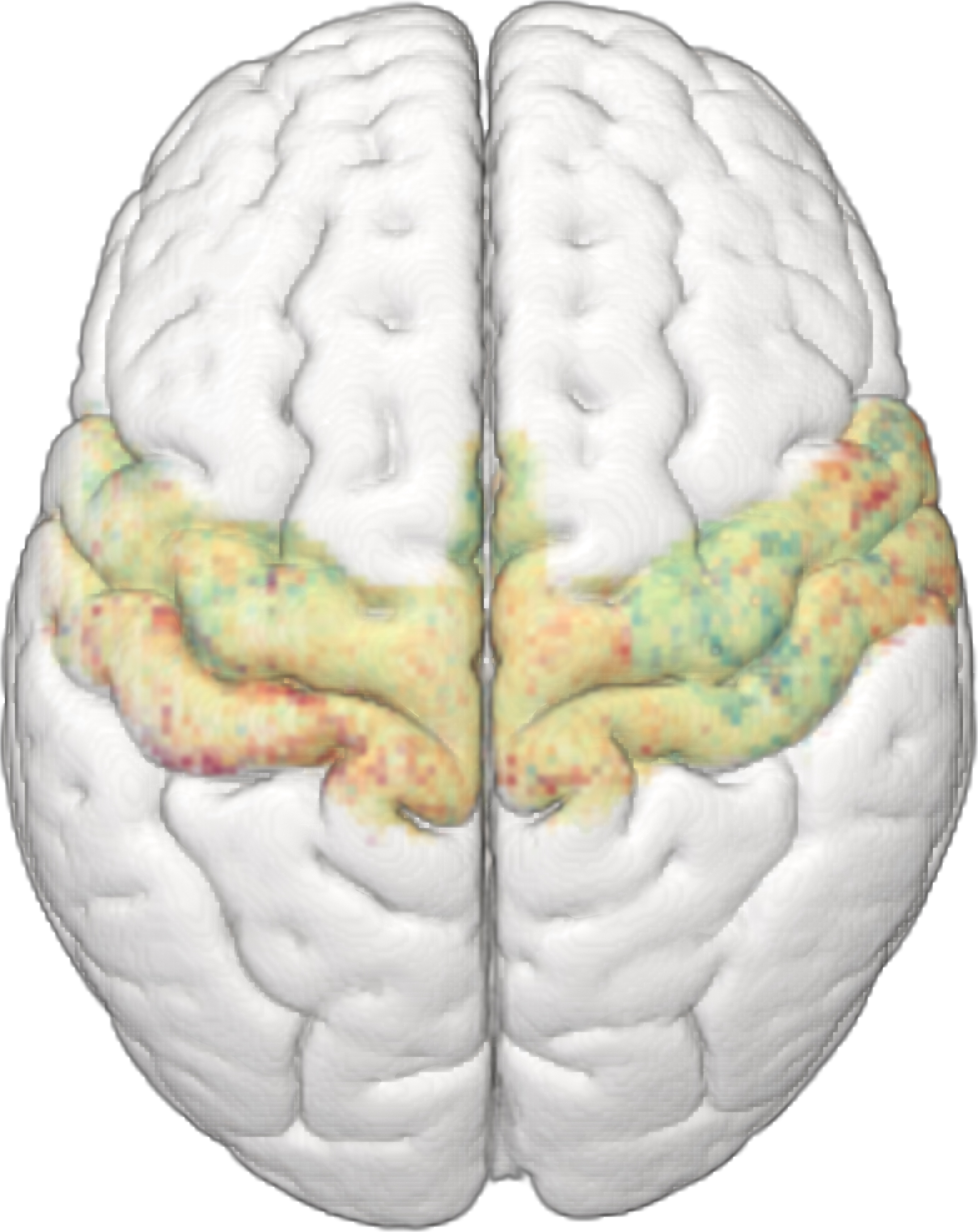}
    \hspace{15pt}
    \includegraphics[width=0.2\textwidth]{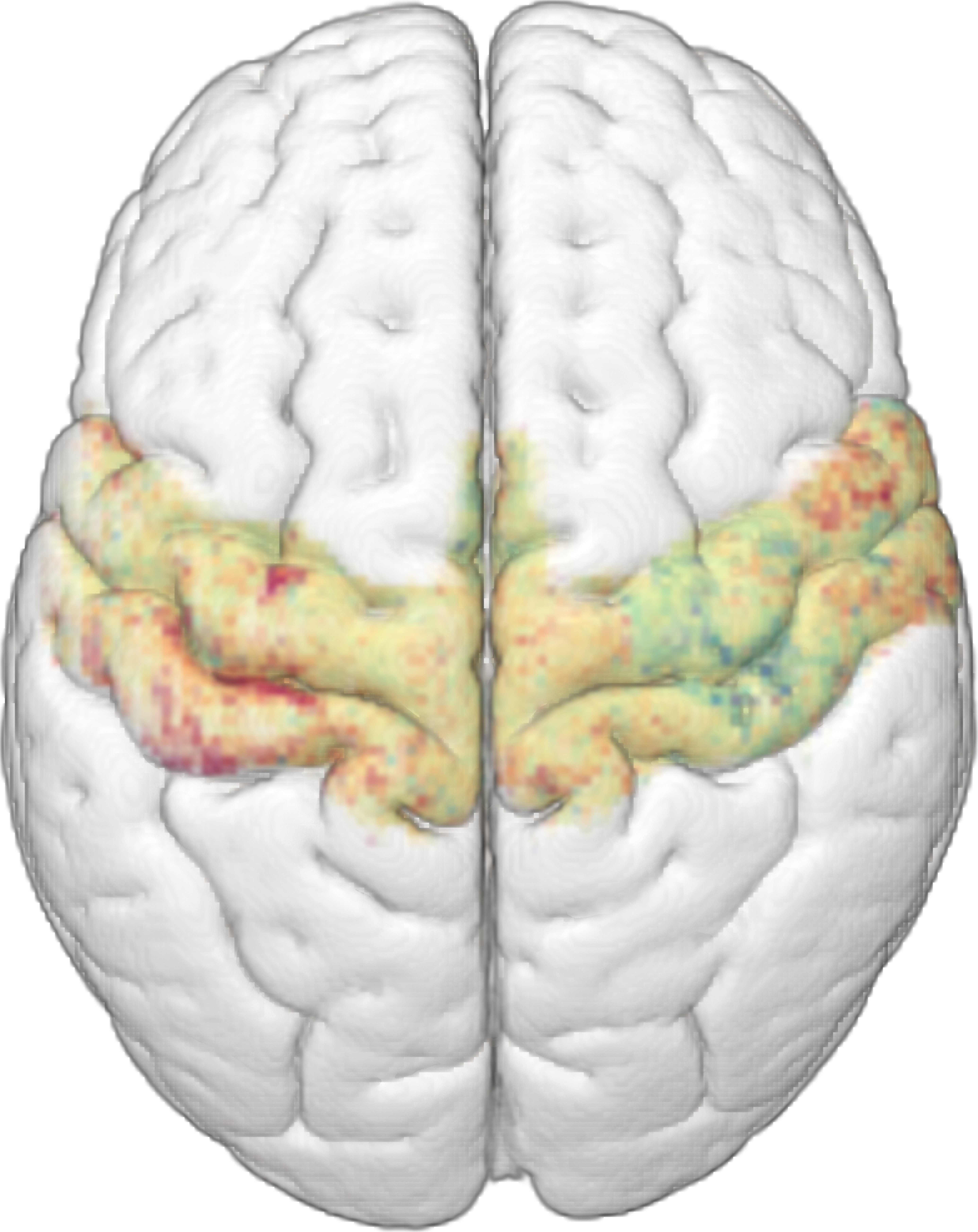}
    \hspace{15pt}
    \includegraphics[width=0.2\textwidth]{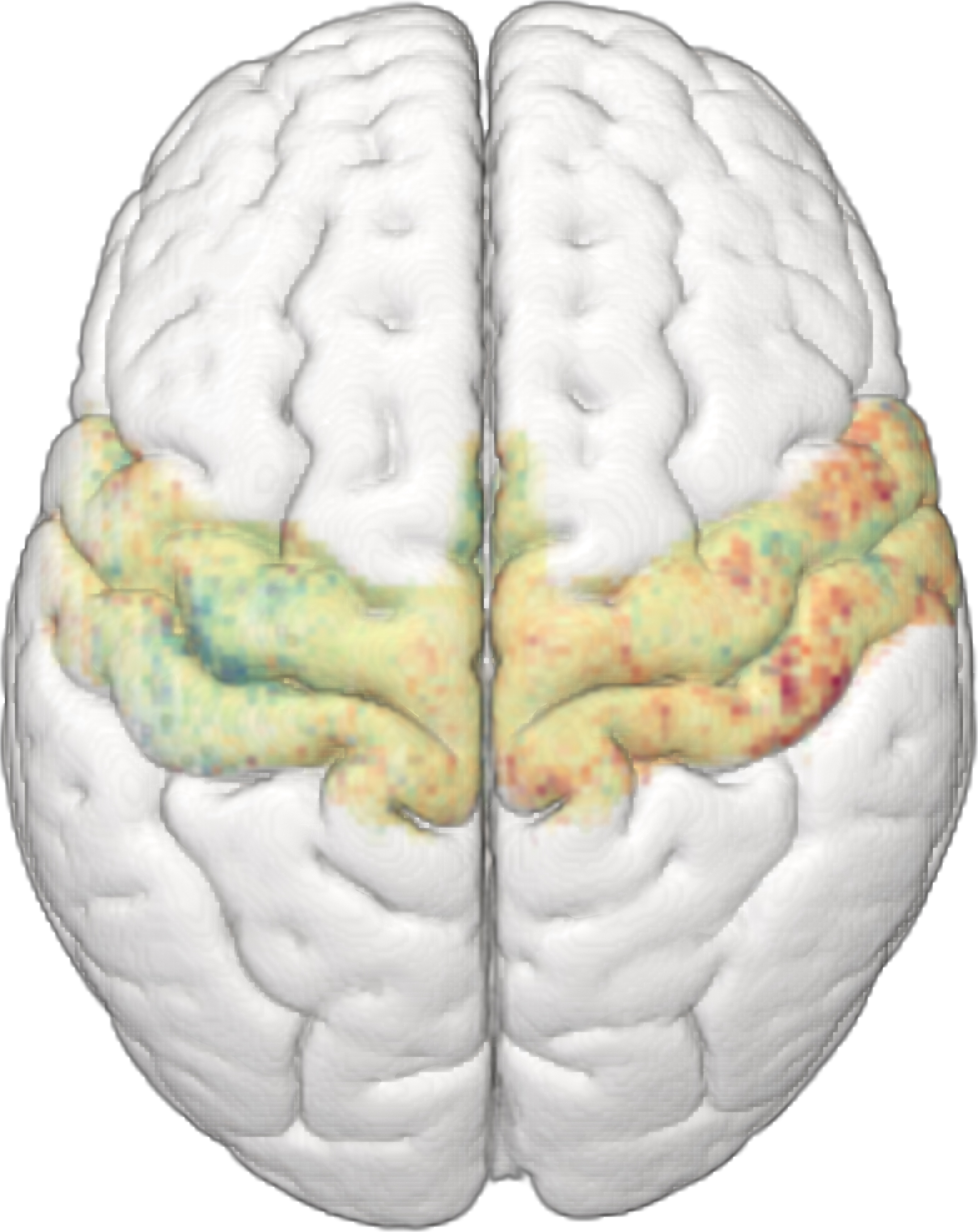}
    \hspace{15pt}
    \includegraphics[width=0.2\textwidth]{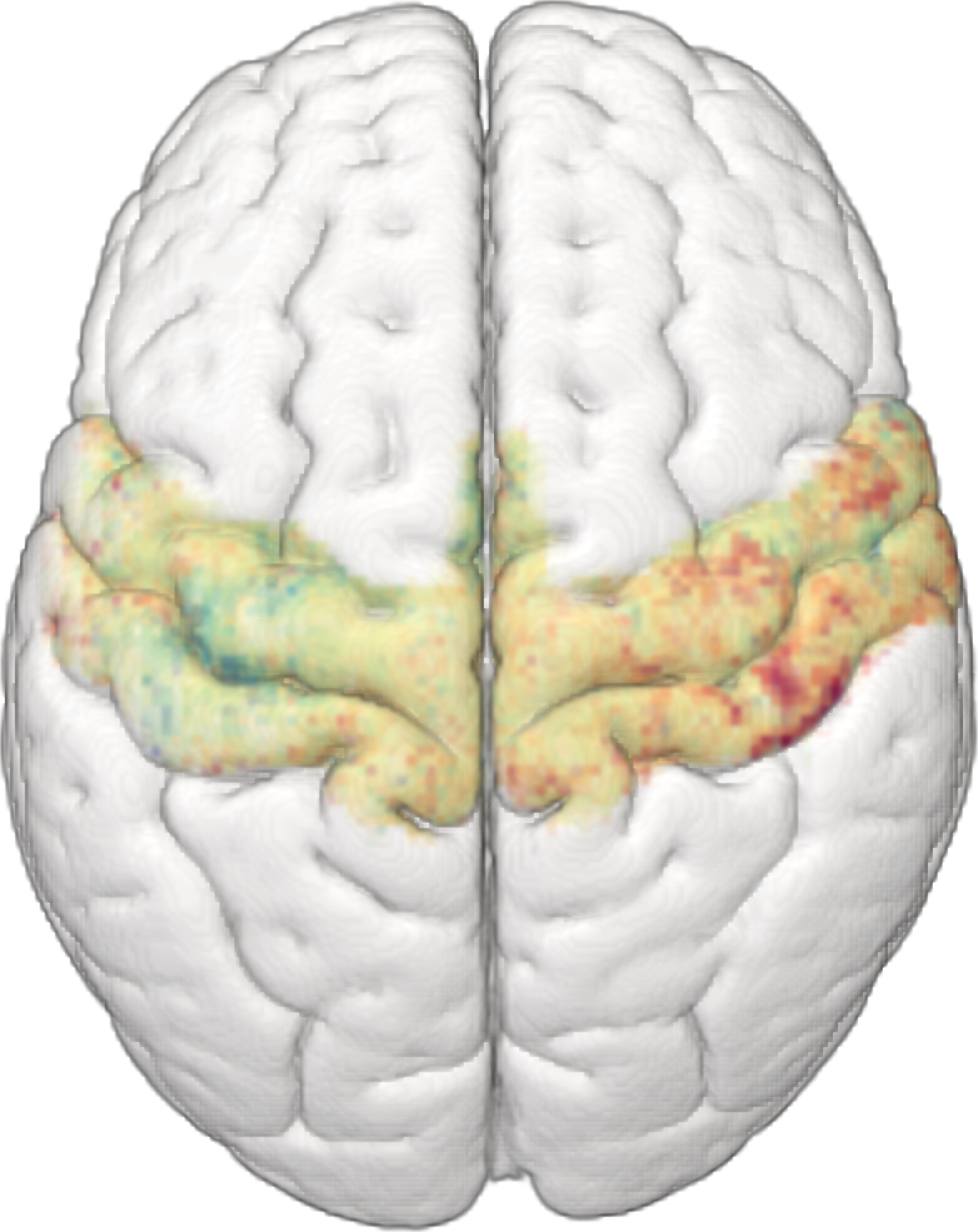}
    \newline
    \vspace{1em}
    \includegraphics[width=0.75\textwidth]{images/pdfs/colorbar.pdf}
    \vspace*{-20pt}
    \caption{One set of synthetic images associated with each of the four
    factor levels.}
    \label{fig:sm synthetic data}
\end{figure}

Using these $\wh{\bbeta}$ values,
$n=6$ synthetic images were generated from a multivariate normal distribution
with variance $\widehat{\bSigma}$,
pulled from the results of the first-level GLM analysis from Section
\ref{sup:firstlevel} above.
Figure \ref{fig:sm synthetic data} shows one such set of synthetic images for
each of the four factor levels.

We fit our model to the synthetic images using the same MCMC algorithm,
initial values, and hyperparameters as in the real data analysis.
Effect estimation error was computed as the residual $\bbeta_{v,
\text{true}}^{\star} - \wh\bbeta_{v}^{\star}$,
where $\wh\bbeta_{v}^{\star}$ is the posterior mean from the MCMC output.
The recovery is shown in Figure 10 in the main paper.

We inferred $\partition_{v}$ using the Bayesian FDR procedure of
\cite{2004Muller},
then reconstructed partitions from the resulting pairwise structure.
Any invalid partitions were resolved by iteratively co-clustering levels with
shared cluster memberships until a valid configuration was obtained.
We compare the latent partition recovery performance of our lhMRPF with two
alternatives, both of which rely on voxel-wise $z$-value maps.
We conducted a spatially independent set of pairwise $z$-tests (treating
$\widehat{\bSigma}$ as known), yielding a $z$-map for each voxel, each run,
and each pairwise comparison.
We then aggregated these values using Stouffer's method
\citep{1949Stouffer} to yield a single $z$-map across all voxels under
consideration for each comparison.
For our spatially independent contrast analysis,
we transformed this $z$-map into a $p$-map and applied an FDR correction at
$\alpha=0.05$.
We then performed pairwise comparisons to reconstruct a partition of factor
levels, assigning two factors to the same group whenever the $p$-value for the
difference did not exceed the critical threshold.
For the pTFCE analysis \citep{2019Spisak},
we passed the pairwise difference $z$-maps as input to the pTFCE algorithm and
constructed the estimated partitions using the same criterion described above.
These procedures mirror our approach with the lhMRPF results following Bayesian
FDR correction.
Note that by design,
lhMRPF always produces valid partitions at every MCMC iteration.
Therefore, if we were to use other types of point estimates,
such as the maximum a posteriori partitions, no corrections would be necessary.
However, because we chose to utilize pairwise properties of the sampled
partitions to facilitate direct comparison with alternative
pairwise-comparison-based methods,
minor adjustments are required to ensure validity in a small proportion of
cases.

Partition error was defined as the number of incorrectly inferred pairwise
equalities or differences.
The final error metric was the distance $\text{d}(\partition_{v},
\partition_{v, \text{true}})$ described in Section 4.2 in the main paper,
with possible values summarized in Table \ref{tab:partition-distance}.
While the mean of this metric for each method is given in Table 3 in the main
paper, we provide additional binary classification metrics in Table
\ref{tab:sim rates} to further elucidate the comparative performance of these
methods.
In Table \ref{tab:sim rates},
we see that the lhMRPF scores highest in accuracy,
while spatially independent tests score highest in precision,
and pTFCE scores highest in recall.
This highlights a few key points:
while independent tests capture the most true positives due to their
sensitivity to highly localized effects,
pTFCE overcompensates by distributing positive decisions across too large a
spatial region.
By balancing both with an informed, partition-based prior,
lhMRPF achieves the highest accuracy and the highest overall F1 score,
a combined measure of precision and recall.

\begin{table}[!ht]
\[\begin{array}{c|ccccccccccccccc}
Partitions & 1 & 2 & 3 & 4 & 5 & 6 & 7 & 8 & 9 & 10 & 11 & 12 & 13 & 14 & 15\\
\hline
1  & 0 & 6 & 6 & 6 & 6 & 8 & 8 & 8 & 10 & 10 & 10 & 10 & 10 & 10 & 12 \\
2  &   & 0 & 8 & 8 & 8 & 6 & 6 & 6 & 4  & 4  & 8  & 4  & 8  & 8  & 6  \\
3  &   &   & 0 & 8 & 8 & 6 & 6 & 6 & 4  & 8  & 4  & 8  & 4  & 8  & 6 \\
4  &   &   &   & 0 & 8 & 6 & 6 & 6 & 8  & 4  & 4  & 8  & 8  & 4  & 6 \\
5  &   &   &   &   & 0 & 6 & 6 & 6 & 8  & 8  & 8  & 4  & 4  & 4  & 6 \\
6  &   &   &   &   &   & 0 & 8 & 8 & 6  & 6  & 2  & 2  & 6  & 6  & 4 \\
7  &   &   &   &   &   &   & 0 & 8 & 6  & 2  & 6  & 6  & 3  & 6  & 4 \\
8  &   &   &   &   &   &   &   & 0 & 2  & 6  & 6  & 6  & 6  & 2  & 4 \\
9  &   &   &   &   &   &   &   &   & 0  & 4  & 4  & 4  & 4  & 4  & 2 \\
10 &   &   &   &   &   &   &   &   &    & 0  & 4  & 4  & 4  & 4  & 2 \\
11 &   &   &   &   &   &   &   &   &    &    & 0  & 4  & 4  & 4  & 2 \\
12 &   &   &   &   &   &   &   &   &    &    &    & 0  & 4  & 4  & 2 \\
13 &   &   &   &   &   &   &   &   &    &    &    &    & 0  & 4  & 2 \\
14 &   &   &   &   &   &   &   &   &    &    &    &    &    & 0  & 2 \\
15 &   &   &   &   &   &   &   &   &    &    &    &    &    &    & 0 \\
\end{array}\]
\caption{Distance matrix for partitions of four elements.
The partitions are labeled according to Figure 1 in the main paper.}
\label{tab:partition-distance}
\end{table}

\begin{figure}[!ht]
    \centering
    {\small Distribution of Partition Recovery Error Across Methods}
    \includegraphics[width=\textwidth]{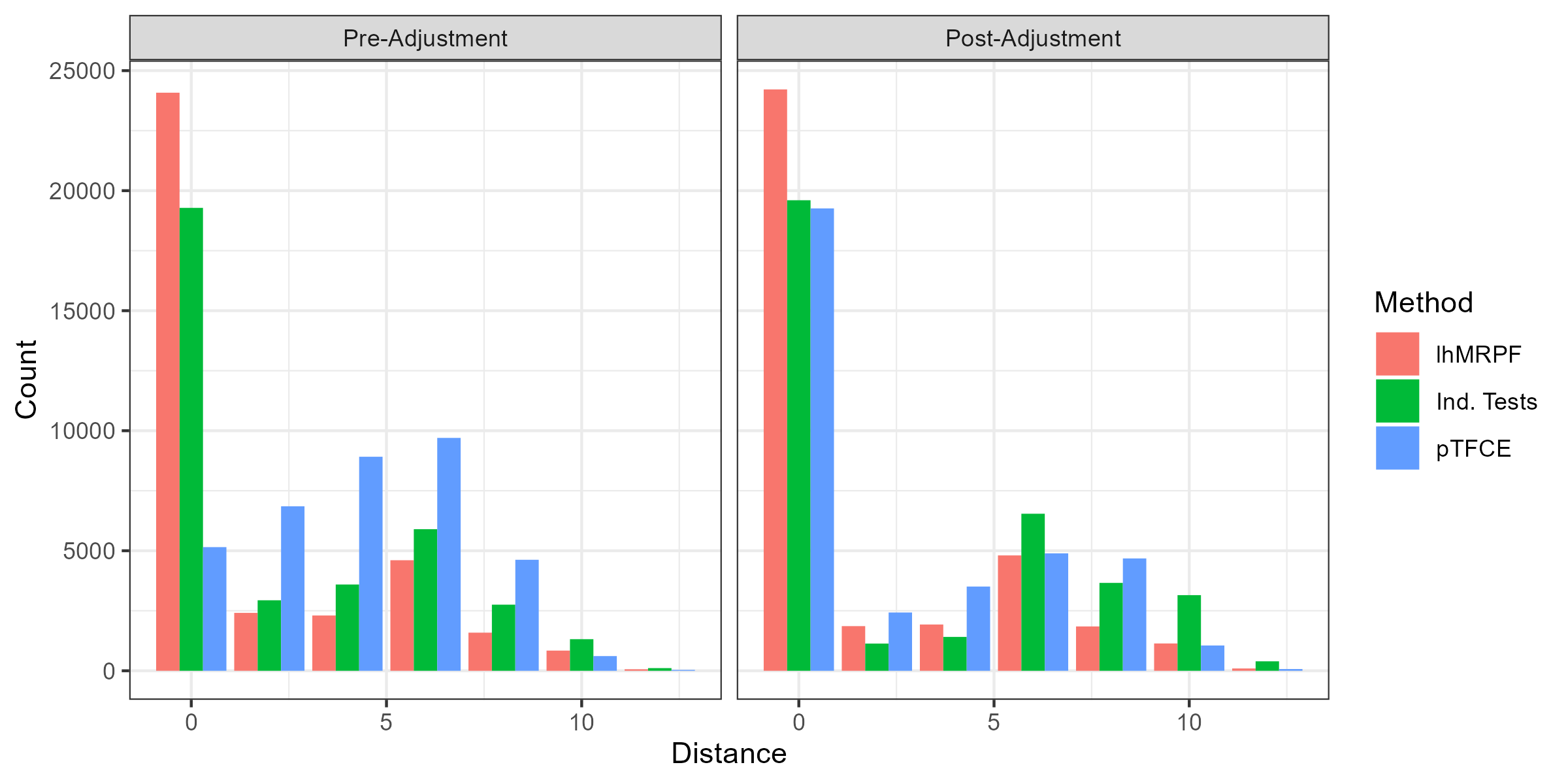}
    \caption{Results for the simulation study:
    Error in estimating the `true' partitions,
    measured by the distance metric given in Section 4.2 in the main paper,
    across different methods.
    Zero error implies correct identification of the underlying partition,
    and error increases by 2 for every incorrectly identified pairwise
    relationship (up to 12) (See Table \ref{tab:partition-distance}).
    Note the difference in the distribution for pTFCE before (right) and after
    (left) adjustment, as coercion to validity using the criterion described in
    Section \ref{sec:sm simulation} in the SM removes many incorrect pairwise
    decisions.}
    \label{fig:sm partitions}
\end{figure}

\begin{table}[!ht]
\centering
\begin{tabular}{|c|ccccccc|}
\hline
\text{Method} & \text{Accuracy} & \text{Precision} & \text{Recall} & \text{F1} & \text{FDR} & \text{FPR} & \text{FNR} \\
\hline
lhMRPF      & $\mathbf{0.852}$ & $0.973$          & $0.613$          & $\mathbf{0.752}$ & $0.027$          & $0.010$          & $0.387$ \\
Ind. Tests & $0.786$          & $\mathbf{0.984}$ & $0.421$          & $0.590$          & $\mathbf{0.016}$ & $\mathbf{0.003}$ & $0.579$ \\
pTFCE      & $0.649$          & $0.512$          & $\mathbf{0.808}$ & $0.627$          & $0.488$          & $0.442$          & $\mathbf{0.192}$ \\
\hline
\end{tabular}
\caption{Results for the simulation study:
Binary classification performance metrics across different methods in terms of
non-partition-based metrics for pairwise comparisons.
All metrics are bounded to $[0, 1]$.
Accuracy is the ratio of correct decisions to all decisions.
Precision and recall are the ratios of correct positive decisions to all
positive decisions and that of correct positive decisions to all true
positives, respectively.
The F1 score is the harmonic mean of precision and recall scores,
and it is thus a summary combination of both. FDR, FPR,
and FNR are the false discovery, false positive,
and false negative rates (i.e.,
the ratios of false positives to all true positives and the ratios of false
positives to true negatives and false negatives to true positives,
respectively). For each metric, the best scores are emphasized,
the highest scores for accuracy, precision, recall, and F1,
and the lowest scores for FDR, FPR, and FNR.}
\label{tab:sim rates}
\end{table}

\clearpage\newpage
\bibliographystyle{natbib}
\bibliography{bib}